%% file: main.tex
\RequirePackage[l2tabu, orthodox]{nag} 

\documentclass[oneside]{um}  

\usepackage{amsbsy}
\DeclareOldFontCommand{\bf}{\normalfont\bfseries}{\mathbf}
\renewcommand\bf{\bfseries}
\usepackage{amsmath}
\newcommand\omicron{o}
\usepackage[linesnumbered,ruled,vlined]{algorithm2e}
\usepackage{csquotes}
\usepackage{siunitx}
\usepackage{multirow}
\usepackage{rotating}
\usepackage{hyperref}
\newsubfloat{table}

\title{Data-Driven and\\Theory-Guided\\Pseudo-Spectral\\Seismic Imaging Using\\Deep Neural Network\\Architectures}  

\tagline{ }                     
\author{Christopher Zerafa}                           
\authorID{425090M}                           
\supervisor{Prof Pauline Galea}                             
\cosupervisor{Prof Cristiana Sebu}                               
\department{Department of Geosciences}                  
\faculty{Faculty of Science}                      
\degree{Ph.D.\ in Geosciences}                      
\doctype{thesis}                              
\degreedate{September, 2021}                        

\graphicspath{
	{./images/}
	{./chap1/images/}
	{./chap2/images/}
	{./chap3/images/}
	{./chap4/images/}
	{./chap5/images/}
    {./chap6/images/}	
    {./appA/images/}	
    {./appB/images/}
    {./appC/images/}	
}   

\makeindex

\begin{document}
\frontmatter 
    \maketitle
    \input{frontmatter/copyright}
    \input{frontmatter/dedication}        
    \input{frontmatter/acknowledgements}   
    \input{frontmatter/abstract}\if@openright\cleardoublepage\else\clearpage\fi
    \tableofcontents*\if@openright\cleardoublepage\else\clearpage\fi
    \listoffigures*\if@openright\cleardoublepage\else\clearpage\fi
    \listoftables*\if@openright\cleardoublepage\else\clearpage\fi
    \input{frontmatter/abbreviations}\if@openright\cleardoublepage\else\clearpage\fi

\pagestyle{umpage}
\floatpagestyle{umpage}
\mainmatter 
   \input{chap1/introduction_main} 
   \input{chap2/lit_overview_main}
   \input{chap3/theory_main}
   \input{chap4/results_main}
   \input{chap5/conclusions_main}   
   \appendix
       \addtocontents{toc}{\protect\setcounter{tocdepth}{0}}
       \input{appA/appendix_a_additional_theory}    
       \input{appB/appendix_b_additional_results}    
       \input{appC/appendix_c_ref_collab} 
    {
       \backmatter
      \if@openright\cleardoublepage\else\clearpage\fi
      \bibliographystyle{apalike}
          {\footnotesize\bibliography{main.bib}}
      \printindex
    }        
   
\end{document}

%% file: frontmatter/copyright.tex
\begin{copyrightenv}
\end{copyrightenv}

%% file: frontmatter/dedication.tex
\begin{dedication}
{\large{When something is important enough,\\you do it even if the odds are not in your favor.}}\\[5mm]
Elon Musk
\end{dedication}

%% file: frontmatter/acknowledgements.tex
\begin{acknowledgements}
Writing a significant scientific thesis is hard work and it would not possible without the support from various people. First of all, I wish to express my greatest appreciation towards my supervisor Prof Pauline Galea and co-supervisor Prof Cristiana Sebu for the intellectual guidance, valuable advice and help that was given to me during my research. The thesis would not have been written without their supervision and support.

My special appreciation to Dr. Godwin Debono who encouraged me to pursue this work and who's support have been invaluable.

I would like to thank Dr. Carlo Giunchi at Istituto Nazionale di Geofisica e Vulcanologia for providing the necessary computational resources and Dr. Elena Cuoco at EGO for the collaborations that we have had, and ones to be held.

I would like to express immense thanks to my wife Rachel, she has been extremely supportive of me throughout this entire process and has made countless sacrifices to help me get to this point. 

Last but not least, gratitude goes to to my parents, Charles and Jane, my brother Daniel, and friends for their support and love. In more ways than one, I am here because of them.
\end{acknowledgements}

%% file: frontmatter/abstract.tex

\begin{abstract}
    Full Waveform Inversion seeks to achieve a high-resolution model of the subsurface through the application of multi-variate optimization to the seismic inverse problem. Although now a mature technology, FWI has limitations related to the choice of the appropriate solver for the forward problem in challenging environments requiring complex assumptions, and very wide angle and multi-azimuth data necessary for full reconstruction are often not available.
    
    Deep Learning techniques have emerged as excellent optimization frameworks. These exist between data and theory-guided methods. Data-driven methods do not impose a wave propagation model and are not exposed to modelling errors. On the contrary, deterministic models are governed by the laws of physics. In between, there are theory-guided methods which have some fixed parameters able mimic physical processes. This enables more intelligibility as compared to purely data driven approach.
    
    Application of seismic FWI has recently started to be investigated within Deep Learning. This has focussed on the time-domain approach, while the pseudo-spectral domain has not been yet explored. However, classical FWI experienced major breakthroughs when pseudo-spectral approaches were employed. This thesis addresses the lacuna that exists in incorporating the pseudo-spectral approach within Deep Learning. This has been done by re-formulating the pseudo-spectral FWI problem as a Deep Learning algorithm for both a data-driven and a theory-guided pseudo-spectral approach. A deep neural network (DNN) and recurrent neural network (RNN) framework are derived. Either was formulated theoretically, qualitatively assessed on synthetic data, applied to a two-dimensional Marmousi dataset and evaluated against deterministic and time-based approaches.
    
    Inversion of data-driven pseudo-spectral DNN was found to outperform classical FWI for deeper and over-thrust areas. This is due to the global approximator nature of the technique and hence not bound by forward-modelling physical constraints from ray-tracing. Pseudo-spectral theory-guided FWI using RNN was shown to be more accurate with only 0.05 error tolerance and 1.45\% relative percentage error. Indeed, this provides more stable convergence, able to identify faults and has more low frequency content than classical FWI. From the comparative analysis of data-driven DNN and theory-guided RNN approaches, DNN was better performing, and recovered more of the velocity contrast, whilst RNN was better at edge definition. In general, RNN was more suited in shallow and deep sections due to cleaner receiver residuals. 
    
    Besides showing the improved performance of FWI formulated as a Deep Learning approach, this thesis highlighted the significant potential of such methods in other fields which have so far not been explored. New research avenues resulting from the shift in the inversion paradigm were identified and the next steps on how to continue developing these two frameworks presented.
 
\end{abstract}

%% file: frontmatter/abbreviations.tex

\chapter*{List of Abbreviations}
\markboth{List of Abbreviations}{List of Abbreviations}
               
\begin{acronym}\itemsep-20pt\parsep-20pt 
\acrodefplural{ANNs}{Artificial Neural Networks}
\acrodefplural{NN}{Neural Networks}
\acrodefplural{DBNs}{Deep Belief Networks}
\acrodefplural{FCLs}{Fully Connected Layers}
\acrodefplural{GANs}{Generative Adversarial Networks}
\acrodefplural{GPUs}{Graphical Processing Units}
\acrodefplural{RBMs}{Restricted Boltzmann Machines}
\acrodefplural{RNNs}{Recurrent Neural Networks}
\acrodefplural{SSEs}{Sum of Squared Errors}
\acrodefplural{UPNs}{Unsupervised Pre-trained Networks}
\acro{CDP}{Common Depth Point}
\acro{CNN}{Convolutional Neural Network}
\acro{CWT}{Continuous Wavelet Transform}
\acro{DBN}{Deep Belief Network}
\acro{DNN}{Deep Neural Network}
\acro{FD}{Finite Difference}
\acro{FFT}{Fast Fourier Transform}
\acro{FWI}{Full Waveform Inversion}
\acro{GAN}{Generative Adversarial Network}
\acro{L-BFGS-B}{Limited-memory Broyden–Fletcher–Goldfarb–Shanno Bound}
\acro{LSTM}{Long Short-Term Memory}
\acro{MLP}{Multi-Layer Perceptron}
\acro{MV}{Moving Variance}
\acro{NN}{Neural Network}
\acro{RBM}{Restricted Boltzmann Machine}
\acro{ReLu}{Rectified Linear Unit}
\acro{RNN}{Recurrent Neural Network}
\acro{RPE}{Relative Percentage Error}
\acro{RTM}{Reverse Time Migration}
\acro{SRC}{Source}
\acro{TTI}{Transverse Transversely Isotropic}
\acro{UPN}{Unsupervised Pre-trained Network}

\end{acronym}

%% file: chap1/introduction_main.tex
\chapter{Introduction}

\section{Motivation}\label{sec:intro_motivation}
The seismic reflection method is by far the most widely used tool in geophysical exploration \citep{Sheriff1985}. It uses artificially generated seismic waves that excite the earth and propagate through the subsurface. They are attenuated by interactions with their medium of propagation, and are partially reflected back and transmitted when coming across a high contrasting acoustic impedance. The reflected data are recorded by receivers (geophones or hydrophones) at or close to the surface. The time required for the waves to travel through the subsurface provides a measurement from which a subsurface model of acoustic media is determined. Geological significance is inferred from the data either directly from the seismic reflection method, or more commonly, through the integration of other methods such as gravitational, magnetics, refraction and other data sources such as well log data, vertical seismic profiles and geological settings of the region. 

The search for new petroleum resources has pushed exploration areas of ever increasing complicated subsurface geology where the success of exploration wells depends heavily on the imaging of seismic data sets. To be able to image this geology, solvers have gradually moved from ray tracing algorithms to one-way wave equation methods and to acoustic and elastic two-way wave equation methods \citep{Nangoo2013}. These imaging techniques have until recently been largely limited to depth mapping of seismic reflections at boundaries between rock formations. The extraction of other information available in seismic data sets was compromised by prohibitive computing processing time and as a result only a fraction of the information contained in seismic datasets could be extracted.

\begin{figure}[!ht]
	\centering
    \subbottom[Velocity model from conventional methods.]{\includegraphics[width=0.45\textwidth]{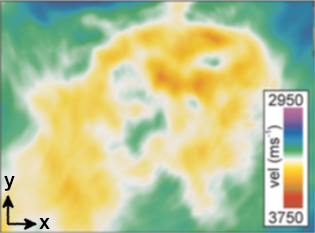}}
    \subbottom[<10Hz FWI velocity model result.]{\includegraphics[width=0.45\textwidth]{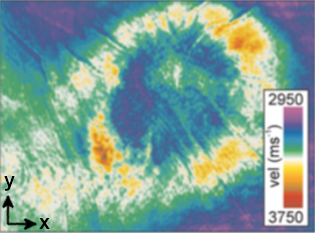}}
	\caption[Horizontal slices though the Samson Dome.]{Horizontal slices though the Samson Dome in the Barent Sea at 1350m showing the uplift in imaging obtained through FWI. Axis indicated extend of the horizontal slice were not present in the original image from \cite{Morgan2013}.} 
	\label{fig:conventional_vs_fwi}
\end{figure}

The availability of super computers and Graphical Processing Units in recent years has enabled seismic processing to implement previously prohibited imaging technology. One such technology is \ac{FWI}, which follows the physics of the wave equation for both phase and amplitude \citep{Tarantola1987}. The technique does not only invert for conventional depth imaging of boundaries but also rock properties such as velocity ($v_p$ - compressional and $v_s$ - shear), lithology and pore-fill thus providing a more comprehensive geology of the subsurface \citep{Plessix2014}. The technique has also gained popularity after it was demonstrated to produce spectacular improvements in imaging subsurface geology beneath a heterogeneous overburden - See Figure~\ref{fig:conventional_vs_fwi}. This is a landmark in seismic exploration and has changed the way seismic data is used and interpreted.

\begin{figure}[!ht]
	\centering
    \subbottom[2004 BP synthetic for FWI.]{\includegraphics[width=0.6\textwidth]{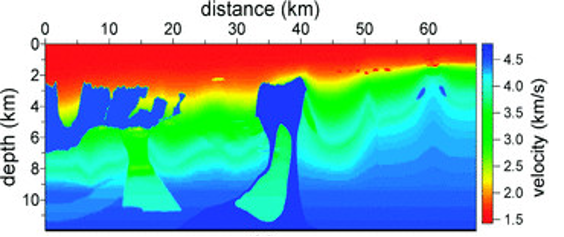}}
    \subbottom[Zoom of complexity at depth.]{\includegraphics[width=0.25\textwidth]{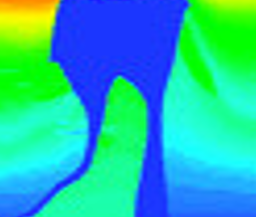}}

    \subbottom[2D FWI result.]{\includegraphics[width=0.6\textwidth]{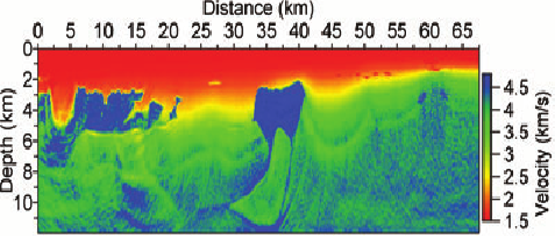}}
    \subbottom[Zoomed section highlighting lack of resolution.]{\includegraphics[width=0.25\textwidth]{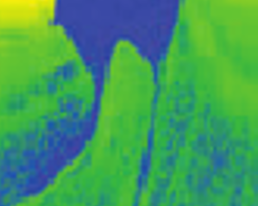}}
	\caption[Limitations of FWI due to poor illumination.]{Limitations of FWI due to poor illumination. From \cite{Shin2010}.} 
	\label{fig:limitations_fwi}
\end{figure}

FWI seeks to achieve a high-resolution model of the subsurface through the application of multivariate optimization to the seismic inverse problem \citep{Virieux2009}. The inversion process begins with a best-guess initial model which is iteratively improved using a sequence of linearized local inversions to solve a fully non-linear problem. Figure~\ref{fig:conventional_vs_fwi} illustrates the image uplift which is achievable with FWI. In situations of more complex structures at depth with convoluted ray-paths in the overburden, the inversion becomes more difficult and more computationally expensive. Figure~\ref{fig:limitations_fwi} illustrates an example of FWI on the 2004 BP synthetic data. The zoomed section (d) illustrates a lack of resolution of FWI.

\section{Aims \& Objectives}
Optimization theory is fundamental to FWI since the parameters of the system under investigation are reconstructed from indirect observations that are subject to a forward modelling process \citep{Tarantola2005}. The accuracy of this forward problem depends on the validity of physical theory that links ground-truth to the measured data \citep{Innanen2014}. Moreover, solving for this inverse problem involves learning the inverse mapping from the measurements to the ground-truth which is based on a subset of degraded or best-estimate data \citep{Tarantola2005, Tikhonov1977}. Thus, two limitations within inverse theory can be identified: (i) the forward problem and (ii) the training data. 

As evidenced throughout the historic development (reviewed in §\ref{sec:lit_rev_fwi}), choice of the forward problem will impact the accuracy of the FWI result. Challenging environments require more complex assumptions to try and better explain the physical link between data and observations, with not necessarily improved levels of accuracies \citep{Morgan2013}. Secondly, the data being used to reconstruct the mapping of measurements for the ground-truth are not optimal. Very wide angle and multi-azimuth data is required to enable full reconstruction of the inverse problem \citep{Morgan2016}; which is not always available. Furthermore, pre-conditioning of data is a necessity prior to FWI to make the inversion better posed \citep{Kumar2012a,Mothi2013,Peng2018,Warner2013}, however if done incorrectly this can degrade the inverse process \citep{Lines2014}.

Recently, deep learning techniques have emerged as excellent models and gained great popularity for their widespread success in pattern recognition tasks \citep{Ciresan2011, Ciresan2012}, speech recognition \citep{Hinton2012} and  computer vision \citep{Krizhevsky2015,Deng2013}. The use of deep neural networks to help solve inverse problems has been explored by \cite{Elshafiey1991}, \cite{Adler2017a}, \cite{Chang2017}, \cite{Wei2017} and has achieved state-of-the-art performance in image reconstruction \citep{Kelly2017, Petersen2017, Adler2017b}, super-resolution \citep{Bruna2015, Galliani2017} and automatic-colourisation \citep{Larsson2016}. These deep learning based waveform inversion processes exist between data and theory guided methods \citep{Sun2019}. Data-driven methods do not impose a wave propagation model. \ac{NN} weights are all trainable and require relatively exhaustive training datasets to invert properly \citep{Sun2019a}. Yet, due to the large number of degrees of freedom, they are not exposed to modelling errors as any conventional FWI algorithm \citep{Wu2018,Li2019}. On the contrary, there are deterministic models, with bases in classical physics. In between, there are theory-guided or physics-informed methods which have some parameters fixed as non-trainable. If non-trainable parameters are built to mimic a physical process, the training model space now consists of only the parameters for that process and the number shrinks drastically. This provides robust training and enables more intelligibility as compared to purely data driven approach \citep{Biswas2019}.

As with most techniques, after decades of development, FWI algorithms now only being improved incrementally, with slight modification to the underlying algorithms. The technique hence requires a fresh injection of ideas and academic pursuits to elevate it to the next phase of development. Application of seismic FWI has recently started to be investigated within the deep learning field and has so far been focused only on the time-domain approach. However, classical FWI experienced pivotal breakthroughs via pseudo-spectral approaches which enabled the technique to go beyond academic experiments and be employed on real datasets (See §~\ref{sec:lit_rev_beyond_academic_exp}). The main aim of the research work presented in this thesis was to investigate whether the same advantages apply when pseudo-spectral FWI is developed within DNN. To current knowledge, there is no prior work investigating the pseudo-spectral inversion within \ac{DNN} frameworks. Specifically, my study was motivated by the following research questions: Are pseudo-spectral approaches possible within a \ac{DNN} framework? Is the shift to a data-driven inversion detrimental? How good can theory-guided inversion be? How do pseudo-spectral DNN compare to deterministic conventional FWI? Are there any particular benefits of data-driven or physics-guided pseudo-spectral DNN approaches? How do pseudo-spectral DNN compare to time-based DNN approaches? Are there any cross-over techniques between Deep Learning and geophysics which would benefit either field? Could this cause the new wave of evolution of the FWI framework?

Hence, the main aim of this work was to develop these two novel approaches in the form of data-driven and theory-guided pseudo-spectral FWI, compare them to traditional approaches and investigate their limitations. To this end, the following steps were followed:
\begin{enumerate}
    \item Building a comprehensive literature review of \ac{DNN} applications within geophysics. This will be focussed particularly on pseudo-spectral FWI approaches and highlights key examples for data-driven and theory-guided approaches.
    \item Re-casting FWI within a \ac{DNN} framework for both a data-driven direct learned inversion and a theory-guided temporal based \ac{DNN} formulation. Both approaches were derived theoretically and assessed on synthetic data. The assessment built up from simple 1D experiments, extended to 2D and evaluated on the standard Marmousi model. The results were validated against classical FWI.
    \item Analysing the limitations of both approaches and discussing future potential developments.
\end{enumerate}

\section{Document Structure}
This dissertation is composed of five chapters, each of them dealing with different aspects of a pseudo-spectral FWI within a Deep Learning framework.
\begin{itemize}
    \item Chapter 1 is introductory and deals with introducing the topic and sets the aims and objectives for this work. 
    \item Chapter 2 gives a literature review of work which highlights trends in FWI, Deep Learning and their combined application respectively. 
    \item Chapter 3 is sub-divided into 3 parts and provides all relevant theory for FWI via DNNs. Part 1 illustrates the key elements within classical FWI. Part 2 presents FWI as a data-driven \ac{DNN} and Part 3 derives RNN as a physics informed framework for FWI. Each of these is compared to classical FWI, with commonalities and differences highlighted.
    \item Chapter 4 builds on the derived theory and evaluates each framework on synthetic data.
    \item Chapter 5 critically analyses the numerical results obtained in Chapter 4 and outlines their limitations. Future work is presented and conclusions addressed.
\end{itemize}

%% file: chap2/lit_overview_main.tex
\chapter{Literature Review}
\textbf{In this chapter the main literature concerning FWI is discussed in the context of pseudo-spectral approaches, with focus on pointing out the general research trends in this area. \acp{NN} are then presented with respect to inverse problems, and their development history discussed. Both these components are put together to identify applications within geophysics, with emphasis on velocity inversion approaches.}

\section{Full Waveform Inversion}\label{sec:lit_rev_fwi}
\ac{FWI} tries to derive the best velocity model and other lithologic properties (as density, anelastic absorption and anisotropy) of the Earth’s subsurface to be consistent with recorded data. An exhaustive search for this ideal model is almost impossible and methods for finding an optimal one describing the data space are necessary. There are two main categories for dealing with this problem: (i) global optimization methods, and (ii) direct solving through linearisation.

Global optimization methods use stochastic processes to try and find the global minimum of the misfit function \citep{Torn1989, Sen1995}. Three most well-known cases of global methods are Monte Carlo methods \citep{Press1968, Biswas2017}, genetic algorithm \citep{Gerstoft1994, Parker1999, Tran2012} and simulated annealing \citep{Rothman1985, Pullammanappallil1994, Tran2011}. Global optimization methods are all very dependent on a fast forward modelling algorithm as they require large amounts of forward modelling calculations. As computers are getting faster and better, the necessity of keeping the parametrization simple might decline. However, current solutions to the seismic inverse problem have to resort to local optimization. The next section reviews direct solving through linearisation for FWI.

\subsection[FWI as Local Optimization]{Formulation of FWI as Local Optimization}\label{sec:fwi_local_opt}
The concept of local optimization for \ac{FWI} was introduced in the 1980’s. \citet{Lailly1983} and \citet{Tarantola1984a} cast the exploding-reflector concept of \citet{Claerbout1971, Claerbout1976} as a local optimization problem which aims to minimise in a least-squares sense the misfit between recorded and modelled data \citep{Virieux2009}. The problem is set in the time-domain as follows: set a forward propagation field to model the observed data, back propagate the misfit between the modelled and the observed data, cross-correlate both fields at each point in space to derive a correction, and do least squares minimization of the residuals iteratively. This outline forms the basis of this technique to this day.

\cite{Gauthier1986} numerically demonstrate a local optimization FWI approach using two-dimensional synthetic data examples. A single diffracting point on a homogeneous model was used to illustrate the importance of proper sampling of the subsurface. Furthermore, this model was used to show that the free surface adds an extra complexity to the problem and increases the non-linearity of the inversion. FWI with or without free-surface multiple modelling is an active area of research to this day \citep{Komatitsch2002,Bergen2019}. 

\subsection{Applications in Time and Frequency}
\begin{figure}[!ht]
	\centering
	\subbottom[Section from real Marmousi velocity model.]{\includegraphics[width=0.45\textwidth]{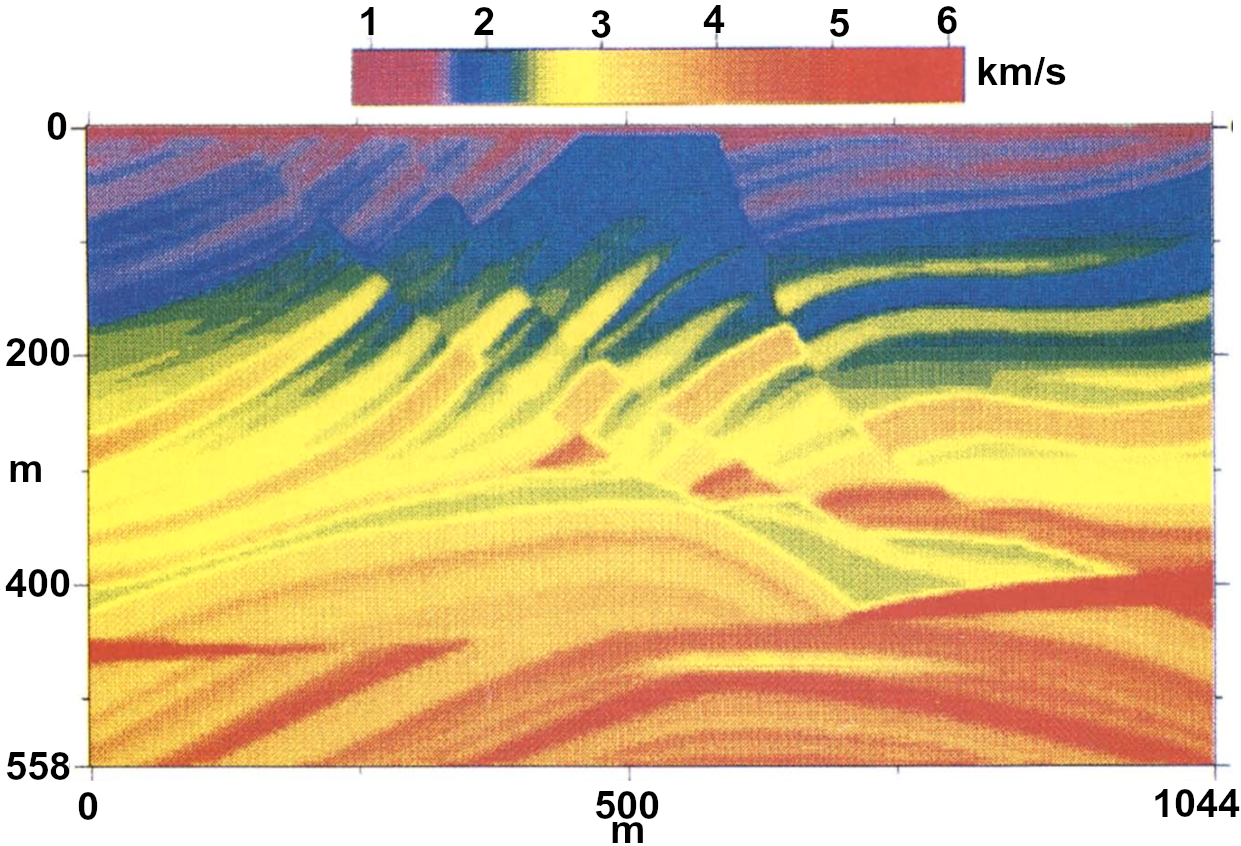}}
	\subbottom[Best estimate using multiscale FWI.]{\includegraphics[width=0.45\textwidth]{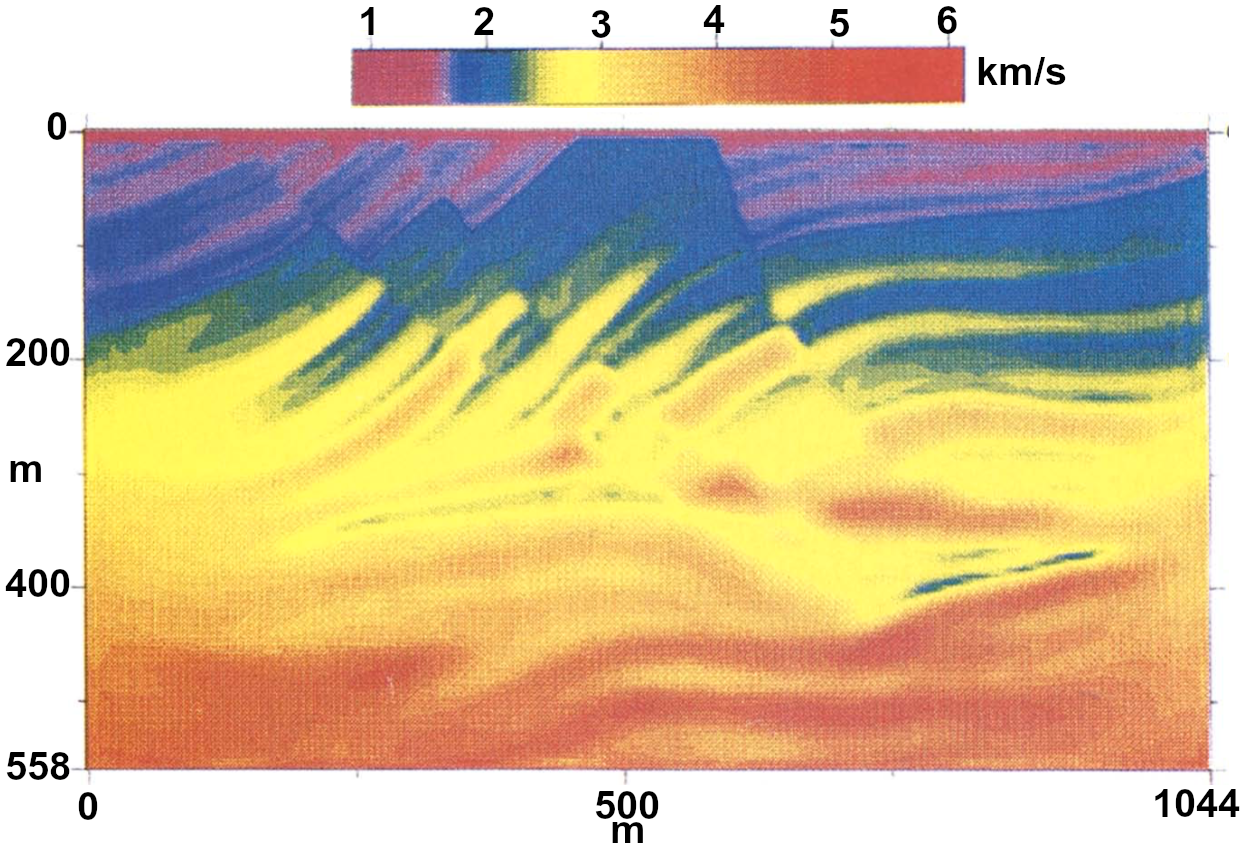}}
	\caption[First practical application of FWI using the Marmousi model]{First practical application of FWI using the Marmousi model. This shows significant improvements for the FWI results as presented by \cite{Bunks1995}.}        
	\label{fig:first_practical_fwi}
\end{figure}

One of the pioneering applications of FWI was presented by \cite{Bunks1995} and is represented in Figure~\ref{fig:first_practical_fwi}. They showed better imaging using a hierarchical multi-scale approach on the Marmousi synthetic model. This strategy initially inverts for low-frequency components where there are fewer local minima, and those that exist are sparser than if for higher frequencies. However, decomposing by scale did not resolve issues of source estimation, source bandwidth and noise \citep{Bunks1995}. In the 1990s, Pratt and his associates proposed FWI via the pseudo-spectral domain \citep{Pratt1990a, Pratt1990b, Pratt1991}. The initial application was to cross-hole data utilizing a finite difference approach and an elastic wave propagator to facilitate the modelling of multi-source data. This was extended to wide-aperture seismic data by \cite{Pratt1996}. Analytically, the time- and frequency-domain problems are equivalent \citep{Virieux2009}. Initial attempts of pseudo-spectral FWI include application to the Marmousi model \citep{Sirgue2004} and land seismic dataset \citep{Operto2004}.

\subsection{Beyond Academic Experiments}\label{sec:lit_rev_beyond_academic_exp}
\begin{figure}[!ht]
	\centering
	\subbottom[Velocitiy model. Top: Conventional, Bottom: FWI.]{\includegraphics[width=0.45\textwidth]{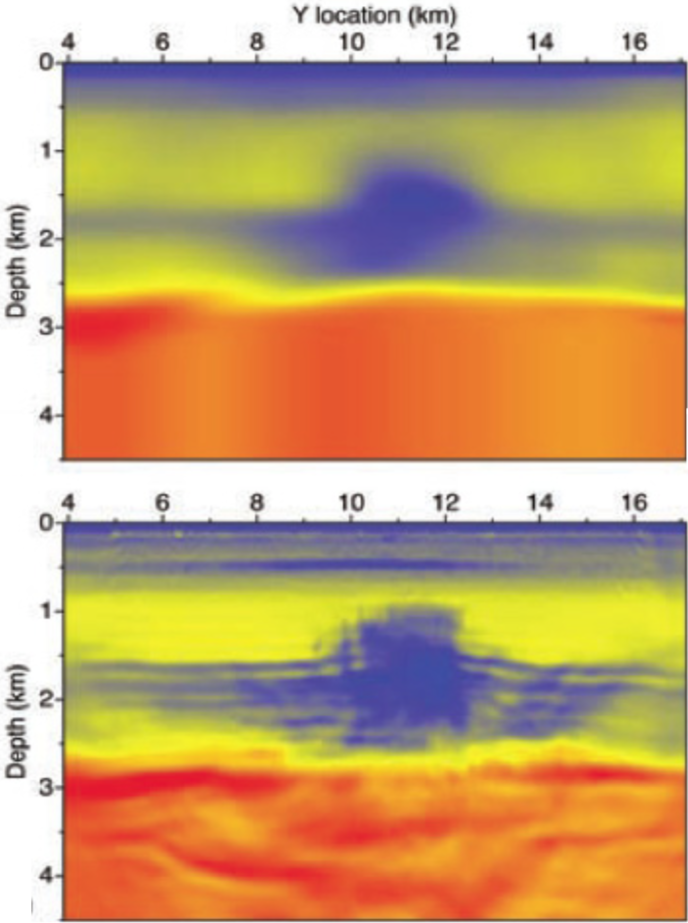}}
	\subbottom[Pre-stack depth migrated section. Top: Conventional, Bottom: FWI.]{\includegraphics[width=0.45\textwidth]{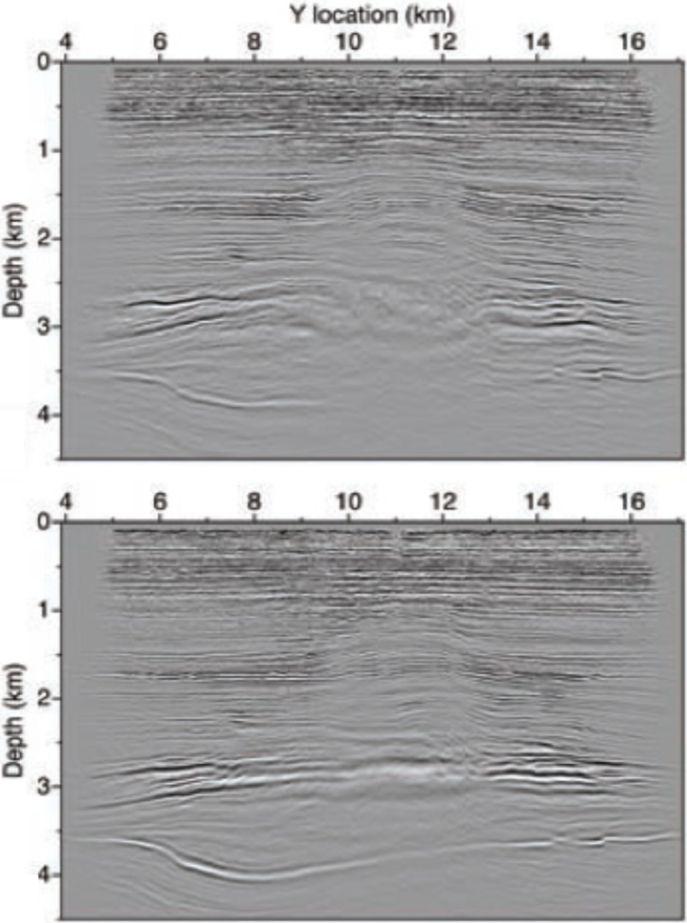}}%
	\caption[Improvements in velocity model and pre-stack depth migrated images obtained through FWI over the Valhall field.]{Improvements in velocity model and pre-stack depth migrated images obtained through FWI over the Valhall field. The FWI updated velocity model demonstrated a network of shallow high-velocity channels and a gas-filled fracture extension from a gas cloud which was not previously identifiable in conventional tomography. The impact is evident in the migrated sections, which show more continues events in otherwise poorly illuminated area. Adapted from \cite{Sirgue2009} and \cite{Sirgue2010}.}        
	\label{fig:fwi_prestack_improvements}
\end{figure}

Theoretically, two-dimensional inversion is only able to explain out-of-plane events by mapping them into in-plane artefacts \citep{Morgan2009}. This meant that FWI restricted to purely academic pursuits \citep{Sirgue2009} and full potential could only be realized if extended to three-dimensions. The first 3D frequency-domain algorithms where developed by \citet{Warner2007} on synthetic datasets, however these used low initial frequencies that are not normally present in real data \citep{Morgan2013}. Examples of this application are demonstrated by \cite{Sirgue2007}, \cite{Ali2007} and \cite{Operto2007}. \cite{Warner2008c} presented the first 3D real data application to a shallow North Sea survey. This improved the resolution of shallow high-velocity channels that resulted in uplifts upon migration. 
In Figure~\ref{fig:fwi_prestack_improvements}, \citet{Sirgue2009} demonstrated successful FWI results for a 3D dataset of the Valhall field, Norway. They inverted wide-azimuth ocean-bottom cable data using a sequence of low frequency bands to generate high-resolution velocity models. The updated velocity model demonstrated a network of shallow high-velocity channels and a gas-filled fracture extension from a gas cloud which was not previously identifiable \citep{Sirgue2010}.

\begin{figure}[!ht]
	\centering
	\subbottom[PSDM stack and CDP gather with original orthorhombic model from tomography.]{\includegraphics[width=0.9\textwidth]{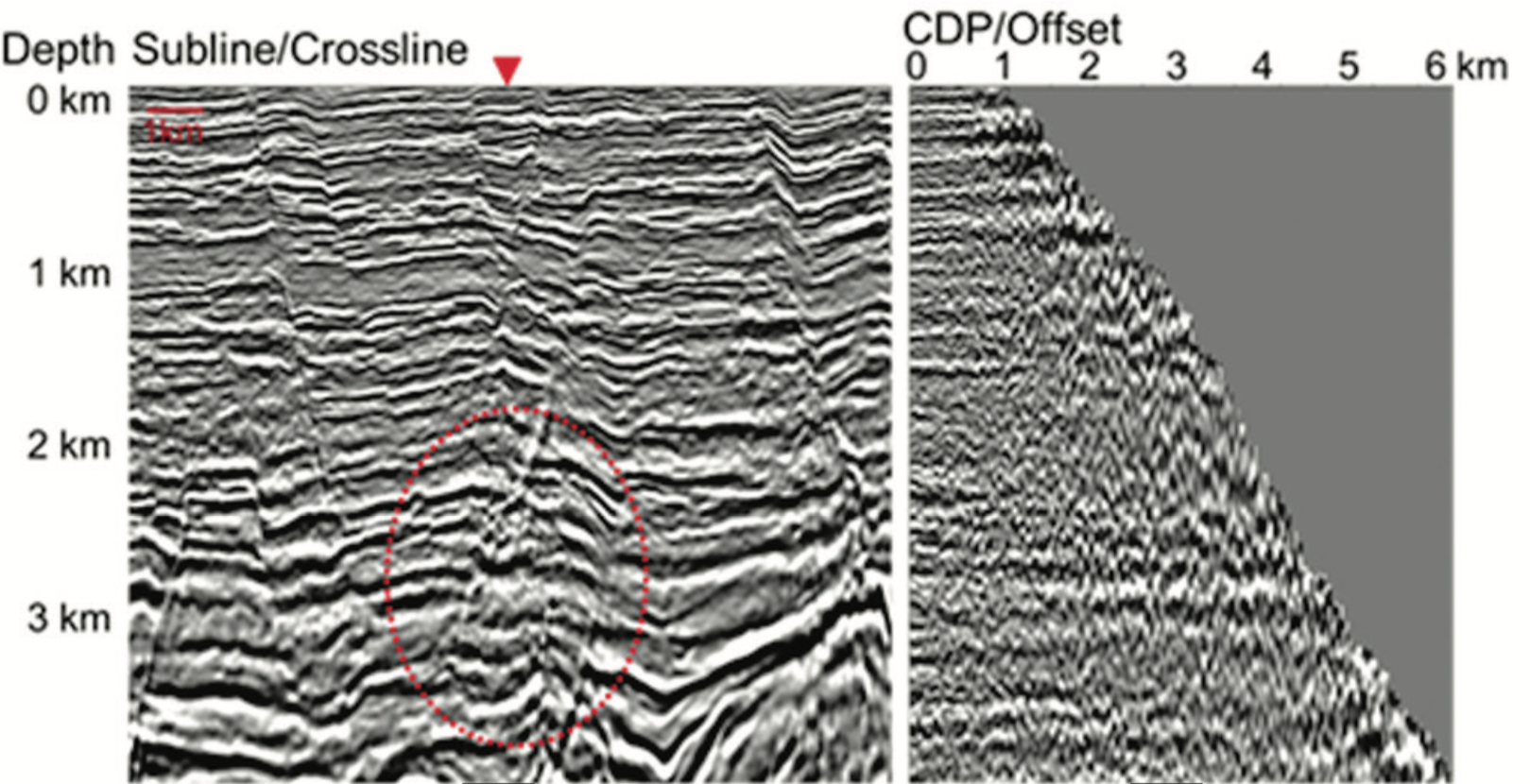}}
	\subbottom[PSDM stack and CDP gather with orthorhombi FWI update.]{\includegraphics[width=0.9\textwidth]{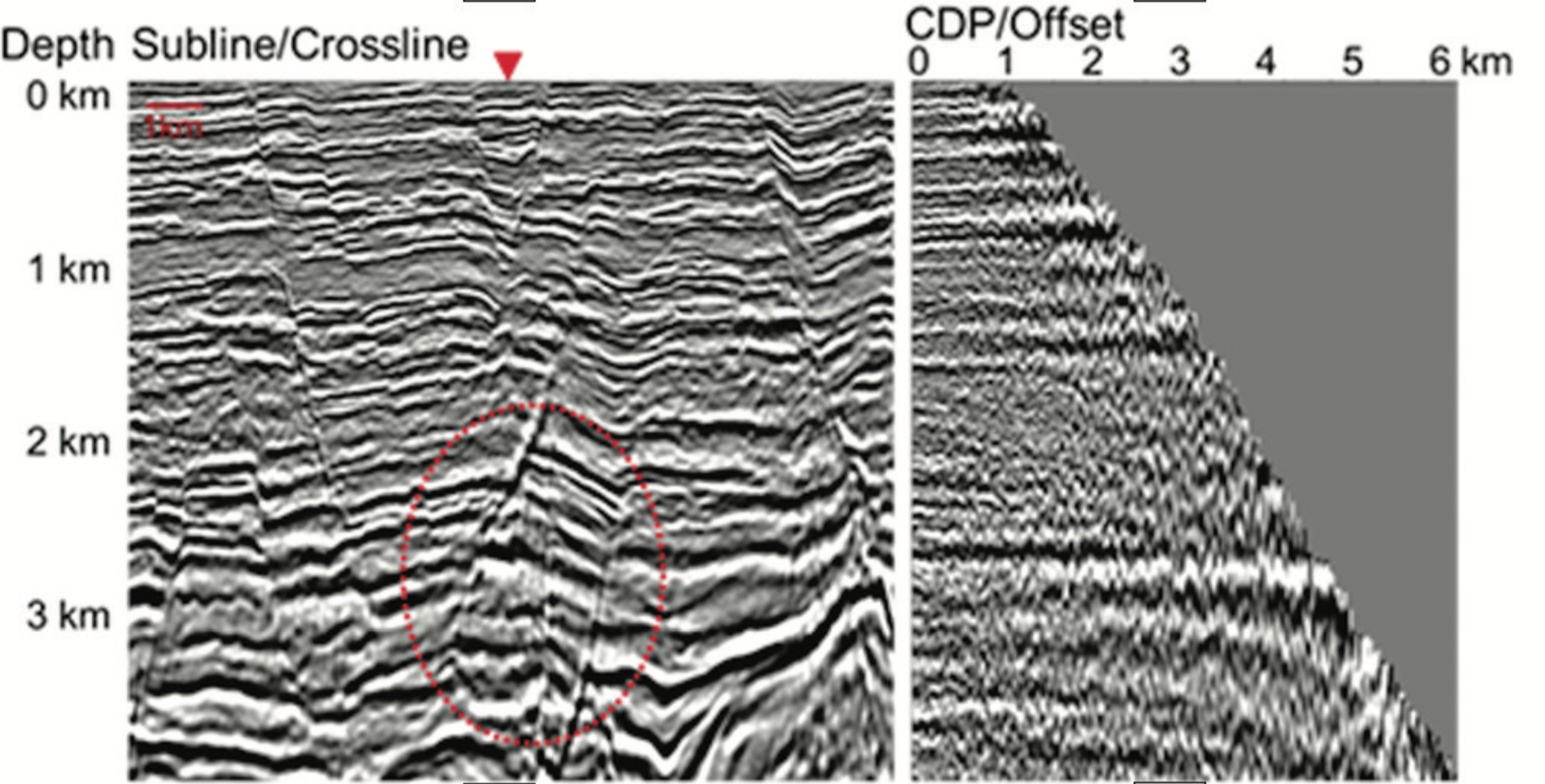}}
	\caption[Imaging improvements obtained through orthorhombic imaging.]{Imaging improvements obtained through orthorhombic imaging. This produces sharp truncations and clearer faults as highlighted by the red dashed ovals, as well as better focussed gathers. Adapted from \cite{Xie2017}.}        
	\label{fig:ortho_fwi}
\end{figure}

\citet{Plessix2010} show results from the application of full waveform inversion to ocean bottom data recorded in the Gulf of Mexico with near-ideal long-offset and wide-azimuth. Their approach was anisotropic and assumed vertical transversely isotropic media with fixed Thomsen’s parameters. The model had better imaging of dips and produced flatter common image gathers in the deep part of the model \citep{Plessix2010}. \cite{Wang2016} developed 3D waveform inversion for orthorhombic media in the acoustic approximation using pseudo-spectral methods. This was found to be stable and produced kinematic accurate pure-mode primary wavefields with an acceptable computational cost. \cite{Xie2017} applied orthorhombic full-waveform inversion for imaging wide-azimuth ocean-bottom-cable data. The results had better azimuthal and polar direction-dependent wave imaging which significantly improved fault imaging -  See Figure~\ref{fig:ortho_fwi}.

A re-occurring theme within this section is the creation of a better approximation to the wavefield propagation within the subsurface; 1D to 2D to 3D discretization, acoustic to anisotropic to elastic to orthorhombic wavefield modelling, with each additional dimension of information resulting in more numerical and computer intensive algorithms \citep{Kumar2012}. Even though computing power has increased dramatically, making FWI more productive, the underlying algorithms are only improving incrementally. Indeed, the next generation of experiments will require changes to acquisition geometry to allow for full-bandwidth and multi-azimuth reconstruction of the wavefield \citep{Morgan2016}.

\section{Deep Neural Networks}

\subsection{Neural Networks for Inverse Problems}
The mathematical formulation of FWI falls under the more general class of variational inverse problems \citep{Tanaka2003}. The aim is to find a function which is the minimal or the maximal value of a specified functional \citep{Dadvand2006}. Indeed, inverse problems attempt to reconstruct an image $x \in X \subseteq \mathbb{R}^d$ from a set of measurements $y \in Y \subseteq \mathbb{R}^m$ of the form
\begin{equation}\label{eq:inverse_theory}
	y=\Gamma(x)+\epsilon
\end{equation}
where $\Gamma: X \mapsto Y, \Gamma \in \mathbb{R}^{m\times d}$ is the discrete operator and $\epsilon \in Y \subseteq \mathbb{R}^{m}$ is the noise. \ac{NN} within Machine Learning can be considered to be a set of algorithms of non-linear functional approximations under weak assumptions \citep{Oktem2018}. Namely, when applied to inverse problems, Equation~\ref{eq:inverse_theory} can be re-phrased as the problem of reconstructing a non-linear mapping $\Gamma_\theta^{\dagger}:Y \mapsto X$ satisfying the pseudo-inverse property
\begin{equation}\label{eq:learned_inverse}
	\Gamma_\theta^{-1}(y) \approx x
\end{equation}
where observations $y\in Y$ are related to $x\in X$ as in Equation~\ref{eq:inverse_theory}, and $\theta$ represents the parametrization of pseudo-inverse by the \ac{NN} learning \citep{Adler2017a}. The loss function defined in Equation~\ref{eq:learned_inverse} is dependent on the type of training data, which is dependent on the learning approach \citep{Adler2017b}.  There are two main classes of learning in Machine Learning: (i) Supervised, and (ii) Unsupervised.

In supervised learning, training data are independent distributed random pairs with input $x\in X$ and labelled output $y \in Y$ \citep{Vito2005}. Estimating $\theta$ for Equation~\ref{eq:learned_inverse} can be formulated as minimizing a loss function $\mathcal{J}(\theta)$ which has the following structure \citep{Adler2017a}:
\begin{equation}\label{eq:supervised_learning}
	\mathcal{J}(\theta) := \mathcal{D}(\Gamma_\theta^{-1}(\boldsymbol{y}), \boldsymbol{x})
\end{equation}
where $\mathcal{D}$ is a distance function quantifying the quality of the reconstruction and $\Gamma_\theta^{-1}:Y\mapsto X$ is the pseudo-inverse to be learned \citep{Adler2017a}. A common metric for the distance function is the sum of squared distances, resulting in:
\begin{equation}
\mathcal{J}(\theta) := \left|\left|\Gamma_\theta^{-1}(y)- x\right|\right|_X^2
\end{equation}
Approaching the inverse problem directly via this approach amounts to learn $\Gamma_\theta^{-1}:Y\mapsto X$ from data such that it approximates an inverse of $\Gamma$. In particular, this has successful applications in medical imaging \citep{Xu2012, Lucas2018}, signal processing \citep{Rusu2017, Dokmanic2016} and regularization theory \citep{Meinhardt2017, Romano2016, DelosReyes2017}.

In unsupervised learning, there exist no input-output labelled pairs and the training data is solely elements of $y\in Y$. The \ac{NN} is required to learn both the forward problem and inverse problem \citep{Andrychowicz2016}. The loss functional for unsupervised learning is given as:
\begin{equation}
	\mathcal{J}(\theta) := \mathcal{L}\left(\Gamma\left(\Gamma_\theta^{-1}(y)\right),x \right)+\mathcal{S}\left(\Gamma_\theta^{-1}(g)\right)
\end{equation}
where $\mathcal{L}:Y\times X \mapsto\mathbb{R}$ is a suitable affine transformation of the data and $\mathcal{S}:X\mapsto \mathbb{R}$ is the regularization function. Main applications of this learning are to inherent structure and have been proven successful in exploratory data analysis applications such as clustering \citep{Sever2015,Gerdova2002} and dimension reduction \citep{Dolenko2015}.


\subsection{Evolution of Neural Networks}\label{sec:evo_NN}
The remaining literature review is restricted to supervised learning approaches using NN as these are more suited for velocity inversion. For a complete review, \cite{Lippmann1987} and \cite{Chentouf1997} provide further detail. 

\subsubsection{Early Neural Nets and the Perceptron}\label{sec:early_nn}
The basic ideas of \ac{NN} date back to the 1940’s and were initially devised by \citet{McCulloch1943} when trying to understand how to map the inner workings of a biological brain into a machine. From a biological aspect, neurons in the brain are interconnected via nerve cells that are involved in the processing and transmitting of chemical and electrical signals \citep{McCulloch1943}. 

\begin{figure}[ht!]
	\centering
	\includegraphics[width=0.75\linewidth]{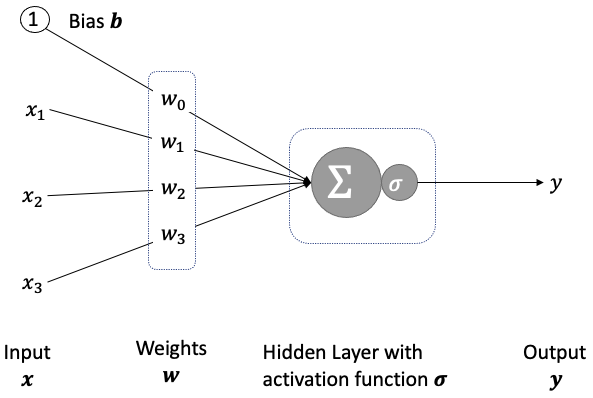}
	\caption[The Single Neuron Perceptron]{The Single Neuron Perceptron. The input values are multiplied by the weights. If the weighted sum of the product satisfies the activation function, the perceptron is activated and ``fires'' a signal. Adapted from \cite{Rosenblatt1958}.}
	\label{fig:perceptron}
\end{figure}
\begin{figure}[ht!]
	\begin{minipage}[c]{0.4\textwidth}
		\centering
		\includegraphics[width=\textwidth]{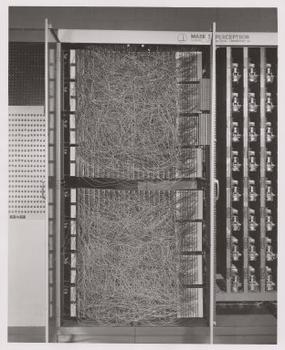}
	\end{minipage}\hfill
	\begin{minipage}[c]{0.55\textwidth}
		\caption[Mark I Perceptron Machine]{The Mark I Perceptron Machine was the first machine used to implement the Perceptron algorithm. The machine was connected to a camera that used 20×20 array of cadmium sulphide photocells to produce a 400-pixel image. To the right is a patch-board that allowed experimentation with different combinations of input features. This was usually wired up randomly to demonstrate the ability of the perceptron to learn. Adapted from \cite{Hecht-Nielsen1990, Bishop2006}.}
		\label{fig:perceptron_mk1}
	\end{minipage}
\end{figure}

Early \acp{NN} with rudimentary architectures did not learn \citep{McCulloch1943} and the notion of self-organized learning only came about in 1949 by \citet{Hebb1949}. \cite{Rosenblatt1958} extended this idea of learning and proposed the first and simplest neural network – the McCullock-Pitts-Perceptron. As shown in Figure~\ref{fig:perceptron}, this consists of a single neuron with weights and an activation function. The weights are the learned component and determine the contribution of either input $x$ to the output $y$. The activation function $\sigma$ adds a non-linear transform, allowing the neuron to decide if the input is relevant for the paired output. Without an activation function, the neuron would be equivalent to a linear regressor \citep{Minsky2017}. \cite{Rosenblatt1958} used this fundamental architecture to reproduce a functional mapping that classifies patterns that are linearly separable. This machine was an analogue computer that was connected to a camera that used 20×20 array of cadmium sulphide photocells to produce a 400-pixel image. Shown in Figure~\ref{fig:perceptron_mk1}, the McCullock-Pitts-Perceptron had a patch-board that allowed experimentation with different combinations of input features wired up randomly to demonstrate the ability of the perceptron to learn \citep{Hecht-Nielsen1990, Bishop2006}.

Rosenblatt’s perceptron was the first application of supervised learning \citep{Russell2008}. However, \citet{Minsky1969} highlight limitations to the applications of a single perceptron. They also point out that Rosenblatt’s claims that the ``perceptron may eventually be able to learn, make decisions, and translate languages'' were exaggerated. There was no other follow ups on this work by Minsky and Papert, and research on perceptron-style learning machines practically halted \citep{Minsky2017}. 

\subsubsection{Back-Propagation and Hidden Layers}
Efficient error back-propagation in NN networks were described in Linnainmaa’s master thesis \citep{Linnainmaa1970}. This minimizes the errors through gradient descent in the parameter space \citep{Hadamard1907} and allows for explicit minimization of the cost function. Back-propagation permits \acp{NN} to learn complicated multidimensional functional mappings \citep{Dreyfus1973}. 

The back-propagation formulation lends itself from major developments in dynamic programming throughout the 1960s and 1970s \citep{Kelley1960, Bryson1961, Linnainmaa1976}. A simplified derivation using the chain rule was derived by \cite{Dreyfus1973} and the first NN-specific application was described by \cite{Werbos81}. It was until the mid-1980s that \cite{Rumelhart1986} made back-propagation mainstream for \acp{NN} through the numerical demonstration of internal representations of the hidden layer. Hidden layers reside in-between input and output layers of the NN.

Back-propagation was no panacea and additional hidden layers did not offer empirical improvements \citep{Schmidhuber2015}. \cite{Kolmogoro1956}, \cite{Hecht-Nielsen1989} and \cite{Hornik1989} pursued development of back-propagation encouraged by the Universal Approximation Theorem. Namely, this theorem states that if enough hidden units are used in a NN layer, this can approximate any multivariate continuous function with arbitrary accuracy \citep{Hecht-Nielsen1989}.
Although back-propagation theoretically allows for deep problems, it was shown not to work on practical problems \citep{Schmidhuber2015}.

\subsubsection{The Vanishing Gradient and Renaissance of Machine Learning}
The major milestone in NN came about in 1991. Hochreiter’s thesis identified that deep \acp{NN} suffer from the vanishing or exploding gradient problem \citep{Hochreiter1991}. Gradients computed by back-propagation become very small or very large with added layers, causing convergence to halt or introduce unstable update steps. Solutions proposed to address this challenge included batch normalization \citep{Ioffe2015}, Hessian-free optimisations \citep{Moller1993, Schraudolph2002, Martens2010}, random weight assignment
\citep{Hochreiter1996}, universal sequential search \citep{Levin1973} and weight pruning \citep{LeCun1990}. 

\begin{figure}[ht!]
	\centering
	\includegraphics[width=0.9\linewidth]{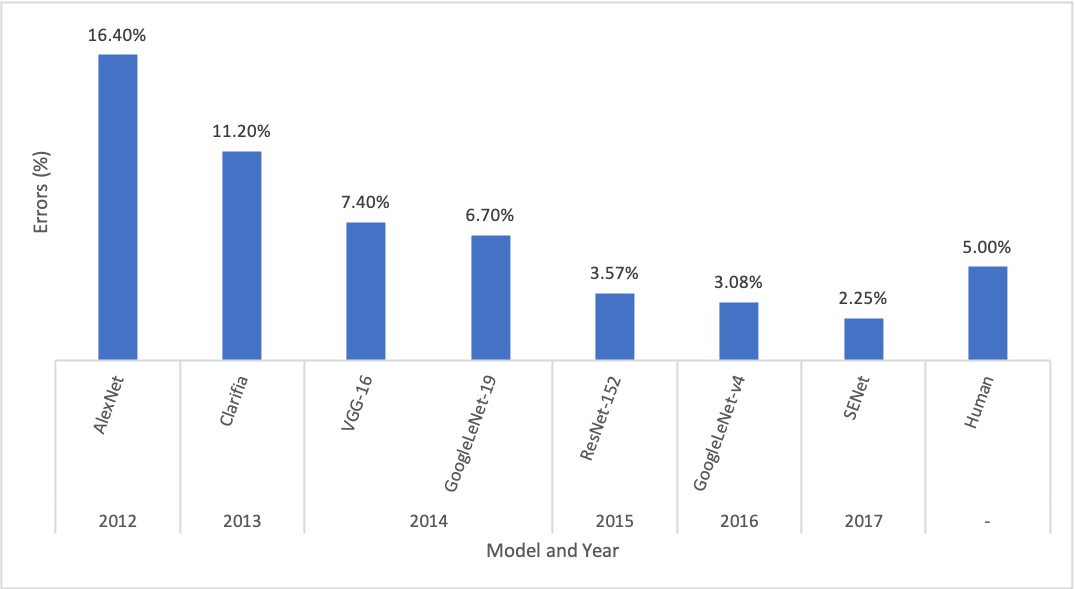}
	\caption[Evolution of accuracy for the ImageNet challenge.]{Evolution of the accuracy for the ImageNet challenge \citep{Krizhevsky2012}. AlexNet won ImageNet in 2012 with 16.4\% error in accuracy. With each year of the competition, the accuracy has been increasing. Sources for these accuracies are Clarifia \citep{Zeiler2014}, VGG-16 \citep{Simonyan2014}, GoogleLeNet-19 \citep{Szegedy2014}, ResNet-152 \citep{He2016}, GoogleLeNet-v4 \citep{Szegedy2016} and SENet \citep{Hu2017}.}
	\label{fig:evolution_imagenet}
\end{figure}

Prior to 2012, NN were apparently an academic pursuit. This changed when AlexNet \citep{Krizhevsky2012} won the ImageNet \citep{Russakovsky2015} visual object recognition by a considerable margin. AlexNet used a deep architecture consisting of eight layers \citep{Krizhevsky2015} and was the only entry employing NN in 2012. All submissions in subsequent years were NN-based \citep{Singh2015} and in 2015, \acp{NN} surpassed human performance in visual object recognition for the first time \citep{Russakovsky2015} - see Figure~\ref{fig:evolution_imagenet}. AlexNet is undoubtedly a pivotal event that ignited the renaissance in interest around deep learning.

\subsection{Deep Neural Network Architecture Landscape}\label{sec:dnn_arch}
According to \citet{Patterson2017}, three of the most common major architectures are (i) Neural Network, (ii) \ac{CNN}, and (iii) \ac{RNN}. 

As discussed in §~\ref{sec:early_nn}, neural networks are non-linear models inspired by the neural architecture of the brain in biological systems. A typical neural network is known as a multi-layer perceptron and consists of a series of layers, composed of neurons and their connections \citep{Goodfellow2016}. 

\begin{figure}[ht!]
	\centering
	\includegraphics[width=0.95\linewidth]{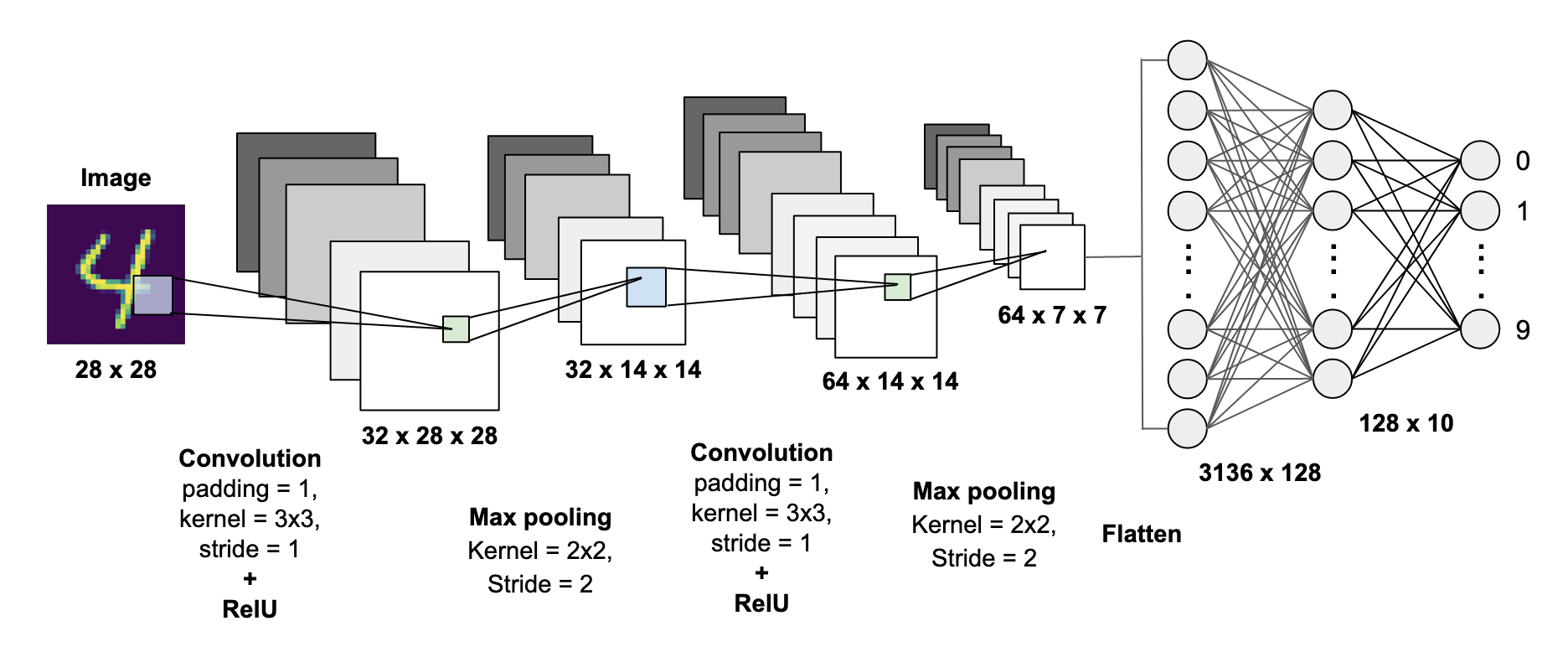}
	\caption[LeNet5 CNN architecture to classify handwritten characters.]{LeNet5 - LeCun et al.’s (1990) CNN architecture used to classify handwritten digits. This consists of two sets of convolutional and max pooling layers, followed by a flattening convolutional layer, then two fully-connected layers and finally a soft-max classifier.}
	\label{fig:CNN_arch}
\end{figure}

\acp{CNN} are regularized version of MLPs with convolution operations in place of general matrix multiplication in at least one of the layers \citep{LeCun1989}. These types of networks find their motivation from work by \citet{Hubel1959, Hubel1962}. Inspired by this work, \citet{Fukushima1982} introduced convolutional layers and downsampling layers, \citet{Zhou1988} developed max pooling layers and \cite{LeCun1990} used back-propagation to derive the kernel coefficients for convolutional layers. The architecture of the NN used by LeCun et al. is known as LeNet5 and is shown in Figure~\ref{fig:CNN_arch} for the classification of hand-written digits. This essentially laid the foundations for modern CNNs. \cite{LeCun2010} gives a comprehensive history up to 2010 and a more recent review is available by \cite{Khan2020}.

\acp{RNN} are in the family of feed-forward neural networks that have recurrent connections or allow for parameter sharing \citep{Lang1988}. These recurrent connection allow for input activations to pass from the hidden nodes in previous states to influence the current input \citep{Waibel1989, Lang1990}. 
\ac{LSTM} networks are one of the most commonly used variations of RNNs \citep{Patterson2017}. These were introduced by \citet{Hochreiter1997} and add the concept of memory gates or cells \citep{Graves2012, Gers1999}. These gates allow for information to be accessible across different time-steps and thus attenuates the vanishing gradient problem present with most RNN models \citep{Patterson2017}.

\subsection{Not Just Algorithms}
\begin{figure}[ht!]
	\centering
	\includegraphics[width=0.95\linewidth]{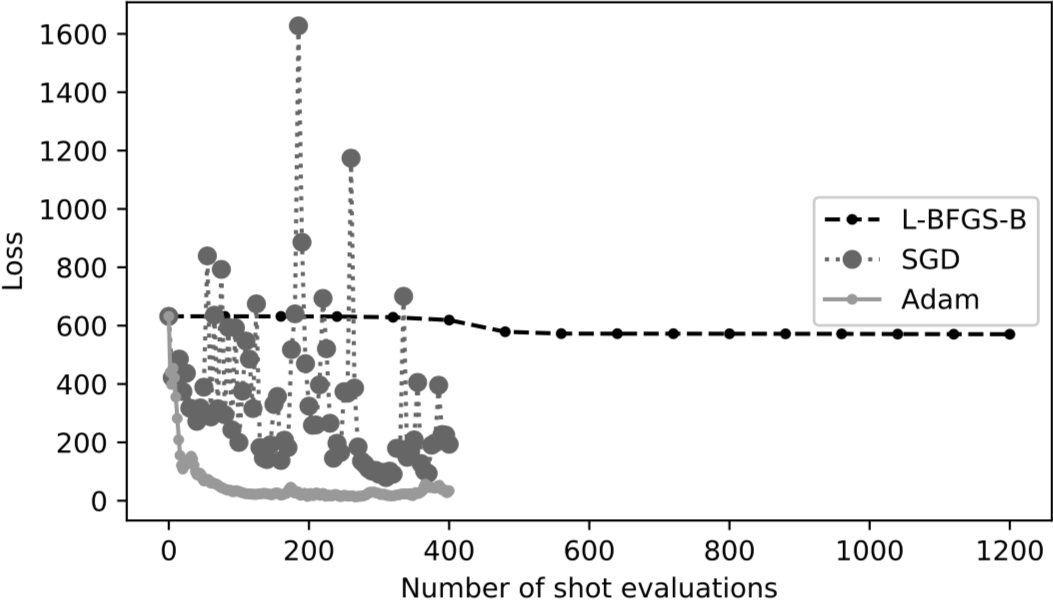}
	\caption[Faster convergence of Adam cost function.]{Adam cost function shown to converge much more rapidly than conventional Stochastic Gradient Descent and L-BFGS-B for FWI. From \cite{Richardson2018}.}
	\label{fig:state_of_art_adam}
\end{figure}

Apart from Machine Learning algorithms, re-interest in DNN has led to software architectures that allow for quick development. The most common include Tensorflow \citep{Abadi2015}, Keras \citep{Chollet2015}, PyTorch \citep{Paszke2017}, Caffe \citep{Jia2014}) and Deeplearning4j \citep{Nicholson2016}. These types of frameworks are facilitating interdisciplinarity between Machine Learning and geophysics. Indeed, \cite{Richardson2018} employed DNN architecture within Tensorflow to solve for FWI. Utilizing a common DNN optimizer - Adam - he shows in Figure~\ref{fig:state_of_art_adam} how the cost function converged quicker in the inversion process as compared to conventional methods in FWI. 

\section{Neural Networks in Geophysics}\label{sec:app_geophysics}
Machine Learning techniques have been utilised across different geophysical applications. Some notable examples include geo-dynamics \citep{Shahnas2018}, geology \citep{Reading2015}, seismology \citep{Shimshoni1998}, paleo-climatology \citep{Dowla1996}, climate change \citep{Anderson2018} and hydrogeology \citep{Hamshaw2018}. Unsupervised algorithms have also been investigated by \citet{Kohler2010} for pattern recognitions of wavefield patterns with minimal domain knowledge. Other geophysical application include seismic deconvolution \citep{Wang1992, CaldeOn-Macias1997}, tomography \citep{Nath1999}, first-break picking \citep{Murat1992}, trace editing \citep{McCormack1993a}, electricity \citep{Poulton1992}, magnetism \citep{Zhang1997}, shear-wave splitting \citep{Dai1994}, event classification \citep{Romeo1994}, petrophysics \citep{Downton2018} and noise attenuation \citep{Li2018a, Halpert2018}. 

\subsection{Legacy Velocity Inversion}
\begin{figure}[ht!]
	\begin{minipage}[c]{0.55\textwidth}
		\centering
		\includegraphics[width=\textwidth]{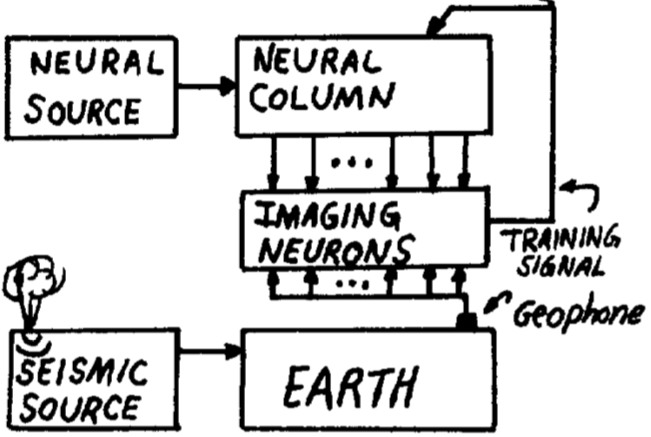}
	\end{minipage}\hfill
	\begin{minipage}[c]{0.4\textwidth}
		\caption[Block diagram for RNN system by Michaels \& Smith (1992)]{Block diagram for RNN system by \citet{Michaels1992}. The difference between a signal and the internal neural signals along a neural column are processed to provide a training signal that modifies neuron weights.}
		\label{fig:Block_RNN_FWI}
	\end{minipage}
\end{figure}

More specific to velocity estimation, the first published investigation for the use of NN was a RNN by \citet{Michaels1992}. Their network architecture represented all components in an elastic FWI experiment with a seismic source, the propagation media and an imaging response. Figure~\ref{fig:Block_RNN_FWI} shows a block diagram representation for their network. The neural column consisted of two 1-layer neuron columns, one for particle displacement and another for particle velocity. 

\begin{figure}[ht!]
	\centering
	\includegraphics[width=0.95\linewidth]{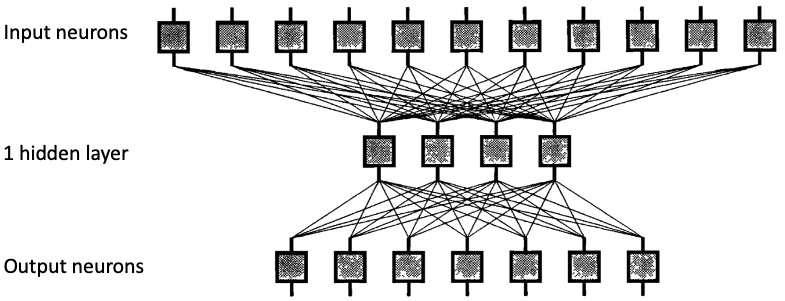}
	\caption[Architecture for first NN application to FWI.]{Architecture for first NN application to FWI. This was a very shallow NN with 1 hidden layer and non-symmetric input-output neurons. Adapted from \cite{Roth1994}.}
	\label{fig:first_NN_FWI}
\end{figure}

\clearpage
\cite{Roth1994} published the first application of NN which estimated 1D velocity functions from shot gathers from a single layer NN in 1994. Figure~\ref{fig:first_NN_FWI} shows their NN architecture. This accepted synthetic common shot gathers from a single source as input and used to compute corresponding 1D large-scale velocity models. The training set used for learning consisted of 450 synthetic models built up of eight strata with constant layer thickness over a homogeneous half-space. Their network was able to approximate a true velocity model sufficiently to act as a starting model for further seismic imaging algorithms. The inferred velocity profiles of the unseen data provided 80\% accuracy levels, and although the network was stable for noise contained data, it was not robust against strong correlated noise. Nonetheless, this investigation sets up NN as possible candidates to solve non-trivial inverse problems.

NNs are not solely limited to creating initial models to FWI. \citet{Langer1996} show how NNs can invert for parameters related to the seismic source and propagation medium using a similar single hidden layer architecture as R{\"o}th \& Tarantola. The difference in approach was two-fold; the NN employed a single seismogram as input as opposed to whole shot gather and pseudo-spectral data was used for training rather than time waveforms directly. The use of a single waveform did allow for improved results, however the use of pseudo-spectral data was instrumental. Transformed NN inference had better accuracies than the conventional time approach. Motivation to use pseudo-spectral data followed the work of \citet{Falsaperla1996} where they identified how the introduced sparsity within pseudo-spectral domain facilitated the learning process for the NN and was more robust to noise.

\subsection{Data-Driven Approaches to Velocity Estimation}
The terminology of data-driven geophysics is not a novel-one. This was first introduced in the literature by \citet{Schultz1994} when estimating rock properties directly from seismic-data through statistical techniques. However, conceptually, this is similar to the deconvolution process within a seismic processing flow \citep{Robinson1957,Robinson1967}. Namely, a filter is derived via autocorrelations and applied as a deconvolution operator \citep{Webster1978}. The term has only recently found a re-invigorated interest. Some modern applications of data-driven geophysical processes include dictionary learning \citep{NazariSiahsar2017}, time series analysis \citep{Wu2018a}, fault identification \citep{Mangalathu2020}, and reservoir characterization \citep{Schakel2014}.

Twenty-one years after \citet{Michaels1992}, \cite{Lewis2017} employed DNN architecture to learn prior models for seismic FWI. Their data driven approach at estimating initial models was applied to salt body reconstruction by learning the probability of salt geo-bodies and use this to regularize the FWI objective function. \citet{Araya-Polo2018} utilised DNN architecture and inverted for 2D high-velocity profiles. For the training process, they generated thousands of random 2D velocity models with up to four faults in them, at various dip angles and positions. Each model had three to eight layers, with velocities ranging from 2000 to 4000 \si{ms^{-1}}, with layer velocity increasing with depth. The DNN architecture is not defined in their paper, however when applied to unseen data with and without salt anomalies, their results achieved accuracies well above 80\% for both cases. This was used to obtain a low-wavenumber starting model then passed to traditional FWI as an initial model. \cite{Wu2018} proposed a convolutional-based network called ``InversionNet'' to directly map the raw seismic data to the corresponding seismic velocity model for a simple fault model with flat or curved subsurface layers. More recently, \cite{Li2019} extended this further and developed a DNN framework called ``SeisInvNet'' to perform the end-to-end velocity inversion mapping with enhanced single-trace seismic data as the input in time domain. 

\subsection{Wave Physics as an Analogue Recurrent Neural Network}
\begin{figure}[ht!]
	\centering
	\includegraphics[width=0.98\linewidth]{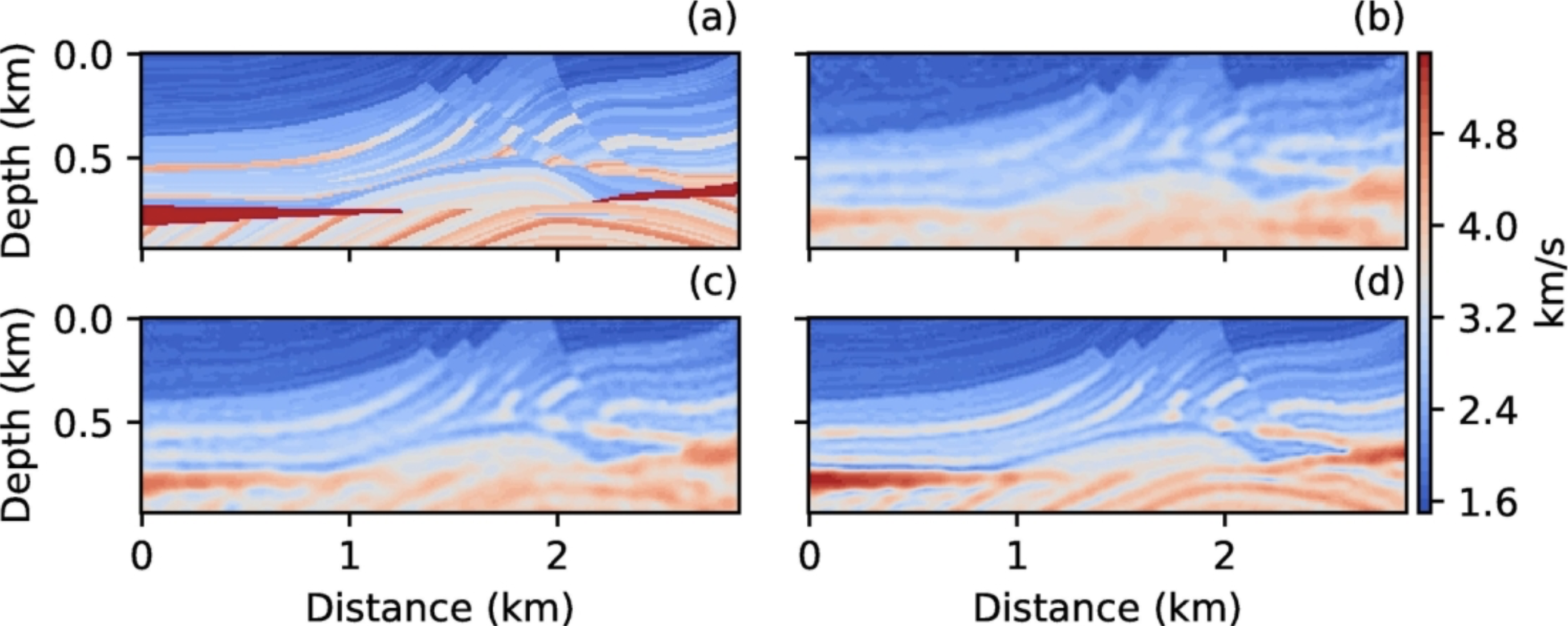}
	\caption[The inversion of Marmousi velocity model using RNN forward modelling.]{The inversion of Marmousi velocity model using RNN forward modelling framework with Adam algorithm optimizer. (a) True Marmousi. (b) 25th iteration. (c) 50th iteration (d) 100th iteration. From \cite{Sun2019}.}
	\label{fig:RNN_FWI_Marmousi}
\end{figure}

Recently, \citet{Raissi2019} and \citet{Hughes2019} derived a function between the physical dynamics of wave phenomena and RNNs. In their work they propose physics-informed neural networks that are trained to solve supervised learning tasks while respecting the laws of physics described by general non-linear partial differential equations. Fundamental to their approach is the ability for DNNs to be universal function approximators. Within this formulation, \cite{Raissi2019} are able to solve non-linear problems without the need to compute a priori assumptions, perform linearisation or employ local time-stepping. Under this new paradigm in modelling, \citet{Raissi2019} show how back-propagation is used ``to differentiate neural networks with respect to their input coordinates and model parameters to obtain physics-informed neural networks. Such \acp{NN} are constrained to respect any symmetries, invariances, or conservation principles originating from the physical laws that govern the observed data, as modelled by general time-dependent and non-linear partial differential equations''. In particular, following up from this work, \cite{Sun2019} recast the forward modelling problem in FWI into a deep learning network by recasting the acoustic wave propagation into a RNN framework. Figure~\ref{fig:RNN_FWI_Marmousi} shows velocity inversion results from \cite{Sun2019} applied to the Marmousi velocity model. These theory-guided inversions still suffer from cycle-skipping, local-minima and high computational cost \citep{Sun2019}. Recent research suggests that Stochastic Gradient Descent algorithms have the capacity to escape local minima to a certain extent \citep{Sun2019}. 


%% file: chap3/theory_main.tex
\chapter{Theoretical Considerations}

\textbf{This chapter reviews and derives the key theoretical approaches used in this dissertation. The first section introduces a classical FWI formulation and summarizes the local optimization problem as a flowchart of elements. Using a combination of ANNs, CNNs and RNNs present in  DNN architectures, FWI is recast as a simulation within a deep learning framework. The next two sections introduce the theoretical setups of data-driven and theory-guided pseudo-spectral FWI and associated theory which can be used to solve FWI as a Deep Learning problem. For a more in-depth assimilation, Appendix~\ref{sec:app_theory} provides complementary material.}

\section{FWI as Local Optimization}\label{sec:theory_theory_fwi_as_local_optimization}
\cite{Lailly1983} and \cite{Tarantola1984a} recast the migration imaging principle introduced by \cite{Claerbout1971} as a local optimization problem \citep{Virieux2009}. For an anisotropic medium, particle motion is based on the wave equation given by:
\begin{equation}\label{eq:wave_equation}
	\frac{1}{c(\boldsymbol{x})^2}\frac{\partial^2p(\boldsymbol{x},t)}{\partial t^2} - \nabla^2 p(\boldsymbol{x},t) = s(\boldsymbol{x},t),
\end{equation}
where $p(\boldsymbol{x},t)$ is the pressure wavefield, $c(\boldsymbol{x})$ is the acoustic p-wave velocity and $s(\boldsymbol{x},t)$ is the source. This can be expressed as a linear operator and solved numerically.

After forward modelling the physical system through the data, the objective is to minimize the difference between the observed data and the modelled data. The difference or misfit between the two datasets is known as the misfit-, objective- or cost- function $\phi$. The most common cost function is given by the $L_2-norm$ of the data residuals:
\begin{equation}\label{eq:l2_norm}
	\phi(x)= \frac{1}{2}\left|\left| \boldsymbol{y}_{\boldsymbol{obs}} - \boldsymbol{y}_{\boldsymbol{cal}}(x) \right|\right|_D^2 =\frac{1}{2}\Delta\boldsymbol{y}^\dagger\Delta\boldsymbol{y},
\end{equation}
where $D$ indicates the data domain given by $n_s$ sources and $n_r$ receivers, $\dagger$ is the transpose and $\boldsymbol{y}_{\boldsymbol{obs}},\boldsymbol{y}_{\boldsymbol{cal}}$ are the observed and calculated data respectively. The misfit function $\phi$ can be minimized with respect to the model parameters $y$ by setting the gradient to zero: 
\begin{equation}
	\nabla\phi=\frac{\partial\phi}{\partial y} = 0.
\end{equation}
To solve for the misfit function, FWI utilizes a linearised and iterative optimisation scheme. Based on the Born approximation in scattering theory \citep{Born1980}, consider the first model calculated to be $\boldsymbol{x}_0$. After the first pass via forward modelling, the model needs to be updated by the model parameter perturbation $\Delta \boldsymbol{x}_0$. This newly updated model is then used to calculate the next update and the procedure continues iteratively until the computed mode is close enough to the true model based on a residual threshold criterion. At each iteration $k$, the misfit function $\phi(\boldsymbol{x}_k )$ is calculated from model $\boldsymbol{x}_{k-1}$ of the previous iteration giving:
\begin{equation}\label{eq:misfit_next_step}
	\phi(\boldsymbol{x}_k)=\phi(\boldsymbol{x}_{k-1} + \Delta\boldsymbol{x}_k).
\end{equation}
Assuming that the model perturbation is small enough with respect to the model, Equation~\ref{eq:misfit_next_step} can be expanded via Taylor expansions up to second orders as:
\begin{equation}\label{eq:misfit_taylor_expansion}
	\phi(\boldsymbol{x}_k) = \phi(\boldsymbol{x}_{k-1}) + \delta\boldsymbol{x}^T\frac{\partial\phi}{\partial x}+\frac{1}{2}\delta\boldsymbol{x}^T\frac{\partial^2\phi}{\partial x^2}\delta\boldsymbol{x}.
\end{equation}
Taking the derivative of Equation~\ref{eq:misfit_taylor_expansion} and minimizing to determine the model update leads to:
\begin{equation}\label{eq:hessian}
	\partial\boldsymbol{x}\approx-\boldsymbol{H}^{-1}\nabla_x\phi,
\end{equation}
where $\boldsymbol{H}=\frac{\partial^2\phi}{\partial x^2}$ is the Hessian matrix and $\nabla_x\phi$ the gradient of misfit function evaluated at $\boldsymbol{x}_0$. The Hessian matrix is symmetric and represents the curvature trend of the misfit function.
\subsection{Model Update}\label{sec:theory_model_update}
Several methods can be utilised for updating for the next iteration. Newton methods update the model directly \citep{Newton1687}, whilst Gauss-Newton use Hessian approximations \citep{Tarantola1984}. The latter is referred to as the step-length \citep{Menke1989} and Equation~\ref{eq:hessian} becomes:
\begin{equation}
	\partial x \approx -\alpha\nabla_x\phi,
\end{equation}
where $\alpha$ is the step-length parameter and the magnitude of $\alpha$ is derived via first perturbing the initial model $\boldsymbol{x}_0$ by a differential change in the direction opposite to the gradient.
\subsection{Regularization}\label{sec:theory_regularization_fwi}
FWI often turns out to be an ill-posed problem given the incomplete parameter space measured in the field. Artefacts and over-fitting might be introduced in the model due to noise or high frequency component of the data \citep{Sirgue2004}. 
Well-posedness can be imposed on the model through \emph{a priori} information \citep{Asnaashari2013}. This is introduced into the misfit function (Equation~\ref{eq:l2_norm}) with the addition of Tikhonov $L_2-norm$ regularization \citep{Tikhonov1977}:
\begin{equation}
	\phi(x)=\frac{1}{2} \left|\left|\boldsymbol{y}_{\boldsymbol{obs}}-\boldsymbol{y}_{\boldsymbol{cal}}\left(\boldsymbol{x}\right)\right|\right|_D^2  + \frac{1}{2}\lambda\left|\left|\boldsymbol{x}-\boldsymbol{x}_{\boldsymbol{\emph{a priori}}}\right|\right|_M^2,
\end{equation}
where $\left|\left|\boldsymbol{x}-\boldsymbol{x}_{\boldsymbol{prior}}\right|\right|_M^2$ is the regularization term and $\lambda$ is the regularisation parameter which controls the trade-off between data and model residuals. Namely, the regularization parameter, $\lambda$, gives relative weight to model optimization term with respect to data optimisation term and acts as a smoother on the modelling \citep{Tikhonov1963}. 

\subsection{Difference between Time and Pseudo-spectral FWI}
Forward modelling for FWI can be done in the pseudo-spectral or time-domain. In case of an attenuating medium, the pseudo-spectral domain is the preferred method as frequency dependent attenuation (quality factor $Q$) is represented by the imaginary component of the velocity \citep{Pratt1999}. Frequency domain application requires the solution of a linear system of equations by a factorization method \citep{Sirgue2004}. This improves the chance for the inversion to locate the global minimum \citep{Sirgue2004}, however it scales poorly with the size of the problem \citep{Operto2007} or else requires assumptions on the physical propagation of the wave equation \citep{BenHadjAli2007}. Time-domain approach has a simpler implementation \citep{Vigh2008}, however it is highly more sensitive to cycle skipping \citep{Vigh2008} and prone to problematic low-wavenumber estimation for the gradient \citep{Sirgue2004}. Time-domain approaches either consider $Q$ to be constant or utilise some relaxation mechanism, thus not fully representing the physics of the underlying problem \citep{Blanch1995}. Either approach has been successfully applied to production datasets, both with merits in different environments. For time implementation reference is made to \cite{Vigh2008}, \cite{Vigh2010}, \cite{Liu2011} and \cite{Cai2015}, and for frequency to \cite{Ben-Hadj-Ali2008}, \cite{Operto2015}, \cite{Plessix2009}. Differences in numerical implementation between time and frequency FWI given in Appendix~\ref{sec:app_theory_Differences_in_numerical_implementation_between_Time_and_Frequency_FWI}.
\subsection{FWI Algorithm Summary}
\begin{figure}[ht]
	\centering
	\includegraphics[width=0.9\linewidth]{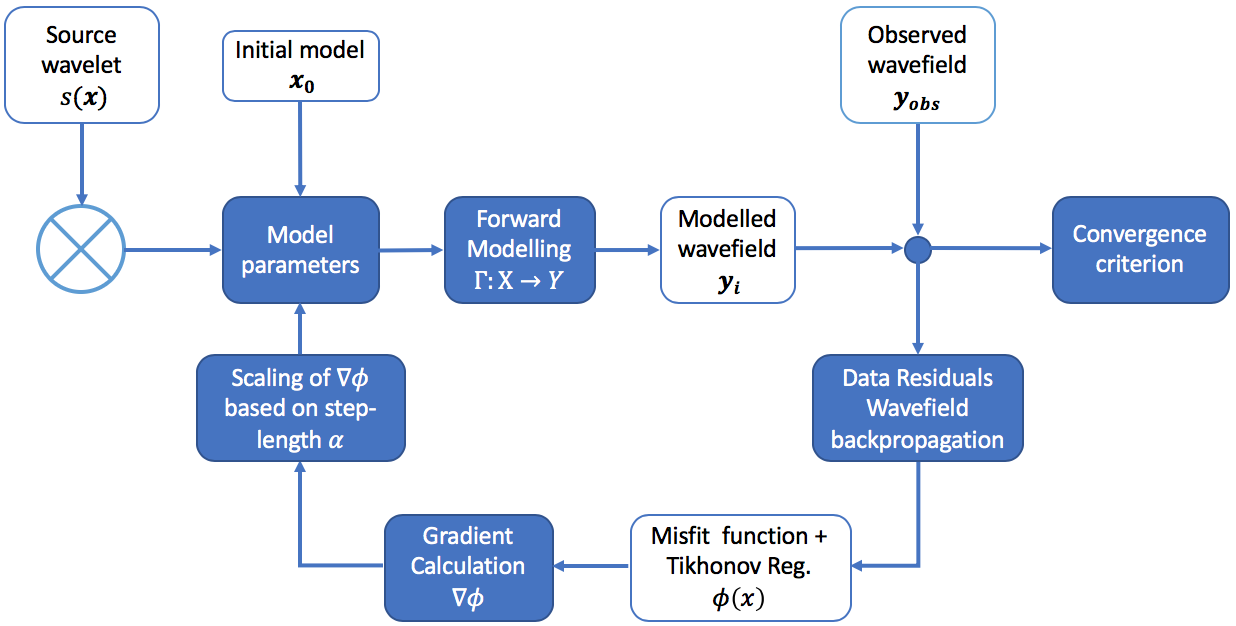}
	\caption[Schematic FWI workflow solved as an iterative optimization process.]{Schematic FWI workflow solved as an iterative optimization process.}
	\label{fig:workflow_FWI}
\end{figure}

Excluding considerations related to the practical implementation of FWI, the local optimization iterative scheme for FWI described in the previous section can be summarised in Algorithm~\ref{algo:fwi_local_optimization} and schematic is illustrated in Figure~\ref{fig:workflow_FWI}.

\begin{algorithm}
Choose an initial model $\boldsymbol{x}_0$ and source wavelet $s(\boldsymbol{x})$.\\
For each source, the forward problem $\Gamma:X\mapsto Y$ is solved everywhere in the model space to get a predicted wavefield $\boldsymbol{y}_i$. This is sampled at receivers $r(\boldsymbol{x})$.\\
At every receiver, data residuals are calculated between the modelled wavefield $\boldsymbol{y}_i$ and the observed data $\boldsymbol{y}_{\boldsymbol{obs}}$.\\
Data residuals are back-propagated to produce a residual wavefield.\\
For each source location, the misfit function $\phi(\boldsymbol{x})$ is applied for the observed data and back-propagated residual wavefield to generate the gradient $\nabla_\phi$ required at every point in the model.\\
The gradient is scaled based on the step-length $\alpha$, applied to the starting model and an updated model is obtained $\boldsymbol{x}_{i+1}$.\\
The process is iteratively repeated from Step 2 until a convergence criterion is satisfied.
\caption{FWI as Local Optimization}
\label{algo:fwi_local_optimization}
\end{algorithm}

\section[FWI as a Data-Driven DNN]{FWI as a Data-Driven DNN}\label{sec:theory_FWI_as_Learned_Direct_Approx}
When applied to inverse problems, \acp{NN} can simulate the non-linear functional of the inverse problem $\Gamma^{-1}:Y\mapsto X$. That is, using a NN, a non-linear mapping can be learned which will minimize
\begin{equation}
	\left|\left|\boldsymbol{x}-g_\Theta(\boldsymbol{y})\right|\right|^2,
\end{equation}
where $g_\Theta$ is the learned formulation of the inverse problem functional $\Gamma^{-1}$, and $\Theta$ the large data set of pairs $(\boldsymbol{x},\boldsymbol{y})$ used for the learning process \citep{Lucas2018}.

\subsection{Artificial Neuron, Perceptron and Multi-Layer Perceptron}
The most elementary component in a NN is a neuron. This receives excitatory input and sums the result to produce an output or activation \citep{Raschka2017}. For a given artificial neuron, consider $n$ inputs with signals $\boldsymbol{x}$ and weights $\boldsymbol{w}$. The output $\boldsymbol{y}$ of a neuron from all input signals is given by:
\begin{equation}
	y=\sigma\left( b + \sum_{j=0}^{n}w_{j}x_j \right),
\end{equation}
where $\sigma$ is the activation function and $b$ is a bias term enabling the activation functions to shift about the origin. Popular activations functions include Binary Step, Linear, Sigmoid, Tanh, Softmax and ReLu functions \citep{Goodfellow2016}. The ReLu or Rectified linear unit is the most widely used activation function. It is non-linear, allowing easily back-propagation of errors. When employed on a network of neurons, the negative component of the function is converted to zero and the neuron is deactivated, introducing sparsity with the network and making it efficient and easy for computation. More information on the other functions is provided in Appendix~\ref{sec:app_theory_Alternative_Activation_Functions}.

\subsection{Feed-forward Architectures and Deep Networks}
\begin{figure}[ht!]
	\centering
	\includegraphics[width=0.98\linewidth]{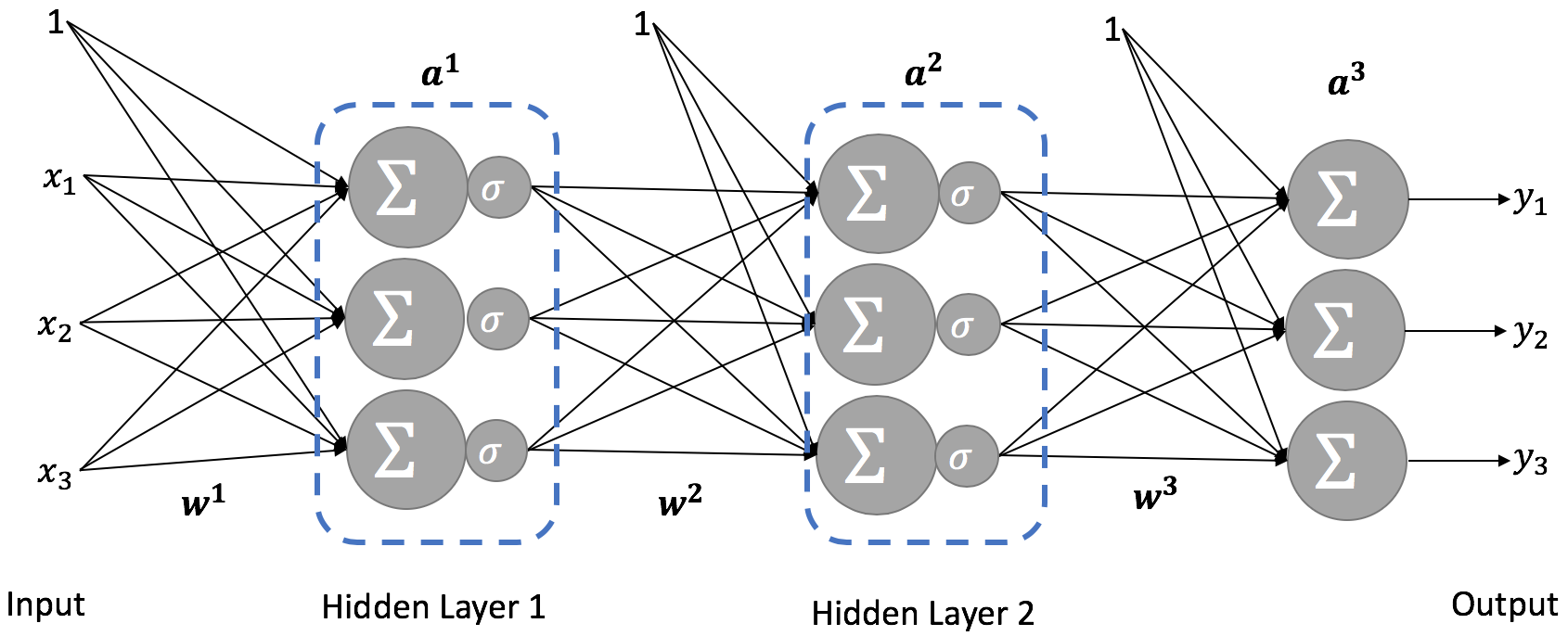}
	\caption[Fully connected NN with 2 hidden layers.]{Fully connected NN with 2 hidden layers. 
	Adapted from \cite{Lucas2018}.}
	\label{fig:MLP_2_layersI}
\end{figure}

The architecture of a NN refers to the number of neurons, their arrangement and connectivity. When a single neuron, the result is a Perceptron. Stacking multiple layers of the simple neurons and fully connecting all the inputs results in the MLP or fully connected layer. Within MLP, the input from the initial nodes $\boldsymbol{x}$ is connected to a hidden layer of neurons and the information is passed layer by layer through the hidden layers until reaching the output layer. Figure~\ref{fig:MLP_2_layersI} shows a fully connected MLP consisting of 2 hidden layers. The output of the units $\boldsymbol{y}$  is the weighted sum of the input units and the application of a non-linear element-wise function \citep{Lucas2018}.

\begin{figure}[!ht]
	\centering
	\subbottom[Deeper networks have been empirically shown to improve the accuracy, since deeper networks are able to achieve higher test accuracies. From \cite{Goodfellow2015}.\label{fig:deep_NN_advantage_METRICS}]{\includegraphics[width=0.9\textwidth]{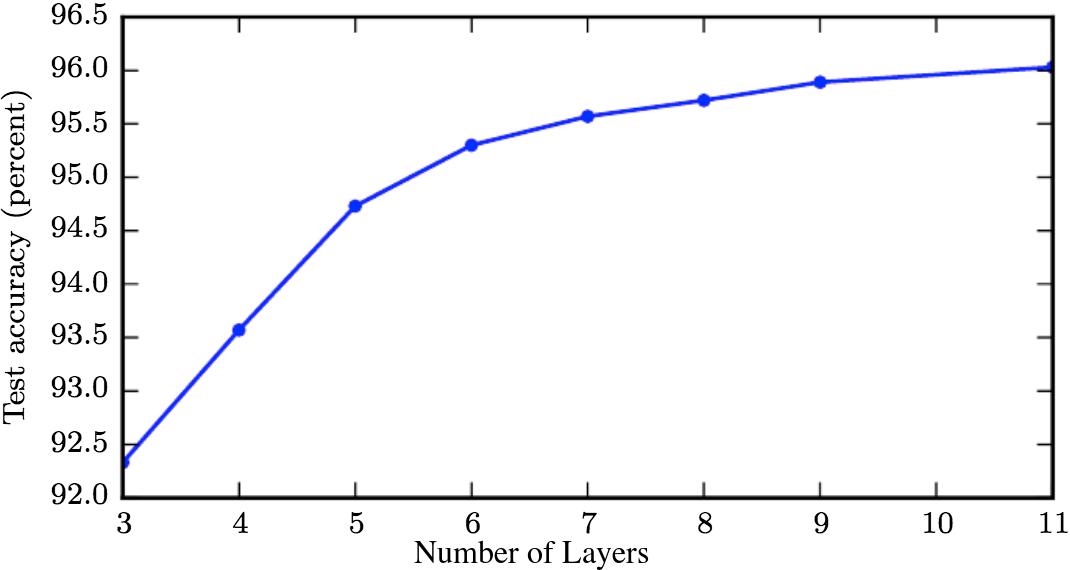}}
	\subbottom[Visual representation of the improved generalization for deep NNs over shallow NNs. The blue and red are the different weight and bias combinations for all the units in a network and the green line represents the decision function explaining the optimal choice for the weights and biases. The dotted grey line represents the line of symmetry which if appled will result in a simpler decision function able to explain more of the the weights and biases.
	\newline \textit{Left:} A shallow NN requires a complicated decision boundary to be able to explain the optimal wieght and bias combinations.
	\newline \textit{Centre:} A simpler boundary decision can be obtained when considering the axis of symmetry. This extra degree of freedom is achievable within a NN by having an additional hidden layer.
	\newline \textit{Right:} An even simpler, almost linear, decision boundary is obtained if an additional axis of symmetry (degree of freedom) is considered, thus giving a deeper network the ability to generalize better. From \cite{Montufar2014}.\label{fig:deep_NN_advantage_REGULARIZE}]{\includegraphics[width=0.98\textwidth]{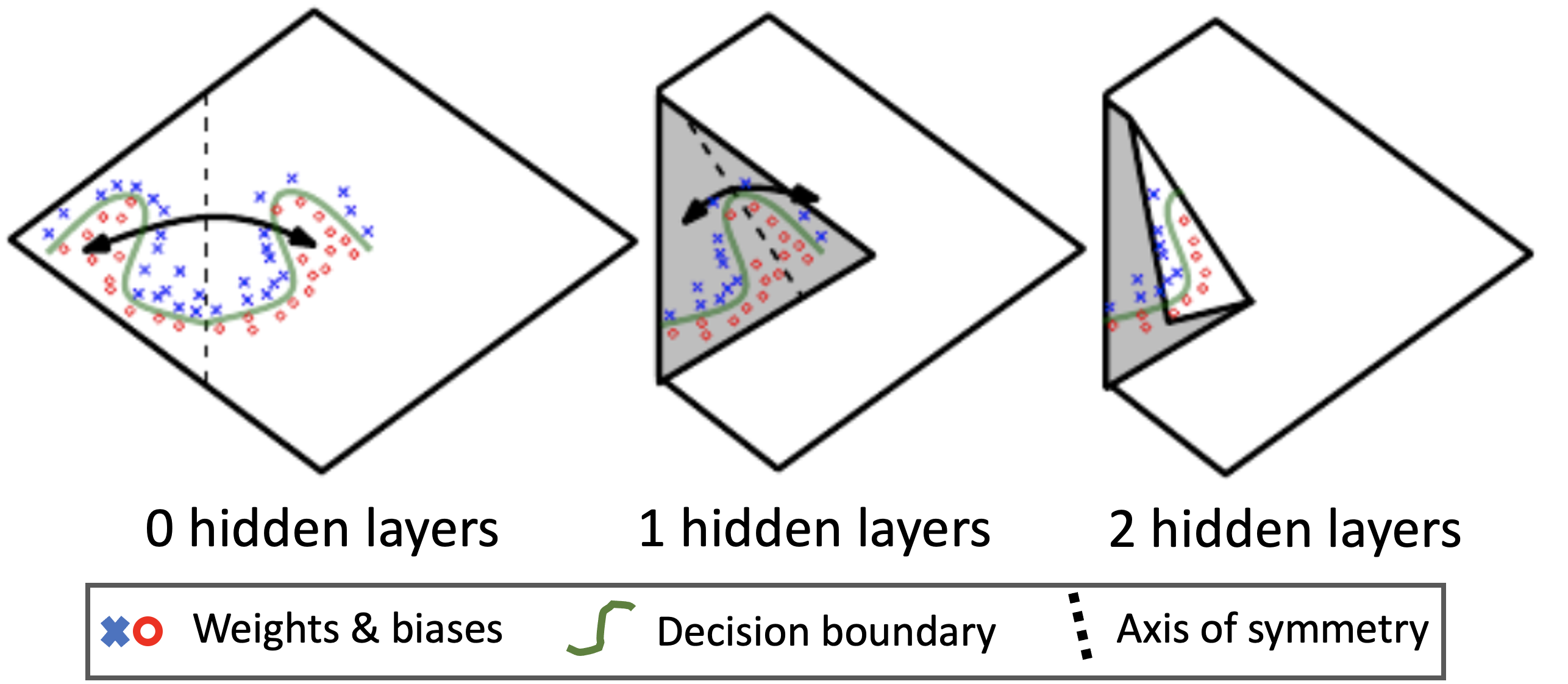}}
	\caption{Advantages of a deep NN over a shallow NN.}        
	\label{fig:deep_NN_advantage}
\end{figure}

When designing a NN, the depth of the network and the width of each layer is key. A network with a single hidden layer is sufficient to fit the training set \citep{Goodfellow2016}. However, deeper NN have better performance metrics (Figure~\ref{fig:deep_NN_advantage_METRICS}) and are able to generalize better to different non-linear functions (see Figure~\ref{fig:deep_NN_advantage_REGULARIZE}), yet harder to optimize \citep{Haykin2009}. Indeed, \cite{Montufar2014} showed that a shallow NN equivalent in terms of performance metrics to a DNN could require an exponential number of hidden units.

\subsection{A Universal Approximation Framework}\label{sec:theory_UAT}
A fundamental attribute to all deep feed-forward NN is that these provide a universal approximation framework \citep{Hornik1989}. In particular, the Universal Approximation Theorem as per \citet{Leshno1993} states that a ``feed-forward network with a linear output layer and at least one hidden layer with an appropriate activation function can approximate any Borel measurable function\footnote{\citet{Stephens2006} state that ``The real-valued function $f$ defined with domain $\mathcal{E}\subset\Omega$, for measurable space $(\Omega,\mathcal{F})$, is Borel measurable with respect to $\mathcal{F}$ if the inverse image of set $\mathcal{B}$, defined as $f^{-1} (\mathcal{B})\equiv\left\{\omega\in\mathcal{E}:f(\omega)\in\mathcal{B}\right\}$ is an element of $\sigma$-algebra $\mathcal{F}$, for all Borel sets $\mathcal{B}$ of $\mathbb{R}$''.} from one finite-dimensional space to another with any desired non-zero amount of error, provided that the network has sufficient hidden layers.''
\clearpage

\subsection{Back-Propagation and Learning with DNN}
When training a DNN, the forward propagation through the hidden layers from input $\boldsymbol{x}$ to output $\boldsymbol{y}$ needs to be measured for its misfit. In terms of regression modelling, the most commonly used cost function is the Sum of Squared Error, defined as:
\begin{equation}\label{eq:dnn_cost_function_J}
	\mathcal{J}\equiv\frac{1}{2}\sum_{j=1}^{\mathcal{J}}\left(\boldsymbol{y}^i-\boldsymbol{y}_{true}\right)^2,
\end{equation}
where $\boldsymbol{y}_{true}$ is the labelled true datasets and $\boldsymbol{y}^{i}$ is the output from the $i^{th}$ pass forward pass through the network. 

\begin{figure}[ht!]
	\centering
	\includegraphics[width=0.9\linewidth]{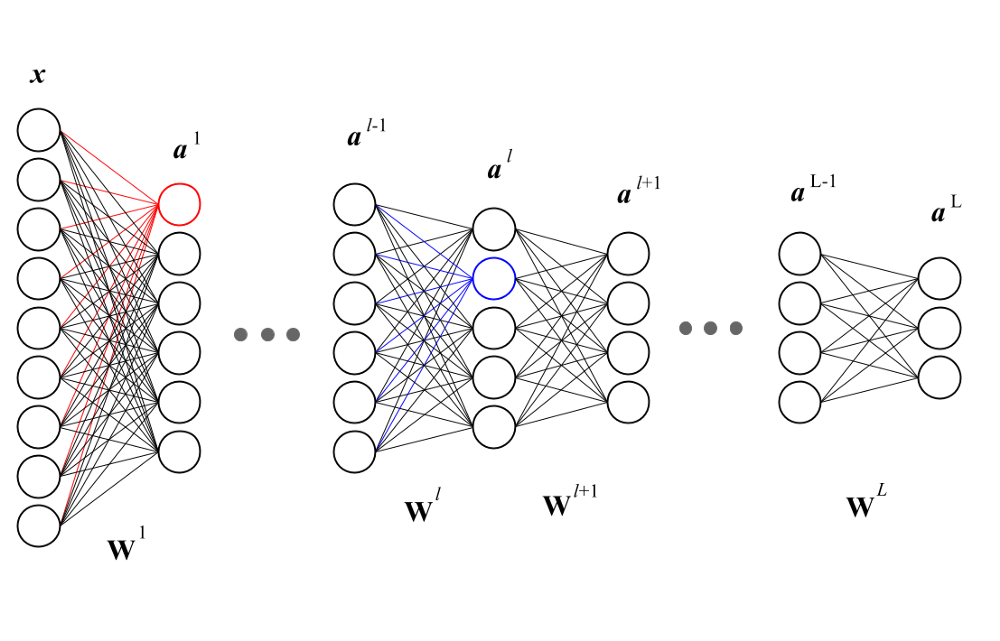}
	\caption[DNN schematic]{Schematic of a DNN with input $\boldsymbol{x}$, $L$ layers, $\boldsymbol{a}^l$ activation function at layer $l$ and weights $\boldsymbol{w}^l$. From \cite{Hallstrom2016}.}
	\label{fig:DNN_schematic}
\end{figure}
Consider the schematic for a fully connected DNN shown in Figure~\ref{fig:DNN_schematic}. Applying forward propagation through the layers of neurons:
\begin{alignat}{3}
&\text{Input sum of neuron $k$ in layer $l$} \qquad && z_k^l &&=b_k^l+\sum_{j}w_{kj}^l a_j^{l-1} \label{eq:input_sum_neurons_k}\\
&\text{Activation function on layer $l$}\qquad &&a_k^l &&=\sigma\left(z_k^l\right)\\
&\text{Input sum of neuron $m$ in layer $l+1$}\qquad &&z_m^{l+1}&&=b_m^l+\sum_{k}w_{mk}^{l+1} a_k^l \label{eq:input_sum_neurons_m}.
\end{alignat}
The objective is to minimize the function $\mathcal{J}$ with respect to the weights. Employing the chain rule, the derivative with respect to a single weight in layer $l$ is given as:
\begin{equation}
	\frac{\partial \mathcal{J}}{\partial w_{kj}^l} = 
	\frac{\partial \mathcal{J}}{\partial z_k^l}\frac{\partial z_k^l}{\partial w_{kj}^l} =
	\frac{\partial \mathcal{J}}{\partial a_k^l}\frac{\partial a_k^l}{\partial z_{k}^l}\frac{\partial z_k^l}{\partial w_{kj}^l}.
	\label{eq:dnn_cost_function}
\end{equation}
Substituting Equations~\ref{eq:input_sum_neurons_k}-\ref{eq:input_sum_neurons_m} in Equation~\ref{eq:dnn_cost_function} yields
\begin{equation}
	\frac{\partial \mathcal{J}}{\partial w_{kj}^l} = 
	\left( \sum_{m} 	\frac{\partial \mathcal{J}}{\partial z_m^{l+1}}\frac{\partial z_m^{l+1}}{\partial a_l^k}  \right)
	\frac{\partial a_k^l}{\partial z_{k}^l}\frac{\partial z_k^l}{\partial w_{kj}^l} =
	\left( \sum_{m}\frac{\partial \mathcal{J}}{\partial z_m^{l+1}} w_{mk}^{l+1} \right) \sigma^{'}\left( z_k^l\right) a_j^{l-1}.
	\label{eq:dnn_cost_function_substitute}
\end{equation}
The error signal of a neuron $k$ in layer $l$ is defined as the total error when the input sum of the neuron is changed and is given by:
\begin{equation}
	\delta_k^l\equiv\frac{\partial\mathcal{J}}{z_k^l}.
\end{equation}
Substituting Equations~\ref{eq:input_sum_neurons_k}-\ref{eq:input_sum_neurons_m} in Equation~\ref{eq:dnn_cost_function_substitute} results in a recursive formulation for the error given by 
\begin{equation}\label{eq:mlp_error}
	\delta_k^l=\left(\sum_{m}\delta_m^{l+1}w_{mk}^{l+1} \right)\sigma^{'}(z_k^l).
\end{equation}
Estimation of the error of the neurons in the final layer $L$ can be a sequential calculation of the error from the previous layer, until all error signals within the network are thus computed. The only derivative to be calculated is the derivative of the cost function $\sigma^{'}$, which is the output of the network. Bias $b_k^l$ for $l^{th}$ layer is a weight to be optimized and the error signal for the bias is given by:
\begin{equation}
	\frac{\partial\mathcal{J}}{\partial b_k^L}=\frac{\partial\mathcal{J}}{\partial z_k^l}\underbrace{\frac{\partial z_k^l}{\partial b_k^l}}_{1} = \delta_k^l.
\end{equation}
Hence, error signals for all neurons in the network can be recursively calculated throughout the network and the derivative of the cost function with respect to all the weights can also be calculated. Training of the DNN is achieved via gradient descent algorithm, referred to as Delta Rule in the machine learning community \citep{Sutton1988}. A small fraction of the derived derivative is subtracted from the weight and updated weight is given as:
\begin{equation}
	w_{kj}^l = w_{kj}^l-\eta\delta_k^l,
\end{equation}
where $\eta$ is a small scalar referred to the learning rate which controls how much of an adjusting is applied to the DNN with respect to the loss gradient. This is synonymous to the model update in FWI presented in §~\ref{sec:theory_model_update}. Equivalence of back-propagation and the classical adjoint method is numerically shown in Appendix~\ref{sec:app_results_equivalence_AD_adjoint}.

\subsection{Optimizing the Loss Function}\label{sec:theory_optimizing_the_loss_function}
Updating of the error gradient in a steepest gradient descent manner might be conceptually straightforward to understand. However, a major drawback with this algorithm is the high risk of getting stuck in local minima \citep{Fletcher1987}. This is an active area of research and \cite{Ruder2016} does an extensive review of optimizers. Four of the most widely used are (i) Adagrad, (ii) Adadelta, (iii) Adam and (iv) RMSprop. Difference in the implementation for these optimizers is given in Appendix~\ref{sec:app_theory_Loss_Optimizers}.

\subsection{Generalization of DNN for “unseen” Data}\label{sec:theory_generalization}
\begin{figure}[ht!]
	\centering
	\includegraphics[width=0.9\linewidth]{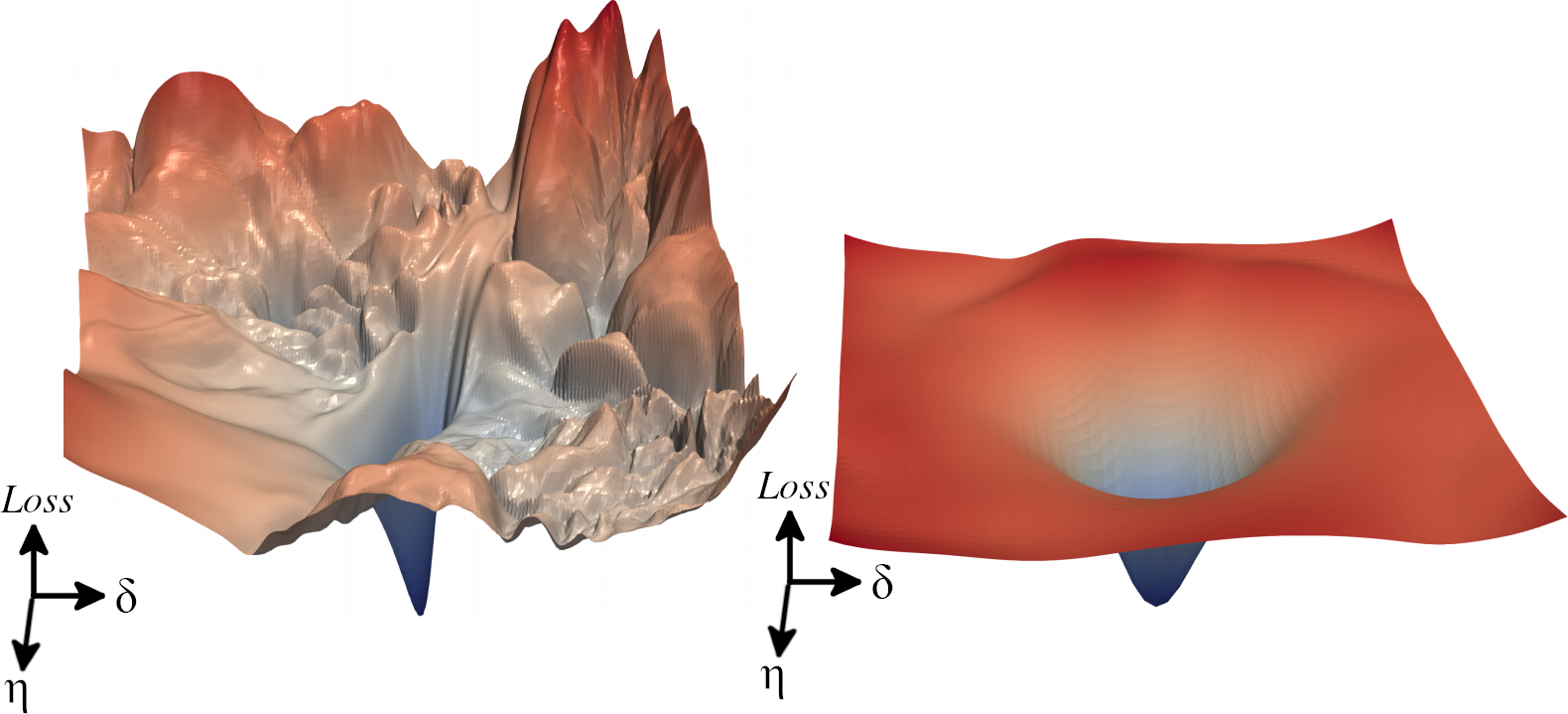}
	\caption[NN loss function without and with regularization]{NN loss function surface without (Left) and with (Right) regularization. Two random vectors $\left(\delta,\eta\right)$ are chosen on the parameter space and values of the loss computed. Vertical axes are in logarithmic scale to show dynamic range. It is highly evident that without regularization, the loss function is susceptible to local minima and at risk to converge to sub-optimal levels. On the other hand, regularization promotes flatness and prevent chaotic behaviour. Adapted from \cite{Li2017}.}
	\label{fig:DNN_regularization}
\end{figure}

Central to DNN is how to make the network able to perform well not just on the training data, but also on new “unseen” inputs. DNN resolve this via two regularization strategies: (i) functional-based, and (ii) NN architecture-based regularizations. These are schemes which update the cost function. On the other hand, architecture-based regularization are alterations to the NN architecture in the form of (i) Dropout, (ii) Data augmentation, and (iii) Early Stopping. Further information on these is available in Appendix~\ref{sec:app_theory_Regularization}. Either of these are essential for DNN as shown in Figure~\ref{fig:DNN_regularization}.

\subsection{Convolutional Neural Networks}
CNNs are feed forward multi-layered networks with layers consisting of convolutions and linear transformers able to perform multiple transformations \citep{LeCun2010}. Apart from Dropout, Activation functions and Fully Connected Layers components, CNNs introduce (i) Convolutional, (ii) Pooling, and (iii) Batch Normalization layers. Convolutional layers add convolutions across the neurons. Pooling layers are dimension reducing layers applied after convolutional layers since location of elements within feature-maps become less important as long as their position relative to others is preserved \citep{Khan2020}. Batch Normalization shifts the covariance of the distribution of hidden units to zero mean and unit variance for each batch \citep{Ioffe2015}, ensuring smooth gradient descent and regularization \citep{Santurkar2018}. Further details on the specific building blocks for CNNs is available in Appendix~\ref{sec:app_theory_CNN_Building_Blocks}.

\subsection{Neural Network Architecture Types}\label{sec:theory_NN_arch_types}
Combining the components introduced in the previous sections, different DNN architectures can be developed. The simplest form for a network would be MLP. This is considered to be a rudimental and basic network. An improvement would be the inclusion of convolutional, max-pooling and batch normalization layers in either one-dimension (1D) or two-dimensions (2D). One such architecture is AlexNet \citep{Krizhevsky2012}. Khan et al.’s 2020 survey provides a detailed overview of the different CNN architectures and how they evolved over time. In particular, two architectures which are considered state-of-the-art as a consensus in the literature are VGG \citep{Simonyan2014} and ResNet \citep{He2016IEEE}. These are only three of a myriad of architectures (Figure~\ref{fig:CNN_history}). More information on these architectures is available in Appendix~\ref{sec:app_theory_Comparison_of_Common_CNN_Architectures}.
\begin{figure}[ht!]
	\centering
	\includegraphics[width=0.98\linewidth]{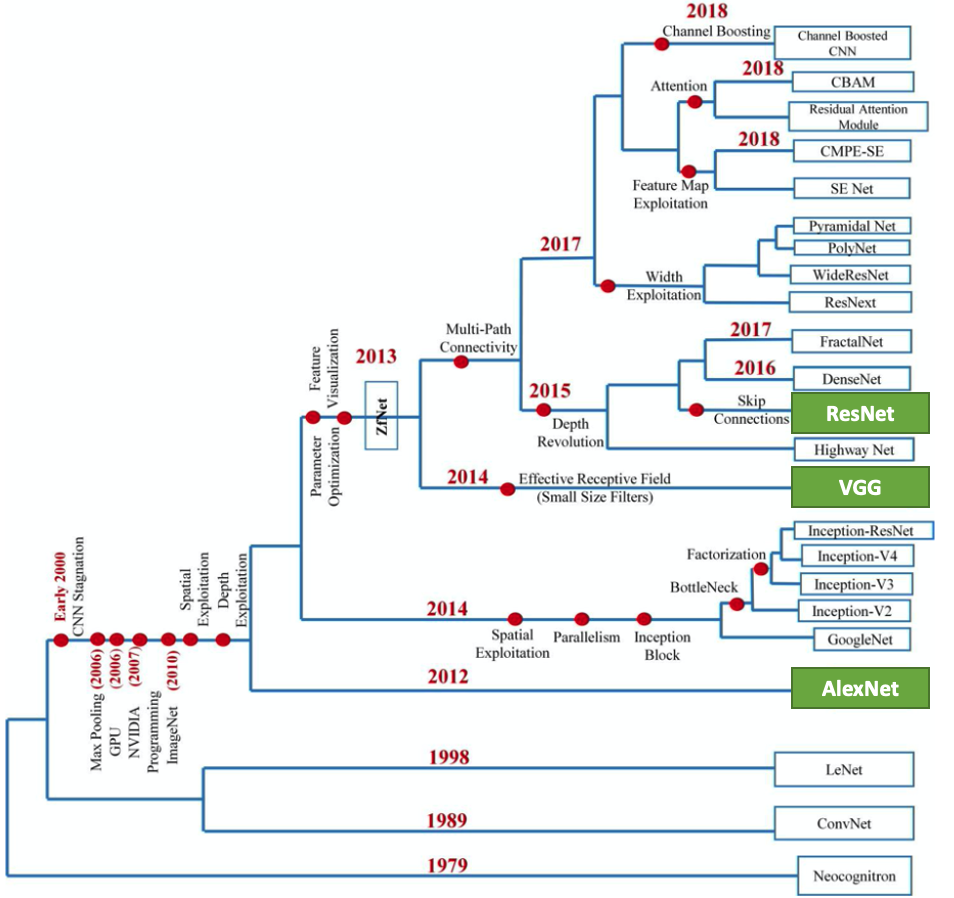}
	\caption[Historic overview of CNN architectures]{Historic overview of CNN architectures, highlighting timeline position for AlexNet, VGG and ResNet. Adapted from \cite{Khan2020}.}
	\label{fig:CNN_history}
\end{figure}

\clearpage
\subsection{Outline for Solving FWI with Data-Driven DNNs}
\begin{figure}[ht!]
	\centering
	\includegraphics[width=0.9\linewidth]{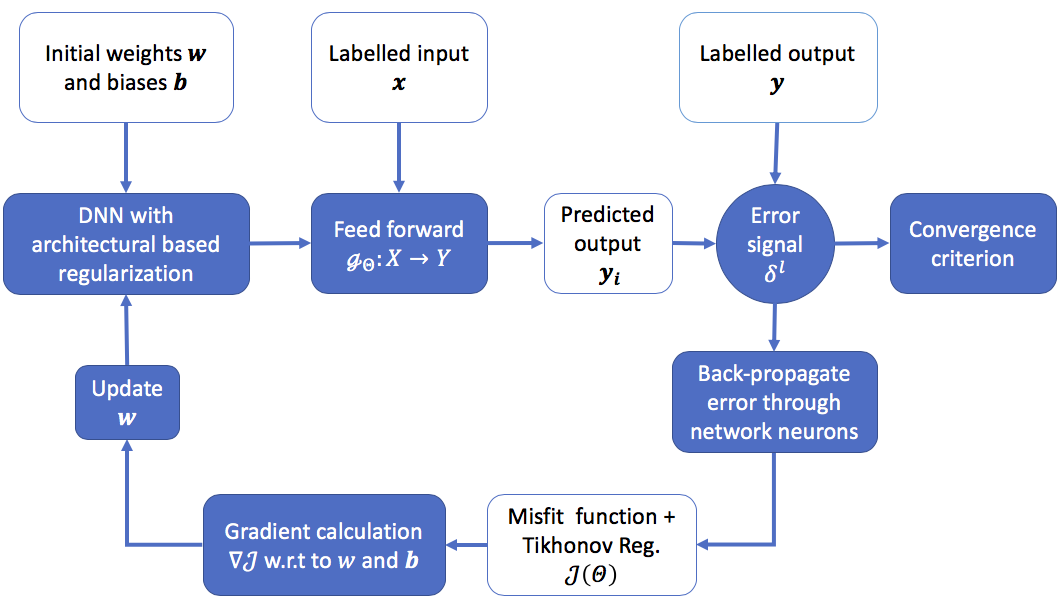}
	\caption[Schematic of DNN workflow]{Schematic of DNN workflow solved via supervised learning with input-output pairs $(\boldsymbol{x}, \boldsymbol{y})$. This is analogous to the FWI algorithm previously shown in Figure~\ref{fig:workflow_FWI}, with the difference that feed forward operator $g_\Theta:X\mapsto Y$ is a learned function.}
	\label{fig:workflow_DNN}
\end{figure}

Utilizing NN architecture and formulae, training of a DNN for FWI can be summarized as in Algorithm~\ref{algo:fwi_dnn}. Schematic of the learning process for DNN is given Figure~\ref{fig:workflow_DNN}.

\begin{algorithm}
Setup a deep architecture from those shown in §~\ref{sec:theory_NN_arch_types}, with regularization measures.\\
Initialise the set of weights $\boldsymbol{w}^l$  and biases $\boldsymbol{b}^l$.\\
Forward propagate through the network connections to calculate input sums and activation function for all neurons and layers.\\
Calculate the error signal for the final layer $\delta^L$ by choosing an appropriate differentiable activation function.\\
Back propagate the errors for all neurons in each $l$ layer $\delta^l$.\\
Differentiate the cost function with respect to biases $\left(\frac{\partial\mathcal{J}}{\partial\boldsymbol{b}^l}\right)$\\
Differentiate the cost function with respect to weights $\left(\frac{\partial\mathcal{J}}{\partial\boldsymbol{w}^l}\right)$\\
Update weights $\boldsymbol{w}^l$ via gradient descent.\\
Recursively repeat from Step 3 until the desired convergence criterion is met.
	\caption{FWI as a data-driven DNN}
	\label{algo:fwi_dnn}
\end{algorithm}

\section[RNN as an Analogue of FWI]{Theory-Guided RNN as an Analogue of FWI}
\begin{figure}[ht!]
	\centering
	\includegraphics[width=0.9\linewidth]{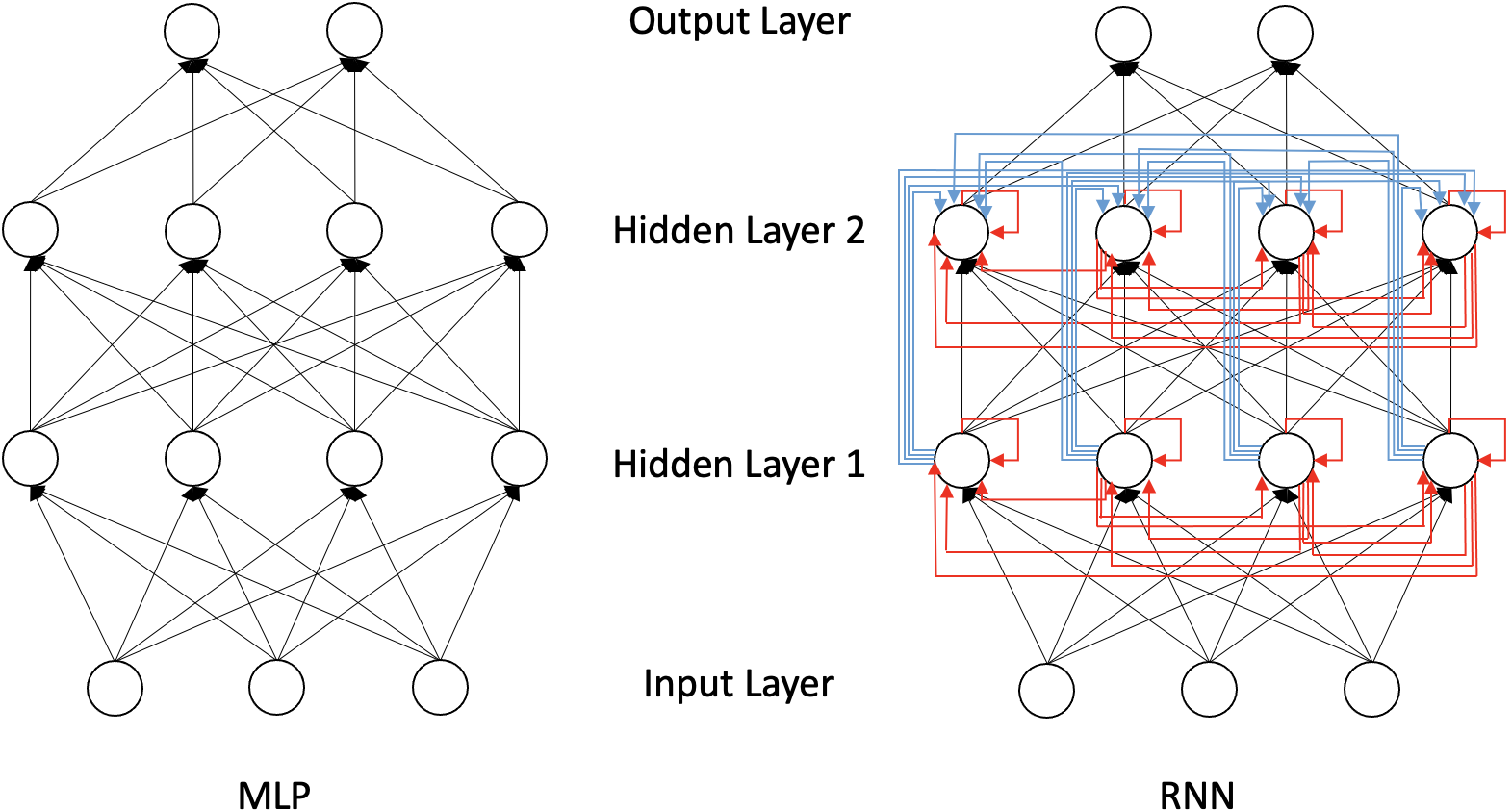}
	\caption[Difference between MLP and RNN architectures.]{Difference between MLP and RNN architectures are Red and Blue cyclical connections. Red are intra-layer connections, whereas Blue are across layers.}
	\label{fig:MLP_vs_RNN}
\end{figure}

Feed forward NNs presented in §~\ref{sec:theory_FWI_as_Learned_Direct_Approx} do not allow for cyclical/recurrent connections between neurons. If this condition is relaxed, we obtain RNNs. This difference is illustrated in Figure~\ref{fig:MLP_vs_RNN}. Although this might seem trivial, recurrent connections allow a ``memory'' of previous inputs to persist in the network’s state \citep{Graves2012}. Through equivalence results of Universal Approximation Theorem for MLP, \cite{Hammer2000} prove that a ``RNN with sufficient number of hidden units can approximate any measurable sequence-to-sequence mapping to arbitrary accuracy''.

\subsection{Forward Pass}
The forward pass of a RNN is identical to that of a MLP, except that the current input and the hidden layer activations from the previous time-step contribute to the current activation function. 

Consider a sequence $x$ of length $T$ and a RNN with input units $I$, hidden units $H$ and output units $K$. Let $x_i^t$ be the input $i$ at time-step $t$, $a_j^t$
the network output to unit $j$ at time-step $t$ and $b_j^t$ the activation of unit $j$ at time-step $t$. For the hidden units we have,
\begin{equation}
	a_h^t=b_h^0+\sum_{i=1}^{I} w_{ih} x_i^t +\sum_{h^{'}=1}^{H}w_{h^{'}h} b_{h^{'}}^{t-1},
\end{equation}
where $b_h^0$ is the initial bias term. \cite{Zimmermann2006} show how initial non-zero bias improve RNN stability and performance. This term is set to always zero to simplify the equation notation. Non-linear, differentiable activation functions $\sigma$ are applied similarly to MLP to get:
\begin{equation}\label{eq:RNN_activation_function}
	b_h^t=\sigma_h \left(a_h^t\right)=\sigma_h \left(\sum_{i=1}^{I}w_{ih} x_i^t +\sum_{h^{'}=1}^{H} w_{h^{'} h} b_{h^{'}}^{t-1} \right).
\end{equation}
Starting at $t = 1$, recursively applying Equation~\ref{eq:RNN_activation_function} will result in the complete sequence of hidden activations. The output $y_k^t$ at the $k^{th}$ neuron at time $t$ is thus given by:
\begin{equation}
	y_k^t=\sum_{h=1}^{H}w_{hk} b_h^t.
\end{equation}
\subsection{Backward Pass}
Given that components of the forward pass and the loss function definition are synonymous with those of MLPs, the remaining components are the derivatives with respect to the model weights. There are two widely used algorithms for this: (i) Real Time Recurrent Learning \citep{Robinson1987}, and (ii) Back-Propagation Through Time \citep{Werbos1988}. Back-Propagation Through Time will be considered going forward. Implementation of Real Time Recurrent Learning is given by \cite{Williams1989} or more recently by \cite{Haykin2009}.

Like standard back propagation, Back-Propagation Through Time consists of repeated application of the chain rule. The only difference is that for RNN, the loss function $\mathcal{J}$ is now dependent on the activation of the hidden layer at the next time-step. Equation~\ref{eq:mlp_error} can be thus re-written as:
\begin{equation}\label{eq:rnn_error}
	\delta_h^t = \frac{\partial\mathcal{L}}{\partial a_j^t}= \theta^{'}(a_h^t)\left(\sum_{k=1}^{K}\delta_k^{t}w_{mk} + \sum_{h^{'}=1}^{H}\delta_{h^{'}}^{t+1}w_{hh^{'}} \right).
\end{equation}
The complete sequence of partial derivate terms can be calculated by starting at time $t=T$ and recursively applying Equation~\ref{eq:rnn_error}. Summing over the whole sequence to get network weights results in:
\begin{equation}
	\frac{\partial\mathcal{L}}{\partial w_{ij}} = \sum_{t=1}^{T}\frac{\partial\mathcal{L}}{\partial a_j^t}\frac{\partial a_j^t}{\partial w_{ij}} = \sum_{t=1}^{T}\delta_j^tb_i^t.
\end{equation}
\subsection{Vanishing Gradient}
\begin{figure}[ht!]
	\centering
	\includegraphics[width=0.8\linewidth]{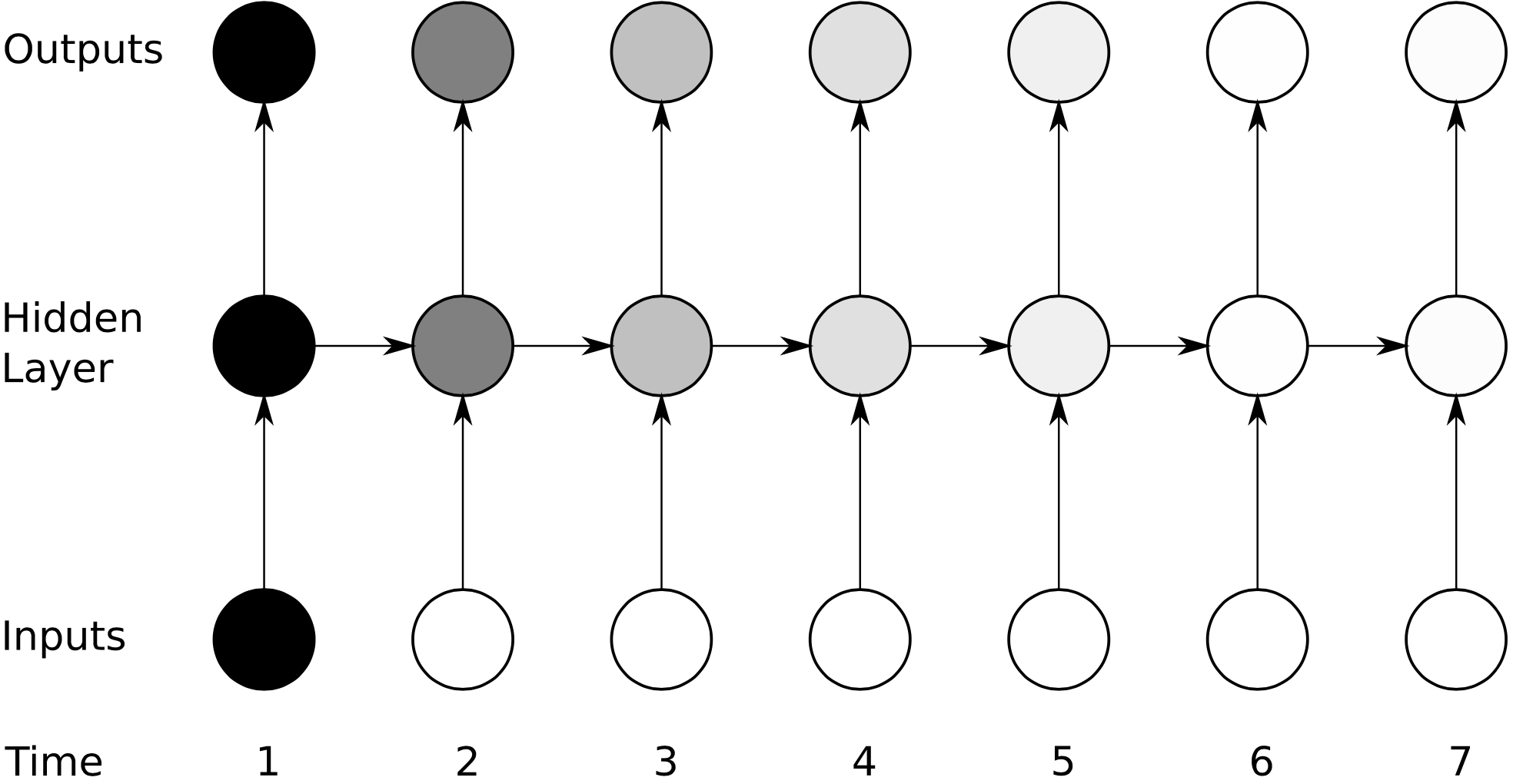}
	\caption[Vanishing gradient problem for RNNs.]{``Vanishing gradient problem for RNNs. The shading of the nodes indicates their sensitivity to the inputs at time one. The darker the shade, the greater the influence. The sensitivity decays over time as new inputs overwrite the activations of the hidden layer, and the network forgets the first inputs.'' From \cite{Graves2012}.}
	\label{fig:RNN_vanishing_gradient}
\end{figure}

Standard RNN architectures suffer from the vanishing gradient problem \citep{Hochreiter1991}. Namely, the sensitivity of deeper neurons either decays or blows up exponentially as it passes through the recurrent connections \citet{Graves2012}. This is schematically illustrated in Figure~\ref{fig:RNN_vanishing_gradient}. Attempts to solve for this included gradient-descent variants \citep{Pearlmutter1989, Williams1989, Elman1990, Fahlman1991, Schmidhuber1992a}, explicitly introduced time delays or time constants \citep{Lang1990,Plate1993,Lin1996,Mozer1992}, simulated annealing and discrete error propagation \citep{Bengio1994}, hierarchical sequence compression \citep{Schmidhuber1992} and Kalman filtering \citep{Puskorius1994}. A comprehensive list is available within \cite{Hochreiter1997}.

\subsection[Long Short-Term Memory]{Long Short-Term Memory (LSTM)}
\begin{figure}[!ht]
	\centering
	\subbottom[The repeating module in a standard RNN contains a single layer.\label{fig:LSTM_module_RNN}]{\includegraphics[width=0.7\textwidth]{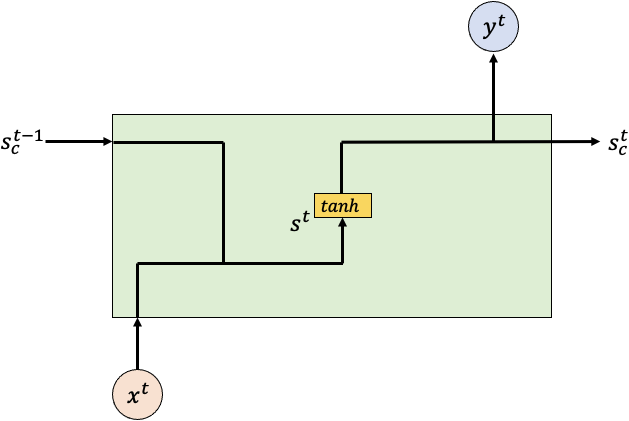}}
	\subbottom[A LSTM memory block has an additional three gates – Input, Output and Forget Gate (red) and a cell state block (blue).\label{fig:LSTM_module_LSTM}]{\includegraphics[width=0.7\textwidth]{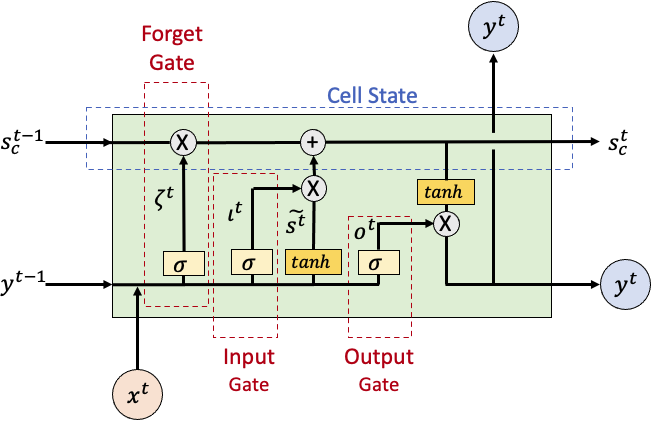}}
	\caption[Comparison between RNN and LSTM blocks.]{Comparison between RNN and LSTM blocks. Adapted from \cite{Olah2015}.}        
	\label{fig:RNN_LSTM_blocks}
\end{figure}

Apart from the subset of algorithms able to handle the vanishing gradient problem in the previous section, \cite{Hochreiter1997} introduce a modified architecture type known as \ac{LSTM}. This NN introduces a set of recurrent connections known as memory blocks (the input, output and forget gates) and a cell state. Figure~\ref{fig:LSTM_module_RNN} shows a standard RNN with a single tanh layer. Figure~\ref{fig:LSTM_module_LSTM} shows the LSTM chain structure but with the additional four interaction layers. Mathematical detail for each of these components is given in Appendix~\ref{sec:app_theory_Individual_RNN_Components}.

As before, $w_{ij}$ is the weight from unit $i$ to unit $j$, $a_j^t$ is the network input to unit $j$ at time step $t$ and $b_j^t$ is the activation of unit $j$ at time step $t$. Given the LSTM memory blocks, the subscripts $\iota$, $\zeta$ and $\omicron$ refer to the input, forget and output gate respectively, the subscript $c$ refers to one of the memory cells $C$ and $s_c^t$ is the state of cell $c$ at time $t$. 

\subsection{Network Training}
The theory reviewed for MLP optimizers in §~\ref{sec:theory_optimizing_the_loss_function} and regularization in §~\ref{sec:theory_generalization} remain unchanged for LSTMs.

\subsection{LSTM as a Substitute for Wave Propagation}
\begin{figure}[!ht]
	\centering
	\subbottom[Unrolled form of acyclic graph of LSTM for FWI.\label{fig:LSTM_unrolled}]{\includegraphics[width=0.7\textwidth]{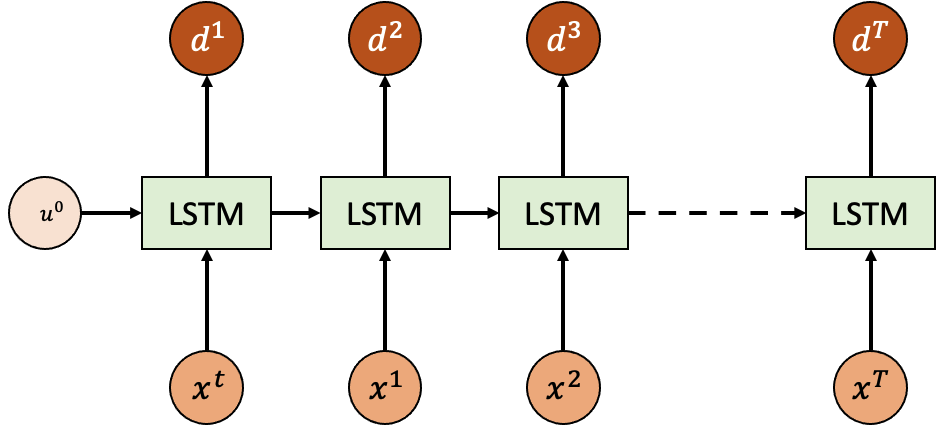}}
	\subbottom[Modified LSTM cell block supporting of forward modelling.\label{fig:LSTM_FWI_components}]{\includegraphics[width=0.9\textwidth]{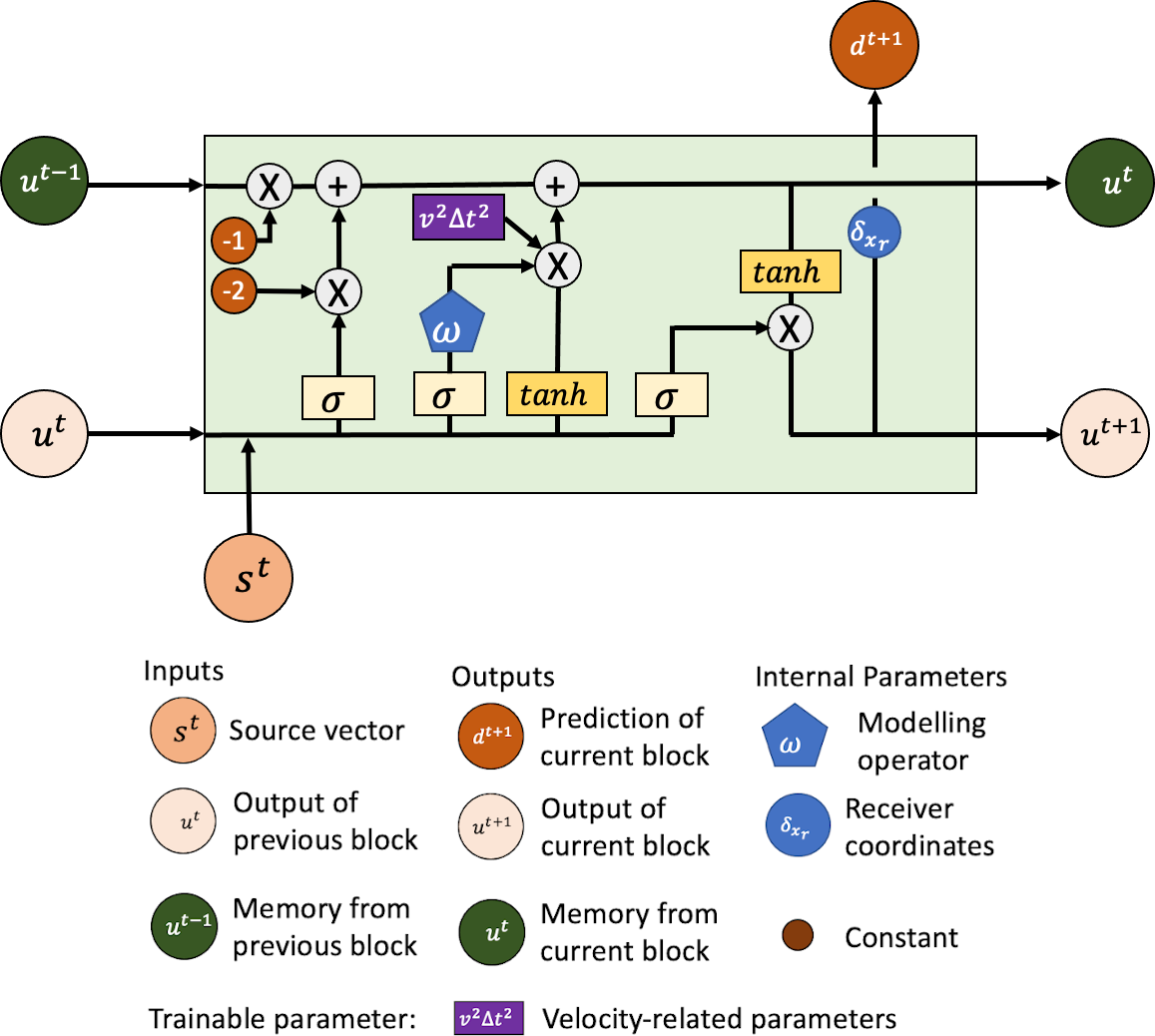}}
	\caption[Recasting of forward modelling within an LSTM]{Recasting of forward modelling of FWI within an LSTM deep learning framework. Adapted from \cite{Sun2019}.}        
	\label{fig:LSTM_FWI}
\end{figure}

Revisiting the discretized FD stencil for wave propagation given as,
\begin{equation}
	p_j^{n+1}=\partial t^2 \left[c_j^2 \mathcal{F}^{-1} \left[k^2 \mathcal{P}_{\nu}^n\right]+s_j^n\right]+2p_j^n-p_j^{n-1},
\end{equation}
it is clear how pressure wave $p$ and source impulse $s$ at current time step $n$ are not affected by the future values $n+1$, but only dependent on the previous state of pressure at $n-1$. This is, by definition, a finite impulse with directed acyclic graph under graph theory definitions \citep{Thulasiraman2011}. With slight modification to the LSTM blueprint in Figure~\ref{fig:LSTM_module_LSTM}, a Deep Learning architecture supporting forward modelling can be cast as a LSTM cell that considers the pressure wave at time $n-1$, produces the modelled shot record at current time $n$ and stores this in memory for the next step $n+1$. Measuring all outputs at each moment in time would equal to the measurements of the wavefield locally at a geophone. This LSTM architecture is shown in Figure~\ref{fig:LSTM_unrolled} as an unrolled graph and Figure~\ref{fig:LSTM_FWI_components} as the building block components within an LSTM. 

The inputs to the LSTM cell are the source term at current time $s^t$, the wavefield at current $u_t$ and previous time step $u_{t-1}$ stored in memory of the LSTM. These wavefields are combined together with untrainable modelling operator $\omega$ and constants $-1$ and $-2$ to replicate the incremental time stepping in forward modelling. Deciding to model in time is equivalent to setting $\omega$ to the Laplacian, whereas setting it to calculate pseudo-spectral second order derivates will lead to pseudo-spectral wavefield modelling. The trainable velocity-related parameter $v^2 \Delta t^2$ is applied to get the current modelled wavefield $u^{t+1}$. This is stored in memory, passed to the forget gate and receiver location discretization $\delta_{x_{r}}$ applied to get the predicted outputs $d^{t+1}$. To train the velocity parameters, seismic shot records are provided as labelled data for training.
\clearpage

\subsection{Outline for Solving FWI with a RNN}
\begin{figure}[ht!]
	\centering
	\includegraphics[width=0.9\linewidth]{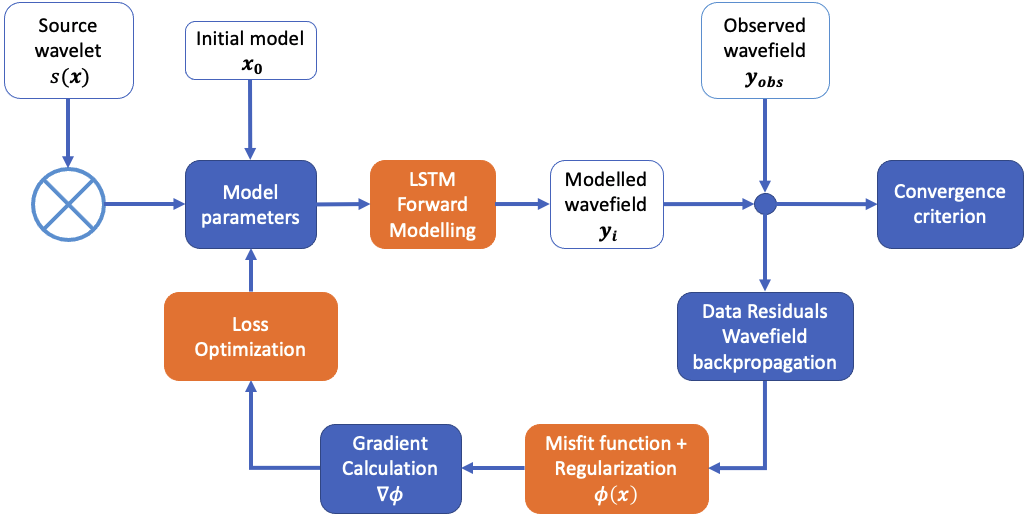}
	\caption[Schematic of RNN workflow]{Schematic of RNN workflow. The orange components are modified from the original FWI workflow in Figure~\ref{fig:workflow_FWI}.}
	\label{fig:workflow_RNN}
\end{figure}

Physics-informed RNNs for FWI is identical to classical FWI, apart from the forward modelling component which is done within an LSTM framework. This is summarized by Algorithm~\ref{algo:FWI_RNN} and schematic shown in Figure~\ref{fig:workflow_RNN}.
\begin{algorithm}
Choose an initial model $\boldsymbol{x}_0$ and source wavelet $s(\boldsymbol{x})$.\\
For each source, solve the forward problem through LSTM time-stepping to get a predicted wavefield $\boldsymbol{y}_i$. This is sampled at receivers $r(\boldsymbol{x})$.\\
At every receiver, data residuals are calculated between the modelled wavefield $\boldsymbol{y}_i$ and the observed data $\boldsymbol{y}_{\boldsymbol{obs}}$.\\
Data residuals are back-propagated to produce a residual wavefield.\\
For each source location, the misfit function $\phi(\boldsymbol{x})$ is applied for the observed data, regularized and back-propagated through residual wavefield to generate the gradient $\nabla_\phi$ required at every point in the model.\\
The gradient is scaled based on loss optimization function, applied to the starting model and an updated model is obtained $\boldsymbol{x}_{i+1}$.\\
The process is iteratively repeated from Step 2 until a convergence criterion is satisfied.
	\caption{FWI as RNN Implementation}
	\label{algo:FWI_RNN}
\end{algorithm}

%% file: chap4/results_main.tex
\chapter{Numerical Results}
\textbf{In the previous section, two formulations of FWI within Deep Learning frameworks were presented. This chapter presents numerical results and additional outcomes of these implementations.}  

\section{Software Setup}\label{sec:results_results_software_setup}
Throughout the work shown in this chapter, Python 3.7 with the Anaconda distribution was used as the primary development language.

\section[FWI as a Data-Driven DNN]{FWI as a Data-Driven DNN}\label{sec:results_FWI_as_a_Learned_Direct_Approximation}

\subsection{Train-Test Data Split}\label{sec:results_Train-Test_Data}
Learning the inversion from time to pseudo-spectral domain requires a training dataset which maps time to pseudo-spectral components and their respective velocity profile. A data generator was designed to create synthetic data on-the-fly for a 2000\si{ms} time window. The steps involved in the data generator are:
\begin{enumerate}[i)]
    \item Randomly create velocity profile $v_p$ for a 2000ms distance, with value ranging from 1450\si{ms^{-1}} and 4500\si{ms^{-1}}. The lower bound of 1400\si{ms^{-1}} was selected for the water column since the observed velocity in normal off-shore seismic exploration conditions ranges from 1440\si{ms^{-1}} to 1540\si{ms^{-1}} \citep{Cochrane1991}. The upper bound of 4000\si{ms^{-1}} was selected as this is the upper limit of velocity in porous and saturated sandstones \citep{Lee1996}. The assumption is made that limestones, carbonates and salt deposits are not present in the subsurface model being inverted since these would have velocity in excess of 4000\si{ms^{-1}} and go beyond the above defined parameters.
    \item Estimate the density $\rho$ using Gardner’s equation \citep{Gardner1974} $\rho=\alpha v_p^{\beta}$ where $\alpha=0.31$ and $\beta=0.25$ are empirically derived constants.
    \item At each interface, calculate the Reflection Coefficient $\mathcal{R}=\frac{\rho_2v_{p_{2}} - \rho_1v_{p_{1}}}{\rho_2v_{p_{2}} + \rho_1v_{p_{1}}}$ where $\rho_i$ is density of medium $i$ and $v_p$ is the p-wave velocity of medium $i$.
	\item For each medium, calculate the Acoustic Impedance $\mathcal{Z}=\rho v_p$.
    \item Define a wavelet $\mathcal{W}$. This was selected to be a Ricker wavelet at 10\si{Hz} \citep{Ryan1994}. The Ricker wavelet is a theoretical waveform that takes into account the effect of Newtonian viscosity and is representative of seismic waves propagating through visco-elastic homogeneous media \citep{wang2015frequencies}, thus making it ideal for this numerical simulation. The central frequency of 10\si{Hz} was chosen as a nominal value based on literature results to be representative of normal FWI conditions \citep{Morgan2013}. Beyond 10\si{Hz} would be considered to be super-high-resolution FWI \citep{mispel2019high}, which goes beyond the scope of this work.
	\item The reflection coefficient time series and wavelet are convolved to produce the seismic trace $\mathcal{T}$.
	\item Fourier coefficients for magnitude $\mathcal{M}(\zeta)$ and phase $\mathcal{M}(\phi)$ are derived based on the \ac{FFT}.
\end{enumerate} 

\begin{figure}[ht!]
	\centering
	\includegraphics[width=0.9\textwidth]{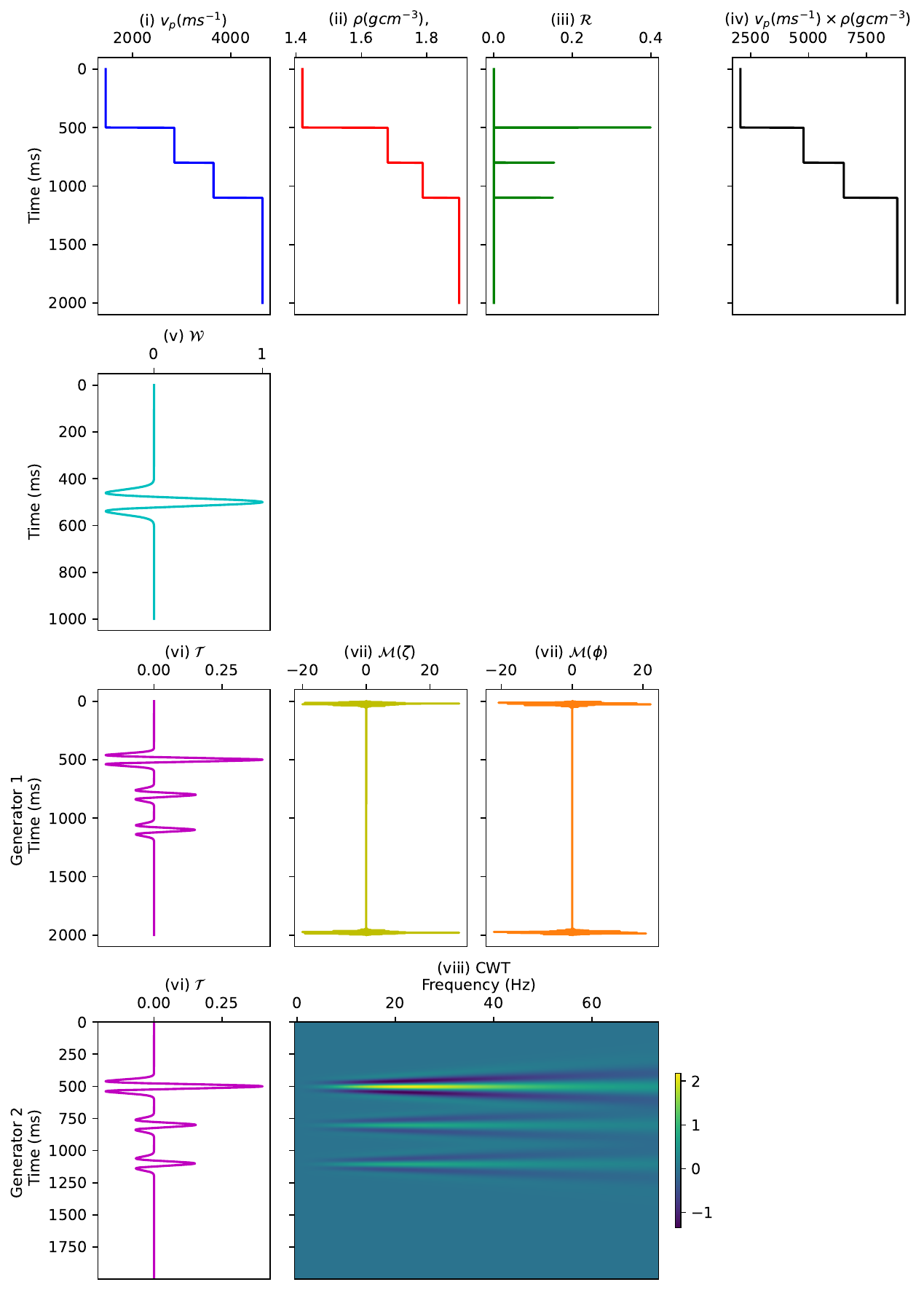}
	\caption[Workflow for creating a pseudo-spectral synthetic trace.]{Workflow for creating a pseudo-spectral synthetic trace.}
	\label{fig:data_generators}
\end{figure}

To exploit higher dimensionality and use 2D CNNs, a secondary generator was designed to perform a \ac{CWT}. This was identical to the previous generator, expect that in Step (vii), produce a CWT with sampling frequencies from 1-75\si{Hz} and wavelet identical to the wavelet given in Step (v). This is referred to as Step (viii). The different steps for these two generator flows are shown in Figure~\ref{fig:data_generators} for a sample velocity profile. These generators will be referred to as Generator 1 and Generator 2 respectively.

\begin{figure}[ht!]
    \centering
    \includegraphics[width=0.95\textwidth]{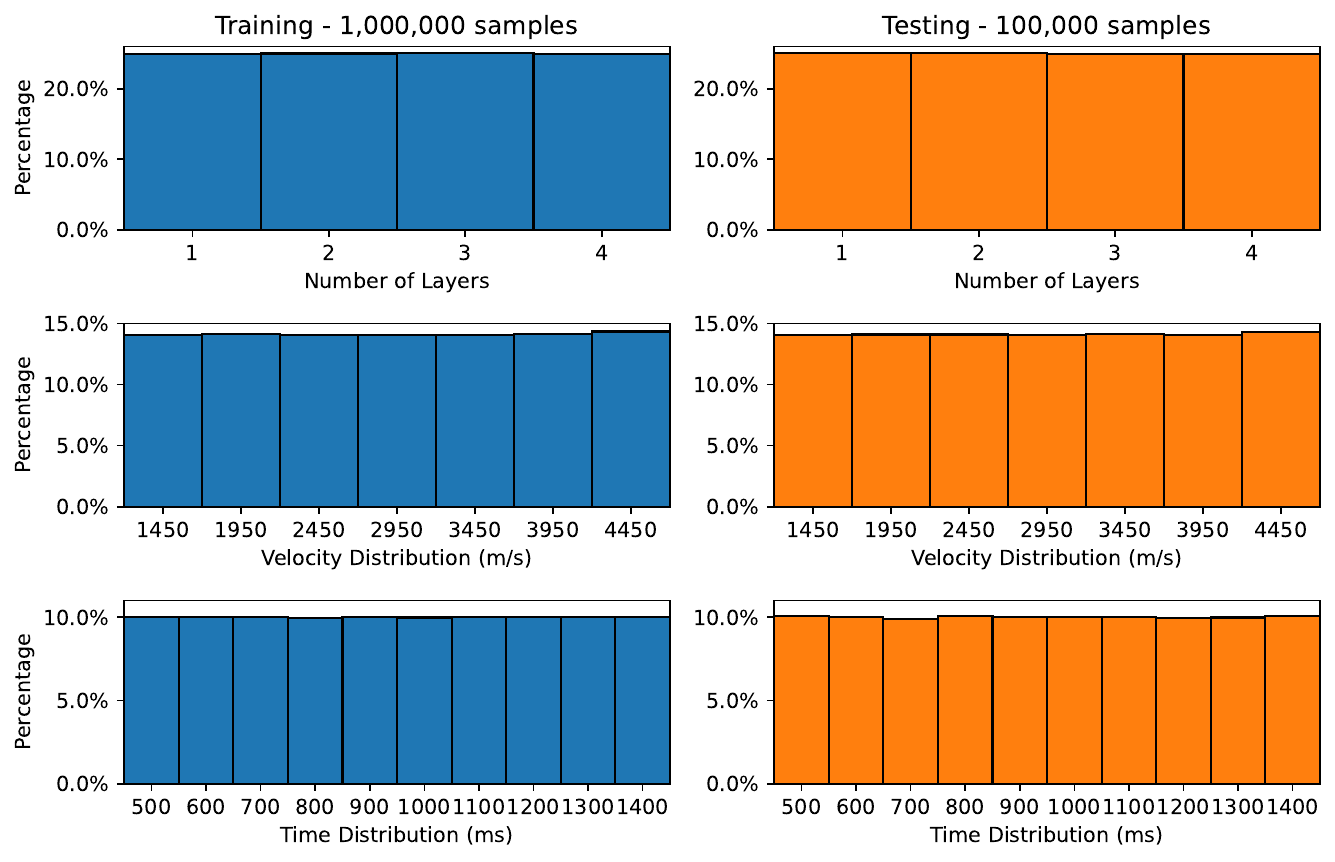}
    \caption[The overall statistics on the training and testing dataset.]{The overall statistics on the training and testing dataset.}
    \label{fig:data_generators_params}
\end{figure}

\clearpage
The parameters assigned with the generators are given in Table~\ref{tab:data_generator_params} and distribution of layers within a training run for 1,000,000 samples and 100,000 testing samples are given in Figure~\ref{fig:data_generators_params}. The \textbf{vel\_min\_separation}, \textbf{time\_min\_separation} are two key parameters as they control the velocity and temporal resolution of the model respectively.

\begin{table*}[ht!]
    \centering
    \begin{tabular}{@{}llc@{}}\toprule
        Parameter                  & Description                 & Value\\ \hline
        length (ms)                & Length of trace             & 2000  \\
        vel\_min (m/s)             & Minimum velocity            & 1450  \\
        vel\_max (m/s)             & Maximum velocity            & 5000  \\
        vel\_min\_separation (m/s) & Minimum velocity separation & 10    \\
        time\_min (ms)             & Minimum time sample         & 500   \\
        time\_max (ms)             & Maximum time sample         & 1500  \\
        time\_min\_separation (ms) & Minimum time separation     & 2     \\
        layers\_min                & Minimum number of layers    & 1     \\
        layers\_max                & Maximum number of layers    & 4     \\
        dominant\_frequency        &\si{Hz} of dominant frequency & 10     \\ \hline
    \end{tabular}
    \caption{Synthetic data generator parameters.}\label{tab:data_generator_params}
\end{table*}

\subsection{DNN Framework}\label{sec:results_DNN_Framework}
\begin{figure}[!ht]
	\centering
	\subbottom[Fourier components.\label{fig:DNN_framework_1}]{\includegraphics[width=0.47\textwidth]{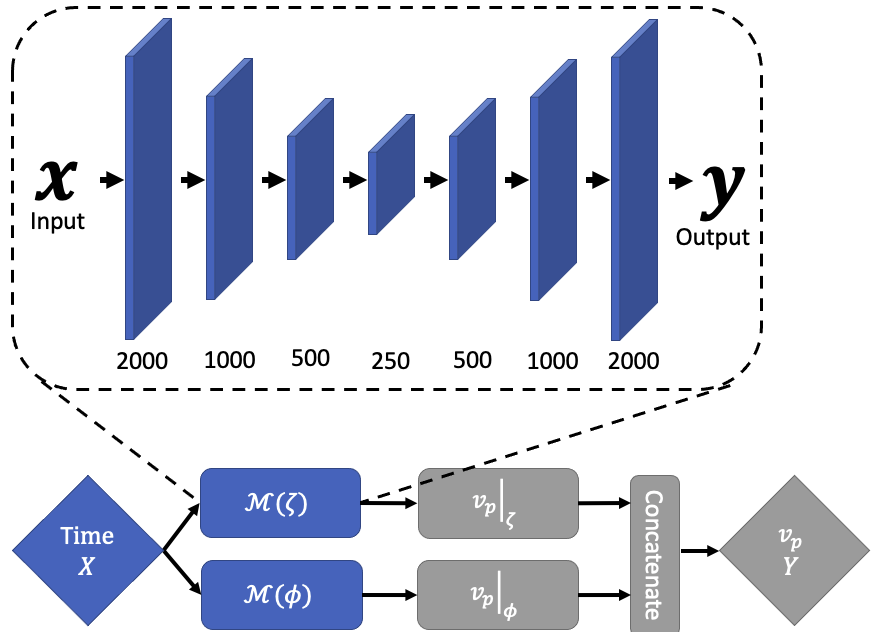}}
	\subbottom[CWT.\label{fig:DNN_framework_2}]{\includegraphics[width=0.47\textwidth]{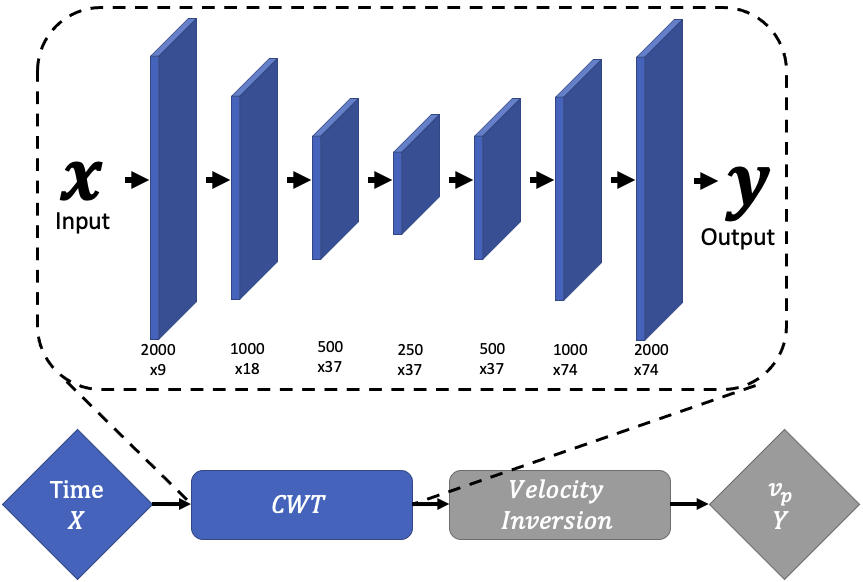}}
	\caption[Pseudo-spectral FWI DNN workflow.]{Pseudo-spectral FWI DNN frameworks to invert for Fourier Transform and CWT. $X$ is the input time domain, $Y$ is the output $v_p$ velocity and $\mathcal{M}$ is the Fourier domain, with magnitude $\zeta$ and phase $\phi$. Each component (blue or grey box) is a network.}         
	\label{fig:DNN_frameworks}
\end{figure}

Figure~\ref{fig:DNN_framework_1} illustrates the DNN framework used to first invert for the Fourier coefficients from the time domain and then invert for velocity profile. The complete workflow has five modules, with each module consisting of a NN with 5 fully connected hidden layers. The layer distributions consisted of an input layer of 2000 neurons, a set of 5 hidden layers of sizes 1000, 500, 250, 500, 1000 neurons, and an output layer of 2000 neurons. This hour-glass design can be considered representative of multi-scale FWI \citep{Bunks1995} since at each hidden layer, the NN learns an abstracted frequency component of the data at a different scale. This is Indeed synonymous with modern DNN approaches such as encoder-decoders and U-Net \citep{ronneberger2015u} and how they extract data representations \citep{Yu2019,Berthelot2018}. The final concatenate network learns the optimal way for combining the outputs. In total, the DNN had 25 hidden layers. In the case of the CWT pseudo-representation, we designed a similar framework to that shown in Figure~\ref{fig:DNN_framework_1}, except that the learned CWT network has an additional dimension to be able to create the CWT. This is shown in Figure~\ref{fig:DNN_framework_2}. The learned CWT network has layers of shape $(2000\times9), (1000\times18), (500\times37), (250\times37), (500\times37), (1000\times74), (2000\times74)$. The velocity inversion DNNs were built to be representatives of Conv1D, Conv2D, VGG, ResNet architectures respectively. A full architectural summary and training process for these network is provided in Appendix~\ref{sec:app_results_architectural_summary_dnn_workflow}.

\clearpage
\subsection{Multi-layer Numerical Results}\label{sec:results_Multi-layer_1D_Numerical_Results}
The first experiment was to assert the validity of the framework in a 1D synthetic case. Using Generator 1 with 1,000,000 training and 100,000 testing samples, with DNN framework A for the Fourier components, loss function was set to be the Sum of Squared Error, stabilized via fixed $L_2-norm$ regularization, data batching, early stopping, and executed for 120 epochs (the number times that the algorithm passes through the entire training dataset). Gradient descent update was optimized via an ADAM optimizer. The DNN was implemented using Keras 2.2.4 and TensorFlow 1.13.1 backend. This was trained on an Intel i7-7800x X-series CPU workstation provided by the Department of Physics at the University of Malta.

\begin{figure*}[ht!]
	\centering
	\includegraphics[width=0.9\textwidth]{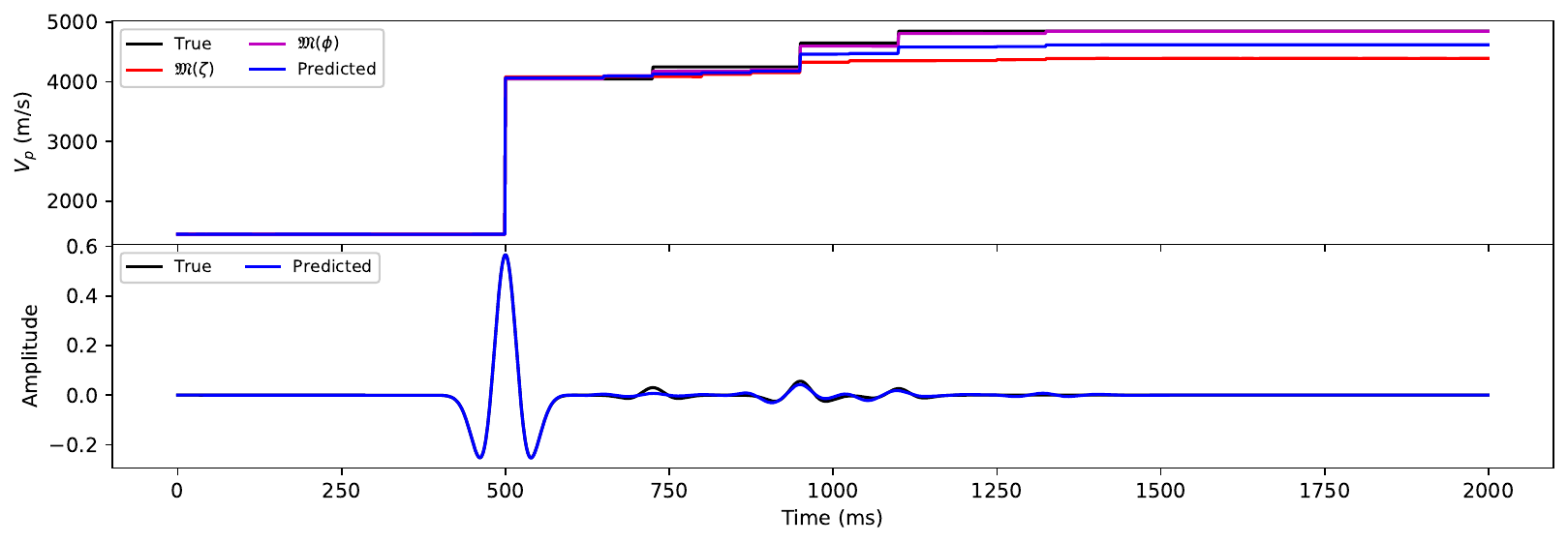}
	\includegraphics[width=0.9\textwidth]{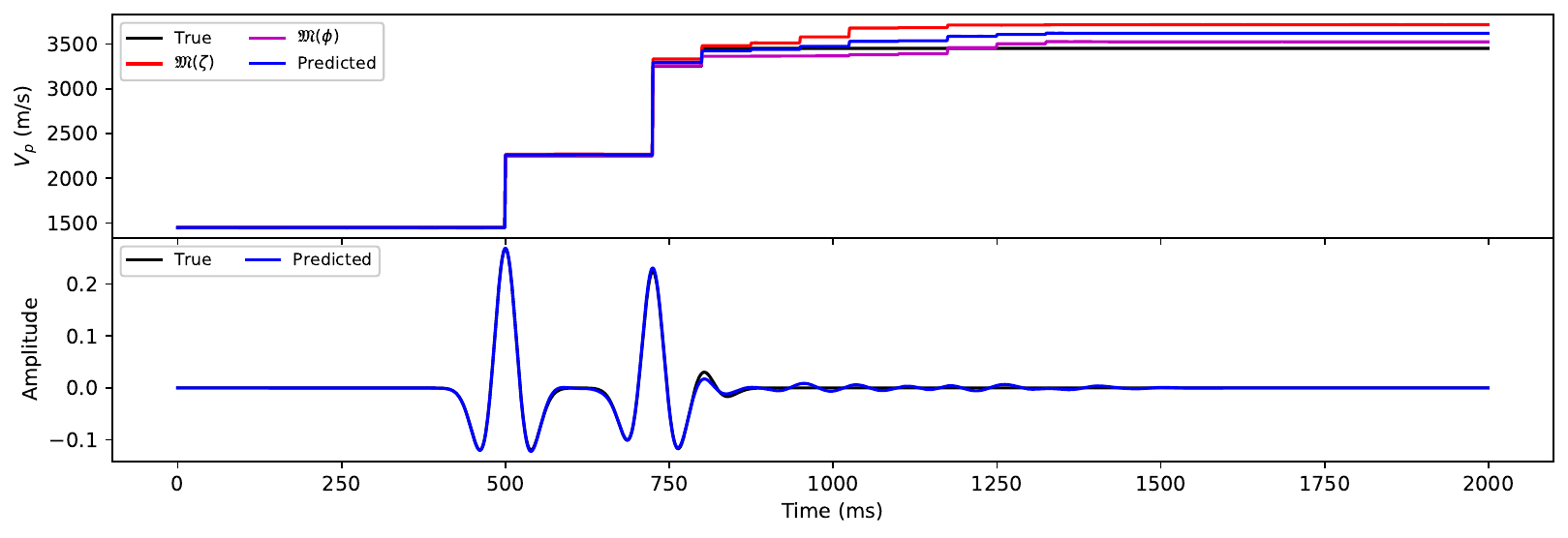}
	\includegraphics[width=0.9\textwidth]{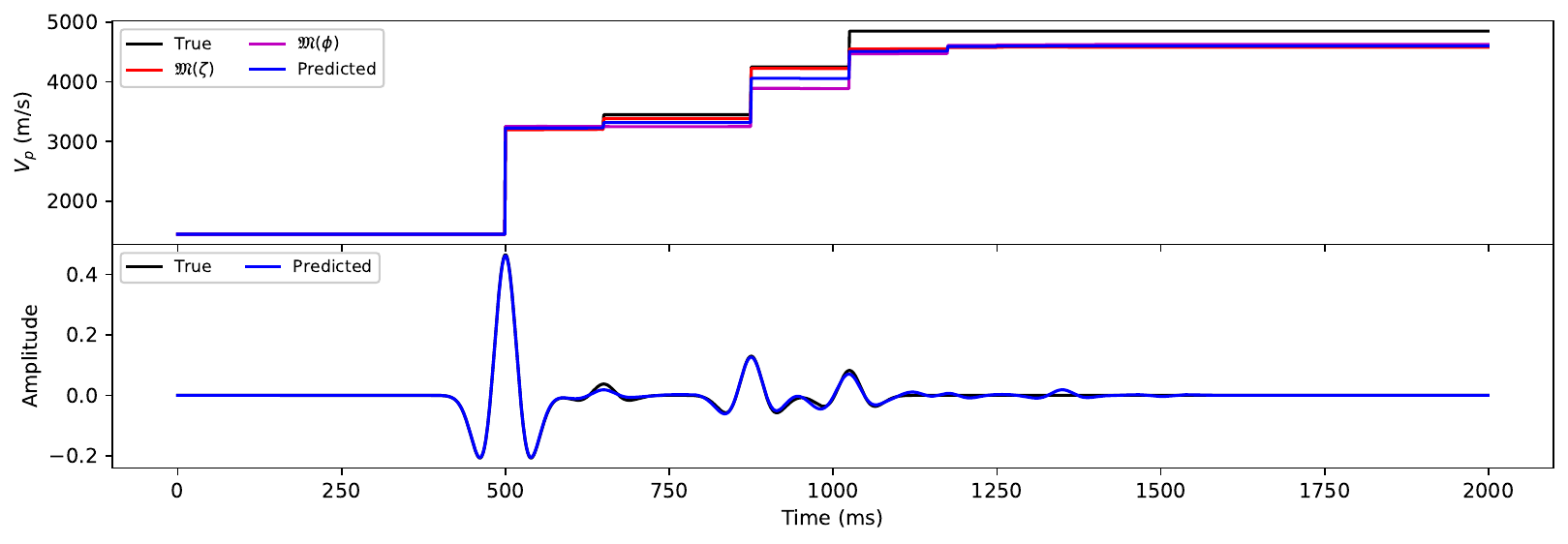}
	\includegraphics[width=0.9\textwidth]{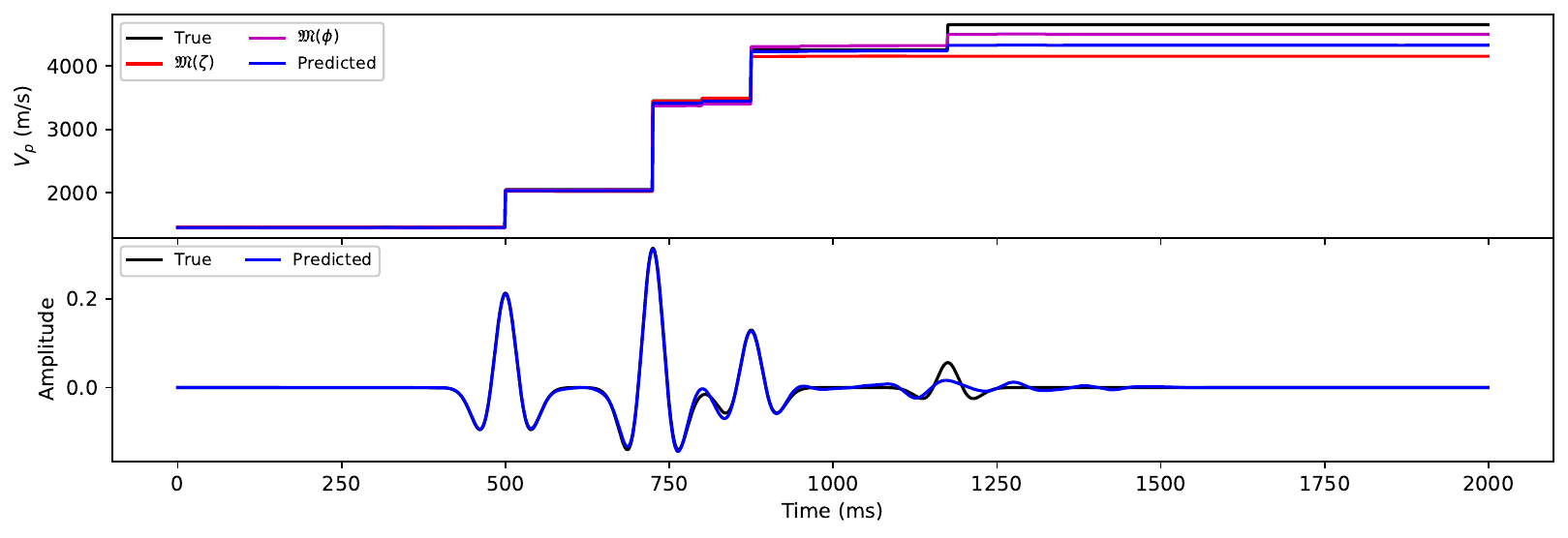}
	\caption[DNN predictions]{Four different predictions obtained from learned weights of the DNN on unseen data. The top panels are the velocity profile reconstructions from the two NN architecture branches ($\mathcal{M}(\zeta)$ and $\mathcal{M}(\phi)$) and the combined result. Bottom panels are the observed and inverted waveforms.}
	\label{fig:dnn_fwi_1d_predictions}
\end{figure*}

Figure~\ref{fig:dnn_fwi_1d_predictions} illustrates the application of DNN architecture for a sample of unseen data and the respective inversion. Inspection of the first 750\si{ms} indicates that the DNN approach is able to reconstruct both the velocity and the waveform profile near perfectly, irrespective of the number of layers and the magnitude of the acoustic difference in this time range. Indeed, these indicate the validity of this approach. Beyond 750\si{ms}, reconstructions start suffering from slight degradation. As illustrated in the velocity reconstruction of the middle figure, the inaccuracy is minimal and ranges $\pm$100\si{ms^{-1}}. This leads to perturbations in the reconstruction and does not allow for perfect matching. Further inspection suggests that the main source of error is due to the magnitude component of the network (red). 

\begin{figure*}[ht!]
    \centering
    \includegraphics[width=0.9\textwidth,  trim = {0.4cm 11.9cm 0.4cm 0.1cm}, clip]{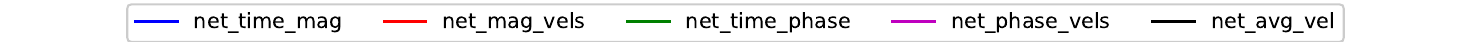}
    \hfill
    \subbottom[Training dataset MSE over the different epochs per DNN component. Overall performance is decreasing per epoch, indicating that the DNN is learning to invert.\label{fig:dnn_perf_metric_train}]
    {\includegraphics[width=0.43\textwidth,  trim = {0.3cm 0.2cm 20cm 0cm}, clip]{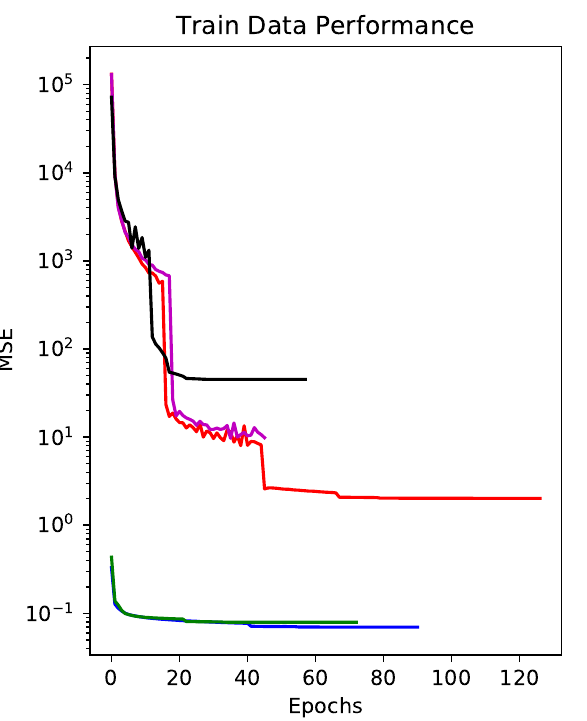}}
    \hspace{10pt}
    \subbottom[Test dataset MSE over the different epochs per DNN component. Overall MSE is decreasing per epoch and there are not signs of over-fitting.\label{fig:dnn_perf_metric_test}]
    {\includegraphics[width=0.43\textwidth,  trim = {10.2cm 0.2cm 10.1cm 0cm}, clip]{2_multilayer_7_xjenza_DNN_perf_train_test-eps-converted-to.pdf}}
    
    \subbottom[Learning Rate performance over the different epochs per DNN component.\label{fig:dnn_perf_lr}]
    {\includegraphics[width=0.43\textwidth,  trim = {20cm 0.2cm 0.15cm 0cm}, clip]{2_multilayer_7_xjenza_DNN_perf_train_test-eps-converted-to.pdf}}
    \caption{DNN training performance metrics.}
    \label{fig:dnn_fwi_1d_performance}
\end{figure*}

Figure~\ref{fig:dnn_fwi_1d_performance} shows the DNN mean squared error performance over the different epochs per DNN component. This graph indicates that the network is learning since mean squared error is overall decreasing at each epoch. The drastic decreases in the mean squared error at different epoch levels can be attributed to the step-wise reductions in learning rate shown in Figure~\ref{fig:dnn_perf_lr}. This varying learning rate allows the network to move to a deeper optimization level and approach a more global minima for the optimization problem. Interestingly, this performance plot indicates that the technique might suffer from a compounding error issue. The two best performing components are the first layer of learning for the inversion, namely Time-to-FFT-Magnitude and Time-to-FFT-Phase, as their mean squared error performance plateaus are at $10^{-1}$. In the second phase of the inversion from the respective FFT components to velocities (FFT-Magnitude-to-Velocity and FFT-Phase-to-Velocity) error plateaus are at $10^1$, which is two orders of magnitude greater. The final network component sits even higher on the scale at $10^2$.

\clearpage

\subsection{Pre-Processing}\label{sec:results_Normalisation}
In classical DNN approaches, it is best practice to normalise or standardize the dataset. An experiment was executed to assess what would happen with and without normalisation of the data for the Magnitude component architecture shown in the previous section. 10,000 training and 1,000 validation traces were generated using Generator 1, stored in memory so the only variance will be the scaling and trained for 200 epochs with early stopping and reducing learning rate monitor. The compute and memory resources necessary for this test were small enough such that this experiment was executed on a 2.6 GHz 6-Core Intel Core i7, 16GB RAM personal computer. The scaling approaches considered are the Standard Scaler and Min-Max Scaler. These are defined as:
\begin{equation}
    x_{MM} = \frac{x-x_{MIN}}{x_{MAX}-x_{MIN}}, 
\end{equation}
\begin{equation}
    x_{SS} = \frac{x-\mu}{\sigma},
\end{equation}
where $x_{MM}$, $x_{SS}$ are the scaled values for Min-Max and Standard Scaler respectively, $x_{MIN}$, $x_{MAX}$ are the minimum and maximum values of the data, $\mu$ is the mean and $\sigma$ is the standard deviation of the training samples.

\begin{figure}[ht!]
    \centering
    \includegraphics[width=0.98\textwidth]{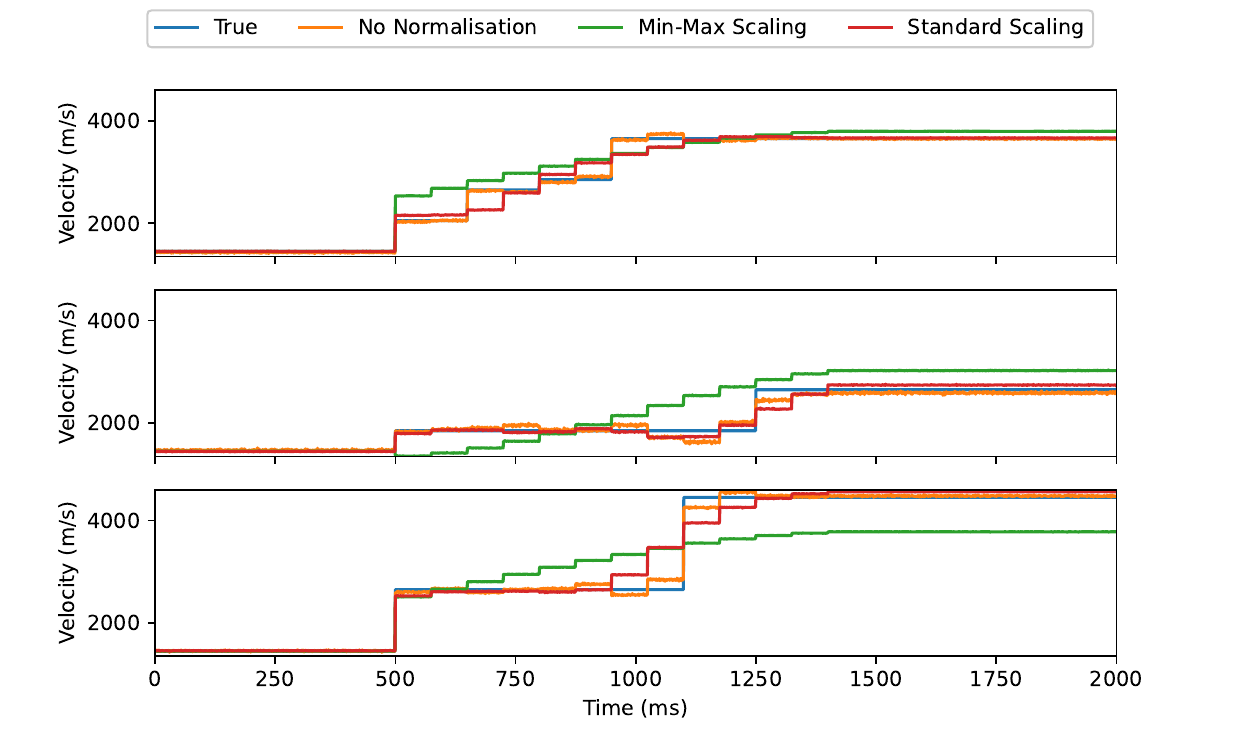}
    \caption[Comparison of velocity profiles with and without scaling.]{Comparison of velocity profiles with and without scaling.}
    \label{fig:scaling_velocity}
\end{figure}

The mean square error for the dataset with-out and with processing was evaluated and is shown in Table~\ref{tab:normalisation_mse}. The value of the mean squared error indicates that pre-processing in the form of normalisation or standardization should not be applied to the problem dataset. The impact of the pre-processing on the inverted velocity profiles is shown in Figure~\ref{fig:scaling_velocity}. In either case, Min-Max scaling was the worst performant, only able to reconstruct the first layer at 500\si{ms}.

\begin{table*}[ht!]
    \footnotesize
    \centering
    \begin{tabular}{@{}cc@{}}\toprule
Pre-Processing & Mean Square Error \\ \hline
No Normalisation          & 4,041 \\
Min Max Normalisation     & 296,672 \\
Standard Scaling          & 17,653 \\\hline
\end{tabular}
\caption{Quantitative assessment on the impact of pre-precessing}\label{tab:normalisation_mse}
\end{table*}

\clearpage

\subsection{Architecture Comparison}\label{sec:results_Architecture_Comparison}
In §~\ref{sec:results_Multi-layer_1D_Numerical_Results}, the validity of the approach was assessed and pitfalls identified for a simple example. This was extended to identify the ideal configuration in terms of architecture, loss, time-to-train and over-fitting. The following computation was made using Python 3.7 and Tensorflow 2.0.0 with a Keras backend. It was executed on an NVIDIA Titan V Graphical Processing Unit with 5120 cores and 12GB ram provided in collaboration with Dr. Carlo Giunchi at Istituto Nazionale di Geofisica e Vulcanologia at Pisa.

\begin{figure}[ht!]
    \centering
    \includegraphics[width=0.98\textwidth]{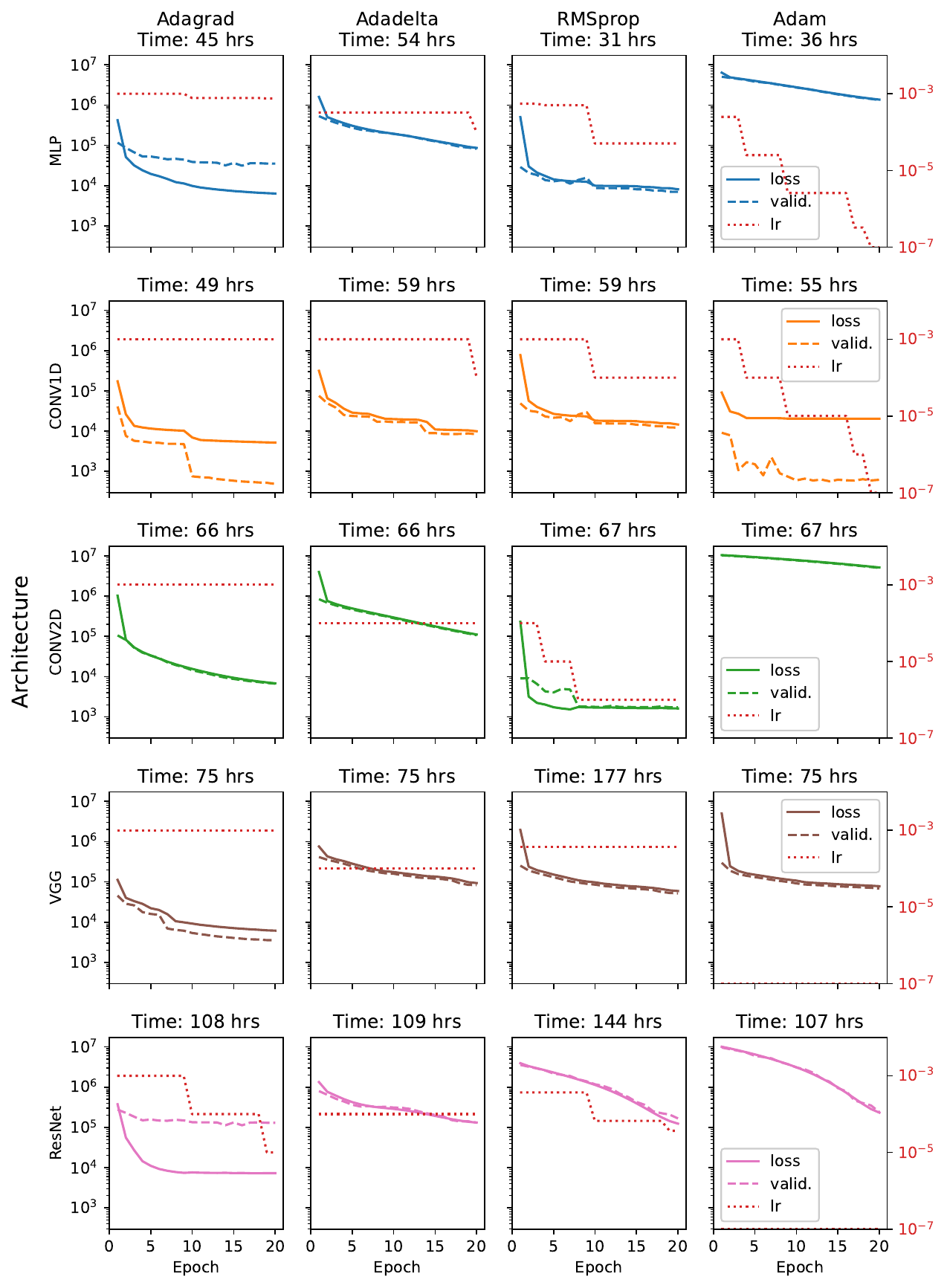}
    \caption[Normalised comparison of DNN architectures]{Normalised comparison of DNN architectures, loss optimizer, duration of training and validation curves. The networks were trained for the same number of epochs without early stopping. The loss is the MSE and lr is the Learning Rate.}
    \label{fig:dnn_fwi_1_comparison_dnn_arch}
\end{figure}

The conversion from time trace to pseudo-spectral representation was fixed for all 1D and 2D networks such that the comparison was done on only the different inversions architectures. This architecture given as ``Time to Pseudo-Spectral 1D'' or ``Time to Pseudo-Spectral 2D'' can be found in Appendix~\ref{sec:app_results_architectural_summary_dnn_workflow}. Figure~\ref{fig:dnn_fwi_1_comparison_dnn_arch} gives a comparison of different DNN architectures, loss optimizers, duration of training and validation curves. The networks were trained for the same number of epochs without early stopping. The training and validation data consisted of 1,000,000 and 100,000 generated traces using Generator 1 and Generator 2 respectively. The loss was fixed to be the MSE and lr represents the learning rate on a secondary axis in Red. 

Considering all loss curves, the best performing setup is that for Conv2D and RMSprop with 20,000 loss and worst performant are MLP-Adam and Conv2D-Adam. The deeper more complex Conv2D, VGG and ResNet architectures in general seem to be experiencing some under-fitting due to the parallel, non-convergent training-validation curves. This would be indicative that the complexity in the current problem is not high and less complex network type such as Conv1D would be more suitable. As evidenced by the gradual monotonic decrease in validation curves, none of the architecture-loss optimizer combinations experienced over-fitting. Conv1D-Adam and MLP, Conv1D and Conv2D for RMSprop seem to indicate increases between epochs 2 and 10 on the validation curves. This would be symptomatic to over-fitting; however, the learning rates move to lower orders of magnitude. This enables networks to continue training and move to lower orders of loss value. The lr for some of the combinations is remaining unvaried, namely Conv1D-Adagrad, Conv2D-Adagrad and Conv2D-Adadelta, all of VGG and ResNet-Adadelta and ResNet-Adam. This is indicative that these networks are not close to reaching a global minimum and would benefit from training for more epochs. Indeed, this would indicate that more complex architectures such as VGG and ResNet require longer training epochs. Indeed, the loss-validation curves further highlight this as, in general, they do not plateau. The more complex and deeper the architectures required longer training times. MLP averaged training time of 41.5 compute hours, whereas ResNet averaged 117 compute hours. These performance metrics should not be considered in isolation and visual inspection of inversion should be equally assessed.

\begin{figure}[!ht]
	\centering
    \subbottom[Trace A\label{fig:dnn_fwi_1_vel_results_A}]{\includegraphics[width=0.9\textwidth]{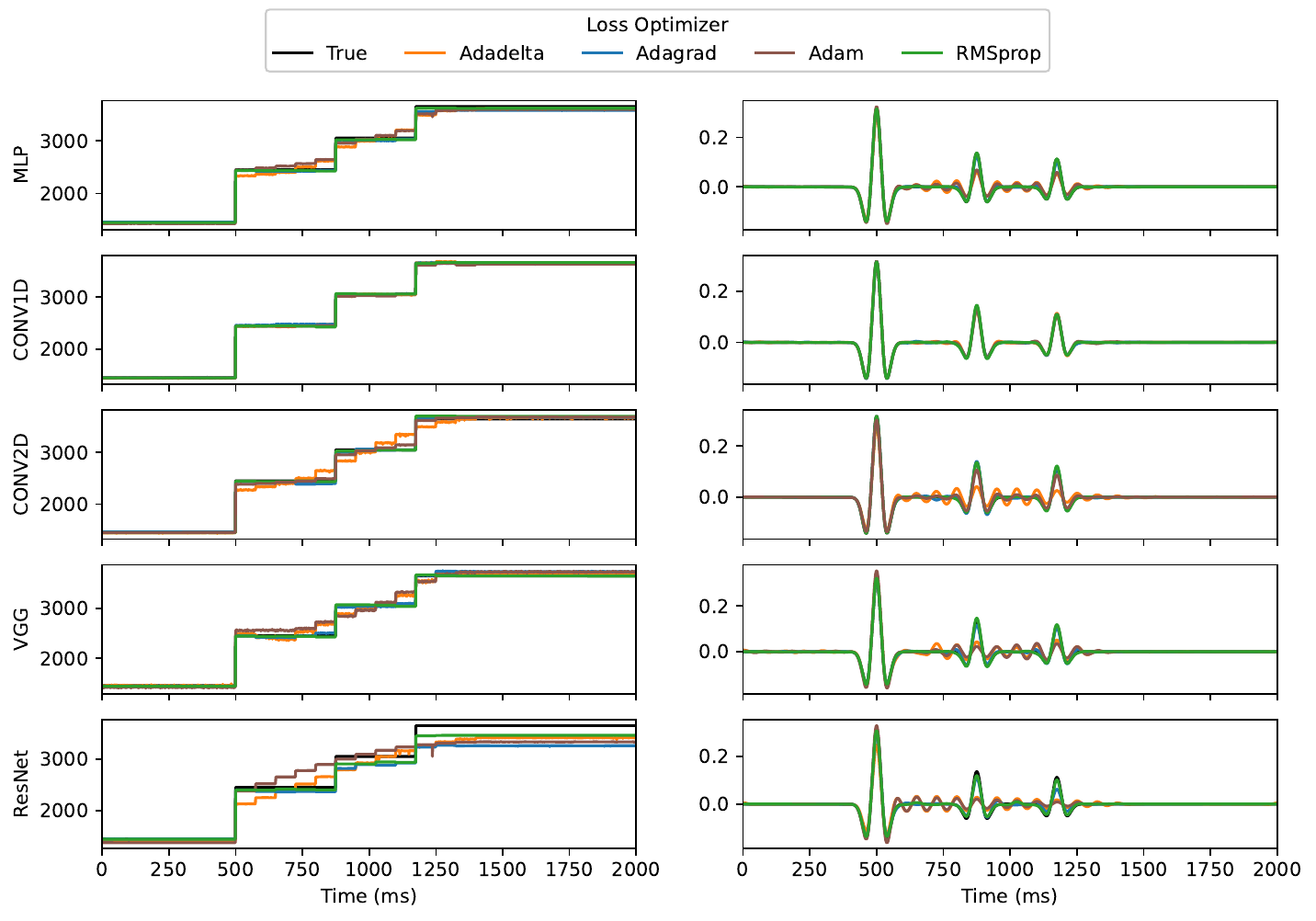}}
	\subbottom[Trace B\label{fig:dnn_fwi_1_vel_results_B}]{\includegraphics[width=0.9\textwidth]{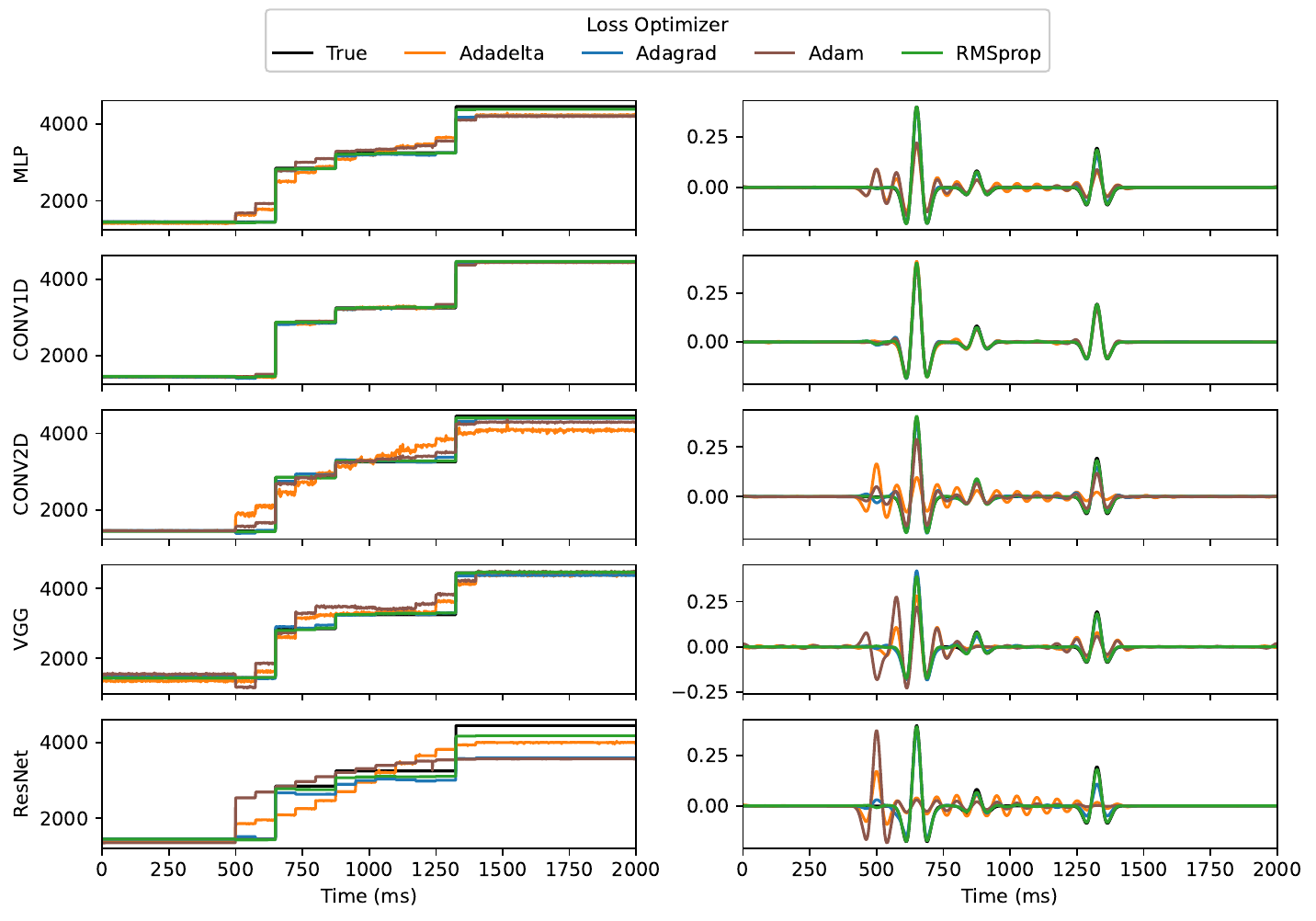}}
    \caption[Velocity inversion for different DNN architectures and losses.]{Velocity inversion for different DNN architectures and losses.}
    \label{fig:dnn_fwi_1_vel_results}
\end{figure}

Figure~\ref{fig:dnn_fwi_1_vel_results} show two sample traces inverted for all these architecture and loss optimizer combinations. Upon initial qualitative inspection, it is clearly evident how Conv1D is the superior architecture. This further confirms the previous assertion that for this given experiment, deeper and more complex architecture types such as Conv2D, VGG and ResNet are not necessary and can be detrimental to overall performance. Common to all architectures is a ringing effect on the time trace inversion. This is due to inadequate inversion for the velocity profile where the initial velocity increase is identified correctly, but beyond this, there is a step-wise incremental velocity profile. This is present in simple MLP and more complex DNNs. Conv1D is the only architecture type which is symptom free. 

\clearpage
\subsection{Architecture-Loss Combination}\label{sec:results_arch_loss_combination}
All results are summarised in Table~\ref{tab:quantitative_assessment_arch_loss} to quantitatively assess the inversion process and the DNN performance metric. The evaluation criteria are:
\begin{itemize}
    \itemsep0em
    \item Duration ($d$): 	Duration of training
    \item Train ($t$): 		Lowest MSE within training
    \item Validation ($v$): Qualitative assess of under-fitting/over-fitting and learning rate performance
    \item Inversion ($i$):	RMSE of 100,000 validation velocities as compared to true velocity
\end{itemize}

These criteria are ranked from 1-20, with 20 being the best result. The score was calculated as
\begin{equation}
    \text{Score} = d+t+v+2i.
\end{equation} 
The formula is arbitrarily chosen and linear in nature, making ideal for interpretation and understanding. The additional weight of 2 for the inversion rank emphasizes the inversion is the most important criteria. The rank criteria ranks all scores, with the highest score being best. The best performing architecture-loss combination is identified as \textbf{Conv1D-Adadelta}. Table~\ref{tab:quantitative_assessment_summary_arch} and Table~\ref{tab:quantitative_assessment_summary_loss} summarize Table~\ref{tab:quantitative_assessment_arch_loss} per architecture and loss optimizer respectively. Table~\ref{tab:quantitative_assessment_summary_arch} further reinforces the choice for ideal setup being of type Conv1D since this architecture ranked in the top four, irrespective of Loss Optimizer. Table~\ref{tab:quantitative_assessment_summary_loss} is in agreement that Adadelta is the better loss optimizer for our setup, however the difference is relatively small and not substantial. Choosing a different loss optimizer would not result in deterioration of our result. Full results used to build these tables is given in Appendix~\ref{sec:app_results_summary_results_exp_1}.

\begin{table}[!htb]
    \centering
    \begin{minipage}[b]{.45\textwidth}
        \centering
            \begin{tabular}{cccc}\toprule
                \multirow{2}{*}{Architecture} & \multicolumn{3}{c}{Score} \\ \cline{2-4} 
                                            & Avg      & Min   & Max   \\ \hline
                Conv1D                        & 74.3 & 70 & 79 \\
                MLP                           & 63.8 & 61 & 66 \\
                Conv2D                        & 51.8 & 48 & 56 \\
                VGG                           & 48.5  & 34 & 55 \\
                ResNet                        & 33.5  & 28 & 42 \\\hline
            \end{tabular}
            \caption{Quantitative assessment for architectures.}
            \label{tab:quantitative_assessment_summary_arch}
    \end{minipage}
    \qquad
    \begin{minipage}[b]{.45\textwidth}
        \centering
            \begin{tabular}{cccc}\toprule
            \multirow{2}{*}{Architecture} & \multicolumn{3}{c}{Score} \\ \cline{2-4} 
                                        & Avg      & Min   & Max   \\ \hline
            Adadelta                      & 58.0   & 42 & 79 \\
            Adagrad                       & 54.6 & 31 & 70 \\
            RMSprop                       & 54.4 & 33 & 72 \\
            Adam                          & 50.4 & 28 & 76 \\\hline
            \end{tabular}
            \caption{Quantitative assessment for loss optimizers.}
            \label{tab:quantitative_assessment_summary_loss}
    \end{minipage}  
\end{table}

\clearpage
\begin{table*}[ht!]
    \footnotesize
    \centering
    \begin{tabular}{@{}cccccccc@{}}\toprule
Architecture & Loss Optimizer & Duration & Train & Validation & Inversion & Score & Overall Rank \\ \hline
MLP          & Adagrad        & 18       & 4     & 20         & 11        & 64    & 6    \\
MLP          & Adadelta       & 16       & 13    & 19         & 8         & 64    & 6    \\
MLP          & RMSprop        & 20       & 7     & 9          & 15        & 66    & 5    \\
MLP          & Adam           & 19       & 19    & 9          & 7         & 61    & 8    \\
Conv1D       & Adagrad        & 17       & 2     & 15         & 18        & 70    & 4    \\
Conv1D       & Adadelta       & 14       & 8     & 19         & 19        & 79    & 1    \\
Conv1D       & RMSprop        & 14       & 9     & 9          & 20        & 72    & 3    \\
Conv1D       & Adam           & 15       & 10    & 17         & 17        & 76    & 2    \\
Conv2D       & Adagrad        & 12       & 5     & 15         & 12        & 56    & 9    \\
Conv2D       & Adadelta       & 12       & 15    & 15         & 4         & 50    & 14   \\
Conv2D       & RMSprop        & 10       & 1     & 9          & 14        & 48    & 15   \\
Conv2D       & Adam           & 10       & 20    & 3          & 10        & 53    & 11   \\
VGG          & Adagrad        & 8        & 3     & 15         & 13        & 52    & 13   \\
VGG          & Adadelta       & 8        & 14    & 15         & 9         & 55    & 10   \\
VGG          & RMSprop        & 1        & 11    & 9          & 16        & 53    & 11   \\
VGG          & Adam           & 8        & 12    & 4          & 5         & 34    & 17   \\
ResNet       & Adagrad        & 4        & 6     & 17         & 2         & 31    & 19   \\
ResNet       & Adadelta       & 4        & 17    & 15         & 3         & 42    & 16   \\
ResNet       & RMSprop        & 2        & 16    & 3          & 6         & 33    & 18   \\
ResNet       & Adam           & 5        & 18    & 3          & 1         & 28    & 20   \\\hline
\end{tabular}
\caption{Quantitative assessment for architecture and loss optimizers.}\label{tab:quantitative_assessment_arch_loss}
\end{table*}  

\subsection{Marmousi Model}\label{sec:results_Numerical_experiment}
\subsubsection{Dataset}\label{sec:results_marmousi_dataset}
\begin{figure}[ht!]
    \centering
    \includegraphics[width=0.99\textwidth]{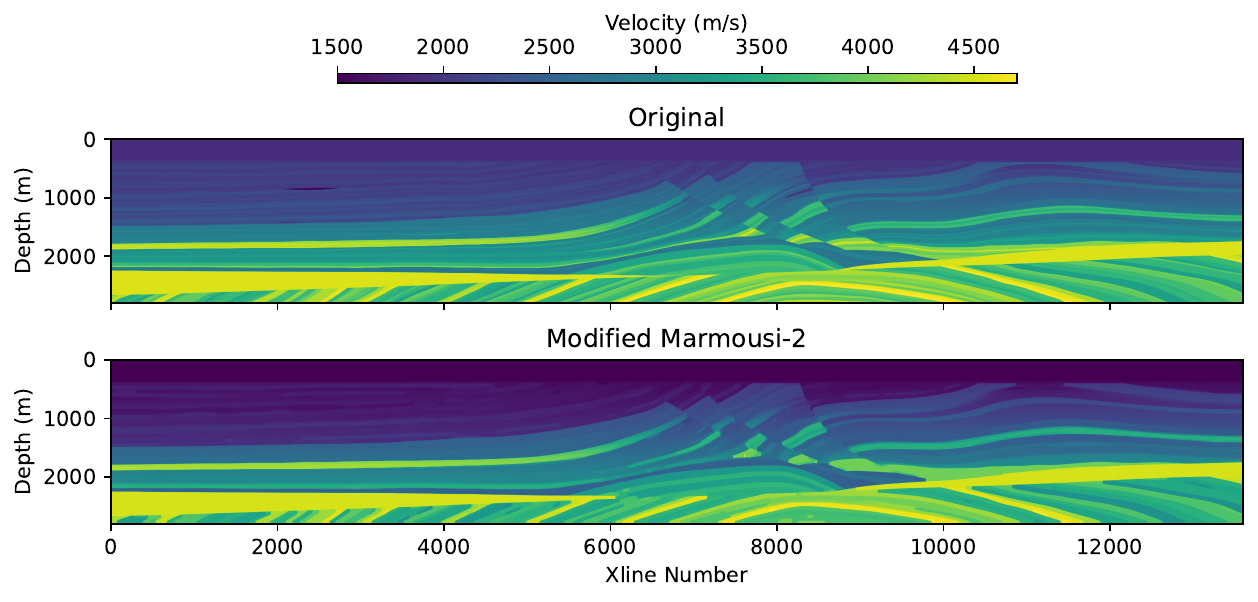}
    \caption[Marmousi-2 and modified Marmousi-2 velocity model.]{Marmousi-2 and modified Marmousi-2 velocity model.}
    \label{fig:marm_modified}
\end{figure}

\begin{figure}[ht!]
    \centering
    \includegraphics[width=0.99\textwidth]{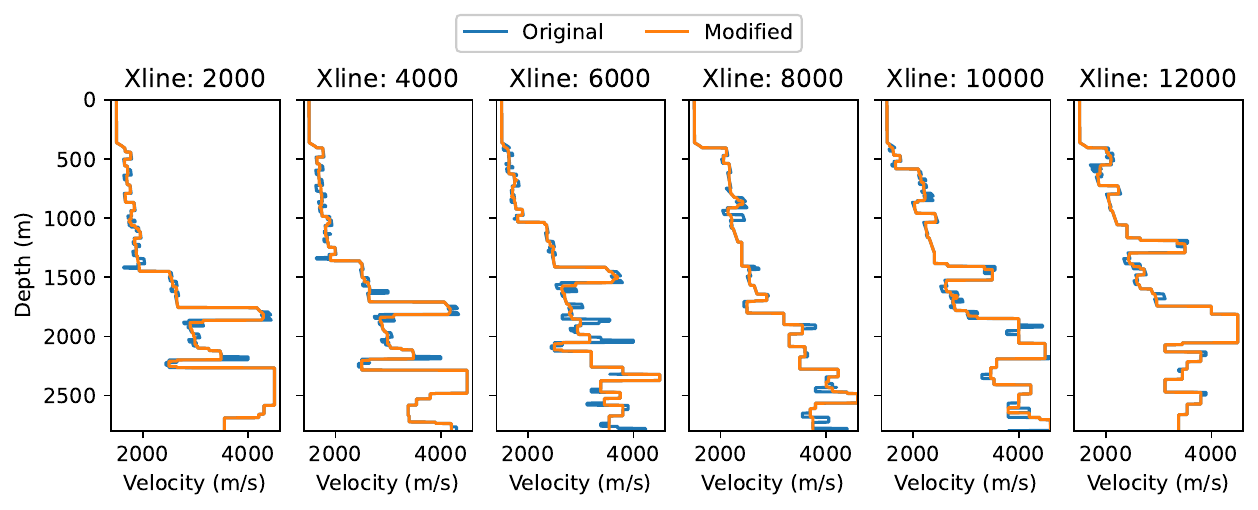}
    \caption[Velocity profiles through crosslines on Marmousi-2 and modified Marmousi-2.]{Velocity profiles through crosslines (Xlines) on Marmousi-2 and modified Marmousi-2 models.}
    \label{fig:marm_modified_velocity}
\end{figure}

The Marmousi-2 model \citep{Martin2002} was used to evaluate the technique on an industry standard dataset. Figure~\ref{fig:marm_modified} and Figure~\ref{fig:marm_modified_velocity} illustrate the Marmousi-2 model and velocity profile respectively. The model has a lateral extension of 17 km and a depth of 3.5 km and includes a total of 199 layers geophysical layers, as well as an extended water layer of 450 m depth to simulate a deep-water setting \citep{Martin2002}. The grid spacing was 10m vertically by 25m laterally, resulting in a 2801 by 13601 grid. The velocity in the model ranges from 1500\si{ms^{-1}} up to 4700\si{ms^{-1}} and after the application of a 150m vertical median filter to reduce the vertical resolution, the number of layers in each velocity profile was analytically calculated to range between 20 to 50 layers. The salt density as taken constant throughout. The generation for this model is provided in Appendix~\ref{sec:app_results_generation_marm}. This will be referred to as the Marmousi model for the rest of the thesis.

\clearpage

\subsubsection{DNN FWI Generator}\label{sec:results_marm_generator}
\begin{figure}[ht!]
    \centering
    \includegraphics[width=0.9\textwidth]{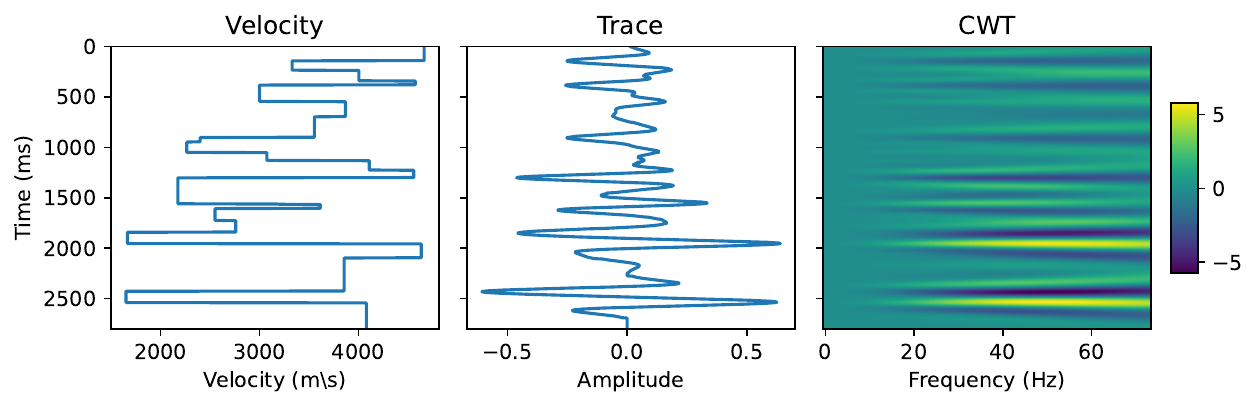}
    \caption[Sample velocity profile, trace and CWT generated by Marmousi generator.]{Sample velocity profile, trace and CWT generated by Marmousi generator.}
    \label{fig:marm_generator}
\end{figure}

Following from the work in Sections~\ref{sec:results_Train-Test_Data}-~\ref{sec:results_arch_loss_combination}, a generator was constructed to be able to invert for the Marmousi model. The generator parameters are given in Table~\ref{tab:marm_data_generator_params}. A sample of the velocity, trace and CWT generated by this generator are available in Figure~\ref{fig:marm_generator}.
\begin{table*}
    \centering
    \begin{tabular}{@{}llc@{}}\toprule
        Parameter                  & Description                   & Value \\ \hline
        length (ms)                & Length of trace               & 2801  \\
        vel\_min (m/s)             & Minimum velocity              & 1450  \\
        vel\_max (m/s)             & Maximum velocity              & 5000  \\
        vel\_min\_separation (m/s) & Minimum velocity separation   & 15    \\
        time\_min (ms)             & Minimum time sample           & 0     \\
        time\_max (ms)             & Maximum time sample           & 2801  \\
        time\_min\_separation (ms) & Minimum time separation       & 5     \\
        layers\_min                & Minimum number of layers      & 20    \\
        layers\_max                & Maximum number of layers      & 50    \\
        dominant\_frequency        & \si{Hz} of dominant frequency & 5     \\ \hline
    \end{tabular}
    \caption[Marmousi data generator parameters.]{Marmousi data generator parameters.}\label{tab:marm_data_generator_params}
\end{table*}

\subsubsection{DNN Training and Architecture Performance}\label{sec:results_marm_DNN_Training_and_Architecture_Performance}
The network was trained for 30 epochs, at 1,000,000 traces and 100,000 traces per training and testing dataset respectively. The network was a slightly modified version of the ideal network in §~\ref{sec:results_arch_loss_combination} due to the longer time length of trace. This network is given as two part network ``Marmousi - Time to Pseudo-Spectral'' and ``Marmousi - Pseudo-Spectral to Velocity'' in Appendix~\ref{tab:app_results_dnn_architectures}.

\begin{figure}[ht!]
    \centering
    \includegraphics[width=0.99\textwidth]{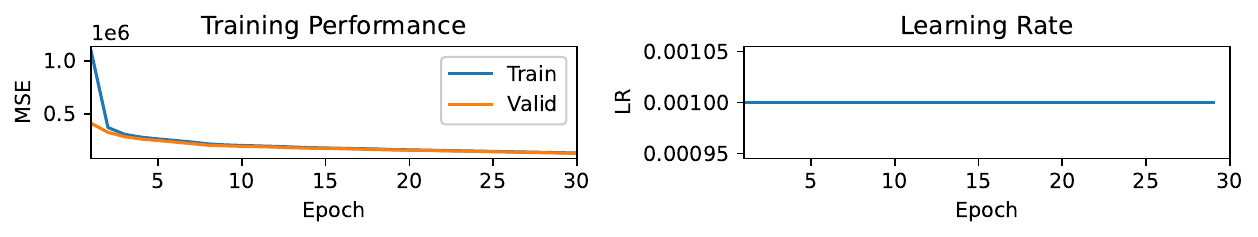}
    \caption[Training and Validation curves for DNN training and Learning Rate values.]{Training and Validation curves for DNN training and Learning Rate values.}
    \label{fig:marm_training_validation_curves}
\end{figure}

\begin{figure}[ht!]
    \centering
    \includegraphics[width=0.99\textwidth]{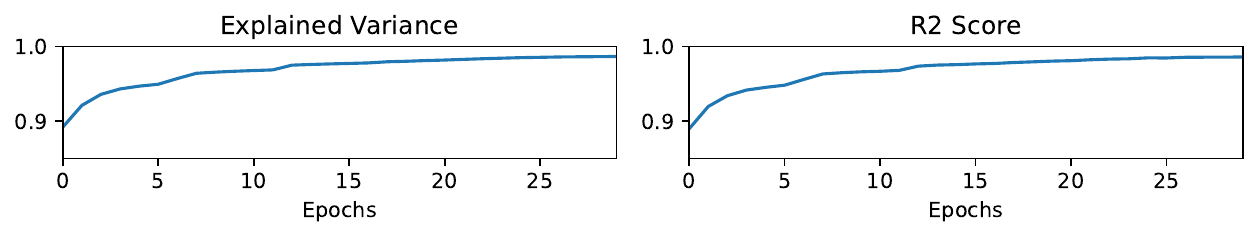}
    \caption[Explained variance and R2 Score metrics calculated per epoch.]{Explained variance and R2 Score metrics calculated per epoch. Both these metrics are approaching one per epoch, suggesting a good overall DNN performance.}
    \label{fig:marm_training_validation_metrics}
\end{figure}

As expected from previous work, the DNN training performance shown in Figure~\ref{fig:marm_training_validation_curves} indicates how this workflow performs well with a monotonically decreasing loss per epoch, with no symptoms of over-fitting or under-fitting. From the Learning Rate plot, the DNN might benefit from a couple more additional training epochs since the automatic reducing learning rate callback within the network was never initiated. Figure~\ref{fig:marm_training_validation_metrics} reinforce this suggestion of good training and with additional metrics calculated per epoch of Explained Variance and R2 Score, both of which are gradually approaching one per epoch.
 
\begin{figure}[ht!]
    \centering
    \includegraphics[width=0.98\textwidth]{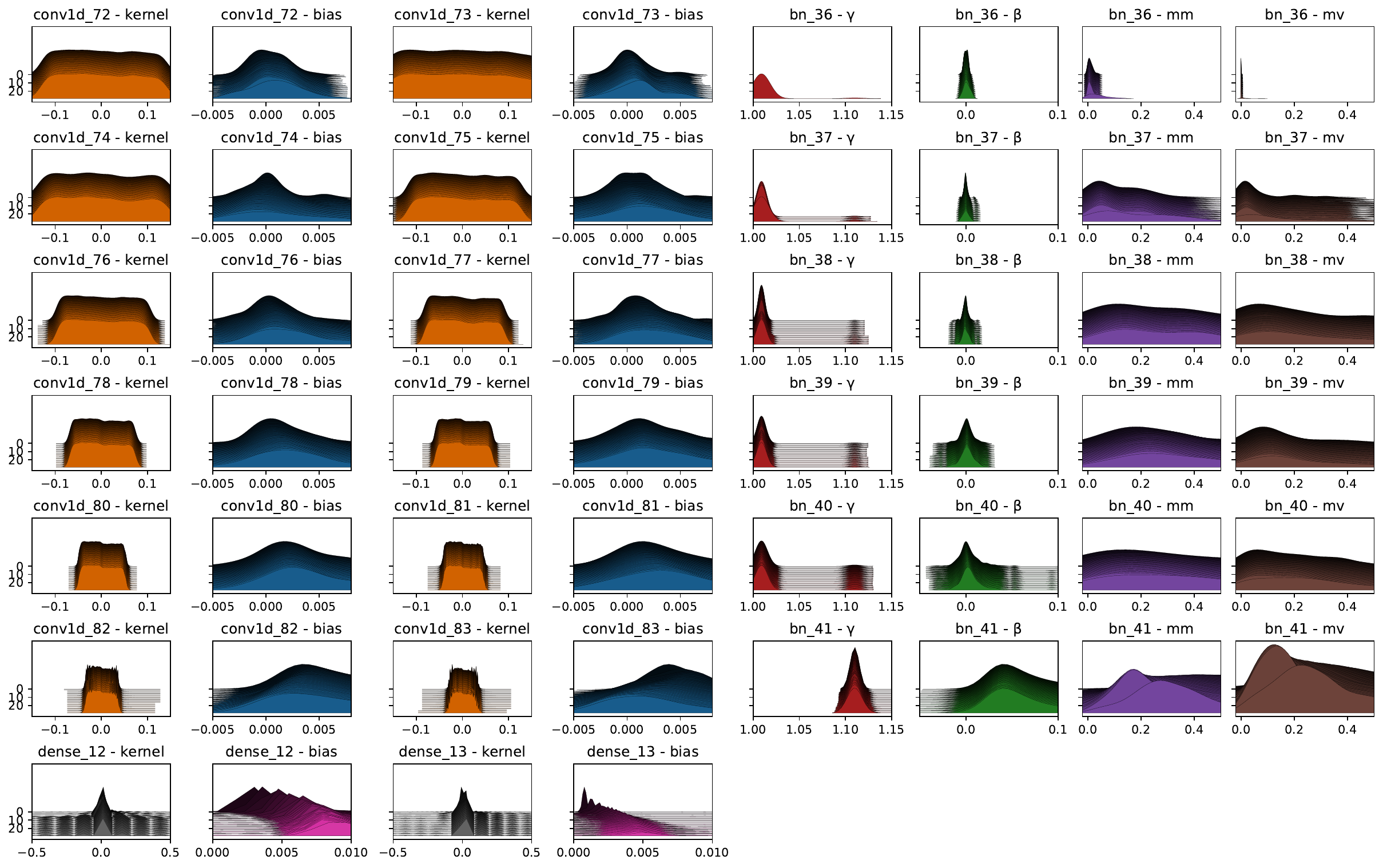}
    \caption[Evolution of network histograms.]{Evolution of network histograms. The depth ($y$-dimension) of these plots are the epochs and the darker saturation indicate the older the epoch value. The $z$-dimension is the density of values represented at $x$-dimension.  MM is the moving average and MV is moving variance.}
    \label{fig:marm_evolution_network_historgrams}
\end{figure}

Figure~\ref{fig:marm_evolution_network_historgrams} show histograms for the evolution of network trainable parameters per epoch. The $y$-axis of these plots are the epochs and the darker saturation indicate the older the epoch value. The $z$-axis is the density of values represented at $x$-axis. The first 6 rows show histograms for the convolution and batch normalisation layers as groups per row. Each group is composed of Convolution – Convolution – Batch Normalisation layers which is getting repeated 6 times in the network. For each convolution, the trainable parameters are the bias and kernel, whilst for Batch Normalisation $\gamma$ and $\beta$ are the per epoch standard deviation and mean respectively, moving variance and moving mean. Convolution kernels are relatively flat, with narrow distribution from $\left[-0.15, 0.15\right]$ to $\left[-0.04, 0.04\right]$. This indicates that the input signal is getting collapsed into smaller probabilities for explanation. Comparing convolution kernels with each row, these are very similar. They both have wider distributions indicating that more abstractions are being included with each pass. The bias starts off with multiple peaks then centre about 0, moving from a left skewed to a more gaussian distribution. This shows that the initial layers are identifying multiple parts of the input signal as being important due to the multiple peaks, but as we go deeper into the network with more epochs, we start seeing that the network “understanding” starts to converge into a gaussian distribution and collapsing all the abstracted information from the upper layers into the desired output. The last row shows histograms for the two dense layers, with trainable parameters for bias and the kernel. The kernel distribution for both dense layers is centred around zero, with a very narrow standard deviation. This indicates how the last two layers a selectively choosing components from the different abstracted feature maps and adding small components to build up the final inversion. The shifting bias from left skewed to more Gaussian, with mean close to 0.004, indicates that the final reconstruction is happening within a Gaussian environment for the DNN. This is an ideal setup as this will facilitate regularizations to unseen data.

\begin{figure}[ht!]
    \centering
    \includegraphics[width=0.98\textwidth]{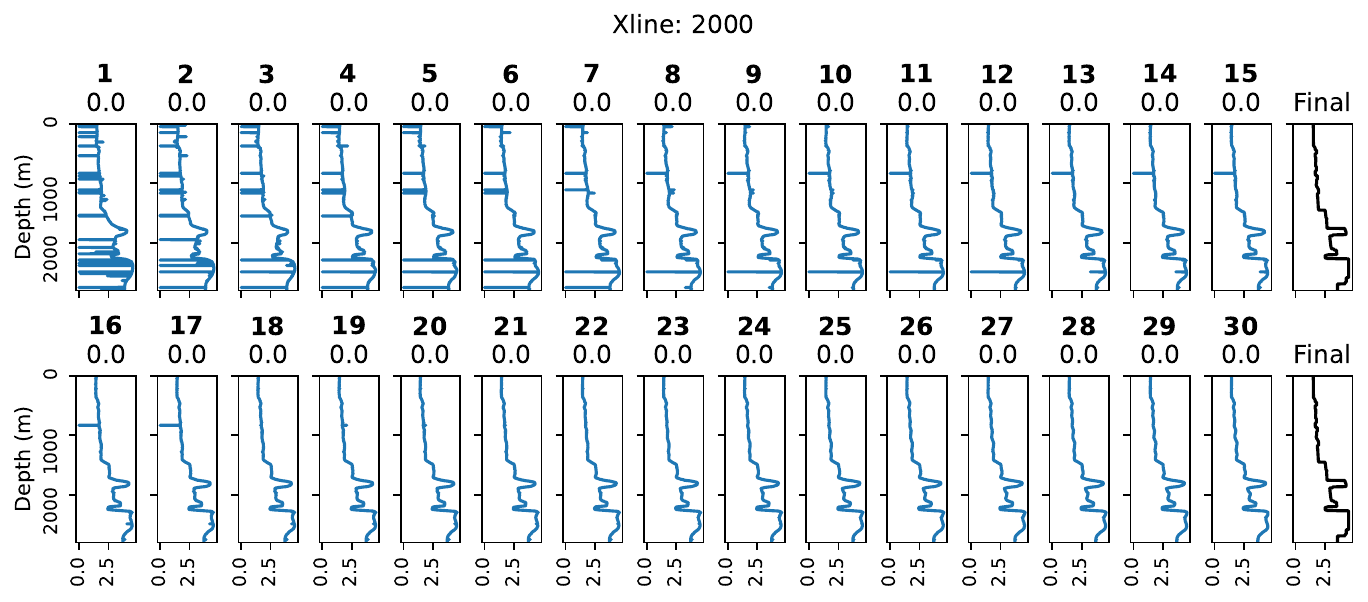}
    \includegraphics[width=0.98\textwidth]{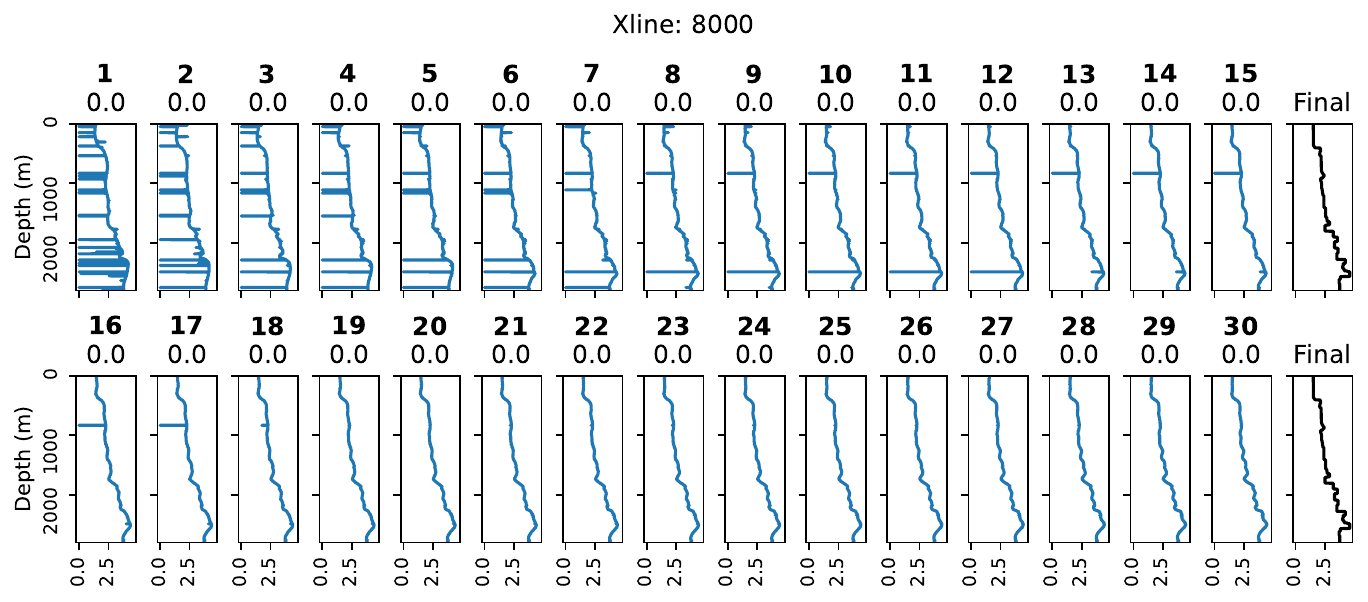}
    \includegraphics[width=0.98\textwidth]{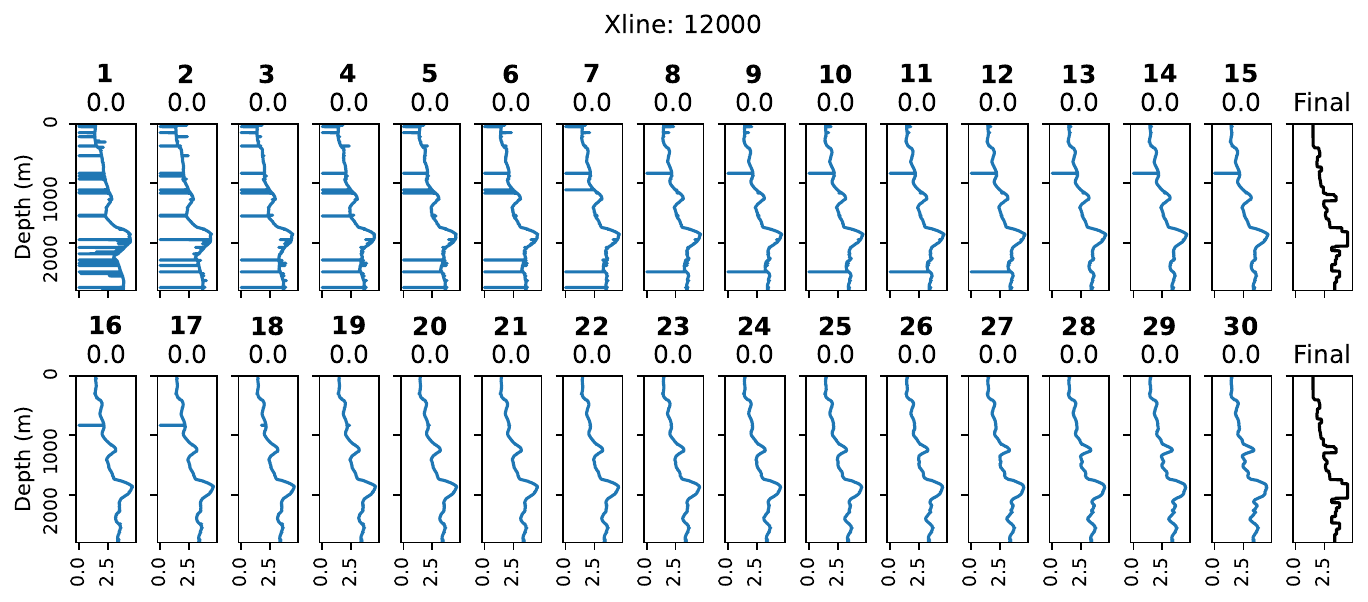}
    \caption[Evolution of velocity profile through different epochs.]{Evolution of velocity profile for Xline 2000, 8000 and 12000 through the different epochs respectively. Above each plot, there is the Epoch number in bold, and the MSE.}
    \label{fig:marm_evolution_velocity_profile}
\end{figure}

\begin{figure}[ht!]
    \centering
    \includegraphics[width=0.98\textwidth]{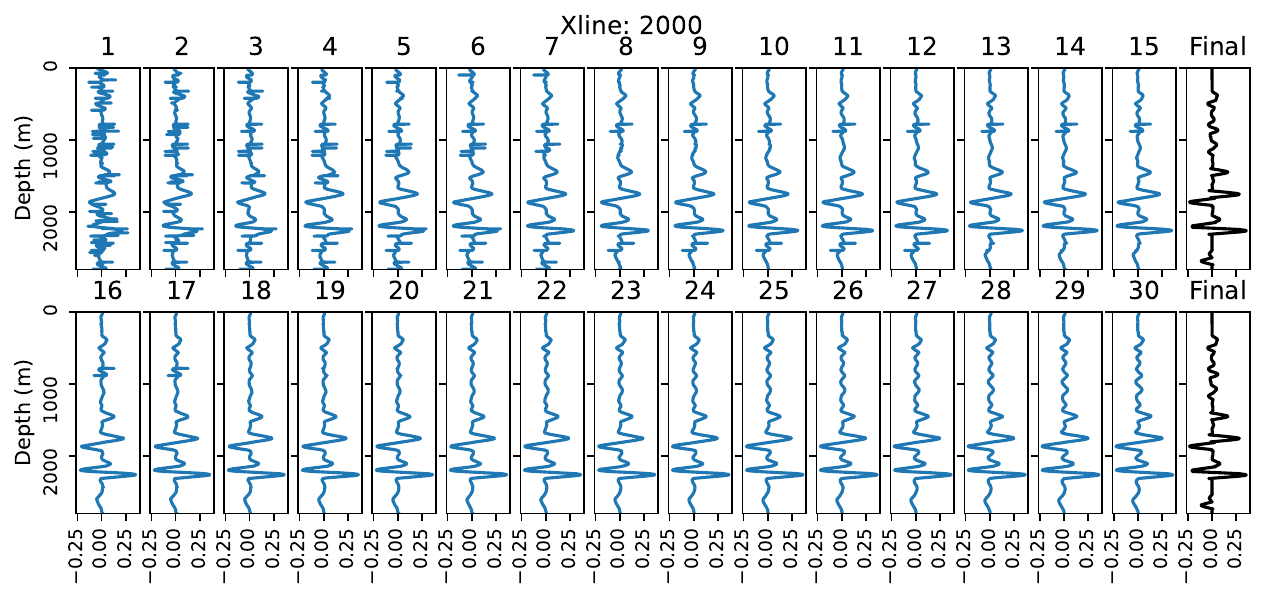}
    \includegraphics[width=0.98\textwidth]{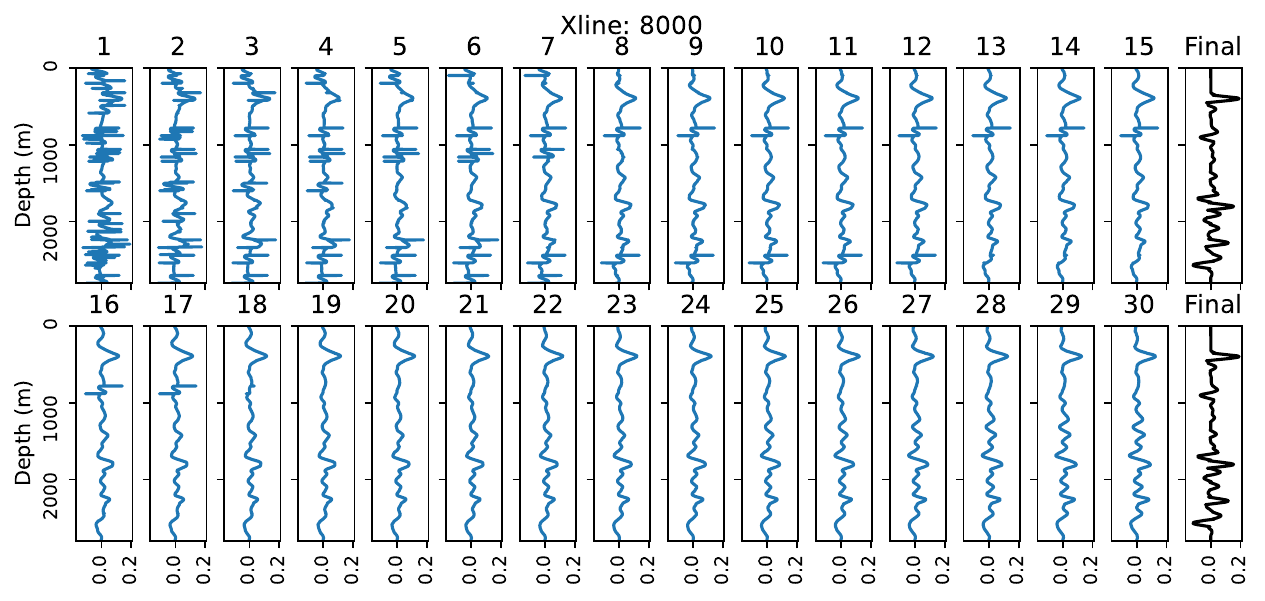}
    \includegraphics[width=0.98\textwidth]{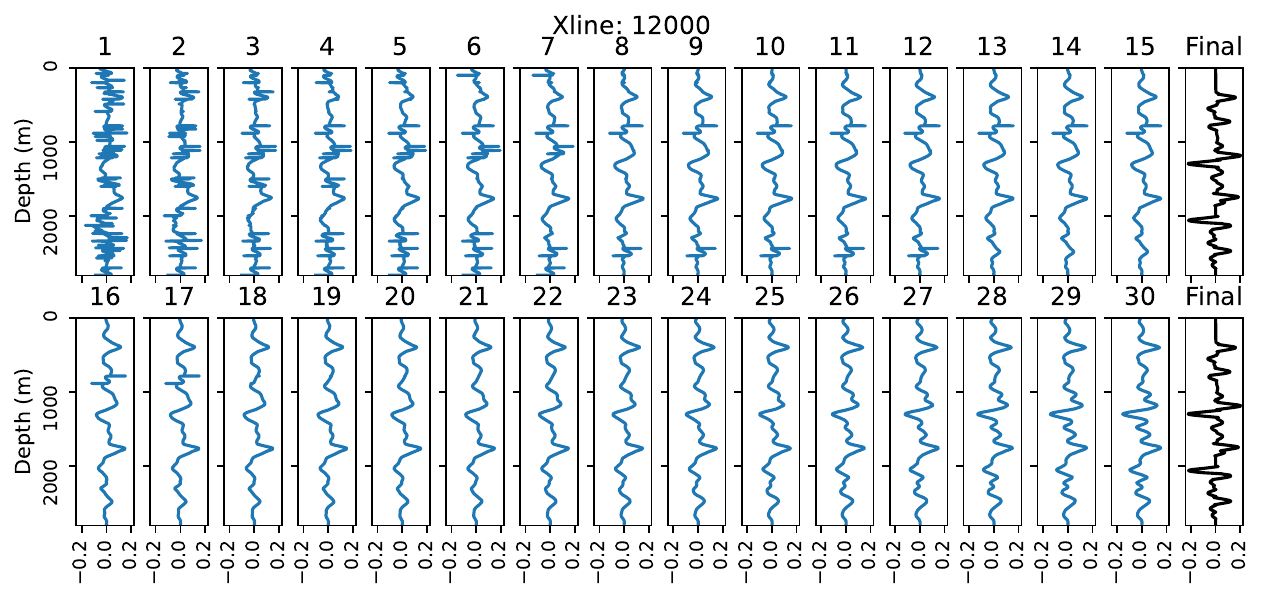}
    \caption[Evolution of trace for Xline 2000, 8000 and 12000 through the different epochs.]{Evolution of trace for Xline 2000, 8000 and 12000 through the different epochs respectively.}
    \label{fig:marm_evolution_trace}
\end{figure}

Figure~\ref{fig:marm_evolution_velocity_profile} and Figure~\ref{fig:marm_evolution_trace} show the velocity inversion and resultant trace for Xline 2000, 8000 and 12000 from that Marmousi model respectively for every training epoch for the DNN. From the initial epoch, the DNN was able to invert most of the information within the velocity profile. The inversion is not perfect, since there is a form of leakage/spikes coming in up to about epoch 10. This is the learning-process of the DNN, since this gets gradually removed from the velocity profile with additional epoch, and at about epoch 20 there is an almost perfect reconstruction. From epoch 20 to epoch 30, the differences are minimal as can be seen by the very small changes to the MSE shown in the plot and as well in the overall DNN loss values in Figure~\ref{fig:marm_training_validation_curves}. 

\clearpage
\subsubsection{Deterministic FWI}\label{sec:results_classical_FWI}
\begin{figure}[ht!]
    \centering
    \includegraphics[width=0.78\textwidth]{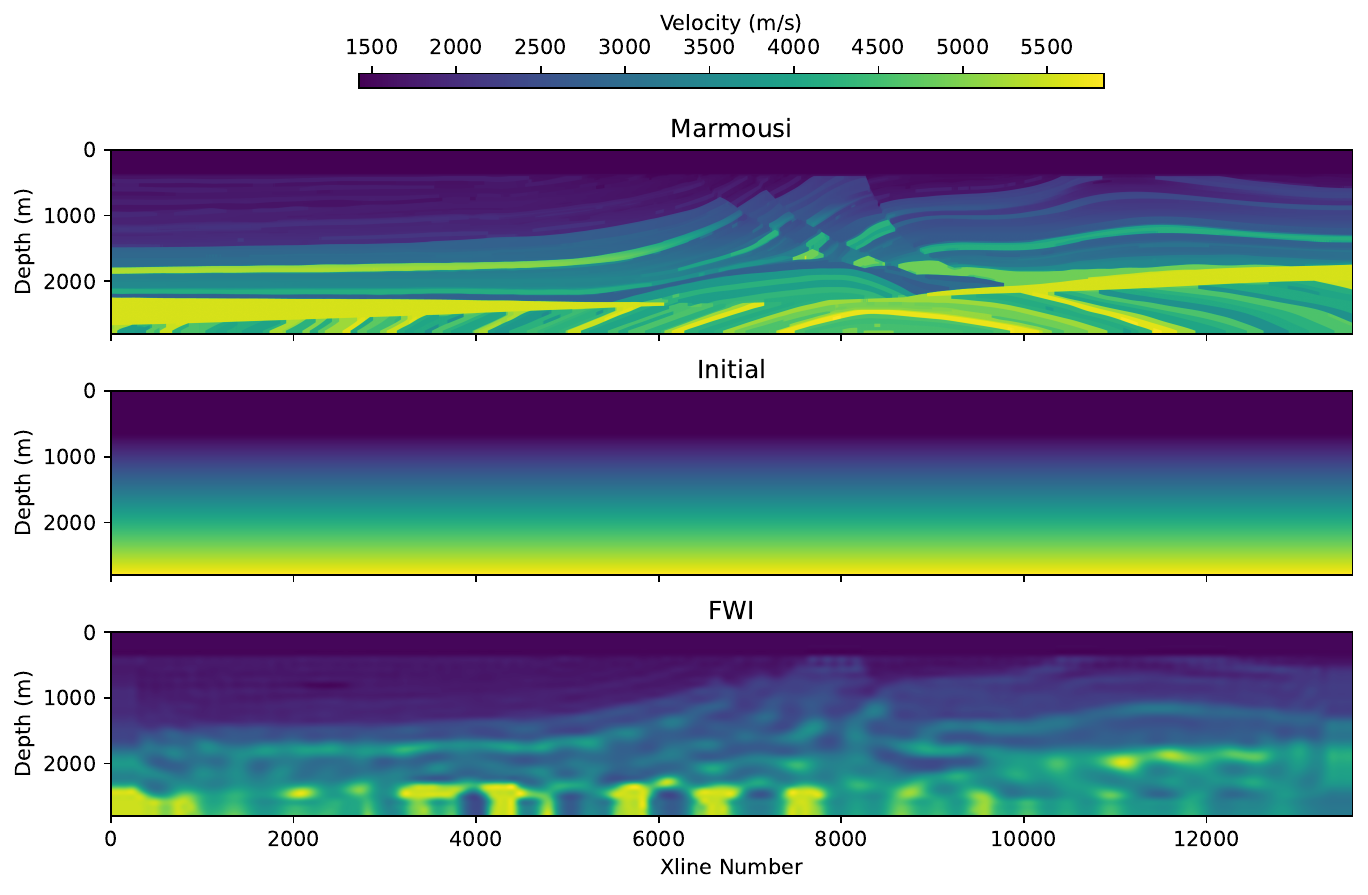}
    \caption[Classical FWI with Sobolev space norm regularization.]{Classical FWI with Sobolev space norm regularization result. Top: Initial modified Marmousi model. Middle: Initial velocity provided for FWI. Bottom: FWI result following from inverting for 3.6\si{Hz}.}
    \label{fig:marm_classical_fwi}
\end{figure}

\begin{figure}[ht!]
    \centering
    \includegraphics[width=0.99\textwidth]{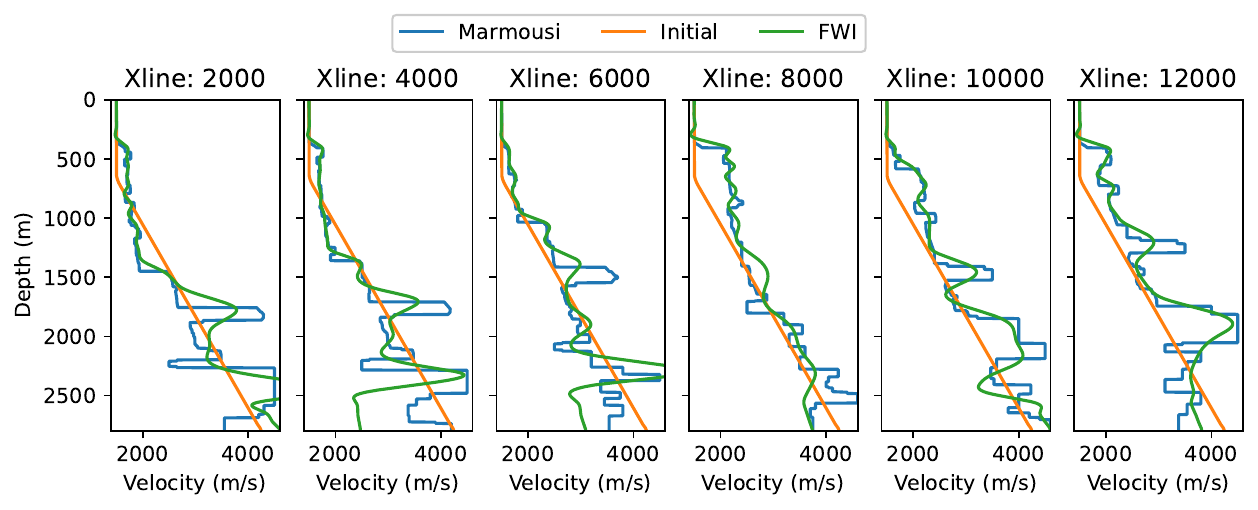}
    \caption[Velocity profiles through Xlines on Marmousi, Initial and FWI results.]{Velocity profiles through Xlines on Marmousi, Initial and FWI results as shown in Figure~\ref{fig:marm_classical_fwi}.}
    \label{fig:marm_classical_fwi_velocity}
\end{figure}

Classical FWI with Sobolev space norm regularization was employed for comparative purposes – see Figure~\ref{fig:marm_classical_fwi} and Figure~\ref{fig:marm_classical_fwi_velocity}. This was a modified version of the FWI optimization framework provided by \cite{Kazei2019}. The maximum frequency of the inversion process was set to be 3.5\si{Hz}. This results in a minimum update resolution of 414m given by $\lambda=\frac{v_{min}}{f_{max}}$ where $v_{min}$ is 1450\si{ms^{-1}} and $f_{max}$ is 3.5\si{Hz}. The dataset was resampled and interpolated by a factor of 10 to enable a faster implementation and still retain the maximum update resolution. The iterative update process started from frequency 1\si{Hz} and iteratively updated by a factor of 1.2 until reaching a maximum frequency of 3.45\si{Hz}. The optimization algorithm was L-BFGS-B \citep{Zhu1997}, with 50 iterations per frequency band in each update. Forward shot modelling was done every 100m, starting from 100m offset, and receivers spaced every 100m. The detailed implementation and inversion parameters are provided in Appendix~\ref{sec:app_results_classical_FWI}. More advanced FWI code could have been implemented to improve lateral continuity and imaging, however results obtained by this implementation provided acceptable run-times and results, thus making feasible for our experimentation. Examples of state-of-the-art code which usually available for consortiums are FULLWAVE\footnote{\url{https://fullwave3d.github.io/}}
or CREWES\footnote{\url{https://www.crewes.org/ResearchLinks/Full_Waveform_Inversion/}}.

\subsubsection{DNN and Classical FWI}\label{sec:results_comparison_dnn_fwi}
\begin{figure}[ht!]
    \centering
    \includegraphics[width=0.99\textwidth]{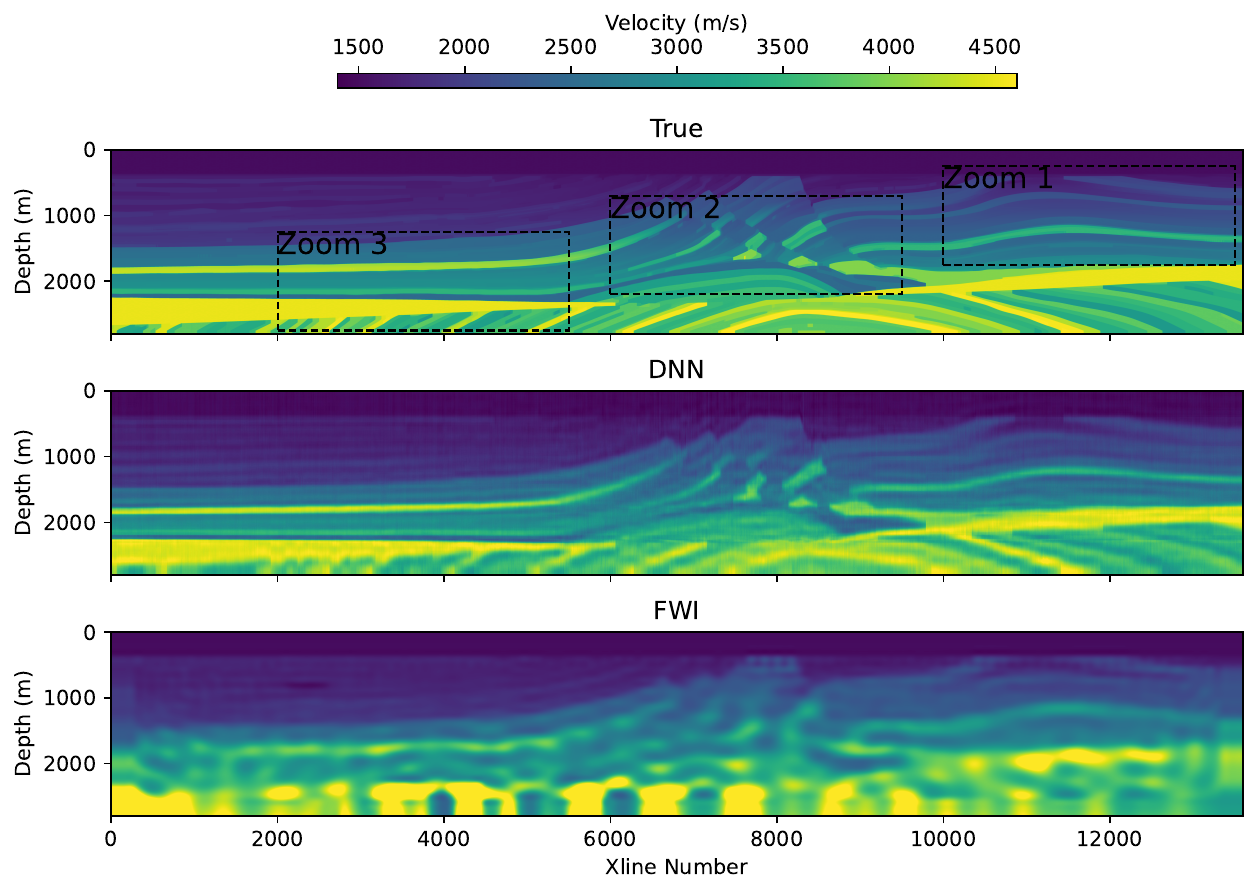}
    \caption[Comparison of DNN and Classical FWI reconstructed velocity models.]{Comparison of DNN and Classical FWI reconstructed velocity models. Top: Initial Marmousi model with highlighted Zoom 1-3 used in Figure 4.22. Middle: DNN FWI result. Bottom: Classical FWI result following from inverting for 3.6 Hz.}
    \label{fig:comparison_dnn_fwi_image}
\end{figure}

\begin{figure}[ht!]
    \centering
    \includegraphics[width=0.98\textwidth]{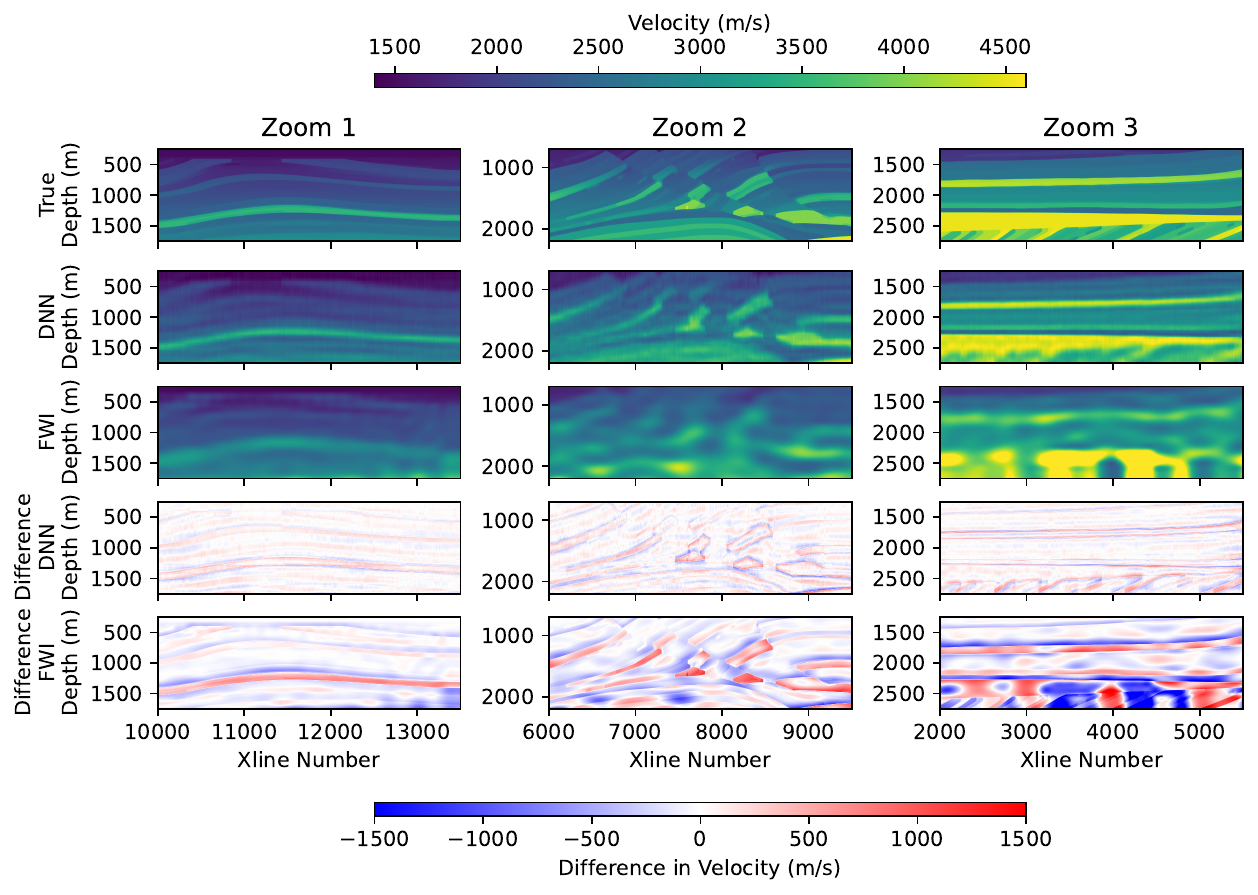}
    \caption[Zoomed comparison of DNN and Classical FWI velocity and errors.]{Zoomed comparison of DNN and Classical FWI reconstructed velocity models and corresponding errors.}
    \label{fig:comparison_dnn_fwi_image_zoom}
\end{figure}

\begin{figure}[ht!]
    \centering
    \includegraphics[width=0.96\textwidth]{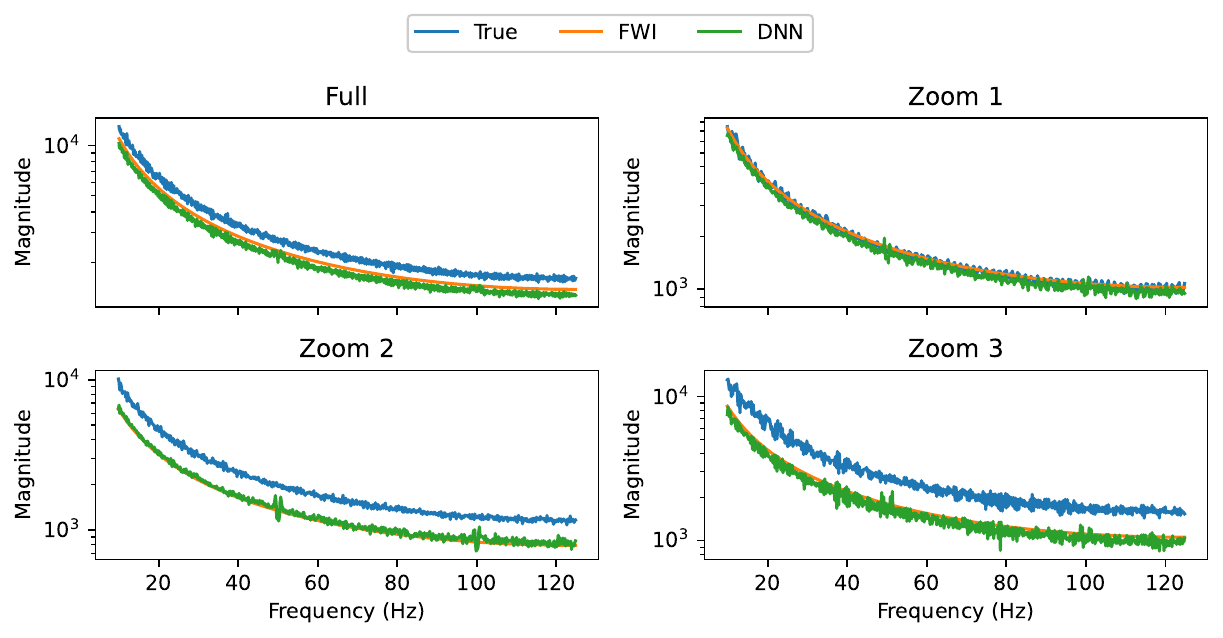}
    \caption[DNN and Classical FWI amplitude spectra.]{DNN and Classical FWI amplitude spectra show that the frequency content is similar in either approach, yet both are lower than the true.}
    \label{fig:comparison_dnn_fwi_spectra}
\end{figure}

\begin{figure}[ht!]
    \centering
    \includegraphics[width=0.99\textwidth]{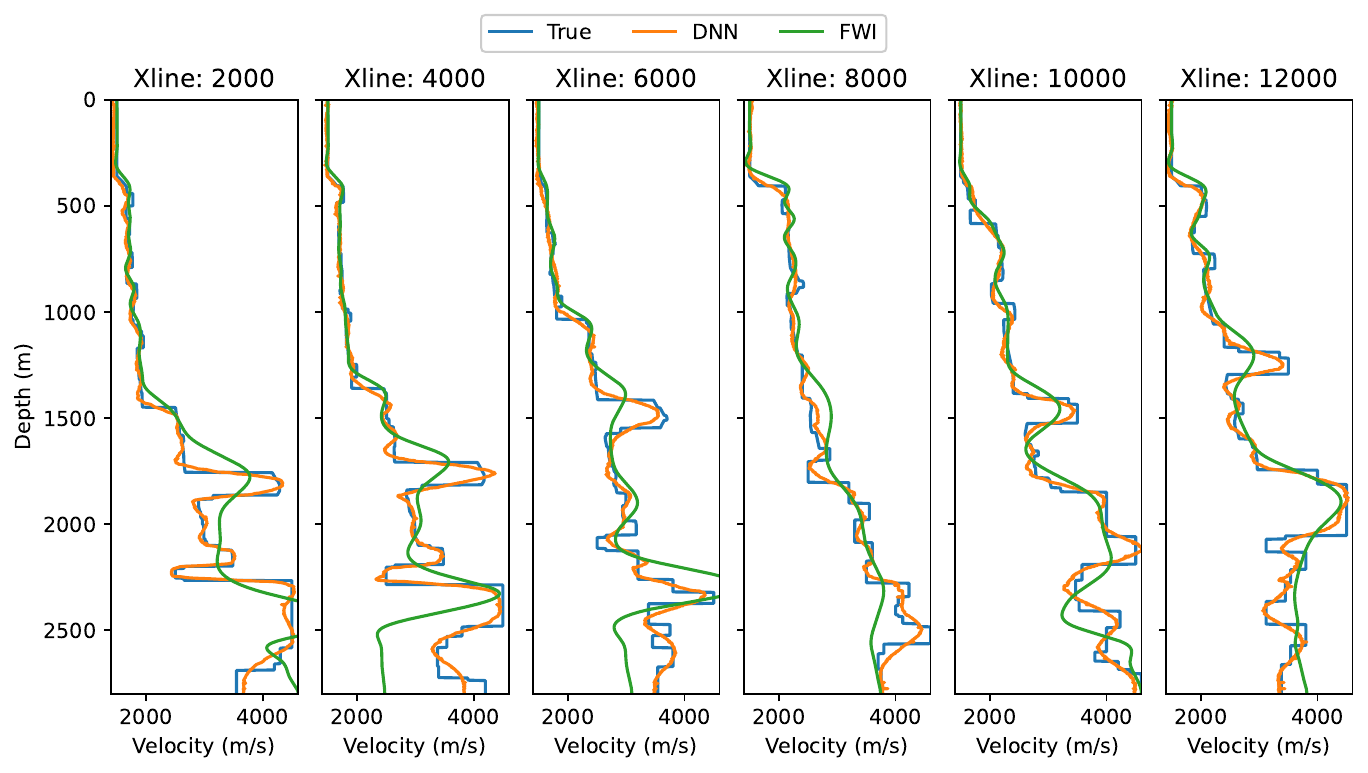}
    \caption[Velocity profiles through Xlines for Initial, DNN and FWI results.]{Velocity profiles through Xlines on Initial, DNN and FWI results as shown in Figure~\ref{fig:comparison_dnn_fwi_image}.}
    \label{fig:comparison_dnn_fwi_velocity}
\end{figure}

To evaluate the performance of our DNN approach, classical FWI and the DNN FWI approach are compared in Figure~\ref{fig:comparison_dnn_fwi_image}. Off the start, it is clearly evident how the DNN approach is producing a lot more uplift than the standard approach. There is improved imaging in the sediment layers, with distinct layers being reconstructed which would otherwise be missed with classical FWI – Zoom 1 in Figure~\ref{fig:comparison_dnn_fwi_image_zoom}. The middle section, with the heavily over-trusted layers shown in Zoom 2, the velocity layers are also being reconstructed to good levels and the small sedimentary pockets at the pinch of the over-thrust are being to be imaged as well. These are being missed completely in Classical FWI. Sub-salt in Zoom 3, DNN is once again producing much better imaging up to the salt and below the salt. Indeed, sub-salt, we are starting to image partially some of the layer coming up into the salt. The inversion process is not perfect as shown by the differences in velocities in Figure~\ref{fig:comparison_dnn_fwi_velocity} for either of the three zoomed sections. Comparison of the error maps, the problematic areas of DNN are also those for classical FWI. In Zoom 1, the amplitude of the large velocity layer coming in at 1400m depth is not being inverted properly. The onset of this layer is not as problematic, but leakage is evidently present. Similarly, for Zoom 3, the salt arrival at 2200m depth, is being imaged by DNN and not by FWI. In Zoom 2, the error hotspot for DNN are similar to those of FWI, however the magnitude of the error is of an order different. This would indicate that DNN is very good performant when it comes to inverting for large velocity packages. Figure~\ref{fig:comparison_dnn_fwi_spectra} gives the amplitude spectra for the full and zoomed velocity models respectively. This show that the frequency content is similar in either approach, yet both are lower than the true.

\begin{figure}[ht!]
    \centering
    \includegraphics[width=0.99\textwidth]{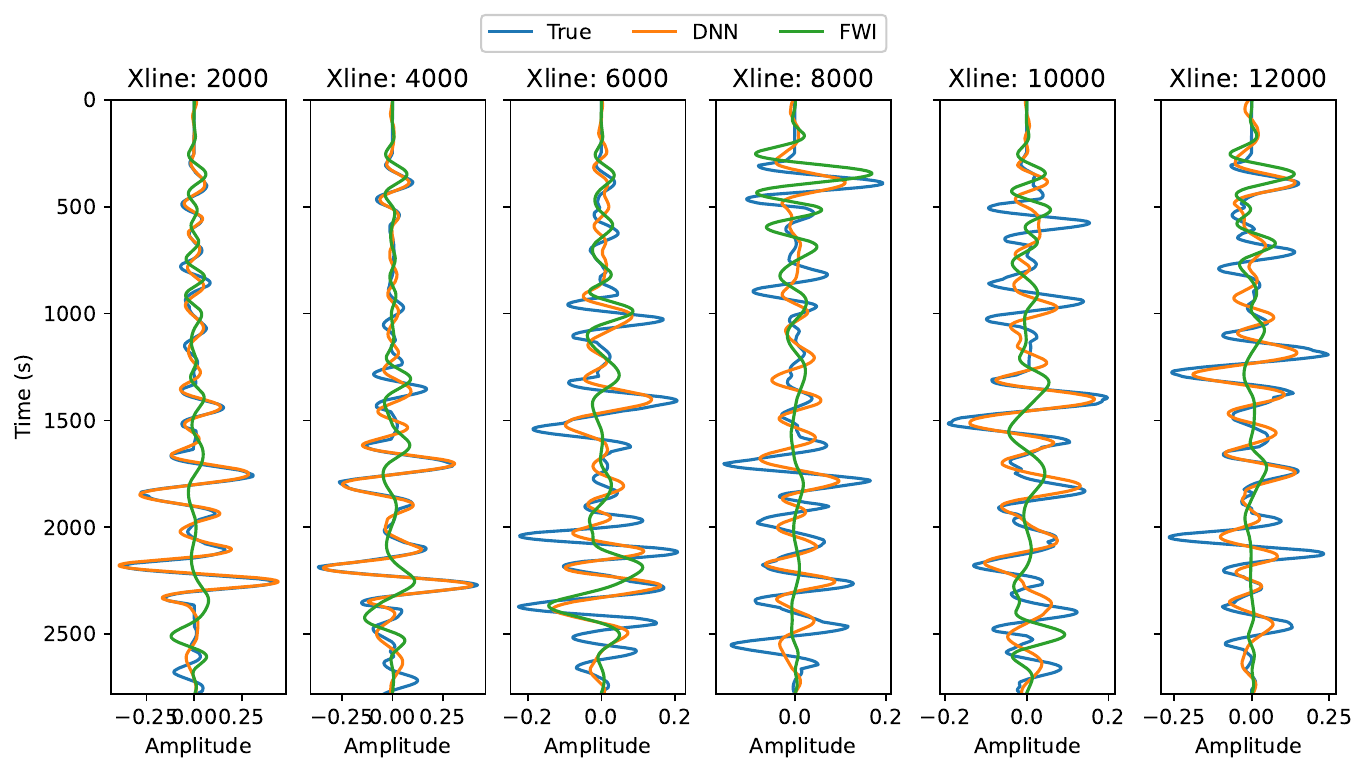}
    \caption[Time inversions of velocity for Initial, DNN and FWI results.]{Time inversions of velocity profiles through Xlines on Initial, DNN and FWI results as shown in Figure~\ref{fig:comparison_dnn_fwi_image}.}
    \label{fig:comparison_dnn_fwi_trace}
\end{figure}

The velocity profiles (Figure~\ref{fig:comparison_dnn_fwi_velocity}) and trace reconstruction (Figure~\ref{fig:comparison_dnn_fwi_trace}) confirm that our DNN approach inverted more of the signal than classical FWI. Upon closer investigation, we are seeing small spikes on the velocity on the salt section of Xlines 2000, 4000 and 6000. Further training would potentially mitigate this, or a median filter could be applied post-inversion to resolve this. From the velocity profiles, we see how FWI is able to update the shallow sections up to 1400m really well, potentially better than DNN as it is able to identify a velocity inversion at depth 500m on Xline 8000 and a pronounced segment layer at depth 800m on Xline 12000. However, beyond 1400m depth, the geometry and forward-modelling physical constraints from ray-tracing come into play and FWI is unable to provide more uplift at deeper velocity packages.

\clearpage
\section[RNN as an Analogue of FWI]{Theory-Guided RNN as an Analogue of FWI}
Based on the formulation presented in the previous Chapter, results for theory-guided RNN as an analogue for FWI are presented in this Section. As indicated in the Literature Review, this idea is not novel. However, the use of pseudo-spectral spatial gradient calculation for the inversion process is novel. Two-dimensional experiment formulation are shown, and confirm the pseudo-spectral forward modelling implementation. This is then applied to synthetic data results.

\subsection{Experiment Setup}
The original code for the time inversion as RNN framework was provided by \cite{Richardson2018} and developed in TensorFlow v1.4. TensorFlow has been in active development since 2018, with a major release of v2.0 in September 2019. This provided much improvement in terms of efficiency and it allowed for easier implementation on GPUs. For this reason, the framework was re-written in v2.0 to enable the use of INGV’s NVIDIA Titan V GPU. This code is published as part of the additional resources to this dissertation in Appendix~\ref{sec:app_results_code_repo}.

\subsection{Forward Modelling using RNNs}\label{sec:results_forward_modelling_using_rnnd}
RNN should be able to model the different wave field components if it is to replace the forward modelling component. This was first tested by considering the 1D case for both Time and Frequency implementations and compared to a 1D Green function solution. This was achievable via a custom RNNcell unit developed in TensorFlow and can be inspected code repository provided in Appendix~\ref{sec:app_results_code_repo}. The 1D implementation provided promising results and is available as part of Appendix~\ref{sec:app_results_rnn_fwi_1d}.

\begin{figure}[!ht]
    \centering
    \subbottom[Constant velocity model.\label{fig:rnn_2d_multisource_vel}]{\includegraphics[width=0.35\textwidth]{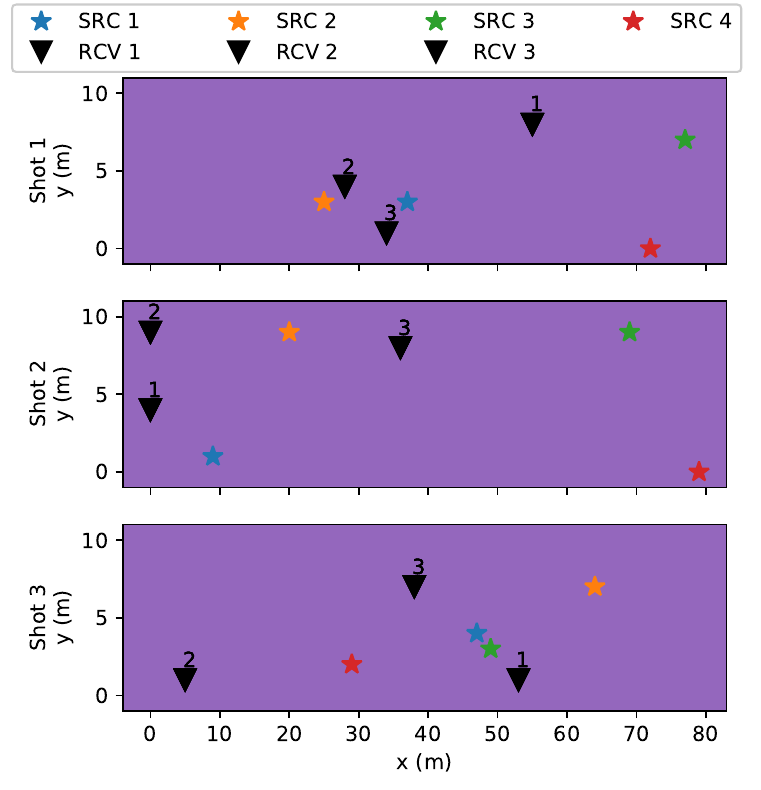}}
    \subbottom[Direct wave RNN forward modelling.\label{fig:rnn_2d_multisource_trace}]{\includegraphics[width=0.5\textwidth]{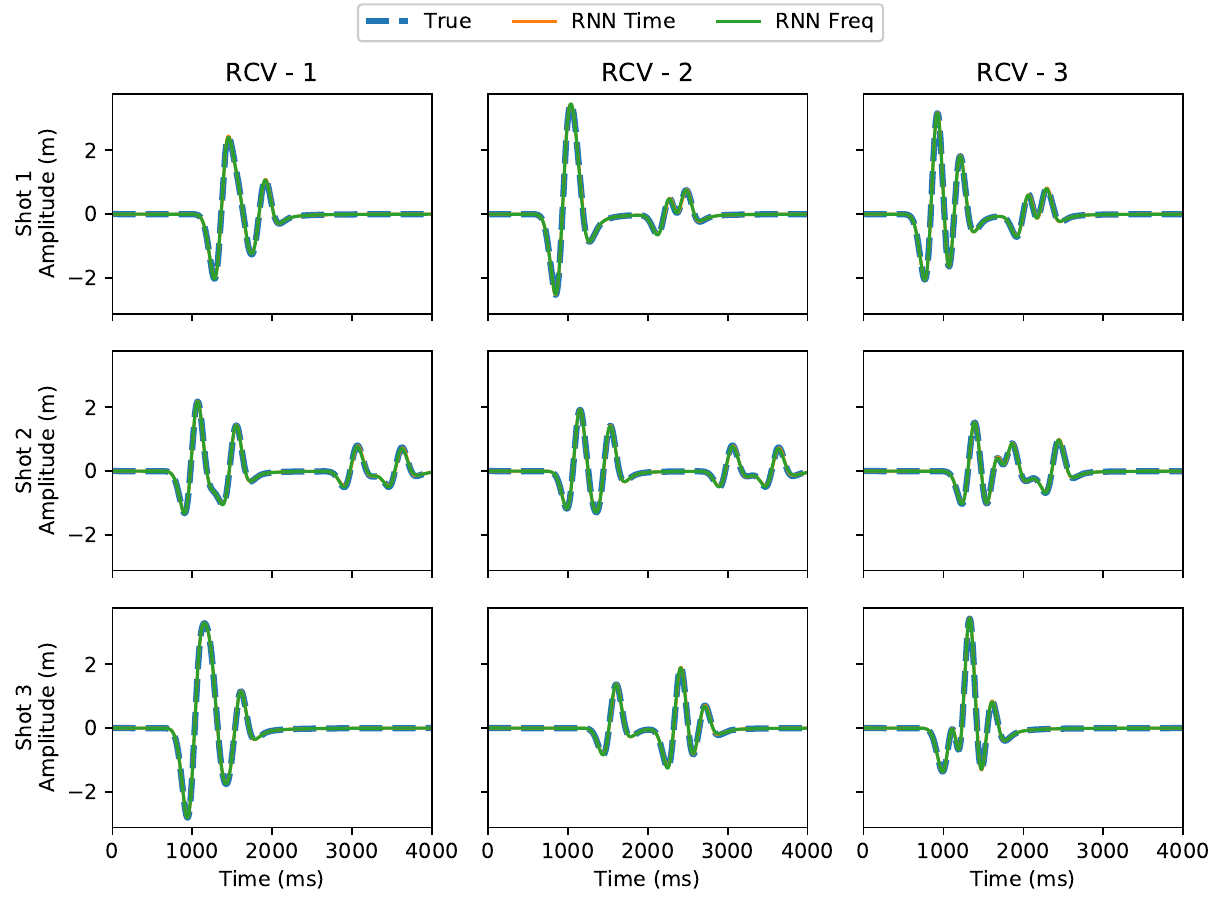}}
    \caption[Direct wave forward modelling for multi-source, multi-receiver geometry.]{Direct wave forward modelling for multi-source, multi-receiver geometry.}        
    \label{fig:rnn_2d_direct_multi}
\end{figure}

The experiment was extended to 2D and tested for all wavefield components. A 25Hz Ricker wavelet was propagated through a 2D 1500\si{ms^{-1}} constant velocity model (Figure~\ref{fig:rnn_2d_multisource_vel}) with a multi-source multi-receiver geometry setup. The 25Hz source wavelet goes into the hyper-resolution realm for FWI and is beyond the resolution that will be investigated on the synthetic model, however this allows for gauging the limit of accuracy. This model setup was forward propagated for 5333 time-steps at 1ms, with a 10m grid spacing. Namely, this implies that 5333 LSTM cells where employed for the forward modelling. The resulting direct waves are illustrated in Figure~\ref{fig:rnn_2d_multisource_trace}, with True being the analytical solution calculated using a 2D Green’s function, RNN Time and RNN Freq are the RNN implementations for forward modelling using Time and Fourier spatial derivatives respectively. Qualitatively, there is no visible difference between either approach.

\begin{figure}[!ht]
	\centering
	\subbottom[Step velocity model.\label{fig:rnn_2d_reflection_vel}]{\includegraphics[width=0.35\textwidth]{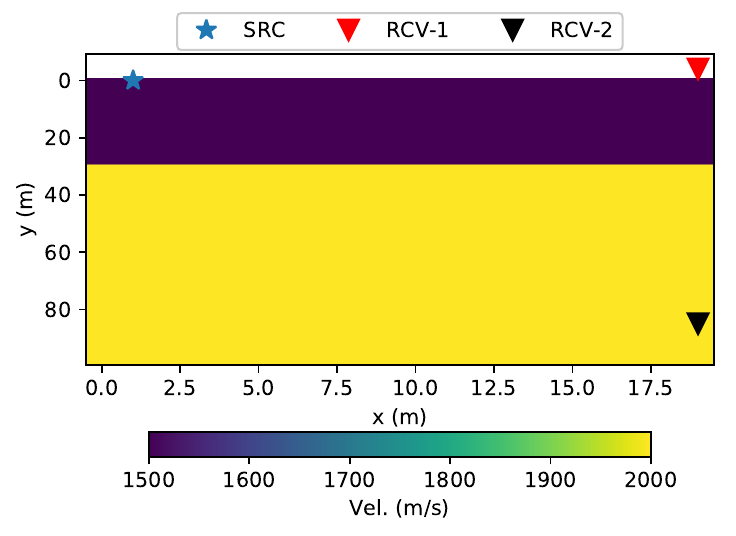}}
    \subbottom[Top: RCV-1 located at ground level reacts to direct arrival at 125ms and reflected arrival at 250ms. \newline Bottom: RCV-2 shows transmitted arrival.\label{fig:rnn_2d_reflection_wave}]{\includegraphics[width=0.5\textwidth]{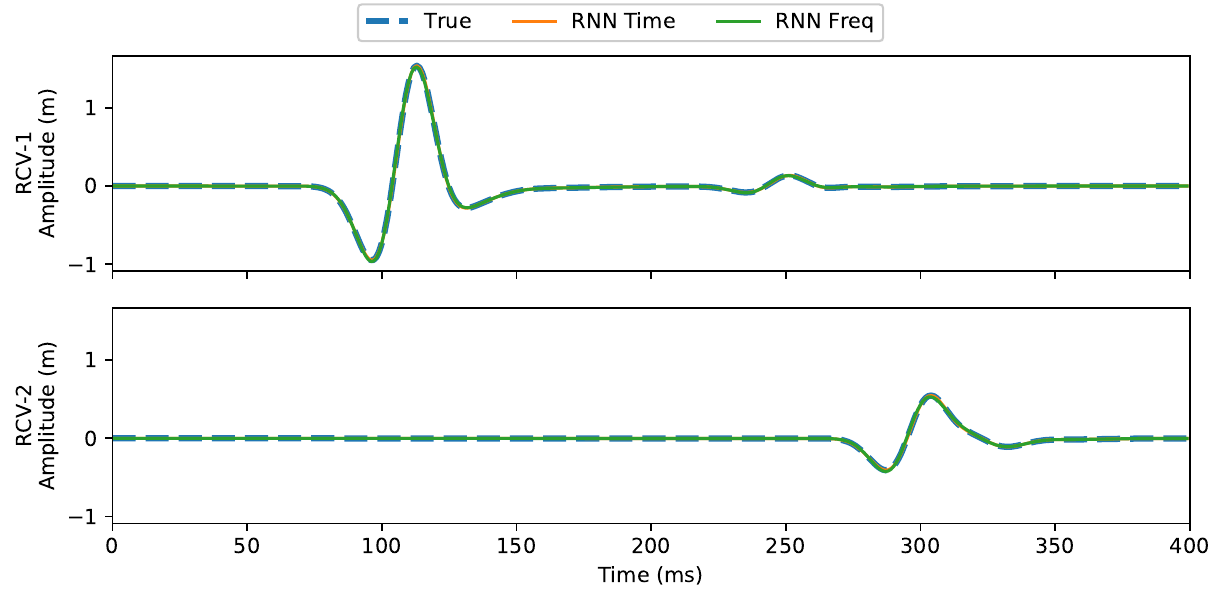}}
	\caption[Reflected and transmitted wave RNN forward modelling.]{Reflected and transmitted wave RNN forward modelling.}        
	\label{fig:rnn_2d_reflection}
\end{figure}

Reflected and transmitted arrivals were tested using a simple step velocity model ranging from 1500\si{ms^{-1}} to 2000\si{ms^{-1}} as shown in Figure~\ref{fig:rnn_2d_reflection_vel}. Figure~\ref{fig:rnn_2d_reflection_wave} is the forward modelled wavefield for the two receiver locations (RCV-1 and RCV-2), top and bottom respectively. RCV-1 at ground level interacts with the direct wave at 125ms and reflected wave at 250ms. RCV-2 is below the acoustic impedance layer at 30m and shows the transmitted wave. Comparing these to the analytical solution, either are able to model the wave components perfectly.

\begin{figure}[!ht]
\centering
\subbottom[Point-scattering velocity model.\label{fig:rnn_2d_scattered_vel}]{\includegraphics[width=0.35\textwidth]{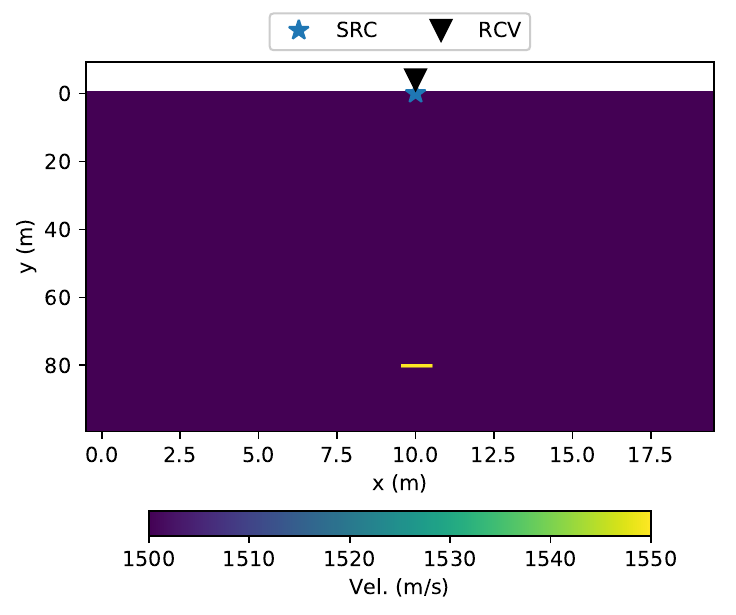}}
\subbottom[Scattering wavefield modelling. Direct wavefield was excluded in the modelling.\label{fig:rnn_2d_scattered_wavefield}]{\includegraphics[width=0.5\textwidth]{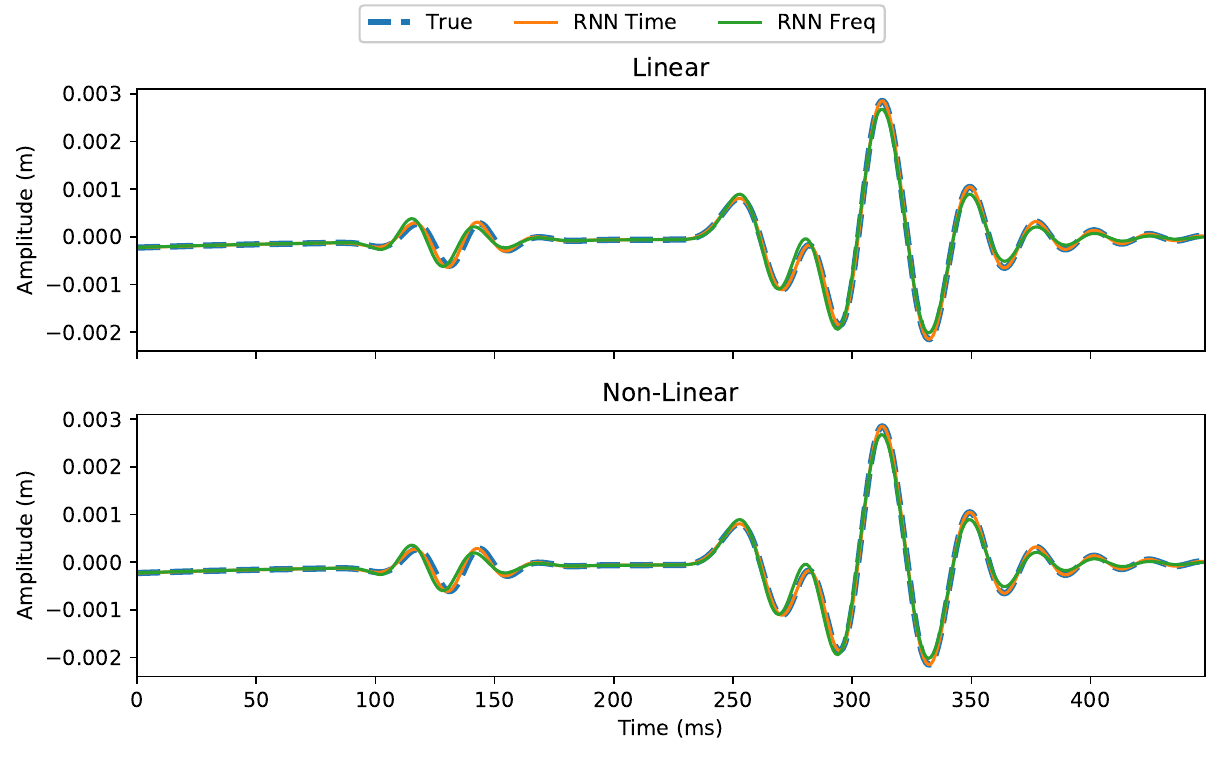}}
\caption[Scattering wave RNN forward modelling.]{Scattering wave RNN forward modelling.}    
\label{fig:rnn_2d_scattering}
\end{figure}

The remaining wavefield components are scattering waves. A constant velocity model of 1500 \si{ms^{-1}} was created with a 1550\si{ms^{-1}} point scatterer (Figure~\ref{fig:rnn_2d_scattered_vel}). RNN implementations were modelled to be depended non-linearly on the scattering amplitude and then approximately linearised. The results are given in Figure~\ref{fig:rnn_2d_scattered_wavefield}. The direct wave was not included in the scattered wavefield reconstruction. Similarly to previous components, scattering are modelled successfully.

Table~\ref{tab:rnn_2d_direct_multi} lists quantitative metrics for the wavefield components. RNN Freq was found to be better for imaging the direct wave (Table~\ref{tab:rnn_2d_direct_multi}), with an improvement of 0.01 in error tolerance and 0.3\% \ac{RPE}. Metrics in Table~\ref{tab:rnn_2d_reflection} and Table~\ref{tab:rnn_2d_scattering} indicate that RNN Time matches the 2D Green’s function near perfectly, whilst RNN Freq introduce error of less than 0.04 and 0.1\% \ac{RPE}. RNN Time is able to model the wavefield within a maximum 0.06 error tolerance and 1.74\% RPE, whilst RNN Freq is overall more accurate with 0.05 and 1.449\% respectively. Given these metrics and the observed models, the discrepancies between the analytical solution and the RNN implementation are deemed acceptable and should be suitable for the modelling process.

\begin{table*}[ht!]
    \footnotesize
    \centering
    \subbottom[Direct wave.\label{tab:rnn_2d_direct_multi}]{
        \begin{tabular}{@{}lcc@{}}\toprule
            Modelling  & Error Tolerance    & RPE (\%) \\ \hline
            RNN Time   & 0.060 & 1.740            \\
            RNN Freq   & 0.050 & 1.449            \\ \hline
        \end{tabular}}\qquad
    \subbottom[Reflected and transmitted wave.\label{tab:rnn_2d_reflection}]{
        \begin{tabular}{@{}lcc@{}}\toprule
            Modelling  & Error Tolerance    & RPE (\%) \\ \hline
            RNN Time   & 0.001 & 0.002                  \\
            RNN Freq   & 0.020 & 0.013                  \\ \hline
        \end{tabular}}
        
    \subbottom[Scattering wave.\label{tab:rnn_2d_scattering}]{
        \begin{tabular}{@{}lcc@{}}\toprule
            Modelling  & Error Tolerance    & RPE (\%) \\ \hline
            RNN Time – Non-linear	& 0.003	& 0.010     \\
            RNN Time – Linear   	& 0.010 & 0.025     \\
            RNN Freq – Non-linear	& 0.030	& 0.076     \\
            RNN Freq – Linear	    & 0.040 & 0.097     \\ \hline
    \end{tabular}}
    \caption[Empirical comparison of 2D wavefield components.]{Empirical comparison of 2D wavefield components.}
\end{table*}

\clearpage
\subsection{Gradient Comparison}
The gradient of the cost function defines the direction in which the model needs to be updated to reach a global minimum (§~\ref{sec:theory_model_update}). Classical FWI approaches generally use the adjoint state method to calculate gradients or the finite differences approach (although computationally expensive), whereas DNN frameworks use automatic differentiation. Theoretical equivalence has been shown in Appendix~\ref{sec:app_results_equivalence_AD_adjoint}, and we now confirm computational equivalence following the approach described by the work of \cite{Richardson2018}.

\begin{figure}[ht!]
    \centering
    \includegraphics[width=0.9\textwidth]{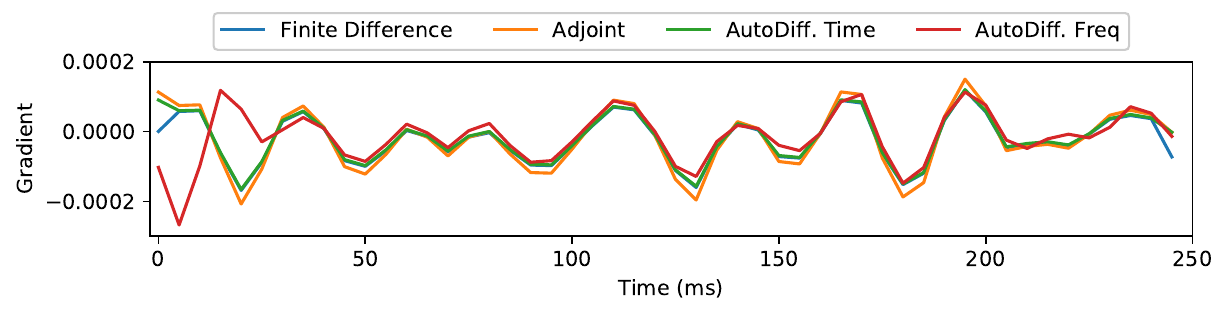}
    \caption[Gradient comparison of RNN implementation with classical approaches.]{Gradient comparison of of RNN implementation with classical approaches. AutoDiff. is the automatic differentiation implementation in Tensorflow v2.0.}
    \label{fig:rnn_gradient_comp}
\end{figure}

A random 1D model was generated, randomly perturbed and gradient of cost function evaluated along the trace. Figure~\ref{fig:rnn_gradient_comp} and Table~\ref{tab:rnn_gradient_comp} compare the gradients at each point for classical finite differences and adjoint techniques to automatic differentiation (AutoDiff.). The adjoint state and AutoDiff. Freq react similarly and slightly over-estimates the gradient, with the pseudo-spectral approach being worse. AutoDiff. Time under-estimates the gradient with an infinitesimal error. Gradients deviate at the edges in either case, with AutoDiff. Freq producing evident perturbation in the initial few time-steps. This is due to the choice of the batch-size within the inversion process and is further discussed in subsection §~\ref{sec:rnn_hp_tunin}. Although this might seem worrying, the scale of this deviation is very minimal and no concerning effects were observed within the previous experimentation leading to this investigation. The other discrepancies are attributed to numerical inaccuracies as per \cite{Richardson2018}.
\begin{table*}[!ht]
    \footnotesize
    \centering
    \begin{tabular}{@{}lccc@{}}\toprule
        \textbf{Finite Difference gradient baseline} & Adjoint & AutoDiff. Time & AutoDiff. Freq \\ \hline 
        Error tolerance                     & $1.000\times10^{-5}$ & $-2.196\times10^{-9}$      & $3.000\times10^{-4}$       \\
        RPE (\%)                            & $0.593$   & $1.302\times10^{-5}$      & $1.779$        \\ \hline 
    \end{tabular}
    \caption{Empirical comparison of gradient calculations.}\label{tab:rnn_gradient_comp}
\end{table*}

\subsection{Hyper-Parameter Tuning}\label{sec:rnn_hp_tunin}
\begin{figure}[ht!]
    \centering
    \includegraphics[width=0.95\textwidth]{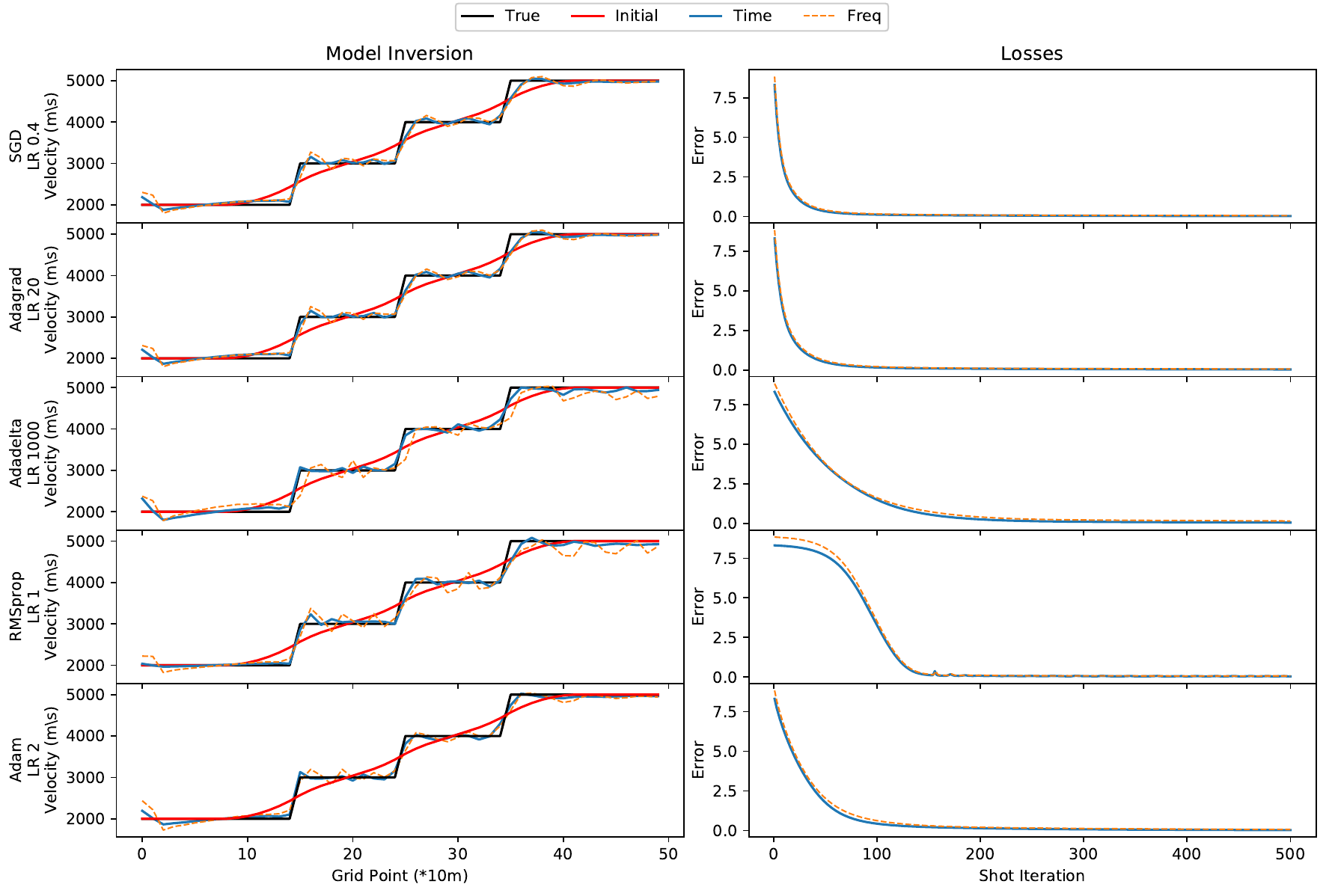}
    \caption[Hyper-Parameter tuning.]{Tuning of hyper-parameters to identify ideal loss optimizer combination.}
    \label{fig:rnn_hp_tuning}
\end{figure}

Similarly to the approach shown in \cite{Sun2019}, a benchmark 1D 4-layer synthetic profile, with velocities [2, 3, 4, 5]\si{kms^{-1}}, was used to identify the ideal parameters for the RNN architecture. This is illustrated as the Black line in Figure~\ref{fig:rnn_hp_tuning}. Classical 1D second-order FD modelling was used to generate the required true receiver data. Multiple learning rates for the different loss optimizers were investigated to try and identify the ideal combination. Figure~\ref{fig:rnn_hp_tuning} shows the best combination for all losses with an ideal batch size of three. The full investigation for this tuning is given in Appendix~\ref{sec:app_results_rnn_hp_tuning}. 

Left side of Figure~\ref{fig:rnn_hp_tuning} shows the inverted velocity profiles, with Red being the initial velocity profile. For Stochastic Gradient Descent, the learning rates was found to be both between zero and one. This is as expected and follows conventional loss optimization. On the other hand, the other loss optimizers had to be scaled to beyond one due to the magnitude differences brought by accumulated squared-norms of the gradients as investigated by \citet{Sun2018}. This is allowed provided the scaling coefficient is between zero and one. For Adagrad, following from \cite{Duchi2011}, the $\beta$ hyper-parameter was fixed at 0.9 and learning rate found to be 20. Adadelta, RMSprop and Adam optimal learning rates were identified at 1000, 1 and 2 respectively. 

The right side of Figure~\ref{fig:rnn_hp_tuning} gives the loss progression. All optimizers iteratively reduce the error with additional shots and on similar scales. Stochastic Gradient Descent and Adagrad do this relatively sooner than the rest, yet the inverted velocity is not as good as the other optimizers. RMSprop follows a rather slow gradual decrease in loss, which then sudden increases. This is expected given that RMSprop updates are derived from a moving average of the square gradients and require an inertial start.

\begin{figure}[ht!]
    \centering
    \includegraphics[width=0.9\textwidth]{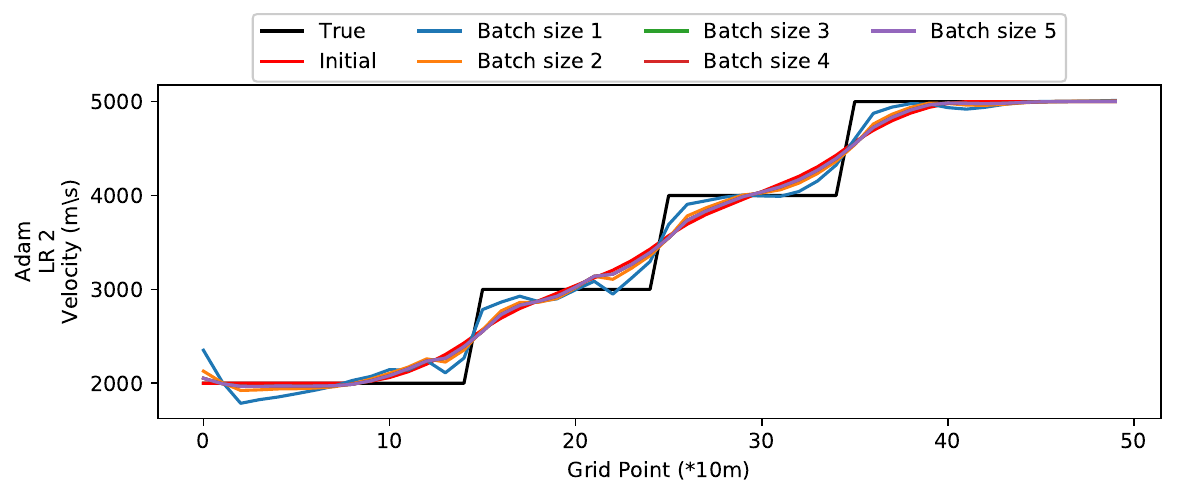}
    \caption[Effect of batch-size on inversion process.]{A smaller batch-size introduces error at the initial part of the velocity profile due to more localized updates. This was derived for a time time implementation for RNN architecture with Adam loss optimizer and learning rate of 2.}
    \label{fig:rnn_hp_tuning_batch_size}
\end{figure}

Based on this investigation, \textbf{Adam} with a learning rate of 2 was identified as the best optimizer. This provided the most stable inversion for either RNN Time or Freq, with the most update and reasonable error loss performance. Mis-match in the shallow part of the velocity is due to the choice of batch-size within the RNN update process. Figure~\ref{fig:rnn_hp_tuning_batch_size} shows the Adam optimizer fixed with learning rate 2 and inverted for batch sizes ranging from one to five. The smaller the batch size, the greater the error since the inversion is more localized and amplifies the gradient onset error shown in Figure~\ref{fig:rnn_gradient_comp}. The larger the batch size, the better the inversion as more data is being used. This poses a limitation since batch size is limited by the Graphical Processing Unit RAM. Given fore-sight that this approach will be used on a large dataset, this was taken as a caveat and batch size fixed at one for the rest of the implementation. 

\clearpage
\subsection{2D Synthetic}
\subsubsection{True and Initial Models}
\begin{figure}[ht!]
    \centering
    \includegraphics[width=0.98\textwidth]{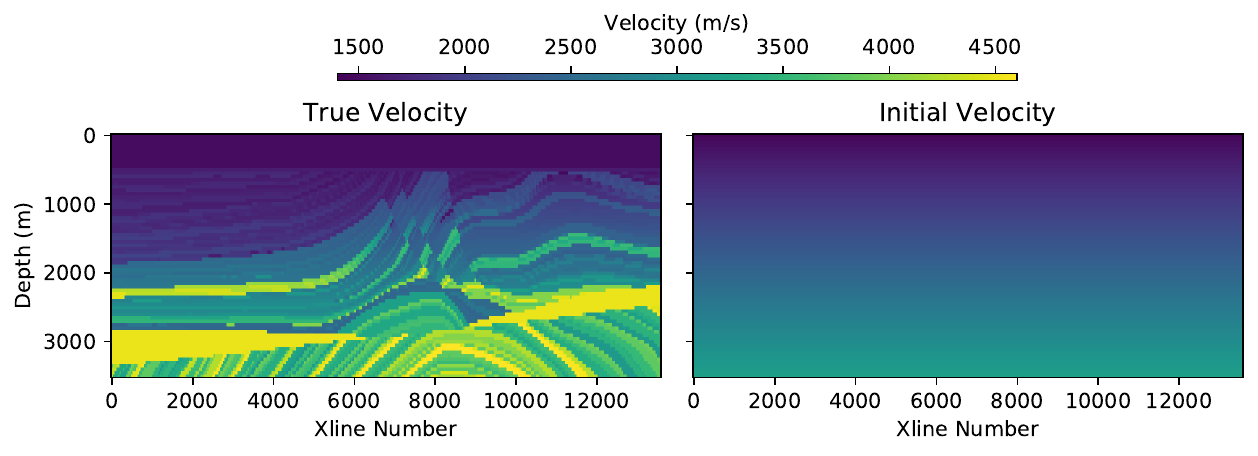}
    \caption[Synthetic 2D Marmousi models for RNN training.]{50\si{m}$\times$50\si{m} grid 2D Marmousi models for RNN training.}
    \label{fig:rnn_marm_models}
\end{figure}

A 2D scalar model of the Marmousi-2 was formulated similar to that in §~\ref{sec:results_marmousi_dataset}. This was re-sampled to a 50\si{m}$\times$50\si{m} grid and smoothed to create the initial model model. These velocity models are plotted in Figure~\ref{fig:rnn_marm_models}. True synthetic receivers were computed by forward modelling through the RNN framework. 56 shots at 300m intervals at depth 200m were generated with a Perfectly Matched Layer at the boundaries. Receivers were set at 50m intervals and modelled for 12\si{s} duration.

\subsubsection{Training of RNN}
\begin{figure}[ht!]
    \centering
    \includegraphics[width=0.9\textwidth]{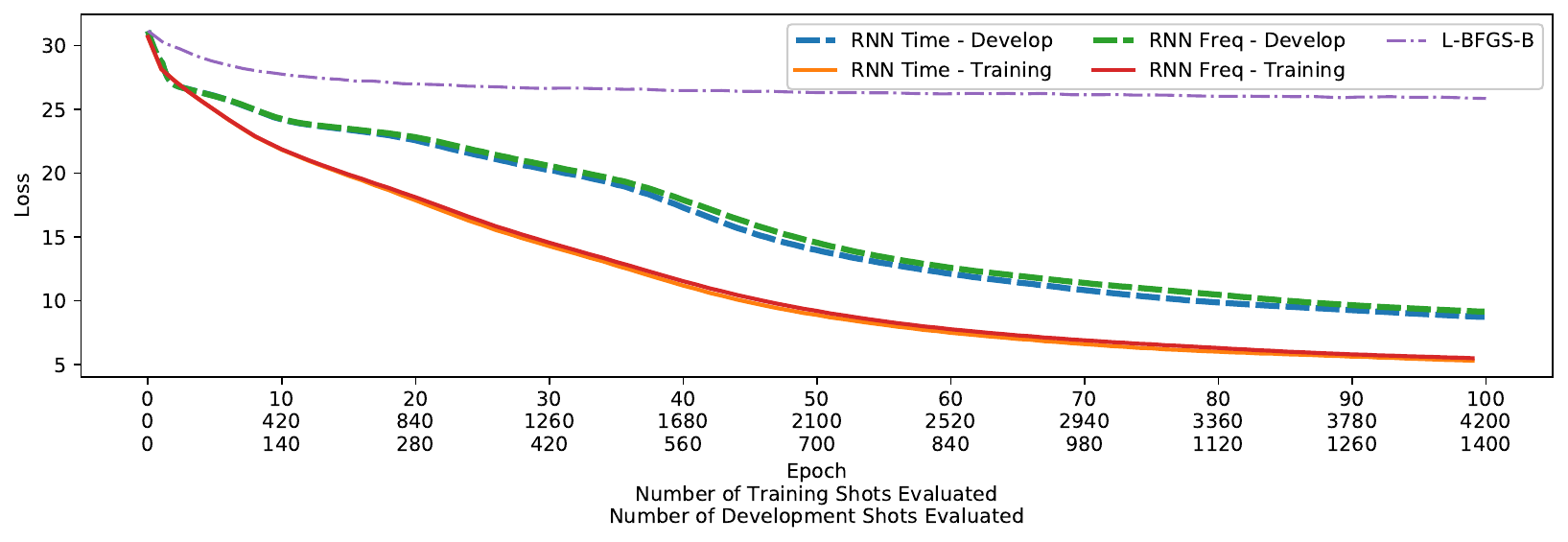}
    \caption[RNN loss performance.]{RNN loss performance for RNN training and development datasets using Adam optimizer with learning rate of 2 and batch size 1. The horizontal labels shows the epoch number and respective number of shots evaluated for training and development. L-BFGS-B is the cost function evaluation for classical FWI plotted on shot number equivalent. Either RNN approach converge quicker than L-BFGS-B, and RNN Freq provides a more stable convergence and better performance then RNN Time.}
    \label{fig:rnn_losses}
\end{figure}

As in standard RNN approaches, the receiver dataset was split into a training and development datasets with at 75\%-25\% split. Training was run for 100 epochs, with early stopping on an NVIDIA Titan V Graphical Processing Unit courtesy of Istituto Nazionale di Geofisica e Vulcanologia. Development loss was calculated every 5th training shot. Figure~\ref{fig:rnn_losses} gives the RNN performance for training and development datasets using Adam optimizer with learning rate of 2.0 and batch size 1. The horizontal labels shows the epoch number and respective number of shots evaluated for training and development. Computational run times are of 14 hours per approach. Both RNN Time and RNN Freq follow similar reductions in loss per epoch and indicate that either implementation converge to an optimal loss. L-BFGS-B loss for classical FWI is shown and is discussed is in the next section.

\subsubsection{Comparison with classical FWI}\label{sec:results_rnn_comparison_to_FWI}
Figure~\ref{fig:rnn_losses} plots the cost function versus the number of shot evaluation equivalent for classical FWI and RNN. The RNN framework is more computationally efficient since either RNN approach converge significantly quicker than L-BFGS-B. RNN Freq provides a more stable convergence and is better performant then RNN Time. The classical FWI is plotted as a shot number equivalent and not the epoch number. The full cost function performance is provided in Appendix~\ref{sec:app_results_classical_FWI}. 

Figure~\ref{fig:rnn_models} compares the inverted velocities and residuals for FWI, together with RNN Time and RNN Freq implementations. Complementary plots showing the model update progressions for this sections are provided as part of Appendix~\ref{sec:app_results_rnn_update_progress}. The true model velocity in Figure~\ref{fig:rnn_models} identifies three zoomed areas which are shown in Figure~\ref{fig:rnn_model_zoomed_in} and Figure~\ref{fig:rnn_velocity_profiles} are velocity profiles taken at 2000 Xline intervals. Figure~\ref{fig:rnn_model_spectra} show the resolution spectra derived via FFT on the velocity models. Comparing FWI and the RNN model in either of these figures, it is clear that the resolution recovery is different. Figure~\ref{fig:rnn_model_spectra} confirms the frequency content in these approaches and shows how RNN models invert more of the lower frequencies in Zoom 2 and Zoom 3. In Zoom 1, FWI is slightly better at frequency recovery beyond 25Hz.

Residual plots (Figure~\ref{fig:rnn_models}-{~\ref{fig:rnn_model_zoomed_in}) and the velocity profiles (Figure~\ref{fig:rnn_velocity_profiles}) show how RNN approaches are able to recover more of the signal in the shallow right side (Zoom 1) and the over-thrust middle area (Zoom 2) of the model. Almost all the signal up to depth 1500m is inverted correctly in Zoom 1 whereas over-thrust faults are near perfectly recovered in Zoom 2 and ~\ref{fig:rnn_model_zoomed_in}F. Zoom 3 is of most interest. The prominent layer at depth circa 2000m is nearly completely missed by RNN models, whereas FWI is able to recover this partially. On the other hand, the deeper 3000m strata are hardly identified with FWI. Residual figures in the full sections show that the RNN model amplitude recover is not as good when compared to FWI (Labels~\ref{fig:rnn_model_zoomed_in}A-B). Indeed, some layers are missed at depth greater than 1500m for Xline number greater than 10,000 (Label~\ref{fig:rnn_model_zoomed_in}C-D). Considering either RNN approach in Figure~\ref{fig:rnn_models}, there is a low-frequency \textit{shadow} artefact introduced till depth 2300m from Xline 0 to 6000 and Xline 8000 to 13900. This is attributed to the practical implementation of batch-size discussed in §~\ref{sec:app_results_rnn_hp_tuning}.

Figure~\ref{fig:rnn_receivers} shows labelled receivers for either model at \ac{CDP} 60, 150 and 300. These CDPs split the model into three sections, representing the different extremities. Label A and B reiterate that the shallow left side is better imaged for FWI, whilst shallow right side is better for RNNs respectively. Label C is the missing high velocity at depth 2000m which has incorrect amplitude for the RNNs, but positioned correctly. Classical FWI has less prominent leakage in this area, yet very evident. Label D is the badly imaged layer at depth between 2000m and 2500m on the right side of the model. Labels E throughout the residuals highlight better low frequency resolution imaging by RNN approaches. Indeed, RNN Freq is able to recover slightly more of these low frequencies and identified by E$^{1}$ and E$^{2}$. Similar improvements are visible throughout the other plots. 

\vspace*{\fill} 
\begin{center}
    \emph{Intentionally left black space.}
\end{center}
\vspace*{\fill}

\begin{figure}[ht!]
    \centering
    \includegraphics[width=0.99\textwidth]{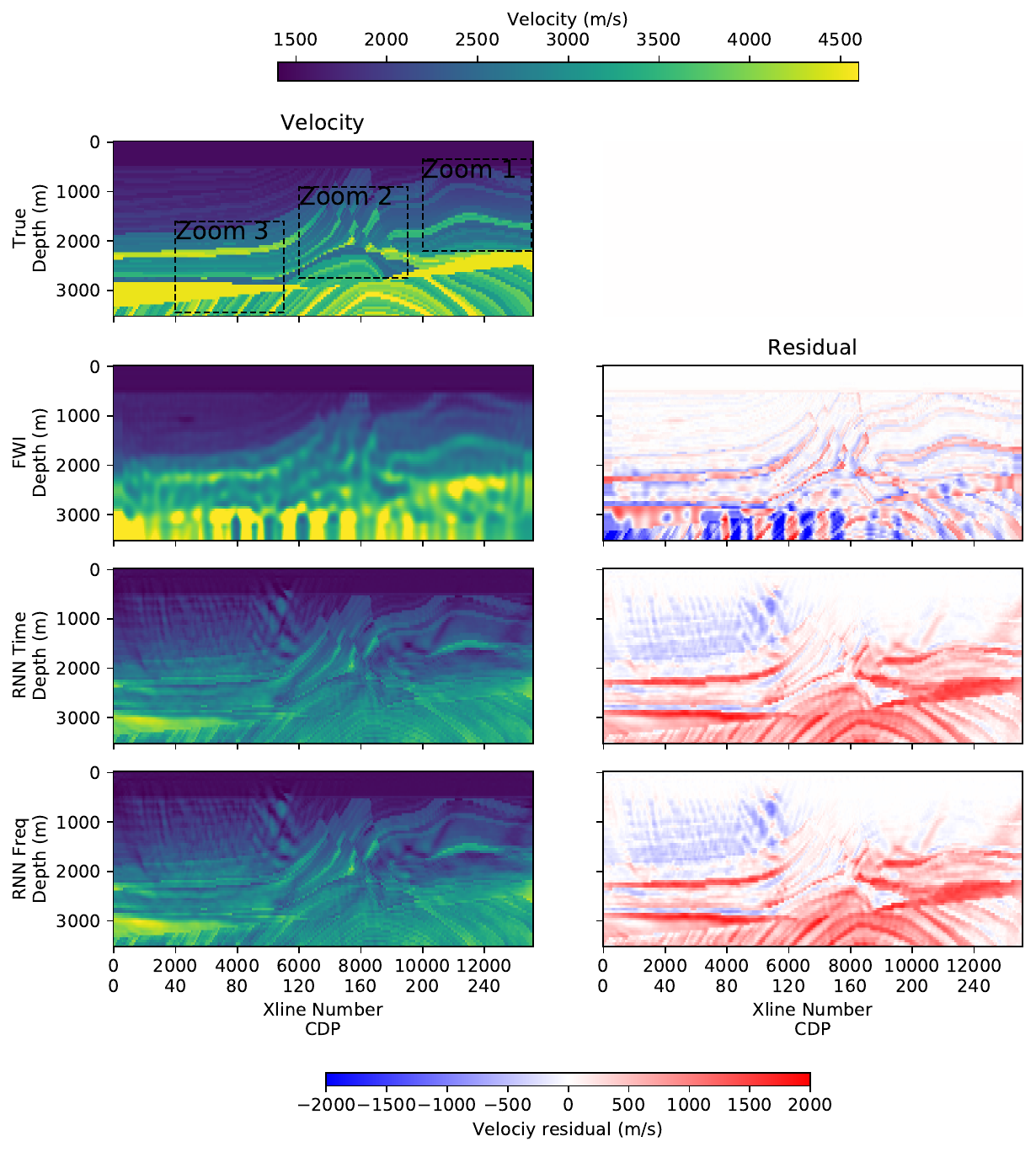}
    \caption[Classical FWI and RNN implementation velocity model inversion.]{Classical FWI and RNN implementation velocity model inversion.}
    \label{fig:rnn_models}
\end{figure}

\begin{figure}[ht!]
    \centering
    \includegraphics[width=0.99\textwidth]{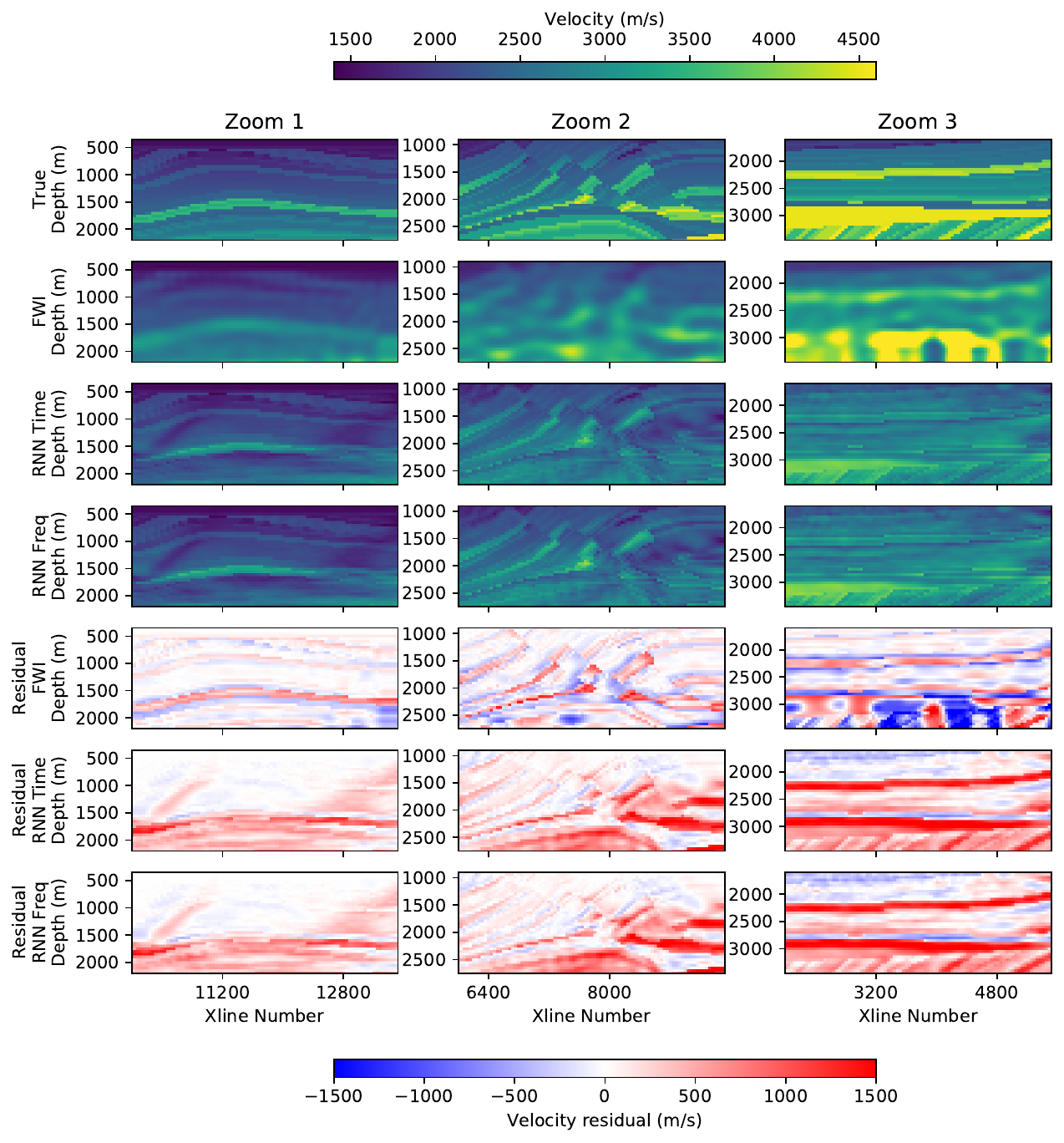}
    \caption[Zoomed In RNN Models]{Zoomed In RNN Models}
    \label{fig:rnn_model_zoomed_in}
\end{figure}

\begin{figure}[ht!]
    \centering
    \includegraphics[width=0.8\textwidth]{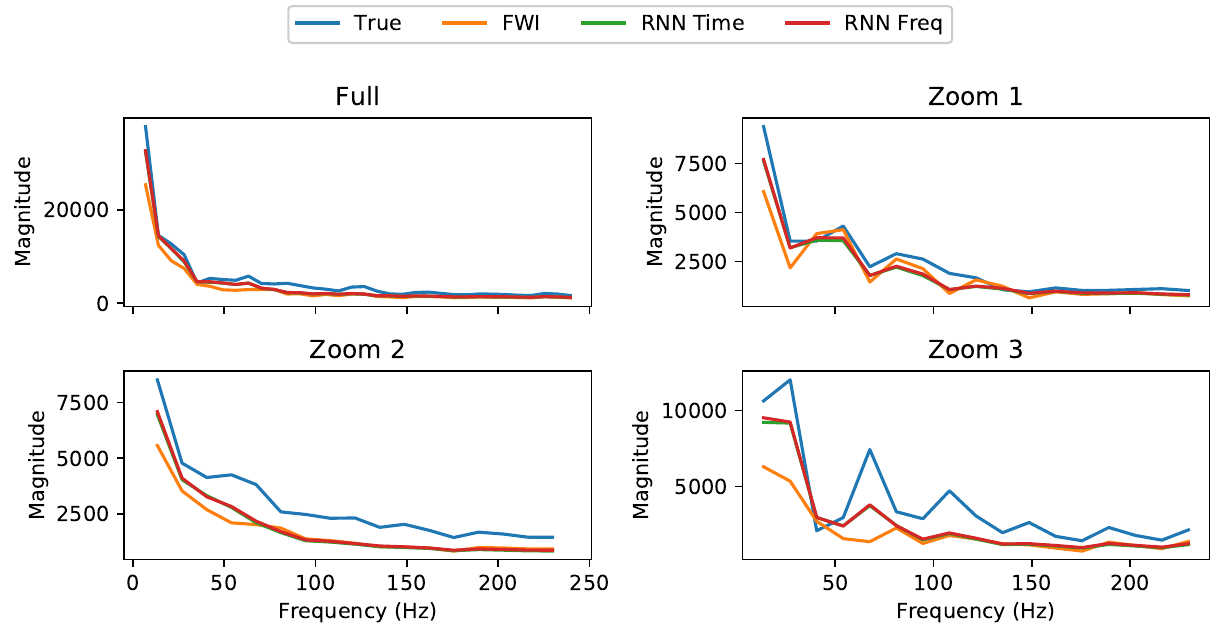}
    \caption[RNN models velocity resolution spectra]{RNN model velocity resolution spectra.}
    \label{fig:rnn_model_spectra}
\end{figure}

\begin{figure}[ht!]
    \centering
    \includegraphics[width=0.9\textwidth]{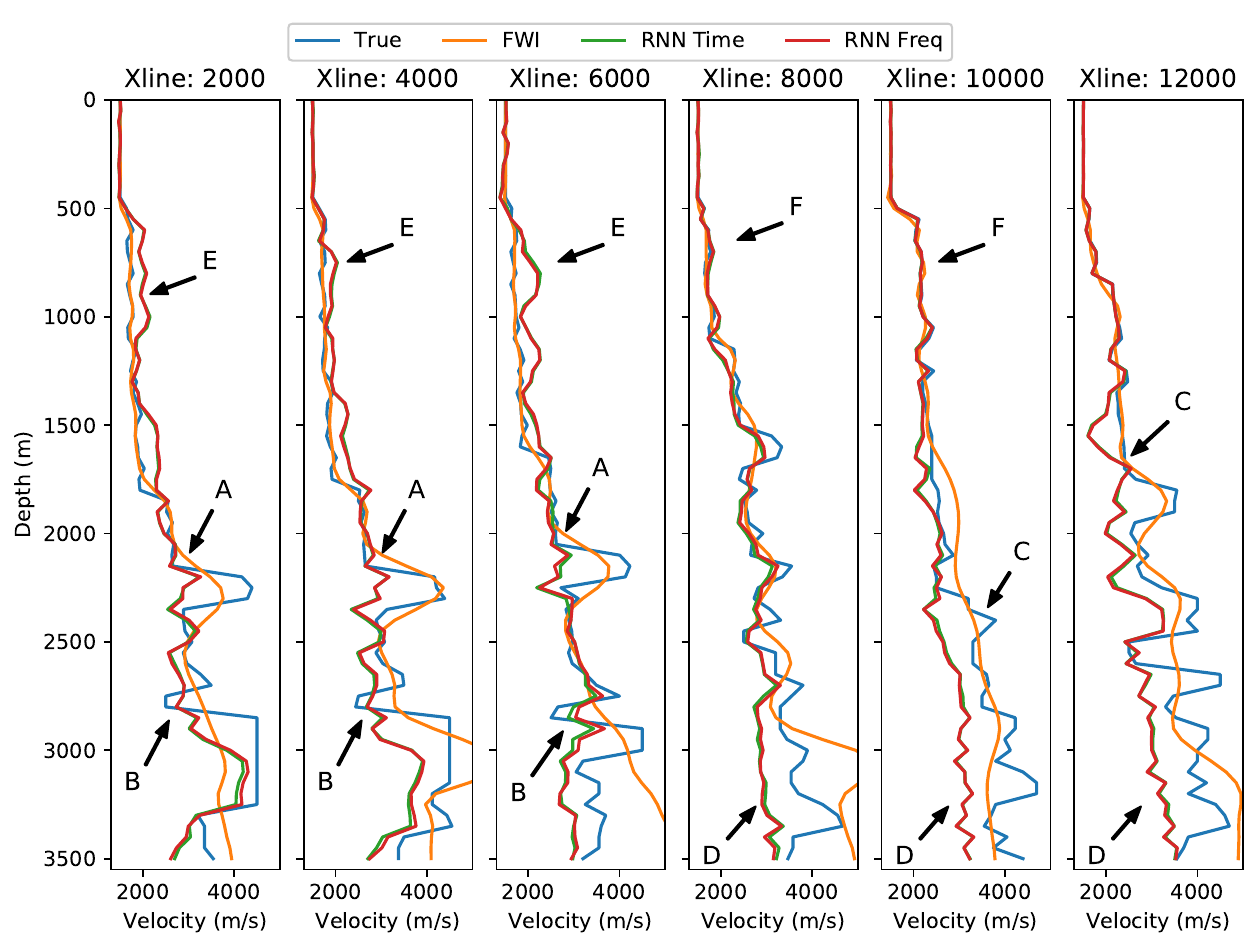}
    \caption[Velocity profiles for RNN and classical FWI.]{Comparison of velocity profiles for RNN and classical FWI. Label \textbf{A-B}: RNN is able to identify strata near perfectly, however unable to inverte the amplitudes values correctly. Label \textbf{C-D}: Missed layers from RNN approaches. Label \textbf{E}: Low frequency artefact for RNN. Label \textbf{F}: Near perfect velocity inversion in the middle Xlines, over shallow depth.}
    \label{fig:rnn_velocity_profiles}
\end{figure}

\begin{figure}[ht!]
    \centering
    \includegraphics[width=0.9\textwidth]{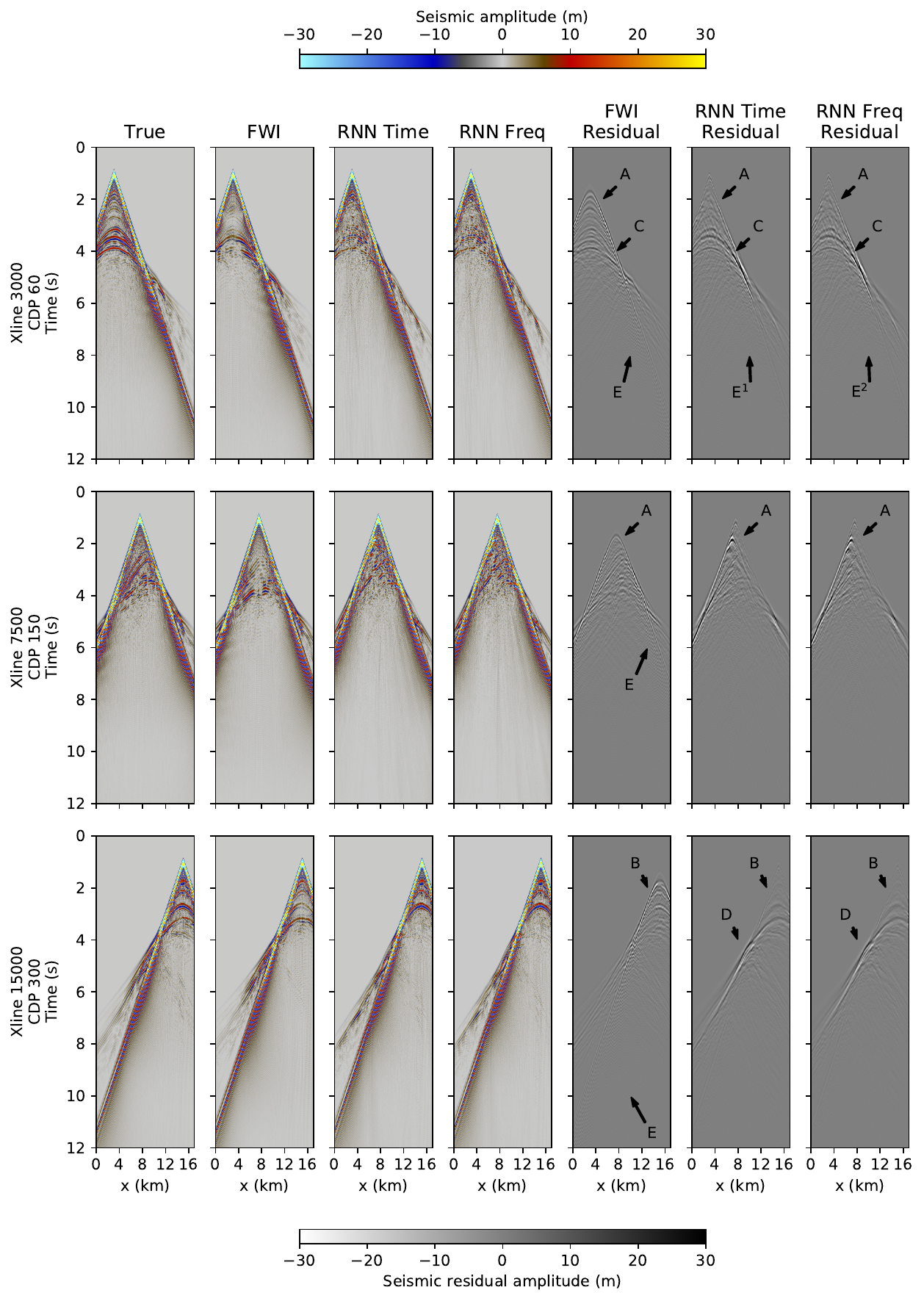}
    \caption[Labelled receivers for True, FWI and RNN models.]{Receivers for True, FWI and RNN models at CDP 60, 150 and 300. Label \textbf{A} and \textbf{B}: Shallow left side is better imaged for FWI. Label \textbf{C}: Missing high velocity with incorrect amplitude but positioned correctly. Label \textbf{D}: Badly imaged layer. Labels \textbf{E}: Better low frequency imaging for either RNN approaches. Labels \textbf{E$^{1}$} and \textbf{E$^{2}$}: RNN Freq is able to recover slightly more low frequencies. 
    }
    \label{fig:rnn_receivers}
\end{figure}

\clearpage
\section{Data-Driven and Theory-Guided NN Frameworks}
In this section, DNN refers to ``Data-Driven FWI'' and RNN to ``Theory-Guided RNN as an Analogue of FWI''.

\subsection{Data Volume}
The volume of data used to train for DNN was 20 orders of magnitude greater. Figure~\ref{fig:dnn_rnn_losses} shows the normalised loss for the two derived frameworks, with respective count for the number of training and validation shot or shot equivalent. Given that the DNN framework used randomly generated traces with each epoch of 1,000,000 and 100,000 trace for training and validation respectively, these were batched into groups of 340 traces to match the number of receivers in the RNN approach.

\begin{figure}[ht!]
    \centering
    \includegraphics[width=0.9\textwidth]{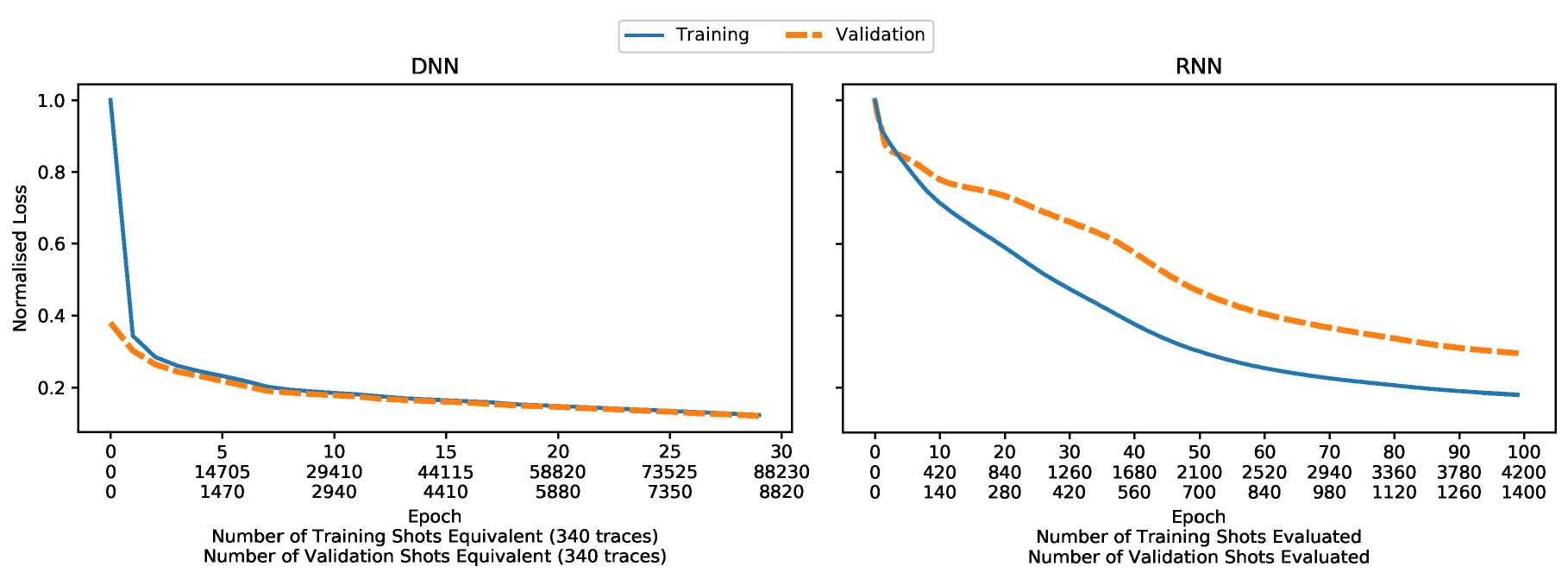}
    \caption[Classical FWI and RNN implementation velocity model inversion.]{Classical FWI and RNN implementation velocity model inversion.}
    \label{fig:dnn_rnn_losses}
\end{figure}

\subsection{Data-Driven Uplift}
DNN produces more imaging uplift since it is not bound by the deterministic forward-modelling physical constraints from ray-tracing. Figure~\ref{fig:fwi_ray_tracing} shows a sample of shots through the Marmousi model. Ray-paths below the high velocity at depth 2-2.5km do not arrive at the receivers at the surface for the current model offsets. Acquisition geometry controlling the offset is a hard-limit within FWI. \citet{Morgan2009} considers large-offsets to be fundamental for successful FWI. Moreover, this hinders seismic imaging in and around salt bodies since, by definition, only one-way ray-paths are considered \citep{Jones2014}. In the case of the data-generators used for the DNN approach, these are free from offset-constraints and do not influence the inversion experiment employed by classical FWI.


\begin{figure}[ht!]
    \centering
    \includegraphics[width=0.95\linewidth]{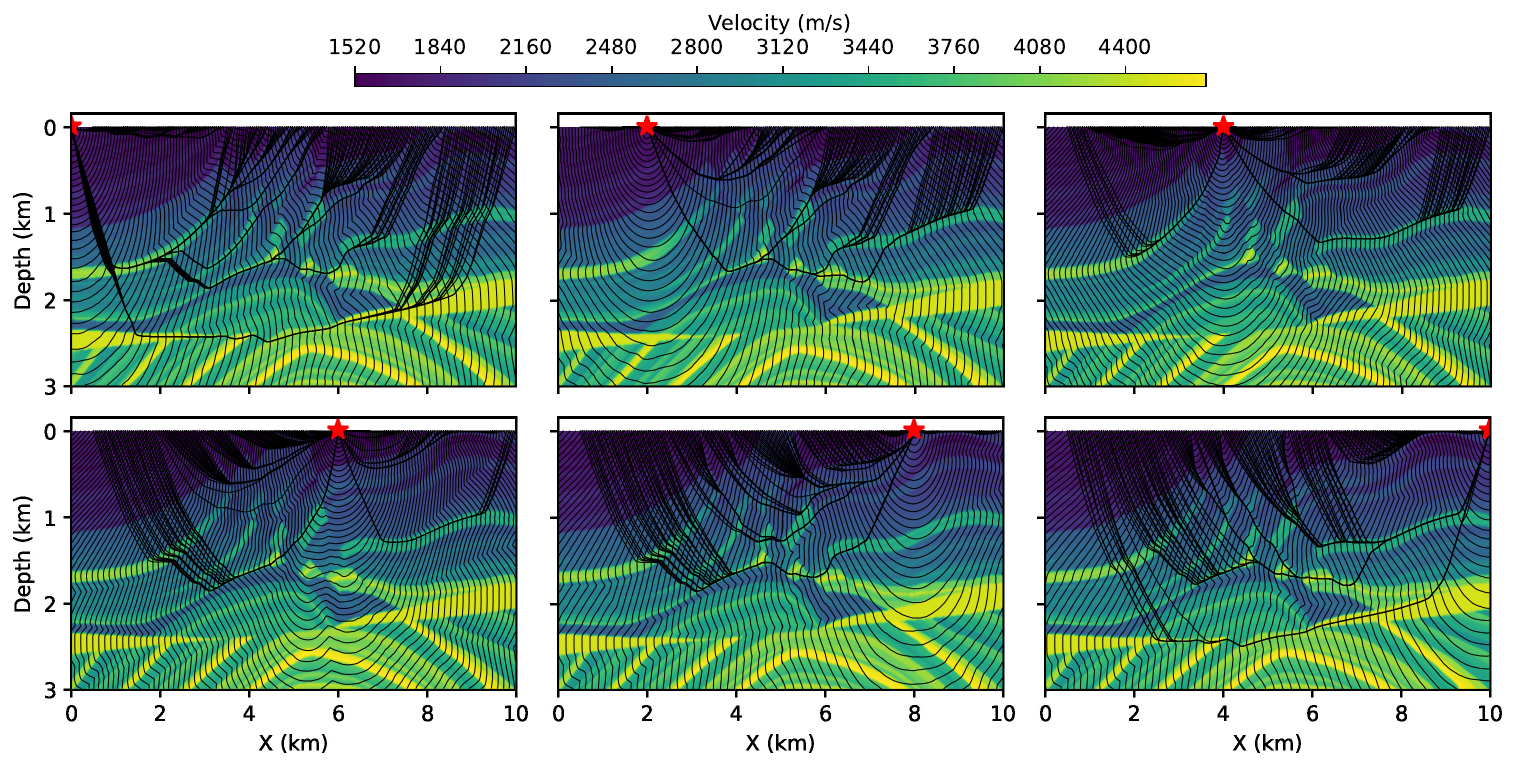}
    \caption[Ray-tracing coverage within forward modelling]{Ray-tracing coverage within forward modelling derived from deterministic geophysics.}
    \label{fig:fwi_ray_tracing}
\end{figure}


\subsection{Inversions}
Figure~\ref{fig:dnn_rnn_models} and~\ref{fig:dnn_rnn_model_zoomed_in} are the full and zoomed-in sections for the inverted models for classical FWI, DNN and RNN respectively. DNN has better layer amplitude continuity as seen by in Zoom 1 and Zoom 3. RNN is better at edge definition than both DNN and FWI as in Zoom 2. 

The velocity profiles in Figure~\ref{fig:dnn_rnn_velocity_profiles} illustrate with Label A how either approach is good in the middle and shallow sections. Indeed, if the RNN artefact labelled A$^{1}$ is excluded in smaller Xlines, this would be valid for either Xline. Label B show how large velocity contrasts within the velocities is well defined for DNN, followed by FWI. In particular, the RNN approach is able to identify the edges, but not able to reconstruct the amplitudes correct. Smaller velocity increases are not an issue as shown by B$^{1}$. A combination of FWI and RNN could potentially exploit the benefit of either. However, not in areas with large velocity contrasts such as salt areas and carbonates. With depth, DNN is a better framework as marked with Label C. Comparing RNN and FWI, RNN is more suited at the edges with Depth as shown with Label C$^{1}$. 

Receivers for pseudo-spectral NN frameworks and true models are shown in Figure~\ref{fig:dnn_rnn_receivers}. Either approach is able to recover different shallow arrivals successfully as marked with Label A. DNN has residual and RNN is able to invert more of the shallow, however FWI recovers direct arrival components nearly completely. Considering CDP 150 and 300, RNN is reinforced as being better performant at the edges from the residuals marked with Label A$^1$. Excluding the gradients artefact of RNN, theses receivers indicate that RNN is the most suited for shallow sections. Both DNN and RNN have leakage on CDP 60 and CDP 300 as shown with Label B. RNN's residual show evidence of signal, possible symptomatic to cycle-skipping. With depth, RNN in general has less residual as shown with Label C.

\begin{figure}[ht!]
    \centering
    \includegraphics[width=0.8\textwidth]{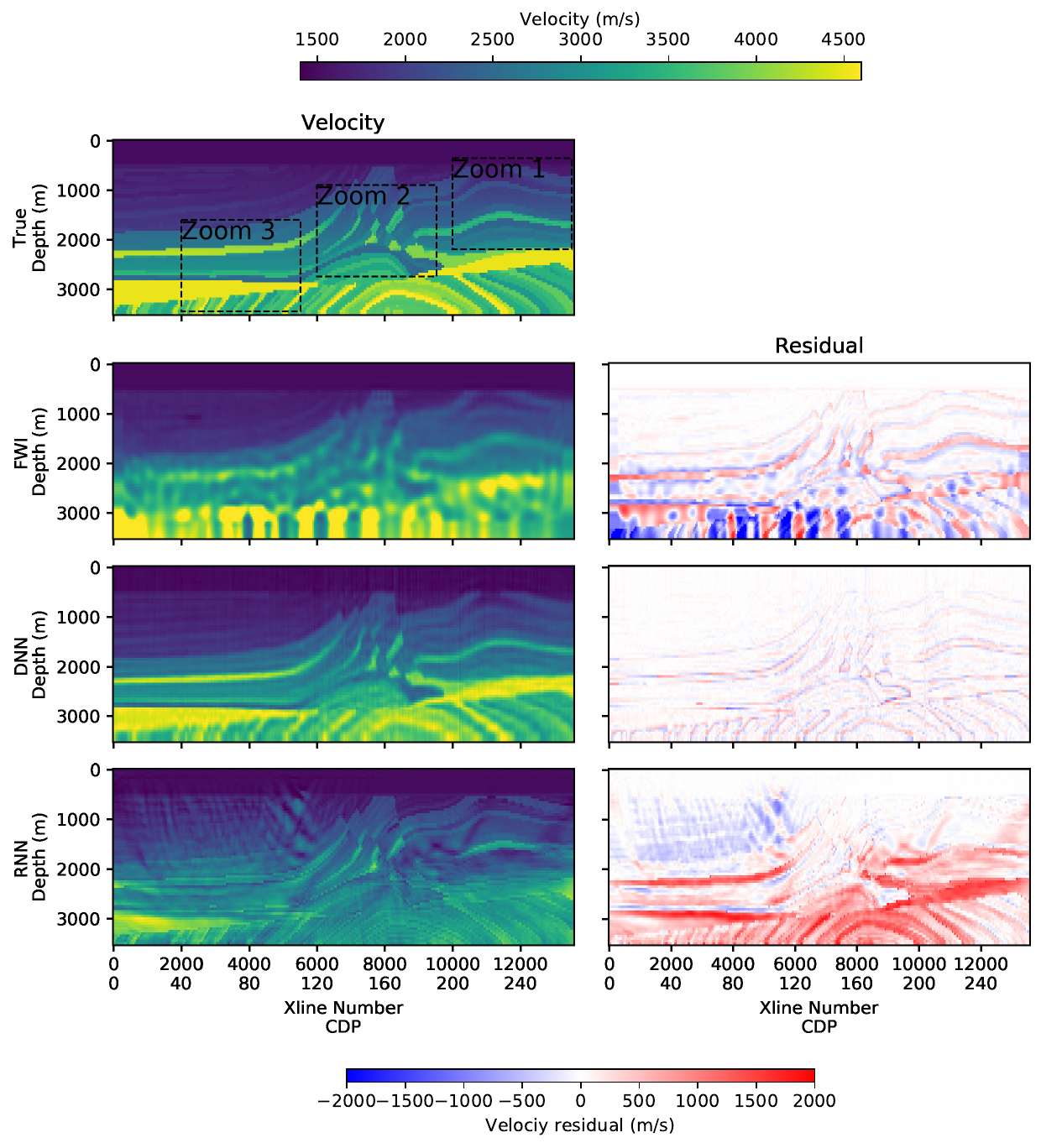}
    \caption[Pseudo-spectral NN framework comparison of velocity inversion.]{Full model showing differences between DNN and RNN velocity inversion as compare to classical FWI, with residual differences.}
    \label{fig:dnn_rnn_models}
\end{figure}

\begin{figure}[ht!]
    \centering
    \includegraphics[width=0.98\textwidth]{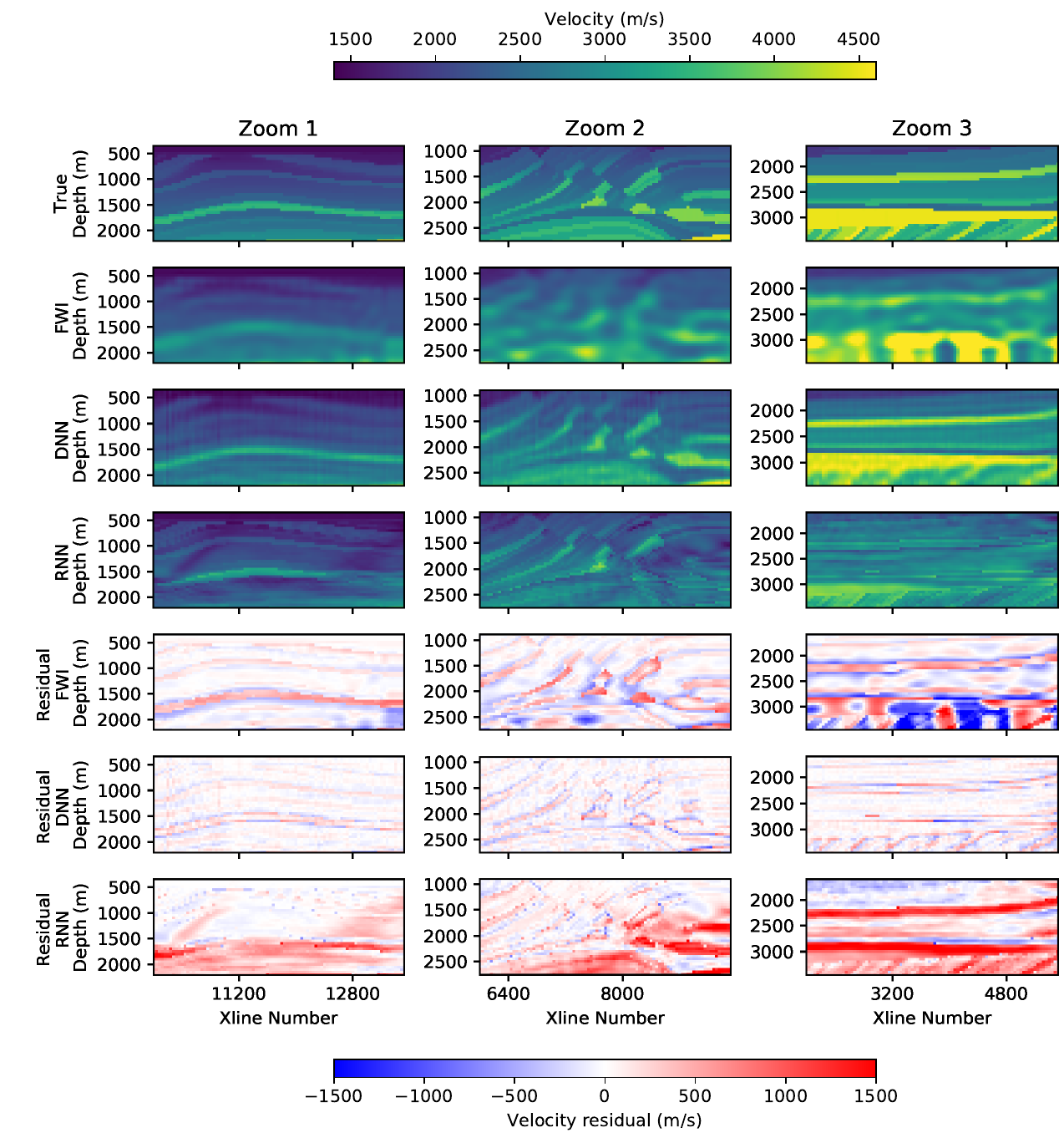}
    \caption[Zoomed In pseudo-spectral NN framework comparison]{Zoomed in sections showing differences between DNN and RNN velocity inversion as compare to classical FWI, with residual differences.}
    \label{fig:dnn_rnn_model_zoomed_in}
\end{figure}

\begin{figure}[ht!]
    \centering
    \includegraphics[width=0.98\textwidth]{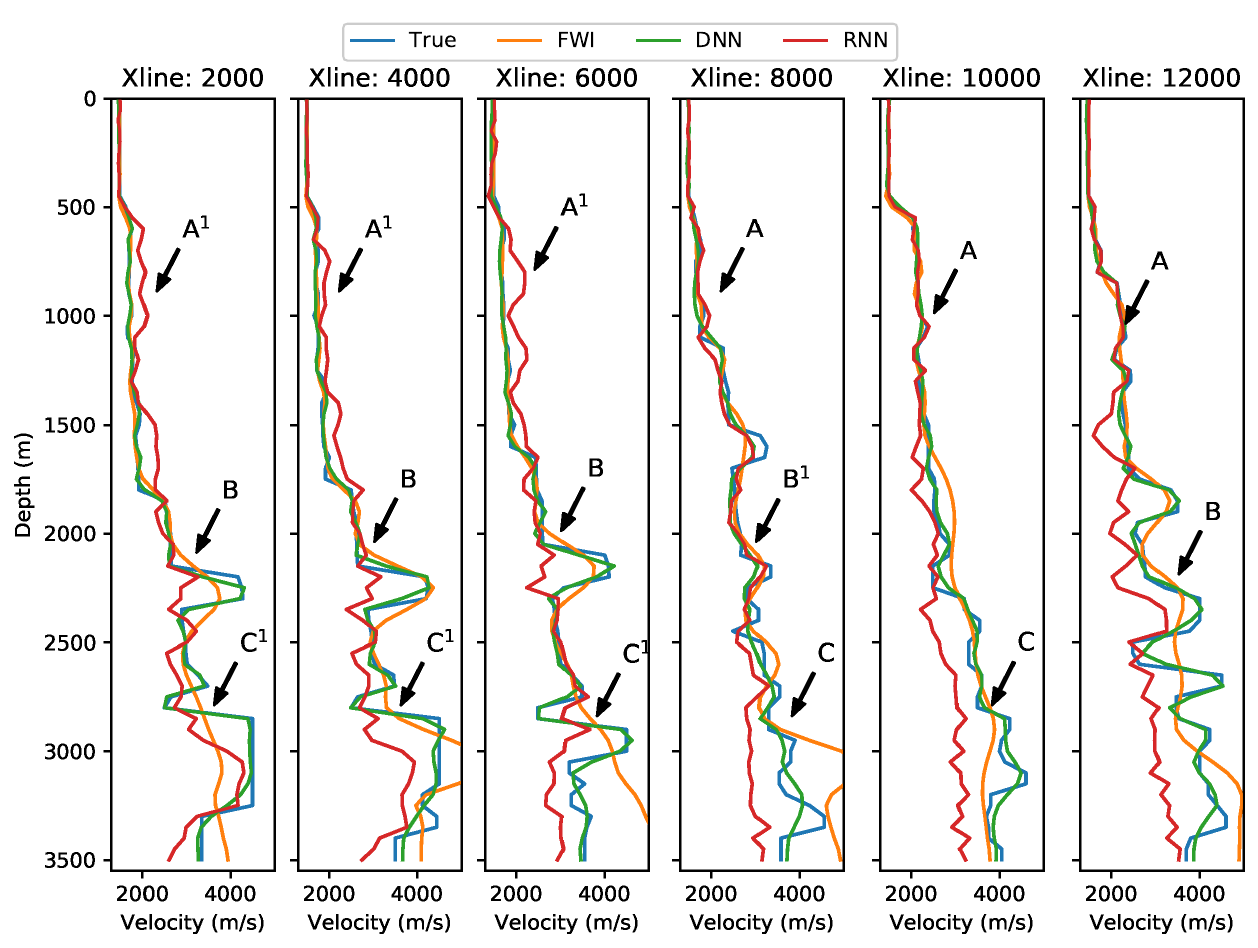}
    \caption[Velocity profiles for pseudo-spectral NN framework comparison.]{Comparison of velocity profiles for DNN, RNN and classical FWI. Label \textbf{A}: Either approach is good in the middle and shallow sections, with the exception of the gradient artefact marked \textbf{A$^{1}$}. Label \textbf{B}: Problematic recover of large velocity contrasts for RNN, whilst smaller velocity contrasts are inverted corrected as marked by \textbf{B$^{1}$}. Label \textbf{C}: DNN is a better framework with depth for velocity, whilst RNN is more suited at the edges as shown with \textbf{C$^{1}$}}
    \label{fig:dnn_rnn_velocity_profiles}
\end{figure}

\begin{figure}[ht!]
    \centering
    \includegraphics[width=0.92\textwidth]{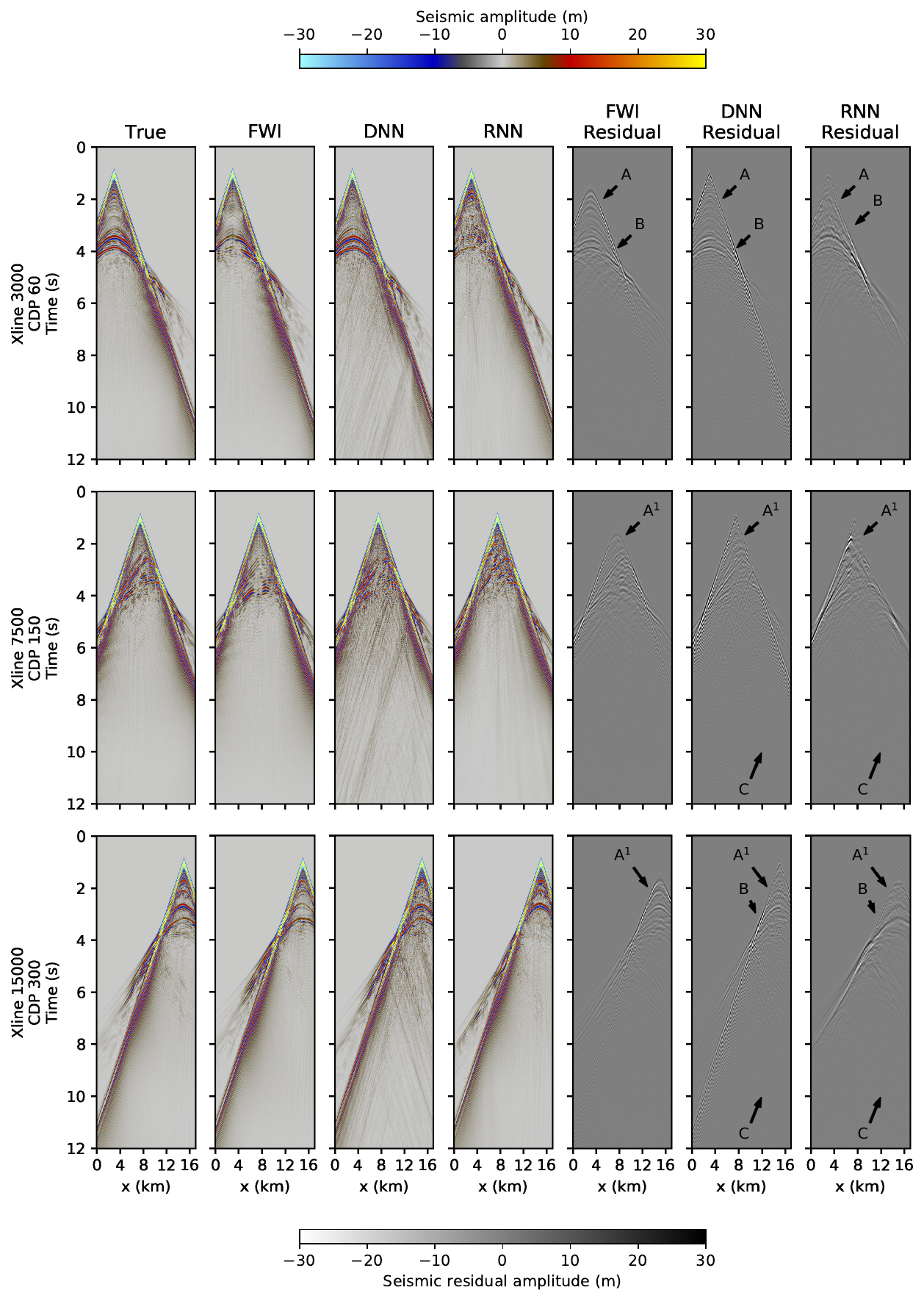}
    \caption[Labelled receivers for pseudo-spectral NN framework comparison.]{Labelled receivers for True, FWI and RNN models at CDP 60, 150 and 300. Label \textbf{A}: NN frameworks recover shallow arrivals successfully, with RNN reinforced as being better performant at the edges from the residuals marked with \textbf{A$^1$}. Label \textbf{B}: Some signal leakage on NN frameworks. Label \textbf{C}: Less leakage from RNN with depth.}
    \label{fig:dnn_rnn_receivers}
\end{figure}

\clearpage
\section{Summary}
The key outcomes from this chapter can be summarised in the following lists. For ``FWI as a Data-Driven DNN'':
\begin{itemize}
    \item Elements within a classical FWI framework can be replaced with a DNN framework successfully. This was practically assessed for multi-strata model using geophysics-generated data.
    \item Normalization is not a necessary component within the framework.
    \item The ideal architecture is a 1D convolutional based (Conv1D), with Adadelta loss optimization.
    \item Kernel and bias distribution plots for the training process confirm that the framework is a learning process.
    \item Most of the update happens within the first few epochs. Additional epochs refine the inverted velocity.
    \item Inversion performance in shallow section was equally good for either classical FWI or DNN approach. DNN framework performs better for deeper and over-thrust areas since DNNs are not bound by forward-modelling physical constraints from ray-tracing.    
\end{itemize}
For ``RNN as an Analogue of FWI'':
\begin{itemize}
    \item Pseudo-spectral RNN frameworks are feasible approaches.
    \item Based on comparisons of a simple model with an analytical Green's function formulation, RNN Time is able to model the wavefield within a maximum 0.06 error tolerance and 1.74\% RPE. RNN Freq is overall more accurate with 0.05 error tolerance and 1.449\% RPE.
    \item Adjoint state and RNN Freq gradients overestimate the Finite Difference, whilst RNN Time under-estimates it with an infinitesimal error. RNN Freq produce a perturbation on the onset of the gradient which is attributed to modelling artefact.
    \item Based on the model size and compute available, the ideal loss was Adam with a learning rate of 2 and batch size of 1. Model batch size proved to be a limitation for practical implementations.
    \item RNN is computationally more efficient than the classical FWI presented in this work. RNN freq shows more stable convergence.
    \item Classical FWI and RNN approaches have merits. RNN frameworks are able to identify faults, but amplitudes are not fully inverted properly. This results in RNN inversions with some missing layers. However, the low frequency content in RNN approaches is better than classical FWI, particularly for RNN Freq.
\end{itemize}
Outcomes of the comparison between the NN approaches:
\begin{itemize}
    \item Well performing DNN was achievable through the use of a very large dataset.
    \item DNN recovers more of the velocity contrast and RNN is better at edge definition.
    \item RNN is the most suited for the shallow sections (exclude the gradient artefact) and at depth due to the cleaner residuals on the receivers. This is only valid in the middle section of the model with the most coverage from ray-tracing. Indeed, RNN approaches have some leakage on the edges which might be symptomatic to cycle-skipping.
    \item RNN Freq is able to recover more low frequencies. 
\end{itemize}

%% file: chap5/conclusions_main.tex
\chapter{Discussion and Conclusions}
\textbf{In this chapter, DNN refers to ``FWI as a Data-Driven DNN'' and RNN as ``Theory-Guided RNN as an Analogue of FWI''. The first section presents a critique to the work, addresses possible pitfalls, identifies areas of potential improvement and suggests directions for future research. This is followed by a final section which presents concluding remarks for this dissertation.}

\section{Critique, Limitations and Future Work}
\subsection{Inversion Paradigm}
DNN, RNN and FWI are three different frameworks which fall within different parts of the inversion spectrum. On one end there is the data-driven approach for DNN, on the other there is purely deterministic geophysics with classical FWI, and in between there is theory-guided RNN.



\begin{figure}[ht!]
    \centering
    \includegraphics[width=0.8\textwidth]{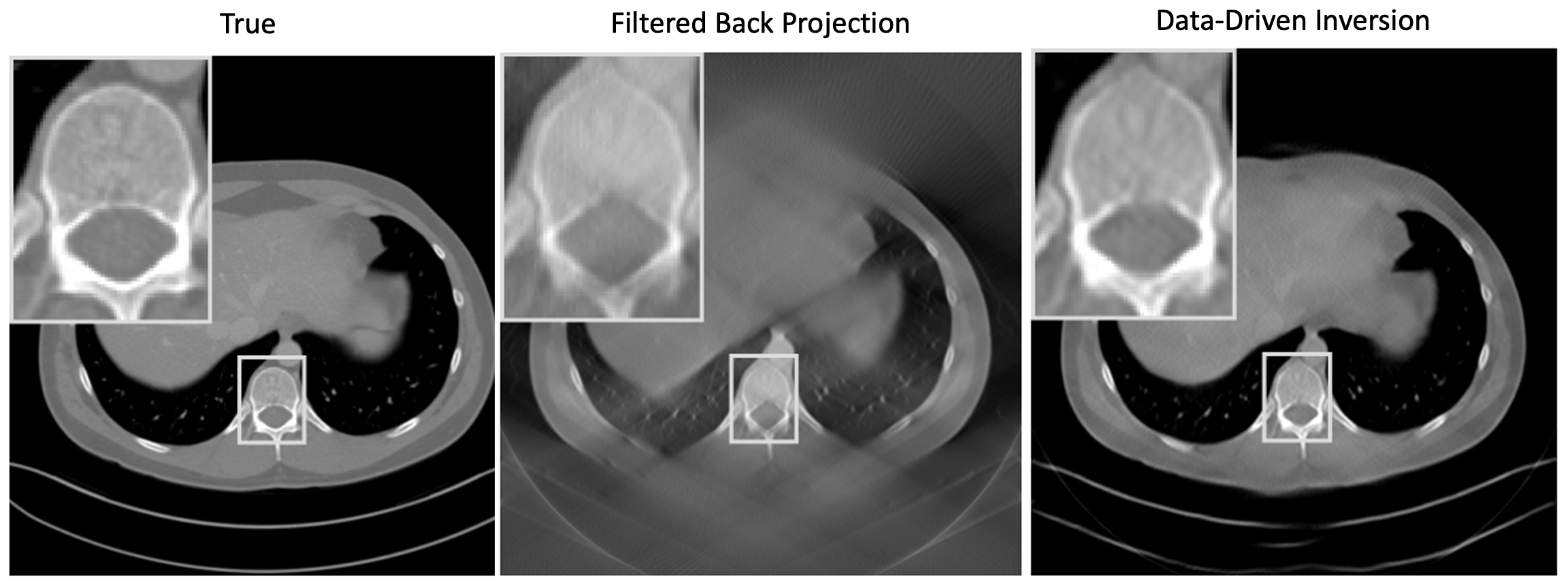}
    \caption[Data-driven inversion for CT-scans.]{Inversion for brain CT-scan, with area of interest. Classical inversion via Filtered Back Projection has areas of poor-illumination due to limited-angle, whilst the data-driven approach offers better inversion. Adapted from \cite{Bubba2019}.}
    \label{fig:fwi_gitta}
\end{figure}
\begin{figure}[ht!]
    \centering
    \includegraphics[width=0.95\textwidth]{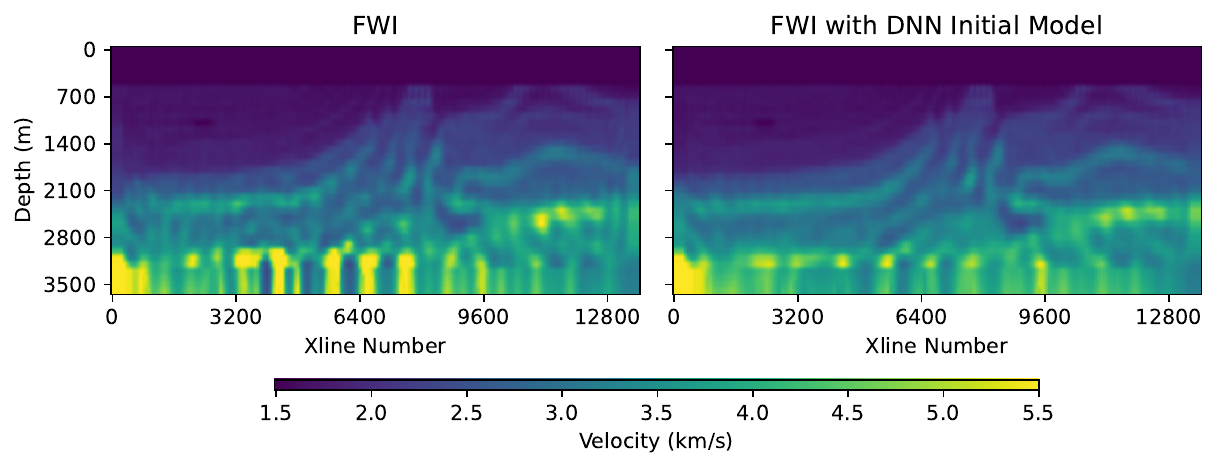}
    \caption[Improved FWI result with DNN as initial model]{Improved FWI result with DNN as initial model.}
    \label{fig:dnn_initial_fwi}
\end{figure}

Within the DNN approach, the inversion component is data-oriented and the data generator is based on geophysics. If data-generators are to include information that might not be deterministically available within a seismic survey, the reconstruction process could invert for this information to \cite{Lewis2017}'s work. A similar approach has already been applied by \cite{Bubba2019} for brain CT-scans (Figure~\ref{fig:fwi_gitta}). These types of data-driven models could be pre-cursors for deterministic models as has been done in the work of \cite{Araya-Polo2018}. Figure~\ref{fig:dnn_initial_fwi} shows inversion for classical FWI without and with DNN as a priori model. The latter approach produces more layer continuity and better imaging at depth. This comes with relatively no extra overhead cost since the DNN would be a pre-trained network. Furthermore, \cite{Grohs2019} analytically determine the upper bounds for the approximation and generalization power for a NN. This could be implemented and provided as a caveat with data-driven inversion by providing confidence maps for the inversion.

Theory-guided inversion inherits advantages and problems from either part of the spectrum. It faces challenges of cycle-skipping and local-minima, whilst it benefits from the use of automatic differentiation to calculate the gradient. This reduces development time as it avoids the need to manually implement and verify the adjoint state method. Furthermore, being at the intersection of physics and computer science, it is inherently strengthened by contributions from two communities of researchers. This opens up possibility of considering other deep learning techniques such as dropout or other acyclic-graph architectures such as directed acyclic graphs \citep{Bogaerts2020}.

In classical FWI, the wavefield at the final time step is affected by the wavefield during the initial time steps. Back-propagation must occur over the entire sequence of time steps for theory-guided RNN. Application of back-propagation through thousands of layers is not a typical application in deep learning applications and automatic differentiation is not designed to efficiently handle such situations. Strategies common to other FWI frameworks to reduce memory requirements could be translated into the field. Examples would include not saving the wavefield at every time step \citep{Nguyen2015}, applying compression to the wavefield time slices \citep{Boehm2015,Kalita2017}, saving wavefields to disk rather then memory \citep{Shen2015}, and regenerating wavefields during back-propagation rather than storing them \citep{Malcolm2016, Yao2020}. Data-driven inversion is gradually proving itself to be a field of its own. An account for some of the main contributions is provided by \cite{Arridge2019}. It is up to the end user to decide to which level of confidence they are willing to base their inversion on data.

\subsection{Training Datasets for Real Data}\label{ref:sec_disc_development_dataset}
\begin{figure}[ht!]
    \centering
    \subbottom[The first two columns are inversions for velocity profiles within the generator limit for the number of layers. For more layers, the inversion tries to generalize for these profiles but misses some components. The inversion is still able to identify some layers, but not to the same resolution as if trained on those layers.\label{fig:dnn_discussion_missing_layer}]{\includegraphics[width=0.9\textwidth]{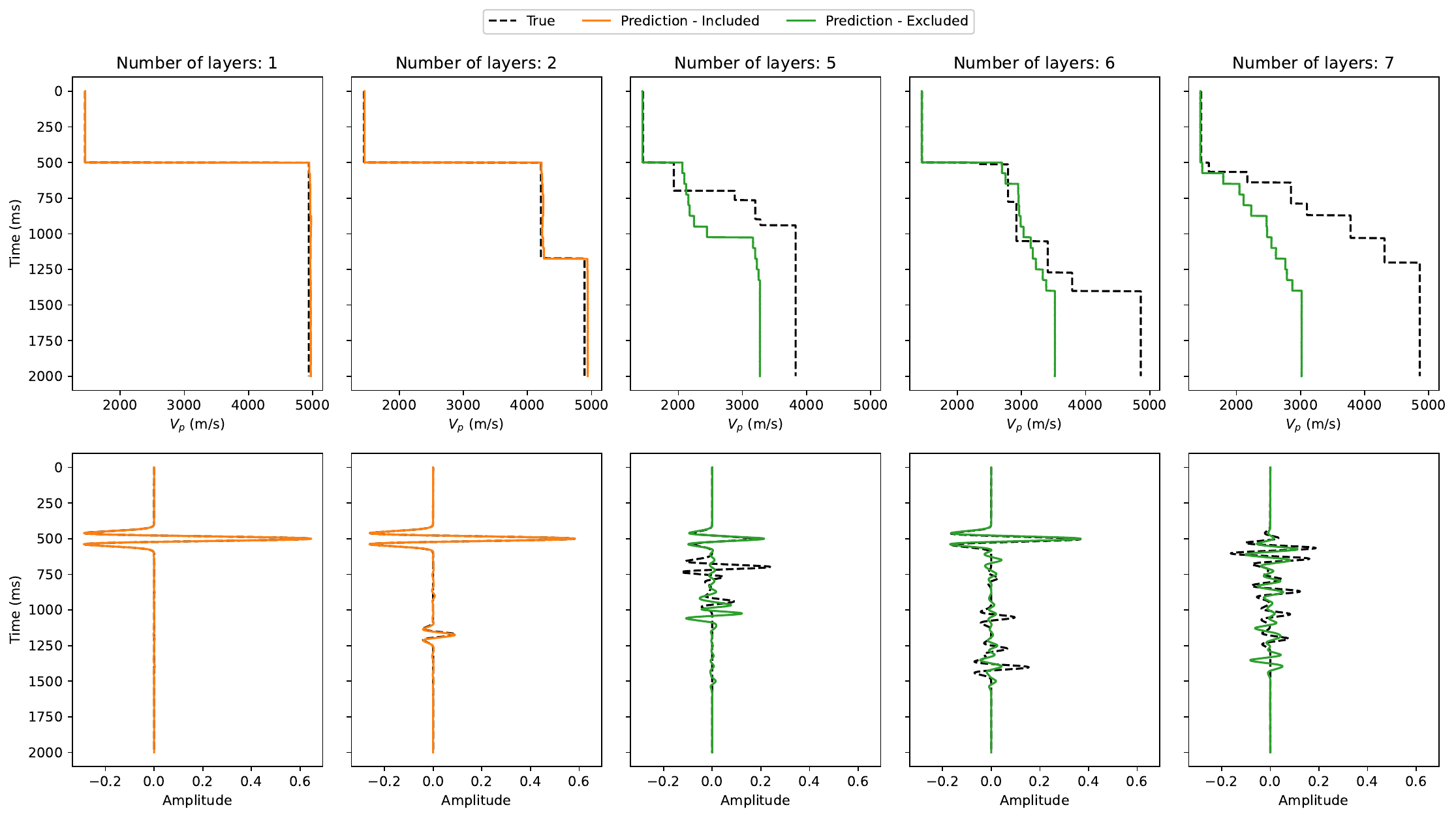}}
    \subbottom[Large velocity inclusions and Velocity inversion would be missed as well.\label{fig:dnn_discussion_Inclusion_Inversion}]{\includegraphics[width=0.9\textwidth]{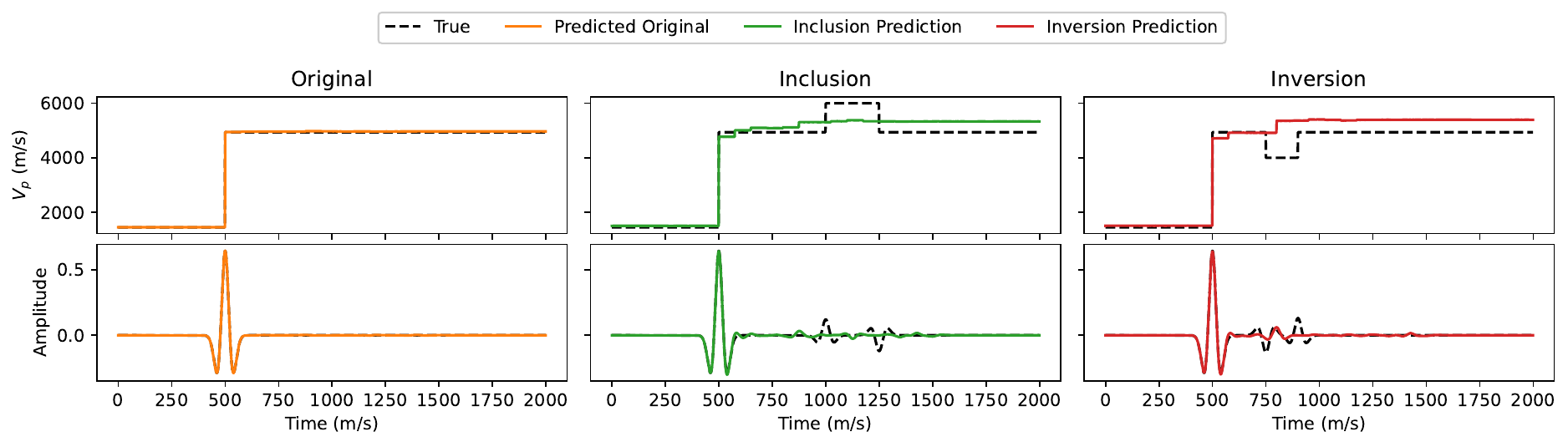}}
    \subbottom[Robustness to noise up to 25\% pink noise contamination.\label{fig:dnn_discussion_Noise_Sensitivity}]{\includegraphics[width=0.9\textwidth]{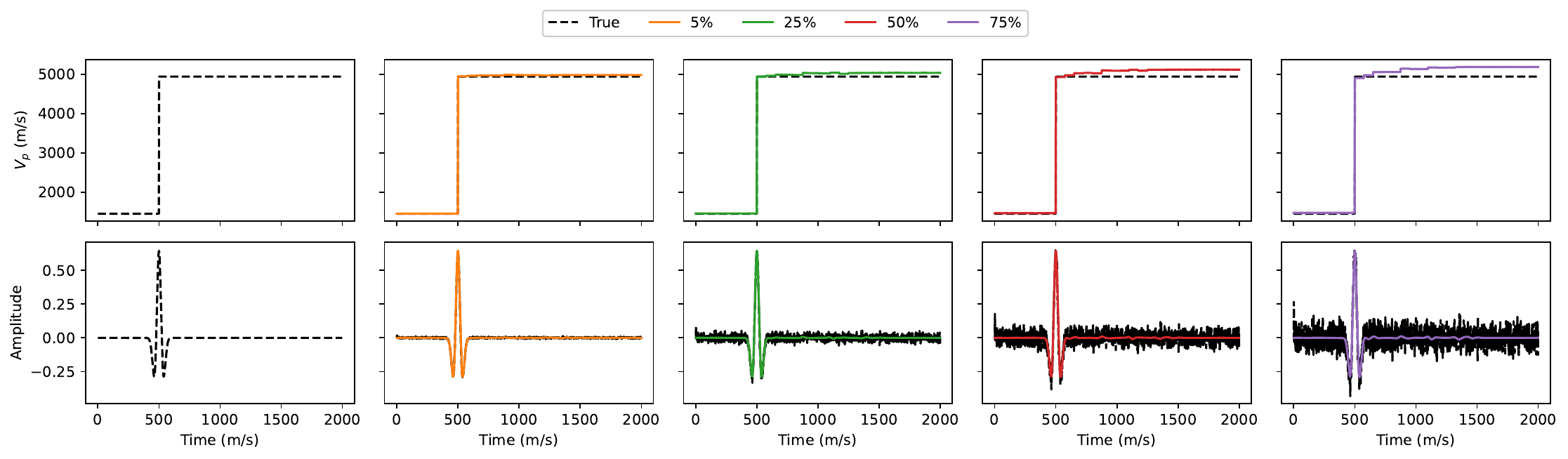}}
	\caption[Development dataset directly influences the quality of inversion.]{Conv1D-Adadelta DNN trained for 30 epochs for a maximum 3 layers for velocity ranging from 1450\si{ms^{-1}} to 5000\si{ms^{-1}}. The quality of the development dataset directly influence the quality of DNN inversion. A: Missing layer. B: Different geophysical models. C: Noise sensitivity. }
\end{figure}

The DNN requires a global model for a real world problem. Consider as example a data generator trained on data that is limited to 3 layers and inversion is carried out for a system made for more than 3 layers. The inversion process will start degradation as shown in Figure~\ref{fig:dnn_discussion_missing_layer}. Conv1D-Adadelta DNN was trained for 30 epochs for a maximum 3 layers for velocity ranging from 1450\si{ms^{-1}} to 5000\si{ms^{-1}}. The first two columns are inversions for velocity profiles within the generator limit for the number of layers. These inversions are of good quality as expected from previous results. For more layers in the velocity profile, the inversion tries to generalize for these profiles but misses some components. The inversion is still able to identify some layers, but not to the same resolution as if trained on those layers. The large velocity contrasts are inverted correctly as shown for the 6 layer velocity profile. Similarly, the geophysics models allowed within the problem need to be considered. Figure~\ref{fig:dnn_discussion_Inclusion_Inversion} illustrates inversion for a large velocity of 6000\si{ms^{-1}} and velocity inversion respectively. Given that these models were not included in the development dataset, the inversion process would never be able to invert for these velocity types correctly. On the other hand, the DNN framework is robust to different noise levels. Figure~\ref{fig:dnn_discussion_Noise_Sensitivity} showcases the inversion for pink noise contaminated data at different levels. Pink noise contamination was chosen due to its prevalence in the low frequency limit \citep{Randall2009}, thus making it more suited to assess the effect on the low frequency component of FWI. The inversion remains relatively unaltered before 25\% and then starts degradation. This inversion process could be used as a de-noising technique given the correct data-generator. 

\clearpage

\begin{figure}[ht!]
    \centering
    \includegraphics[width=0.95\textwidth]{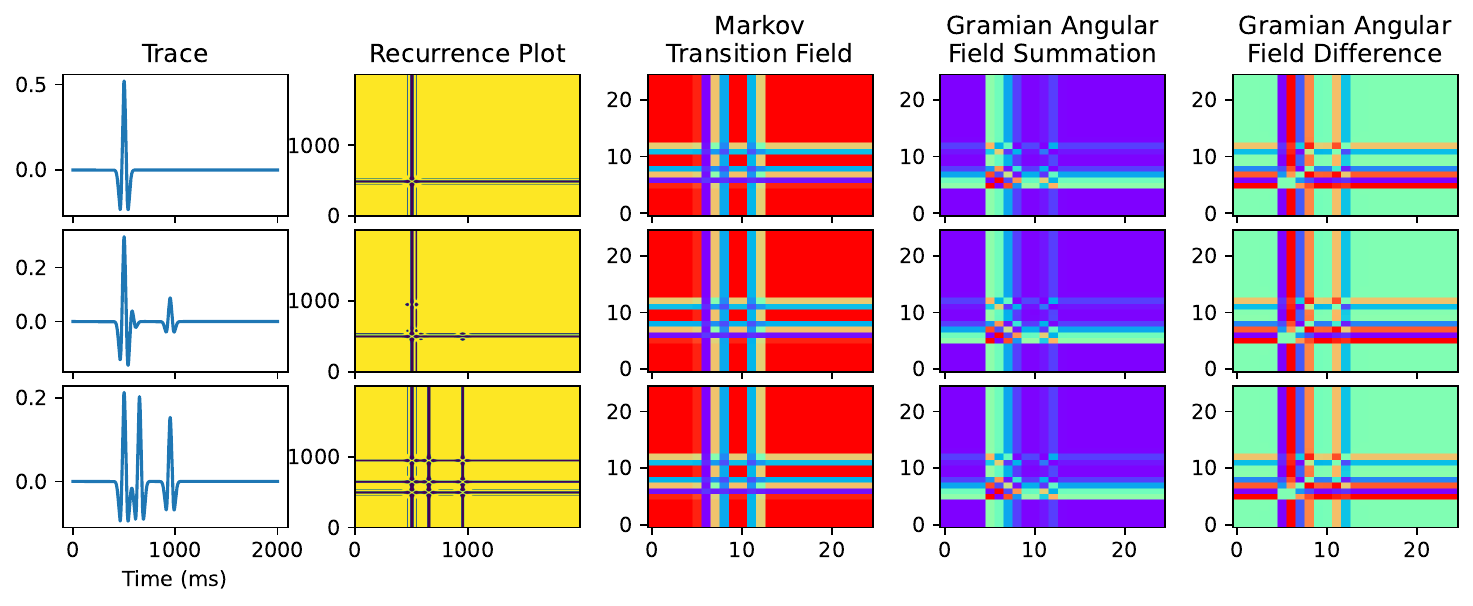}
    \caption[Alternative representations for a trace.]{Alternative representations for a trace.}
    \label{fig:alt_representations_trace}
\end{figure}

\clearpage
Modelling techniques and transform spaces could provide an alternative approach for the DNN. \cite{Jozinovic2020} use Wigner-Ville distributions for pseudo-spectral representations of the seismograms for prediction of intensity measurements of ground shaking, or other representations such as Recurrence Plots \citep{Kamphorst1987}, Markov Transition Fields \citep{Wang2015} and Gramian Angular Fields \citep{Wang2015} - See Figure~\ref{fig:alt_representations_trace}.

For theory-guided RNN, excluding part of the data from training for use as a development dataset is standard practice in deep learning, but not within classical FWI. For a real-world problem, the size of the seismic dataset relative to the model parameters generally has fewer data samples and could potentially prove problematic. Hyper-parameter tuning for the optimal parameters for RNN demonstrate that practice can result in convergence to a good model, yet this does not prove a similar result is achievable when using the entire dataset. 


\subsection{Forward Modelling and Multiples}
All shots considered within the forward problem for either FWI and RNN framework were within the water column for the Marmousi model. This implies that receiver data have surface-related multiples, together with all other inter-layer multiple components. Undergoing forward problem solving with and without multiples is a decision that the literature is still unable to resolve. 

Multiples travel longer paths and are reflected at small angles in contrast to the primaries and are able to illuminate shadow zones where primary reflections cannot reach \citep{Bergen2019}. Inclusion of these wavefield components can lead to improvement within the inversion process as multiples can contain more subsurface information compared to primary and diving wave \citep{Komatitsch2002}. \cite{Bleibinhaus2009} investigated these effects of surface scattering in FWI and concluded that velocity models resulting from neglecting the free surface in the inversion show artifacts and suffered from a loss of resolution. \cite{Liu2020} employ a combination of lower-order multiple as the source and the higher-order multiple to invert, whilst \cite{Zhang2013} transform each hydrophone into a virtual point source with a time history equal to that of the recorded data to help their inversion and are able to produce methods utilizing multiples to improve velocity updates. 

\cite{Hicks2001} and \cite{Operto2006} show how traditional FWI would become unstable when inverting with free surface related waves. This said, removing of multiples introduce additional processing steps which are subject to error and could lead to removal of signal. The consensus is that the choice of multiple inclusion is per different use case. Indeed, the work presented could be revisited for the sensitivity of multiples within the inversion.

\subsection{Implications of Data Volume and Computational Power}
More data is directly correlated with better modelling for NN frameworks, and this ability is limited by the resources available. Similar to classical FWI, computational power is a limitation within the frameworks presented. This was already identified within the RNN approach with the limit from the Graphical Processing Unit RAM, constraining the model size and batch processing. A larger batch-size for RNN processing would intuitively imply that the optimization is less likely to get stuck with local minimum and reduce the probability of cycle-skipping. Workaround for this could be multi-Graphical Processing Unit systems, such as NVIDIA's DGX station\footnote{\url{https://www.nvidia.com/en-us/data-center/dgx-systems/}$\textsuperscript{\ref{ref:footnote_no_affiliation}}$} and  Lambda Lab's Vector station\footnote{\url{https://lambdalabs.com/gpu-workstations/}$\textsuperscript{\ref{ref:footnote_no_affiliation}}$}, or cloud computing such as Amazon Web Services \footnote{\url{https://aws.amazon.com/nvidia/}$\textsuperscript{\ref{ref:footnote_no_affiliation}}$} and Google Cloud\footnote{\url{https://cloud.google.com/gpu/}$\textsuperscript{\ref{ref:footnote_no_affiliation}}$}\footnote{\label{ref:footnote_no_affiliation}These are just samples of resources and there is no affiliation.}. This cost on memory requirements for \acp{NN} is a common issue with solving optimization of large-scale neural networks \citep{Bottou2018} and efforts have been made into mitigating this via alternating gradient direction methods and Bregman iteration training methods \citep{Boyd2011, Taylor2016}.


\subsection{Maturity of the Frameworks}
\begin{figure}[!ht]
	\centering
	\subbottom[Initial circular model.]{\includegraphics[width=0.32\textwidth]{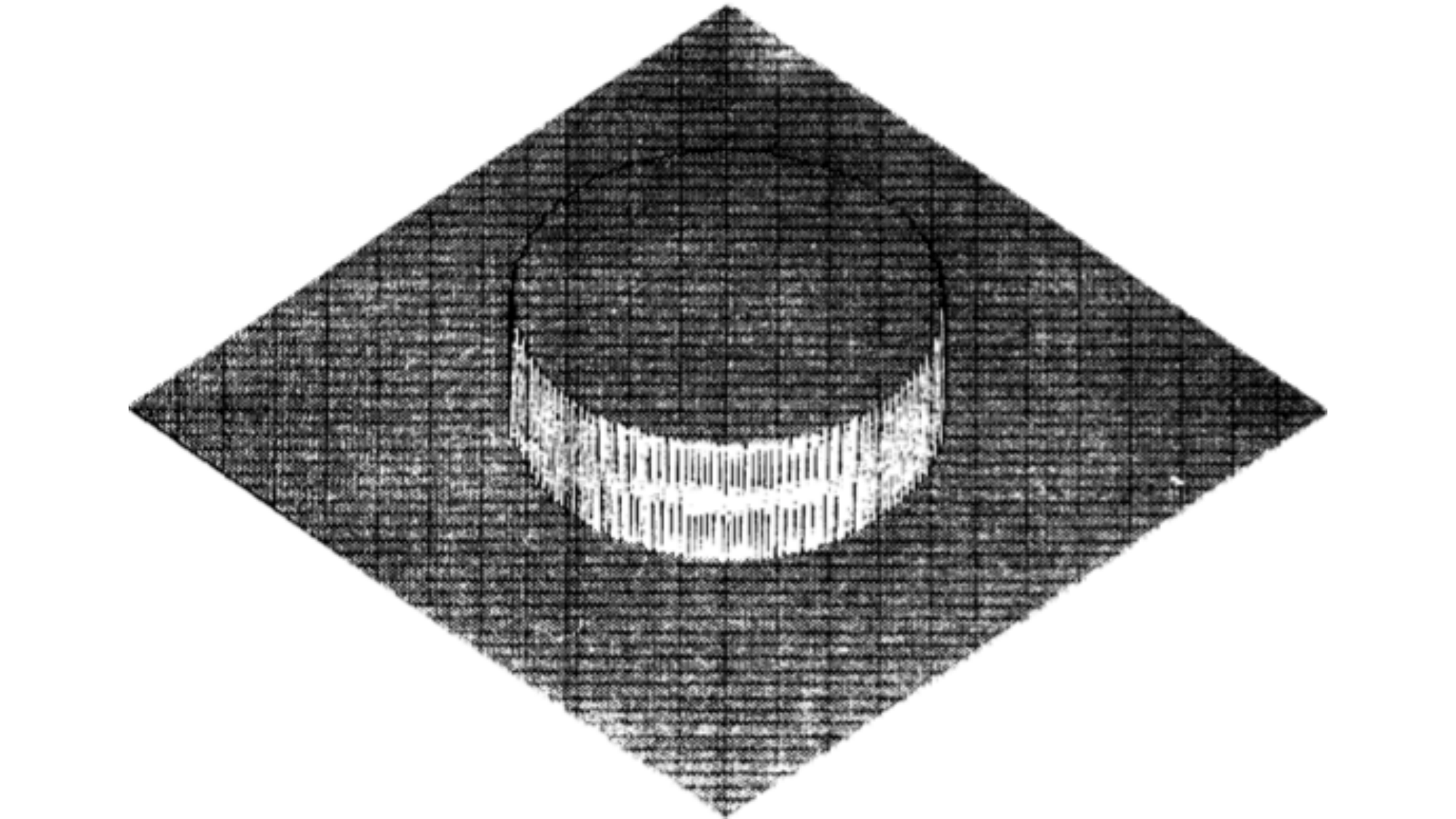}}
    \subbottom[Legacy FWI from 1986.\label{fig:gauthier_fwi_legacy}]{\includegraphics[width=0.32\textwidth]{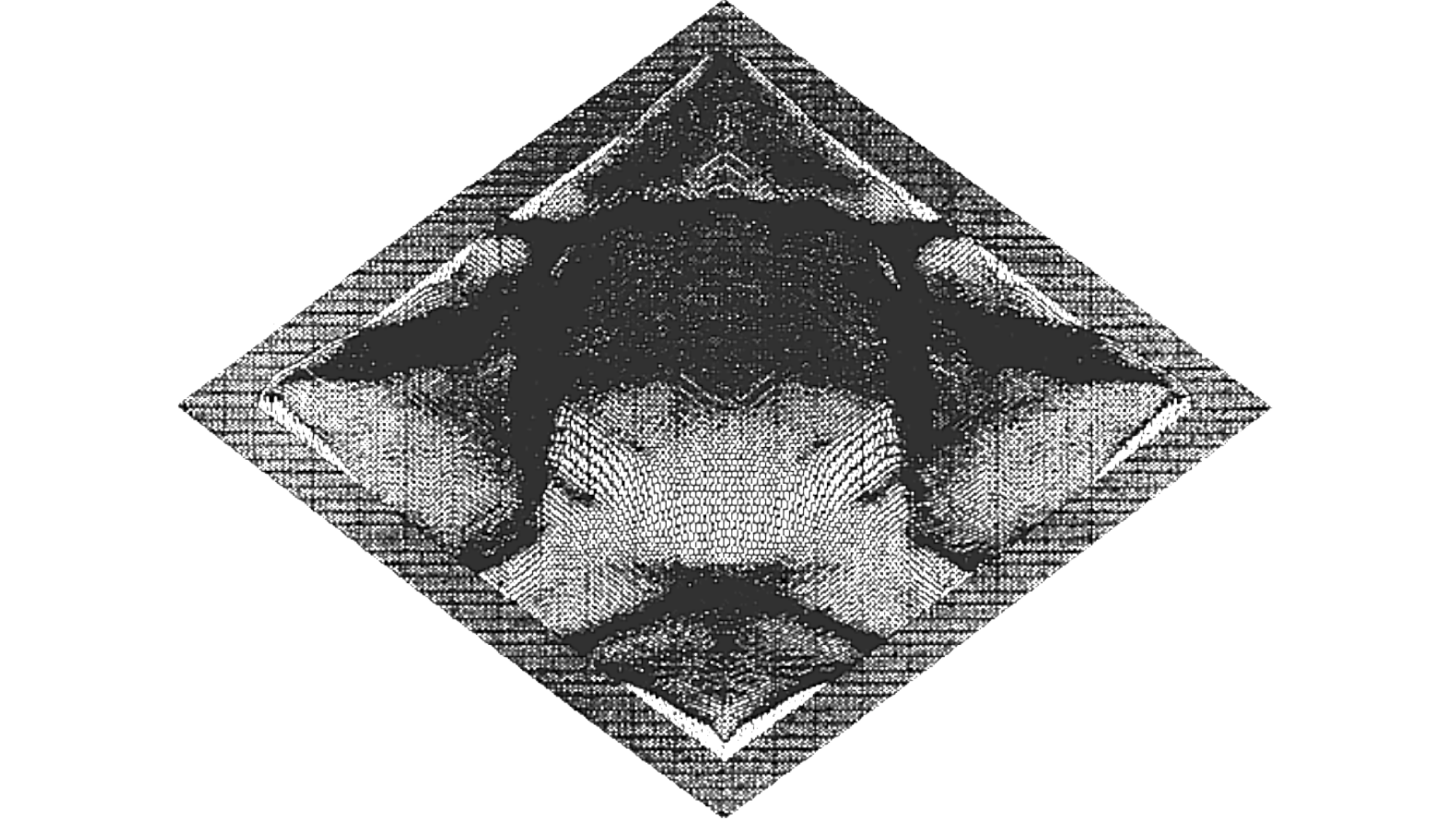}}
    \subbottom[Modern FWI.\label{fig:gauthier_fwi_modern}]{\includegraphics[width=0.32\textwidth]{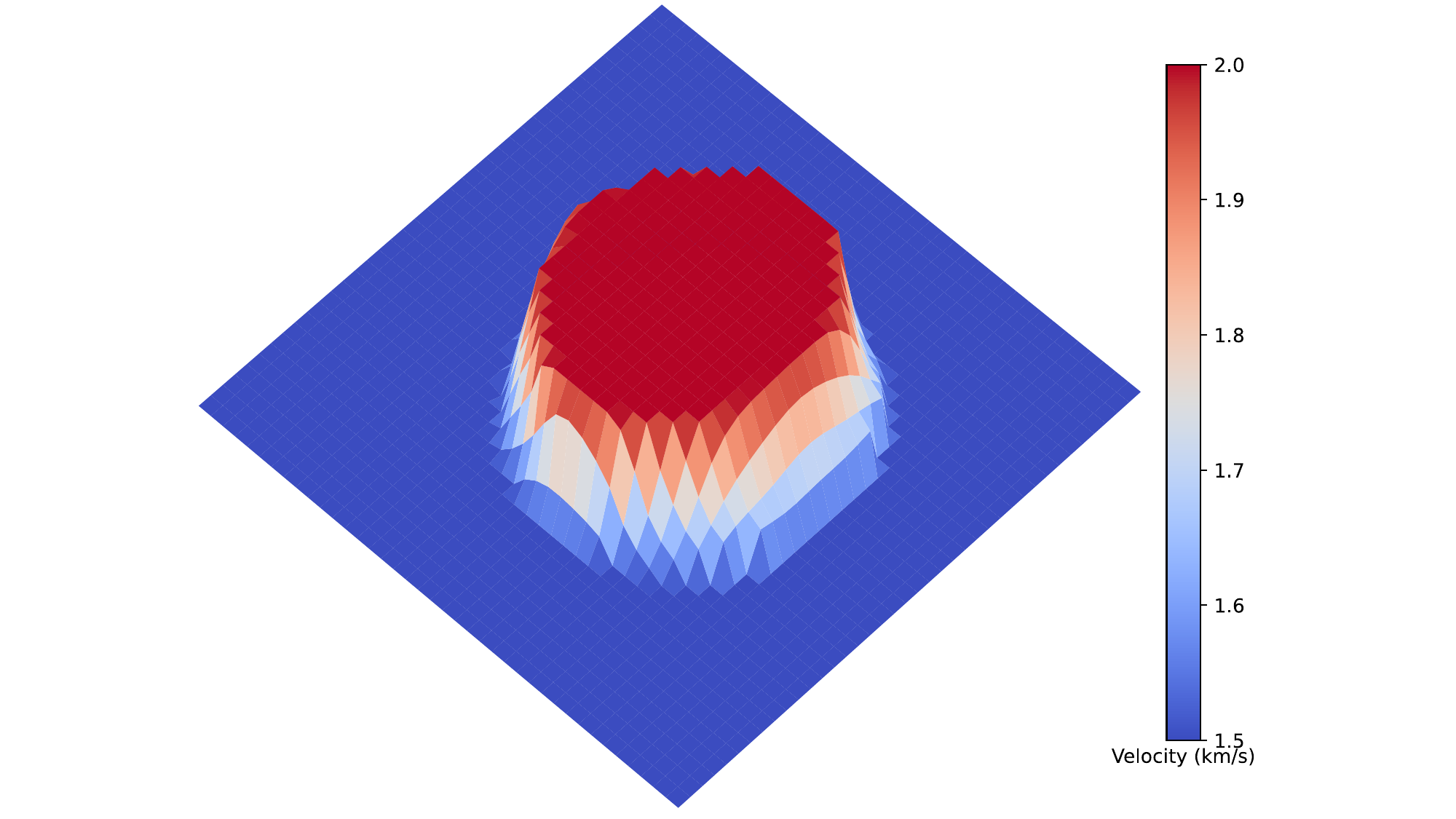}}
	\caption[Comparison between inversion for legacy FWI and modern FWI.]{Comparison between inversion results for legacy FWI and modern FWI for a circular model. 35 years of advances in hardware, computational modelling, mathematics, geophysics and optimization theory produce major uplift at the fraction of time.}
	\label{fig:gauthier_fwi}
\end{figure}

The NN framework results presented in this work should be considered within the context of maturity of the technique. As indicated in the literature review, FWI was originally an academic pursuit and initial inversions results were naive. Figure~\ref{fig:gauthier_fwi} shows results for FWI inversion obtained by \citet{Gauthier1986} for a circular model. Inverting for 8 shots for 5 iterations, with 400 receiver locations, the model is able to recover some structure within the velocity - Figure~\ref{fig:gauthier_fwi_legacy}. Inverting a similar model within a modern FWI framework would produce better global inversion as shown in Figure~\ref{fig:gauthier_fwi_modern}. The legacy inversion by Gauthier et al. took approximately 1 hour to execute on a CRAY-1S supercomputer \citep{Kolodzey1981}, whilst modern FWI was run on a personal computer for 2 minutes. This is a major uplift, which is backed by thirty-fives years of advances in hardware, computational modelling, mathematics, geophysics and optimization theory. Taking a similar time consideration for the NN framework, the potential for data-driven pseudo-spectral approach is still in infancy, yet able to produce improvements to a mature technique.

Both frameworks are still to be implemented to higher dimensions. The DNN approach would follow similarly to the work by \cite{Liu2020}. 2D receivers are modelled for an arbitrary velocity models and 2D convolutional architectures are trained to invert from the seismic data to the velocity model. Naturally, the next step would be 3D geometry. In the case of theory-guided approaches, addition of this axis is expected to provided better imaging.



\subsection{Other Areas of Deep Learning}
The intersection of Deep Learning and FWI is just starting to flourish and offers ample opportunity for new research avenues. Alternative architectures are readily available to be implemented. Some examples for DNN are shown in Figure~\ref{fig:CNN_history}. Transformers \citep{Vaswani2017} are currently state-of-the-art sequence learners and \cite{Shalova2020} developed the initial theoretical work for implementation for FWI. Alternatively, Fourier Recurrent Units \citep{Zhang2018} could be used within theory-guided FWI. These units stabilize training gradients along the temporal dimension with Fourier basis functions and would potentially resolve the gradient anomaly within our work.

The technique of Transfer Learning within Deep Learning has just starting to be considered in geophysics. Indeed, given that DNN generators are stochastic global engines, a pre-trained NN for a survey could easily be employed on another seismic survey with minimal training. \cite{Siahkoohi2019} successfully used the probability distributions of nearby surveys around an area of interest to fine-tune a pre-trained Generative Adversarial Network and be able to map low-cost, low-fidelity solutions to remove surface-related multiples and ghost from shot records and numerical dispersion from receivers.

\cite{Yadav2015} presented the topic of solving differential equations directly using NN architectures and bypassing the finite-difference approach. This framework has been found to be advantageous since (i) the solutions obtained are integrable and differentiable, thus have better interpretability, (ii) solutions are highly generalized and preserves accuracy despite very few points, and (iii) no modification is needed for different kinds of boundary conditions. Indeed, \cite{Zhu2020} developed a neural-network-based full waveform inversion method that integrates deep neural networks with FWI by representing the velocity model with a generative neural network. The velocity model generated by NN is input to conventional FWI partial differential equation solvers. The gradients of both the NN and PDEs are calculated using automatic differentiation, which back-propagates gradients through the acoustic/elastic PDEs and NN layers to update the weights and biases of the generative neural network \citep{Zhu2020}.



\section{Conclusion}
Data-driven and theory-guided approaches for FWI have been reviewed with a comprehensive study of literature. 
The pseudo-spectral approach via deep learning framework was identified to be lacking in any previous work and this proved to be an opportunity for development. FWI was re-cast within a DNN framework for both a data-driven and a theory-guided based formulation. Both approaches were developed theoretically, qualitatively assessed on synthetic data and tested on the Marmousi dataset. 

Elements within a classical FWI framework were shown to be substitutable with DNN components. The base architecture for the network was set to be an hour-glass neuron design. This is representative of multi-scale FWI and modern DNN approaches. Two data-generators were used to validate the framework on multi-layer models. This was tested for normalization and concluded not applicable. Fully connected layers, 1D and 2D convolutions, VGG and ResNet type architectures for Adagrad, Adadelta, RMSprop and Adam optimizer were quantitatively evaluated for computational hours, DNN training performance, validation and learning rate performance, and inversion RMSE. The best performing architecture-loss combination was identified as Conv1D-Adadelta. Conv1D architecture ranked the highest in all tests, whilst the differences in optimizer were superficial. The choice of architecture was the most important aspect as choosing a different loss optimizer would not result in deterioration of the result.

The Conv1D-Adadelta network was trained for inversion of the Marmousi model by using a 20-50 layer generator, with velocities ranging from 1450\si{ms^{-1}} to 5000 \si{ms^{-1}}. It was trained for 30 epochs, at 1,000,000 and 100,000 trace per epoch per training and testing dataset respectively. Analysis of the network kernel and bias distribution for the training and velocity and trace update per epoch confirmed that the framework is a learning process. Most of the update happen within the first few epochs, whilst additional epochs refined the inverted velocity. Multi-scale 3.5Hz classical FWI with Sobolev space norm regularization was compared to this DNN inversion. Inversion performance in shallow sections was equally good for either classical FWI or DNN approach. DNN framework performs better for deeper and over-thrust areas since DNNs are not bound by forward-modelling physical constraints from ray-tracing.

Theory-guided RNN as an analogue of FWI was implemented for 2D experiments and different wavefield components compared to an analytical 2D Green's function and time implementation. Based on these results, RNN Time is able to model the wavefield within a maximum 0.06 error tolerance and 1.74\% RPE. RNN Freq is overall more accurate with 0.05 error tolerance and 1.449\% RPE. Assessment on the gradients indicates how the adjoint state and RNN Freq gradients in general overestimate finite difference calculation, whilst RNN Time under-estimates it with an infinitesimal error. RNN Freq produced a perturbation on the onset of the gradient which was attributed to modelling artefact and could be mitigated in future versions of this approach. Based on the model size and compute available, the ideal loss was Adam with a learning rate of 2 and batch size of 1. Model batch size proved to be a limitation for practical implementations, yet RNN is computationally more efficient than the classical FWI presented in this work. RNN freq provides more stable convergence and is better performant. Overall, RNN frameworks are able to identify faults, but amplitudes are not fully inverted properly.

From the comparative analysis of the NN approaches, DNN was better performing. DNN networks recovered more of the velocity contrast, whilst RNN was better at edge definition. RNN was more suited for the shallow and depth sections due to the cleaner receiver residuals. This was only valid in the middle section of the model with the most coverage from ray-tracing. Indeed, RNN approaches had some leakage on the edges.

A critique addressing benefits and limitations of these two approaches was also included. The impact of the shift in the inversion paradigm was reviewed for both approaches. Data-driven should be considered within global approximation approaches and has potential to be used as \emph{a priori} to deterministic FWI. Research which can be easily translated was addressed in the form of probabilistic maps for the inversions and analytical accuracy upper bounds per iteration. The RNN approach benefits from the wider community of active researchers. The reduction in development time is a direct integration from Computer Science to geophysics. Vice-versa, Deep Learning frameworks can adopt strategies common to FWI. 

Data-driven inversion is still at the very early stages of development, and more research opportunity is still available. However, to truly assess the applicability and relevance of these frameworks, these approaches will have to be applied to real data in the future. The DNN model generators were shown to work within the boundaries of their parameters. Extra layers, velocity inversions and inclusions could be missed altogether if the network is not prepared correctly. Application of pre-trained networks is relatively easy and thus different geophysical model hypothesis could be assessed quickly. In either case, the DNN is robust to noise and future work would involve implementation as a de-noising techniques. Alternative modelling methods and transform spaces were also provided as alternative approaches for DNN. 

The forward modelling approach used through this work was critiqued for the use of multiples. Whether to use or not to use multiples within forward modelling is model dependent and should be evaluated for RNN Freq. Similar to classical FWI, computational power was identifiable as a limitation within these DNN frameworks. Although this is currently a limitation, it will not be in the near future due to the relative quick development of GPUs. A corollary to the whole approach was addressed in the form of the maturity of the approach. 35 years of advances applied to these frameworks would be expected to yield very good results. Finally, other areas of DNN that can be applied to FWI were presented. Alternative architectures such as Transformers and use of Fourier Recurrent Units are readily available. Potential of transfer learning and solving differential equations using NN were presented as future directions of research for these frameworks.

%% file: appA/appendix_a_additional_theory.tex
\chapter[Theoretical Tools]{Additional Theoretical Tools}\label{sec:app_theory}
\section[Difference between Time and Frequency FWI]{Difference between Time and Frequency FWI}\label{sec:app_theory_Differences_in_numerical_implementation_between_Time_and_Frequency_FWI}
To illustrate the differences within numerical implementations between time and frequency FWI, let us consider the 1D finite difference formulation of the acoustic wave equation given by:
\begin{equation}\label{eq:wave_equation_2}
    \frac{1}{c(x)^2} \frac{\partial^2 p(x,t)}{\partial t^2} - \nabla^2p(x,t) = s(x,t).
\end{equation}
Following from \cite{Igel2016}, Equation~\ref{eq:wave_equation_2} is re-arranged and re-written in dense notation as:
\begin{equation}\label{eq:wave_equation_derivs}
    \partial_t^2p = c^2\partial_x^2p + s.
\end{equation}
Discretize space and time with constant increments $dx$ and $dt$ such that
\begin{align}
    x_j &= jdx, \qquad j=\left[0,l_{max}\right]\\
    t_n &= ndt, \qquad t=\left[0,n_{max}\right].
\end{align}
\begin{figure}[ht!]
    \centering
    \includegraphics[width=0.3\linewidth]{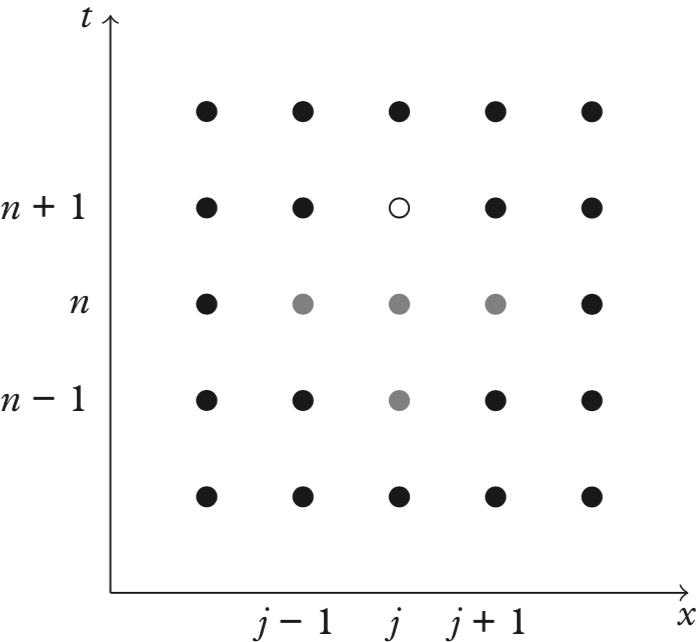}
    \caption[Space-Time FD discretization stencil for the 1D acoustic wave equation.]{Space-Time FD discretization stencil for the 1D acoustic wave equation. The $x$-axis corresponds to space and the $y$-axis to time. The open circle denotes the point $p(x_j,t_{n+1})$ to which the state of the pressure field is extrapolated. From \cite{Igel2016}.}
    \label{fig:4_point_stencil}
\end{figure}
Implementing a FD stencil as shown in Figure~\ref{fig:4_point_stencil}, the derivatives approximated via central finite differences and Equation~\ref{eq:wave_equation_derivs} can be written as:
\begin{equation}
    \frac{p_j^{n+1} - 2p_j^{n}+p_j^{n-1}}{dt^2} = c_j^2\left[ \frac{p_j^{n+1} - 2p_j^{n}+p_j^{n-1}}{dx^2} \right] + s_j^n,
\end{equation}
where the upper index corresponds to time discretization and the lower index corresponds to spatial discretization. Pressure at location $n+1$ based on current location $n$ and previous $n-1$ location is given by:
\begin{equation}\label{eq:wave_equation_time_step}
    p_j^{n+1} = c_j^2\left[p_j^{n+1} - 2p_j^{n}+p_j^{n-1} \right] + 2p_j^n-p_j^{n-1}+ dt^2s_j^n.
\end{equation}
Given a source wavelet $s_j^n$ at location $j$ and time $n$, an initial wavefield $p_j^0$ and initial conditions such that everything is at rest at time $t=0$, all components on the right-hand side of Equation~\ref{eq:wave_equation_time_step} are known and can be used to propagate particle motion through an acoustic medium.

Considering the Fourier pseudo-spectral implementation \citep{Gazdag1981}, Fourier theory seeks to approximate a given function by a finite sum over some $N$ orthogonal basis functions $\Phi_i$, namely:
\begin{equation}
    f(x)\approx \sum_{i=0}^N a_i \Phi_i (x),
\end{equation}
where $a_i$ are the Fourier coefficients. The main benefit of using this approximation is that given the discrete Fourier transform of functions defined on a regular grid, exact machine precision derivatives up to the Nyquist wavenumber $k_N=\frac{\pi}{dx}$ can be calculated. In particular, it can be shown that the Fourier transform operator $\mathcal{F}$ can obtain the exact $n^{th}$ derivate as:
\begin{equation}\label{eq:fourier_deriv}
    f^n(x)=\mathcal{F}^{-1} \left[(ik)^n\mathcal{F}\left[f(x)\right]\right].
\end{equation}
Substituting Equation~\ref{eq:fourier_deriv} in the $2^{nd}$ order spatial derivative of the 1D acoustic wave equation in Equation~\ref{eq:wave_equation_derivs} gives:
\begin{equation}\label{eq:2nd_derivs}
    \partial_x^2p_j^n=\mathcal{F}^{-1}\left[ (ik)^2 P_\nu^n \right]=\mathcal{F}^{-1}\left[-k^2 P_\nu^n\right],
\end{equation}
where $P_\nu^n$ is the discrete complex wavenumber spectrum at time $n$. As such, the main overall error in the numerical solution comes from only the time integration scheme. Substituting Equation~\ref{eq:2nd_derivs} in Equation~\ref{eq:wave_equation_derivs}, discretize on a FD stencil and rearranging yields:
\begin{equation}
    p_j^{n+1}=dt^2 \left[c_j^2 \mathcal{F}^{-1}\left[k^2P_\nu^n\right]+s_j^n\right]+2p_j^n-p_j^{n-1}.
\end{equation}
\section{Activation Functions}\label{sec:app_theory_Alternative_Activation_Functions}
\begin{figure}[!ht]
	\centering
	\subbottom[Binary Step]{\includegraphics[width=0.32\textwidth]{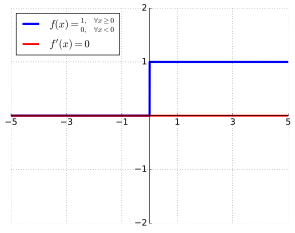}}
    \subbottom[Linear]{\includegraphics[width=0.32\textwidth]{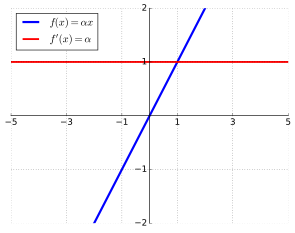}}
    \subbottom[Sigmoid]{\includegraphics[width=0.32\textwidth]{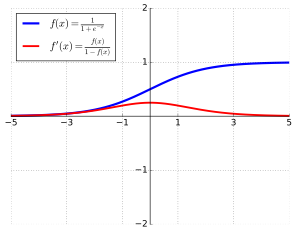}} 

    \subbottom[Tanh]{\includegraphics[width=0.32\textwidth]{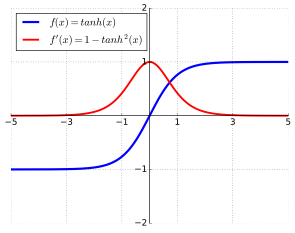}}
    \subbottom[ReLU]{\includegraphics[width=0.32\textwidth]{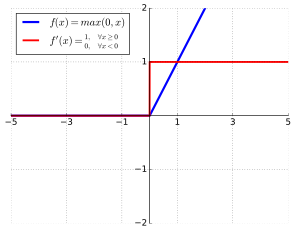}}
	\caption[Activation functions $f(x)$ and their derivative $f^{'}(x)$.]{Activation functions $f(x)$ and their derivative $f^{'}(x)$.}        
	\label{fig:alternate_activation_functions}
\end{figure}

Below are popular activations functions based on \cite{Goodfellow2016}. These are represented graphically in Figure~\ref{fig:alternate_activation_functions}.
\begin{itemize}
    \item {\bfseries Binary Step}: $f(x)=\begin{cases} 
        1, & \forall x \geq 0 \\  
        0, & \forall x < 0  
        \end{cases} $. This a simple switch on-off type function ideal for a binary classification. It is more theoretical than practical as data classifications tasks usually classify into multiple classes. Furthermore, a binary step function has a vanishing gradient, thus making it not usable with back-propagation. 
	\item {\bfseries Linear}: $f(x)=\alpha x,\forall x\in\mathbb{R}$. This is proportional to the input $x$, based on the scalar $\alpha$. The gradient is constant at $\alpha$, resulting in a linear transformation.
    \item {\bfseries Sigmoid}: $f(x)=\frac{1}{1+e^{-x}}$. This is widely used function since it is a non-linear smooth continuously differentiable function. The non-linearity enables neurons to approximate non-linear functions. The gradient is always positive, however it still has a vanishing gradient problem as $f^{'}(x)$ approaches zero for larger $x$. Furthermore, the sigmoid function is non-symmetric and connectivity between neurons is not necessarily always positive. Mitigation is achieved via scaling of the sigmoid, resulting in the Tanh function.
    \item {\bfseries Tanh}: $f(x)= \frac{e^{x}-e^{-x}}{e^{x}+e^{-x}}$. This function caries the same properties of non-linearity, smoothness and continuous differentiability of the sigmoid function, however it has the added benefit of being symmetric over the origin. This mitigates the problem of having all values of the same sign. Nonetheless, Tanh still retains the vanishing gradient problem as the function is flat with small gradients for large $x$.
    \item {\bfseries Softmax}: $f(x_i)=\frac{e^{x_i}}{\sum_{j=1}^{n}e^{x_j}}$. This function is the sigmoid function extended to multiple classes and provides a probability distribution over the predicted classes. 
	\item {\bfseries ReLU}: $f(x)=max(0, x)$. The Rectified Linear Unit is non-linear, allowing easy back-propagation of errors. When employed on a network of neurons, the negative component of the function is converted to zero and the neuron is deactivated, thus introducing sparsity with the network and making it efficient and easy for computation.
\end{itemize}

\section{Loss Optimizers}\label{sec:app_theory_Loss_Optimizers}
To simplify notation in this section, consider objective function $\mathcal{J}(\Theta)$ parameterized by model parameters $\Theta\in\mathbb{R}^d$ to optimize gradient $\nabla_\Theta\mathcal{J}(\Theta)$ with respect to the model parameters for a learning rate $\eta$. This can be recast as:
\begin{equation}
    \Theta\equiv\Theta-\eta\nabla_\Theta\mathcal{J}(\Theta).
\end{equation}
\subsection{Adagrad}\label{sec:app_theory_Adagrad}
The Adaptive Gradient or Adagrad algorithm \citep{Duchi2011} performs learning rate updates with training based on the frequency of distribution in the data. It performs smaller updates (smaller learning rates) on parameters associated with frequently occurring features and larger updates vice-versa. 

As Adagrad uses a different learning rate for every parameter $\Theta_i$ at time step $t$, the gradient of the objective function is given by
\begin{equation}
    g_t=\sum_{i}\nabla_{\Theta_{t}} \mathcal{J}(\Theta_{t,i}) .
\end{equation}
Model updates at time $t+1$ become:
\begin{equation}\label{eq:loss_optimizer_adagrad}
    \Theta_{t+1}=\Theta_{t}-\frac{\eta}{\sqrt{G_t+\epsilon}} g_t,
\end{equation}
where $G_t$ is the sum of squares of the gradients up to time step $t$ and $\epsilon$ is a stabilizing term to avoid division by zero. This optimizer eliminates the need to manually tune the learning rate. However, the accumulation of the squared gradients in the denominator keeps growing with every update and will eventually cause the learning rate to shrink up to the point the algorithm is no longer able to acquire additional knowledge \citep{Ruder2016}.

\subsection{Adadelta}\label{sec:app_theory_Adadelta}
Adadelta \citep{Zeiler2012} extends Adagrad and aims to avoid the continual decay of the learning rates by using a decaying average of past gradients to some fixed size $w$. Equation~\ref{eq:loss_optimizer_adagrad} becomes
\begin{equation}
    \Theta_{t+1}=\Theta_t-\frac{\text{RMS}\left[\Delta\Theta\right]_{t-1}}{\text{RMS}\left[g\right]_t}g_t,
\end{equation}
where $RMS\left[g\right]_t$ is the Root Mean Squared for the gradient, $RMS\left[\Delta\Theta\right]_{t-1}=\sqrt{E\left[\Delta\Theta^2\right]_{t-1}+\epsilon}$ is the Root Mean Squared of parameter updates. Since $RMS\left[\Delta\Theta\right]_{t}$ is unknown, it is approximated until the previous time step $t-1$. This algorithm removes the need for a default learning rate.

\subsection{RMSprop}\label{sec:app_theory_RMSprop}
Similar to Adadelta, RMSprop \citep{Hinton2012} attempts to resolve Adagrad’s vanishing learning rates by maintaining a moving average of the square of gradients and divides the gradient by the root of this average. Equation~\ref{eq:loss_optimizer_adagrad} becomes:
\begin{align}
    E\left[g^2\right]_t &= 0.9E\left[g^2\right]_{t-1}+0.1g_t^2 \\
    \Theta_{t+1} &= \Theta_{t}-\frac{\eta}{\sqrt{E\left[g^2\right]_t+\epsilon}}g_t.
\end{align}

\subsection{Adam}\label{sec:app_theory_Adam}
Adaptive Moment Estimation or Adam \citep{Kingma2014} computes adaptive learning rates for each parameter based on estimations of first-order and second-order moments. 
The algorithm updates exponential moving averages of the gradient $(\hat{m}_t)$ and the squared gradient $(\hat{v}_t)$ by
\begin{align}
    \hat{m}_t &= \frac{m_t}{1-\beta_1^t} \\
    \hat{v}_t &= \frac{v_t}{1-\beta_2^t},
\end{align}
where the hyper-parameters $\beta_1, \beta_2 \in [0, 1)$ control the exponential decay rates of these moving averages. Equation~\ref{eq:loss_optimizer_adagrad} for Adam becomes:
\begin{equation}
    \Theta_{t+1}=\Theta_t - \frac{\eta}{\sqrt{\hat{v}_t}+\epsilon}\hat{m}_t.
\end{equation}

\section{Regularization}\label{sec:app_theory_Regularization}
DNN regularization is available via two strategies: (i) functional-based, and (ii) NN architecture-based. 
\subsection{Functional-based}\label{sec:app_theory_Functional-based_NN_regularization}
These type of schemes update the cost function (Equation~\ref{eq:dnn_cost_function_J}) with a regularization term, namely:
\begin{equation}
    \hat{\mathcal{J}} = \mathcal{J} + \alpha \Omega,
\end{equation}
where $\mathcal{J}$ is the original cost function, $\Omega$ is the regularization term or norm penalty and $\alpha\in\left[0,\inf\right)$ is a hyperparameter that weights the relative contribution of $\mathcal{J}$ to $\Omega$. The most commonly used norm penalty is $L^2-norm$ defined as:
\begin{equation}
    \Omega_{L^{2}}=\frac{1}{2}\left|\left|\boldsymbol{w}\right|\right|_2^2,
\end{equation}
where $\boldsymbol{w}$ are the weights throughout the DNN. This lends itself from the broader subject of regularization theory and was previously identified for FWI in §~\ref{sec:theory_regularization_fwi}. Alternative names for this style of regularization are Ridge Regression or Tikhonov regularization.

\subsection{Architecture-based}\label{sec:app_theory_NN_architecture-based_regularization}
DNN can be regularized via alterations to the NN architecture through (i) Dropout, (ii) Data augmentation, and (iii) Early Stopping.
\subsubsection{Dropout}
\begin{figure}[ht!]
    \centering
    \includegraphics[width=0.5\linewidth]{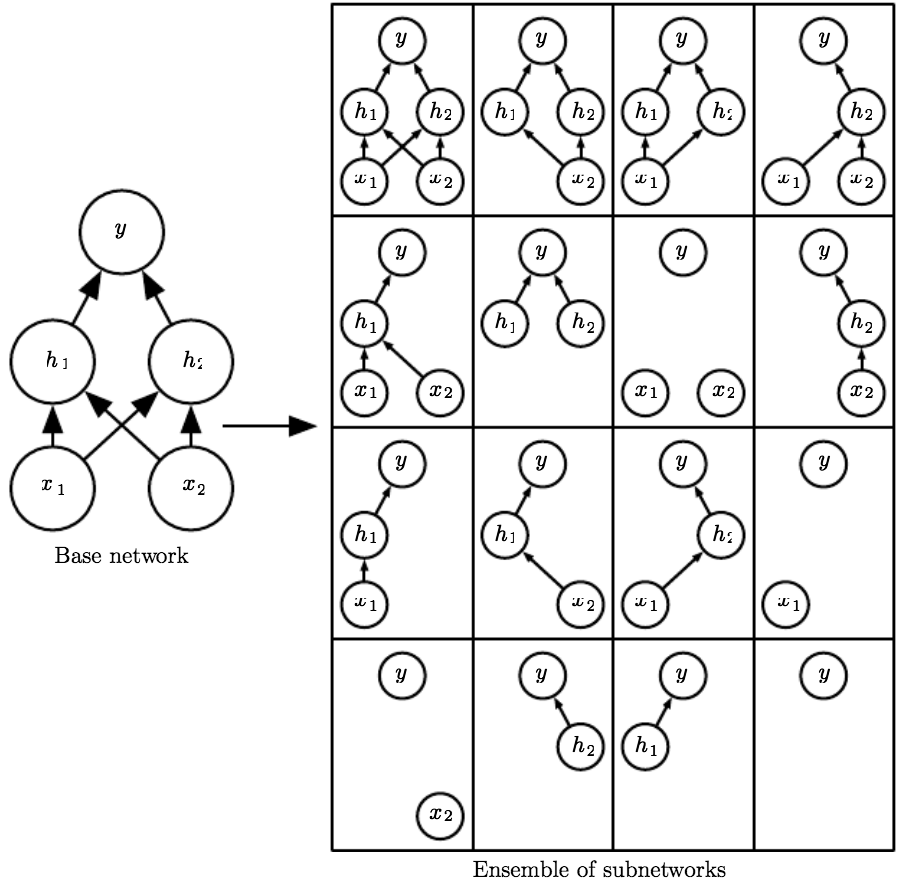}
    \caption[All possible sub-networks produced from dropout.]{All 16 possible sub-networks are produced from a simple base network. A large proportion of these do not have sufficient input-to-output connections and are ignored. For wide and deeper DNNs, these types of sub-networks become insignificant since it would be unlikely to drop all paths from inputs to outputs. If such an occurrence does happen, it would only effect a single epoch and this would be mitigated with long enough training. From \cite{Goodfellow2016}.}
    \label{fig:dropout}
\end{figure}

Dropout is a very simple ensemble-based technique \citep{Srivastava2014}. It is the process of allow only some parts of the network to be updated. Figure~\ref{fig:dropout} illustrates all sub-networks that can be formed from a simple NN that allow weight update within an epoch. Either of these sub-networks is randomly considered and trained at a given epoch and inherently remove nodal dependencies \citep{Srivastava2014}. 

\subsubsection{Data Augmentation}
The best way to make a DNN model generalize better for “unseen” data is to train on more data \citep{Goodfellow2016}. The amount of data available in real-world application is limited and it is reasonably straight forward to create new “fake” data. This is achieved via transformations of the original $x$ input and mapping to the correct $y$ \citep{Lee2018}. \cite{Ekbatani2017} and \cite{Le2017} show successful training of a DNN using synthetic created input-output pairs in practice.

\subsubsection{Early Stopping}
When DNN model complexity is sufficient to represent over-fitting, it is likely that the training error decreases steadily over time, but validation set error begins to rise again and makes the model worse off with any additional iteration \citep{Yao2007}. Early stopping is the technique which terminates training at a given epoch $\mathcal{E}^n$ if the errors on the validation sets at $\mathcal{E}^{n+1}$ are greater than those of $\mathcal{E}^n$. Early stopping regularization is the most widely used regularization technique as it is unobtrusive to the learning dynamics for DNN \citep{Caruana2000}.

\section{CNN Building Blocks}\label{sec:app_theory_CNN_Building_Blocks}
This section reviews the elements for CNNs referenced through-out the dissertation.
\subsection{Convolutional Layer}\label{sec:app_theory_Convolutional_Layer}
\begin{figure}[ht!]
    \centering
    \includegraphics[width=0.7\linewidth]{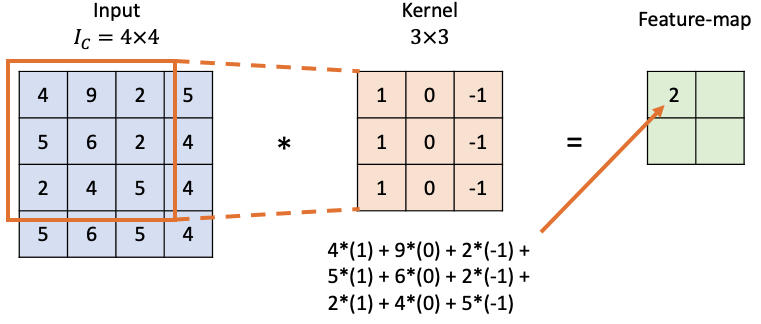}
    \caption[Example of 2D convolution.]{Example of 2D convolution.}
    \label{fig:CNN_calculation}
\end{figure}

A convolutional layer differs from a normal fully connected layer through the introduction of convolutions across the neurons. The window across which convolutions are applied is referred to as a kernel or filter. Mathematically, the convolutional operation on an input image tensor $I_c$ for the $l^{th}$ layer is given by:
\begin{equation}
    f_l^k(p,q)=\sum_c\sum_{x,y} i_c(x,y) e_l^k(u,v),
\end{equation}
where $i_c (x,y)$ is an element of the input image tensor, $e_l^k (u,v)$ is the index of the $k^{th}$ convolutional kernel $k_l$, and the output or feature-map of the $k^{th}$ convolutional kernel is given by $F_l^k=\left[f_l^k(1,1),\cdots,f_l^k(P,Q)\right]$ with $P$ and $Q$ being the total number of rows and columns in $I_C$. A graphical representation of the convolutional operation is given in Figure~\ref{fig:CNN_calculation}.

\subsection{Pooling Layer}\label{sec:app_theory_Pooling_Layer}
\begin{figure}[ht!]
    \centering
    \includegraphics[width=0.4\linewidth]{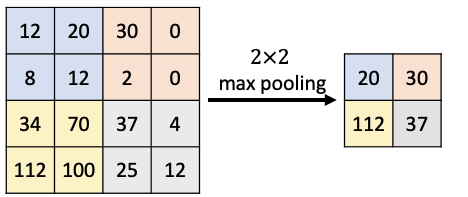}
    \caption[Example of 2D max-pooling.]{Example of 2D max-pooling.}
    \label{fig:CNN_calculation_max_pooling}
\end{figure}

Once a feature-map is extracted from a convolutional layer, the exact location of one element becomes less important as long as its relative position to others is preserved \citep{Khan2020}. Sub-regions of interest can thus be inferred via reduction of dimensionality \citep{Lee2016}. There are two main types mechanisms, max and min pooling. Max pooling picks the maximum value and min pooling picks the minimum value from the assigned region. Figure~\ref{fig:CNN_calculation_max_pooling} illustrates an example of max pooling.

Either of these pooling operations support feature extraction which is invariant to translational shifts and small distortions \citep{Ranzato2007}. These invariant features regulate the complexity of the network and reduce over-fitting. Alternative formulations include average \citep{Boureau2010}, L2 \citep{Wang2012}, overlapping and spatial pyramid pooling \citep{He2015}.

\subsection{Batch Normalization}\label{sec:app_theory_Batch_Normalization}
Input distributions to the layers in deep network may vary when weights are updated during training. Batch Normalization is used to counteract the internal covariance shift in the distribution of hidden unit values by standardising the input to zero mean and unit variance for each batch \citep{Ioffe2015}. Batch normalization for a feature-map $F_l^k$ is given by
\begin{equation}
    \mathcal{N}_l^k=\frac{\mathcal{F}_l^k - \mu_B}{\sqrt{\sigma_B^2+\epsilon}},
\end{equation}
where $\mathcal{N}_l^k$ is the normalized feature map, $\mathcal{F}_l^k$ is the input feature map, $\mu_B$ is the mean, $\sigma_B^2$ is the variance for a batch and $\epsilon$ is added for numerical stability \citep{Ioffe2015}.

\section{Common CNN Architectures}\label{sec:app_theory_Comparison_of_Common_CNN_Architectures}
\subsection{AlexNet}\label{sec:app_theory_AlexNet}
AlexNet \citep{Krizhevsky2012} is considered as the first deep neural network with 8 layers, compared to its predecessor LeNet with 5 layers \citep{LeCun1990}. In particular, it has large filters $(11\times 11, 5\times 5)$ at the initial layers, uses dropout during training, ReLU to improve convergence rates and overlapping sub-sampling and local response normalization to improve generalization. The basic architectural blueprint is given in Figure~\ref{fig:dnn_arch_comparison_alexnet}.

\subsection{VGG}\label{sec:app_theory_VGG}
VGG \citep{Simonyan2014} is a rather simplistic, homogenous network. However, it is 19 layers deep. It stacks $(3\times 3)$ filters to reduce the computational complexity, uses max pooling after the convolutional layers and padding is applied to maintain the spatial resolution. The only problem associated with VGG network are the 138 million trainable parameters, which make it computationally expensive for network training and deploying on system with low resources. The architectural blueprint is given in Figure~\ref{fig:dnn_arch_comparison_vgg}.

\subsection{ResNet}\label{sec:app_theory_ResNet}
\begin{figure}[ht!]
	\centering
	\subbottom[AlexNet: Five convolutional and three fully connected layers.\label{fig:dnn_arch_comparison_alexnet}]{\includegraphics[angle=-90, origin=c, width=0.31\textwidth, trim={8.6cm 5cm 8.6cm 5cm}, clip]{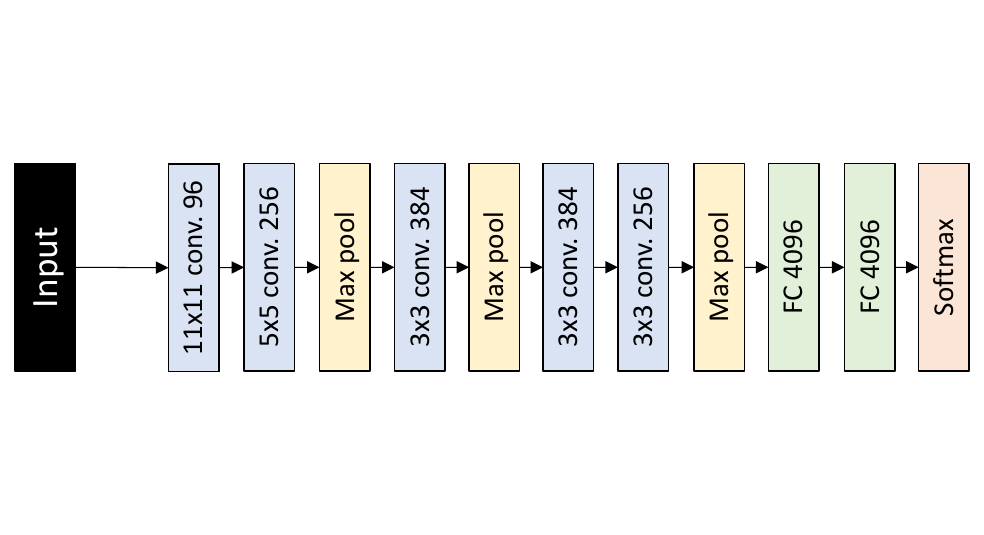}}\quad
    \subbottom[VGG: Five stacked conv. layers with max pooling and 3 fully connected layers.\label{fig:dnn_arch_comparison_vgg}]{\includegraphics[angle=-90, origin=c, width=0.31\textwidth, trim={2cm 4cm 1.7cm 4cm}, clip]{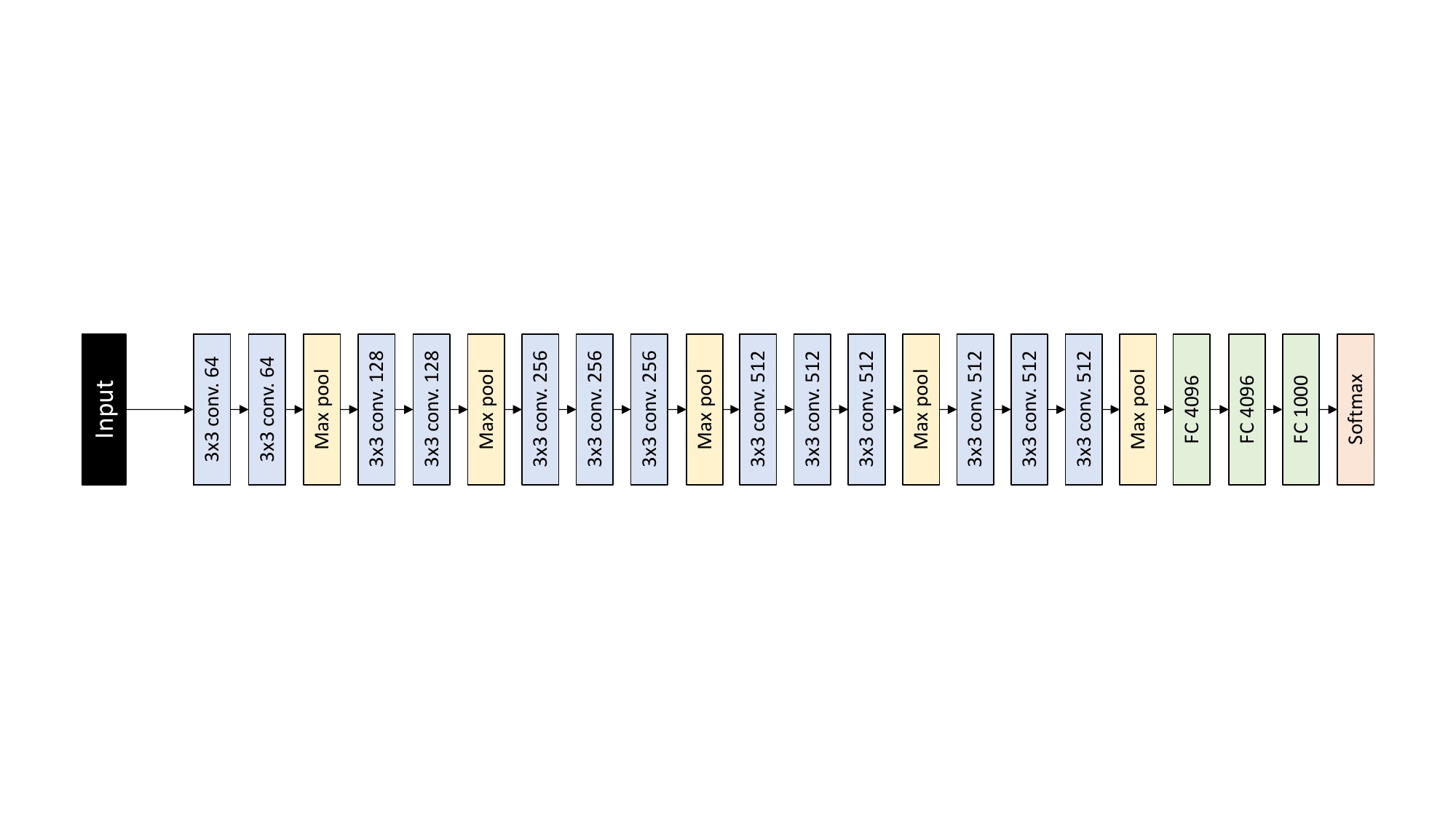}}\quad
    \subbottom[ResNet: 152 layer deep network with residual blocks.\label{fig:dnn_arch_comparison_ResNet}]{\includegraphics[angle=-90, origin=c, width=0.31\textwidth, trim={2cm 4cm 1.7cm 4cm}, clip]{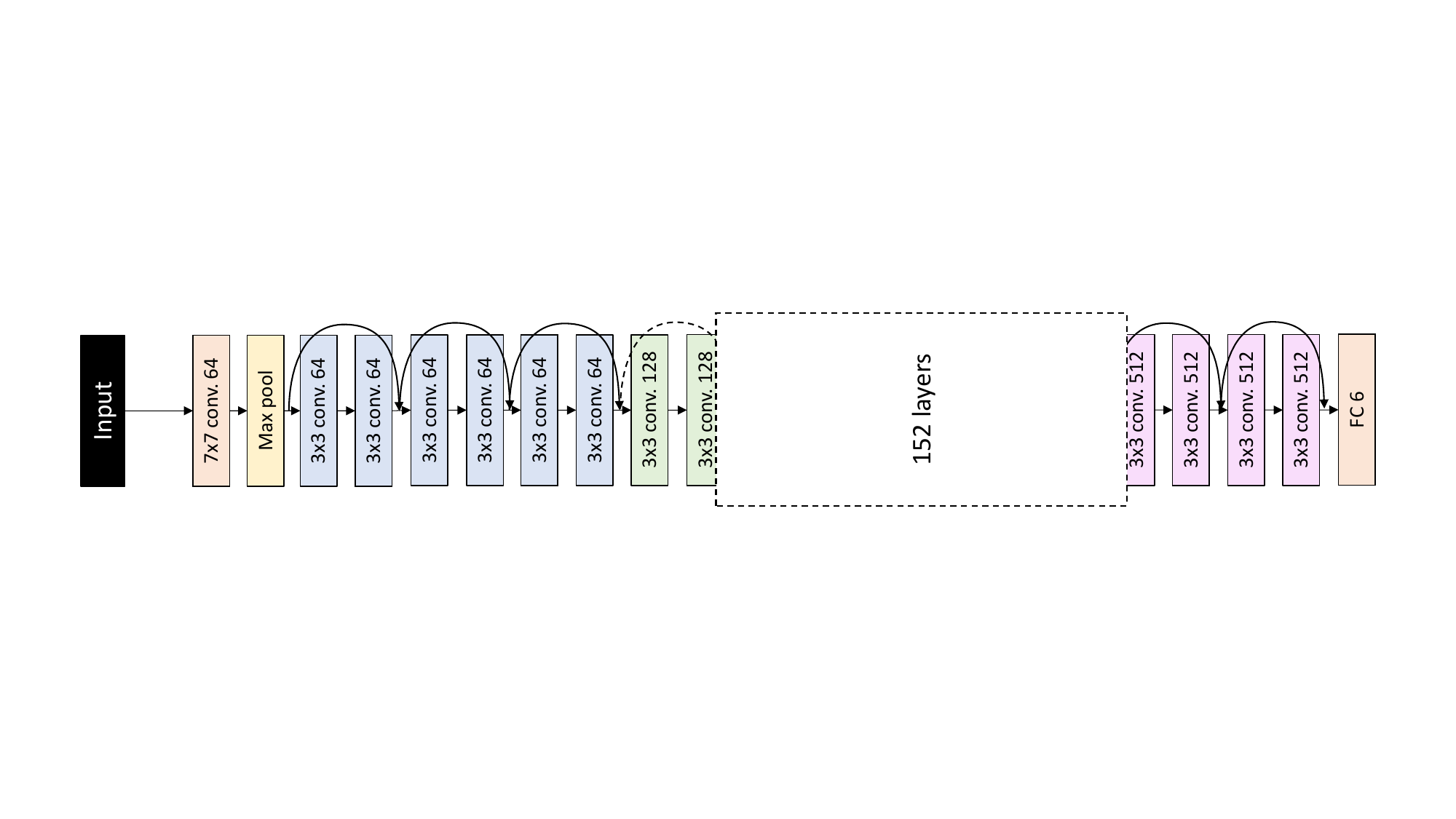}}
    
    \subbottom[Residual block introduces an “identity map” or “skip connection” which skips one or more layers.\label{fig:dnn_arch_comparison_resnet_identity}]{\includegraphics[width=0.5\textwidth, trim={10.5cm 6.4cm 10.5cm 6.5cm}, clip]{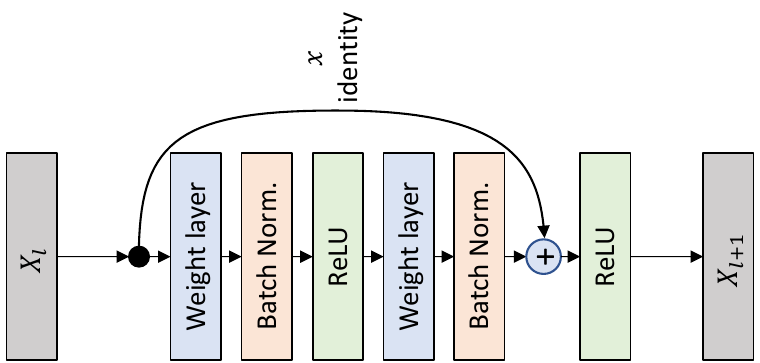}}
	\caption[Activation functions $f(x)$ and their derivative $f^{'}(x)$.]{Architectural blueprints for (a) AlexNet, (b) VGG, (c) ResNet and (d) Residual Block for ResNet. The numbers in each layer in (a)-(c) indicate the number of trainable parameters. Adapted from \cite{Krizhevsky2012}, \cite{Simonyan2014} and \cite{He2016}.}        
	\label{fig:dnn_architecture_Comparison}
\end{figure}

ResNet \citep{He2016} is a 152-layers deep network shown in Figure~\ref{fig:dnn_arch_comparison_ResNet}. ResNet is 20 times deeper than AlexNet and 8 times deeper VGG. Pivotal to this network is the concept of a residual block, shown in Figure~\ref{fig:dnn_arch_comparison_resnet_identity}. This introduces the “identity map” or “skip connection” that skips one or more layers. This propagates the error deep through the network layers and allows it to learn how to minimize the residual.

\clearpage
\section{RNN Graphs and Unfolding}\label{sec:app_theory_Graphs_and_Unfolding}
RNNs are capable of learning sequences representing dynamic temporal behaviour \citep{Rumelhart1986}. These can be categorized into two broad classes: (i) finite impulse with directed acyclic graphs, and (ii) infinite impulse with directed cyclic graphs \citep{Sun2019}. A finite impulse RNN is a directed acyclic graph that can be unrolled and replaced with a strictly feed-forward NN, whilst an infinite impulse RNN is a directed acyclic graph which cannot be unrolled \citep{Sherstinsky2020}. 

\begin{figure}[ht!]
	\centering
	\includegraphics[width=0.98\linewidth]{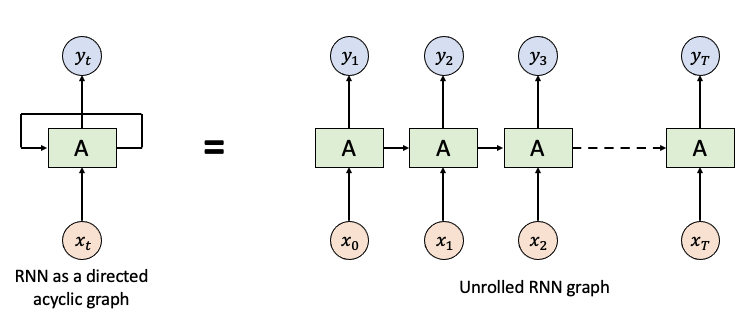}
	\caption[RNNs can be represented as directed acyclic graphs]{RNNs can be represented as directed acyclic graphs. Chuck of the NN $A$ looks at some input $x_t$ and outputs a value $y_t$. A loop allows information to be passed from one step of the network to the next. Bias weights are omitted for clarity. Adapted from \cite{Olah2015}.}
	\label{fig:RNN_Unfolding}
\end{figure}

When doing wavefield modelling, the wavefield at time $t$ is derived from components of the previous time $t-1$ and not future time $t+1$. Given this directionality in wave propagation, forward modelling can be mapped to a finite impulse directed acyclic graph. ``Unfolding'' of a RNN is a useful technique to visualize directed acyclic graphs. Figure~\ref{fig:RNN_Unfolding} shows part of an unfolded RNN. 

\section{LSTM Components}\label{sec:app_theory_Individual_RNN_Components}
\subsection{Forget gate}
The forget gate uses a sigmoid function to decide what information should be passed between hidden states. Values from this gate range between 0 and 1, indicating the level of information to be forgotten.

\subsection{Input gate}
The input gate uses a sigmoid activation function, accepts the previous hidden state and current input and decides which values will be updated. The current input and previous hidden state are passed into the tanh function to squeeze values between -1 and 1 and get a potential new candidate.
 	
\subsection{Cell state}
\begin{figure}[ht!]
    \centering
    \includegraphics[width=0.8\linewidth]{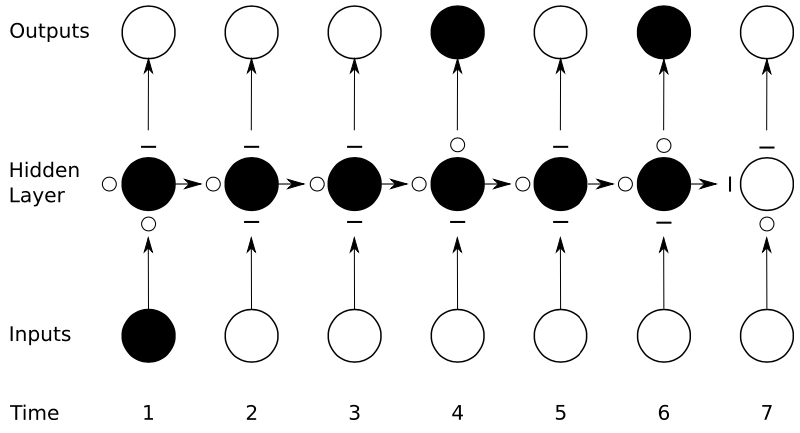}
    \caption[LSTM preserves the gradient to deeper layers of a NN.]{``Preservation of gradient with LSTM structure. The shading of the nodes indicate the influence of the inputs at a particular point in time. The black nodes indicate maximum sensitive and the white nodes are entirely insensitive. The state of the input, forget, and output gates are displayed below, to the left and above the hidden layer respectively. In this example, all gates are either entirely open (‘O’) or closed (‘—’)." From \cite{Graves2012}.}
    \label{fig:LSTM_gradient}
\end{figure}

The cell state acts as a mechanism to transfers information through the sequence. This enables information from earlier time steps to be available at later time steps, thus reducing the effects of vanishing gradient. The preservation of gradient information by LSTM is illustrated in Figure~\ref{fig:LSTM_gradient}.

\subsection{Output gate}
The output gate determines the next hidden state. It 
 uses a sigmoid activation on the current state and previous hidden state, and multiples this new cell state with a tanh to decide which part of the data should be pushed forward through the sequence.

\section[Equivalence of Automatic Differentiation]{Automatic Differentiation and Adjoint State}\label{sec:app_results_equivalence_AD_adjoint}
Following from \cite{Richardson2018}, consider the 1D scalar wave equation consisting of only one shot with one receiver. Using cost function in Equation~\ref{eq:dnn_cost_function_J}, the gradient with respect to one wavefield time step is
\begin{equation}
    \frac{\partial J}{\partial \mathbf{u}_t} = \frac{\partial J}{\partial \mathbf{u}_t} + \frac{\partial J}{\partial \mathbf{u}_{t+1}}\frac{\partial \mathbf{u}_{t+1}}{\partial \mathbf{u}_t}+\frac{\partial J}{\partial \mathbf{u}_{t+2}}\frac{\partial \mathbf{u}_{t+2}}{\partial \mathbf{u}_t},
\end{equation}
where $\frac{\partial \mathcal{J}}{\partial \mathbf{u}_t}$ indicates the row vector of partial derivatives of $\mathcal{J}$ with respect to elements of $\mathbf{u}_t$, while the wave speed $\mathbf{c}$ and the wavefields from other time steps $\mathbf{u_{t' \neq t}}$ are held constant. $\frac{\partial \mathcal{J}}{\partial \mathbf{u}_{t+1}}$ includes all changes to $\mathcal{J}$ caused by changes of the wavefield at a previous $\mathbf{u}_{t+1}$ and a later time step $\mathbf{u}_{t+1}$.

Considering Automatic differentiation, the gradient of the cost function with respect to the wave speed is obtained by chaining different derivates at different time steps via the chain rule. Namely, 
\begin{equation}
        \frac{\partial \mathcal{J}}{\partial \mathbf{c}} = \sum_{t=1}^{N_t}\frac{\partial \mathcal{J}}{\partial \mathbf{u}_t}\frac{\partial\mathbf{u}_t}{\partial \mathbf{c}}
        \label{eqn:accum_djdc}.
\end{equation}
Consider $N_t=4$ as an example. The steps to calculate the gradient of $\mathcal{J}$ would be
\begin{equation}
        \frac{\partial \mathcal{J}}{\partial \mathbf{u}_4}=\frac{\partial \mathcal{J}}{\partial \mathbf{u}_4},
\end{equation}
\begin{equation}
        \frac{\partial \mathcal{J}}{\partial \mathbf{u}_3}=\frac{\partial \mathcal{J}}{\partial \mathbf{u}_3}+\frac{\partial \mathcal{J}}{\partial \mathbf{u}_{4}}\frac{\partial \mathbf{u}_{4}}{\partial \mathbf{u}_3},
\end{equation}
\begin{equation}
        \frac{\partial \mathcal{J}}{\partial \mathbf{u}_2}=\frac{\partial \mathcal{J}}{\partial \mathbf{u}_2}+\frac{\partial \mathcal{J}}{\partial \mathbf{u}_{3}}\frac{\partial \mathbf{u}_{3}}{\partial \mathbf{u}_2}+\frac{\partial \mathcal{J}}{\partial \mathbf{u}_{4}}\frac{\partial \mathbf{u}_{4}}{\partial \mathbf{u}_2},
\end{equation}
\begin{equation}
        \frac{\partial \mathcal{J}}{\partial \mathbf{u}_1} =\frac{\partial \mathcal{J}}{\partial \mathbf{u}_1}+\frac{\partial \mathcal{J}}{\partial \mathbf{u}_{2}}\frac{\partial \mathbf{u}_{2}}{\partial \mathbf{u}_1}+\frac{\partial \mathcal{J}}{\partial \mathbf{u}_{3}}\frac{\partial \mathbf{u}_{3}}{\partial \mathbf{u}_1}.
\end{equation}
The required partial derivatives are given by
\begin{equation}
        \frac{\partial \mathcal{J}}{\partial \mathbf{u}_t}= 2(\delta_{x_r}^T\mathbf{u}_t - d_t)\delta_{x_r}^T,
\end{equation}
\begin{equation}
        \frac{\partial \mathbf{u}_{t+1}}{\partial \mathbf{u}_t}= \mathbf{c}^2 \Delta_t^2 \mathbf{D_x}^2 + 2,
\end{equation}
\begin{equation}
\frac{\partial \mathbf{u}_{t+2}}{\partial \mathbf{u}_t} = -1,
\end{equation}
\begin{equation}
        \frac{\partial\mathbf{u}_t}{\partial \mathbf{c}}= 2 \mathbf{c} \Delta_t^2 \left( \mathbf{D_x}^2 \mathbf{u}_{t-1} - \mathbf{f}_{t-1}\right).
\end{equation}
Substituting into Equation \ref{eqn:accum_djdc} yields
\begin{align}
        \frac{\partial \mathcal{J}}{\partial \mathbf{c}} &= \sum_{t=1}^{N_t} \left(2(\delta_{x_r}^T\mathbf{u}_t - d_t)\delta_{x_r}^T + \frac{\partial \mathcal{J}}{\partial \mathbf{u}_{t+1}}(\mathbf{c}^2 \Delta_t^2 \mathbf{D_x}^2 + 2)
        - \frac{\partial \mathcal{J}}{\partial \mathbf{u}_{t+2}}\right) \\
        &\qquad\times 2 \mathbf{c} \Delta_t^2 \left( \mathbf{D_x}^2 \mathbf{u}_{t-1} - \mathbf{f}_{t-1}\right)\nonumber\\
        &= \sum_{t=1}^{N_t} \left(\mathbf{c}^2 \Delta_t^2 \left(\mathbf{D_x}^2\frac{\partial \mathcal{J}}{\partial \mathbf{u}_{t+1}} + \frac{2}{\mathbf{c}^2 \Delta_t^2} (\delta_{x_r}^T\mathbf{u}_t - d_t)\delta_{x_r}^T\right)\right. \\
        &\left.\qquad + 2\frac{\partial \mathcal{J}}{\partial \mathbf{u}_{t+1}}- \frac{\partial \mathcal{J}}{\partial \mathbf{u}_{t+2}}\right)2 \mathbf{c} \Delta_t^2 \left( \mathbf{D_x}^2 \mathbf{u}_{t-1} - \mathbf{f}_{t-1}\right)\\
        &= \sum_{t=1}^{N_t} -F^{-1}\left(\frac{2}{\mathbf{c}^2 \Delta_t^2}(\delta_{x_r}^T\mathbf{u}_t - d_t)\delta_{x_r}^T\right)
        2 \mathbf{c} \Delta_t^2 \left( \mathbf{D_x}^2 \mathbf{u}_{t-1} - \mathbf{f}_{t-1}\right),
\end{align}
where $F^{-1}(\mathbf{g}_t)$ is wave propagation backward in time of the source amplitude $\mathbf{g}_t$. Recognize that the wave equation can be used to express the factor on the right as the second time derivative of the forward propagated source wavefield,
\begin{align}
        \frac{\partial \mathcal{J}}{\partial \mathbf{c}} &=\sum_{t=1}^{N_t}-F^{-1}\left(\frac{2}{\mathbf{c}^2 \Delta_t^2} (\delta_{x_r}^T\mathbf{u}_t - d_t)\delta_{x_r}^T\right)\frac{2\Delta_t^2}{\mathbf{c}}\frac{\partial^2\mathbf{u}_{t-1}}{\partial t^2}\\
        &=\sum_{t=1}^{N_t}-F^{-1}\left(2 (\delta_{x_r}^T\mathbf{u}_t - d_t)\delta_{x_r}^T\right)\frac{2}{\mathbf{c}^3}\frac{\partial^2\mathbf{u}_{t-1}}{\partial t^2},
\end{align}
where the linearity of wave propagation is used to take factors out of the source amplitude of the left term. The result is the same equation that is used in the adjoint state method.

%% file: appB/appendix_b_additional_results.tex
\chapter[Code and Additional Results]{Code Repository and Additional Results}

\section{Code Repository}\label{sec:app_results_code_repo}
A git repository with all code used in this dissertation is available at \url{https://gitfront.io/r/zerafachris/52df30fb666ba880749c8e951a3d056ce628a6cd/PhD/}. In the following chapter, any reference made to code will refer to this repository.

\section{Marmousi-2 Model}\label{sec:app_results_generation_marm}
\begin{figure}[!ht]
    \centering
    \includegraphics[width=0.4\textwidth]{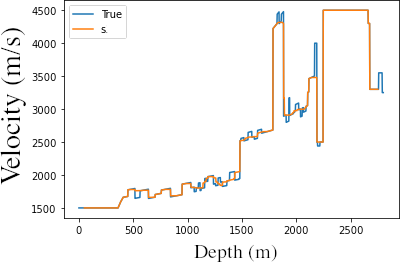}
    \caption[Sample velocity profile through original and modified Marmousi-2 model.]{Sample velocity through original and modified Marmousi-2 model. $s$ is the 150m median filtered modified Marmousi-2.}
    \label{fig:app_smooth_marm_vel_profile}
\end{figure}

The original Marmousi-2 model has been made available by \cite{Martin2006} under a Creative Commons Attribution 4.0 International License. This was modified with a 150m median filter. Figure~\ref{fig:app_smooth_marm_vel_profile} shows the impact of the 150m median filter on the vertical resolution of the model. The code used for this modification is available at \href{https://gitfront.io/r/zerafachris/52df30fb666ba880749c8e951a3d056ce628a6cd/PhD/blob/code/appendix/Marmousi_2_generator.ipynb}{\url{./code/appendix/Marmousi\_2\_generator.ipynb}}.

\section{Classical FWI}\label{sec:app_results_classical_FWI}
\subsection{Inversion}
\begin{figure}[!ht]
        \centering
        \includegraphics[width=0.5\textwidth]{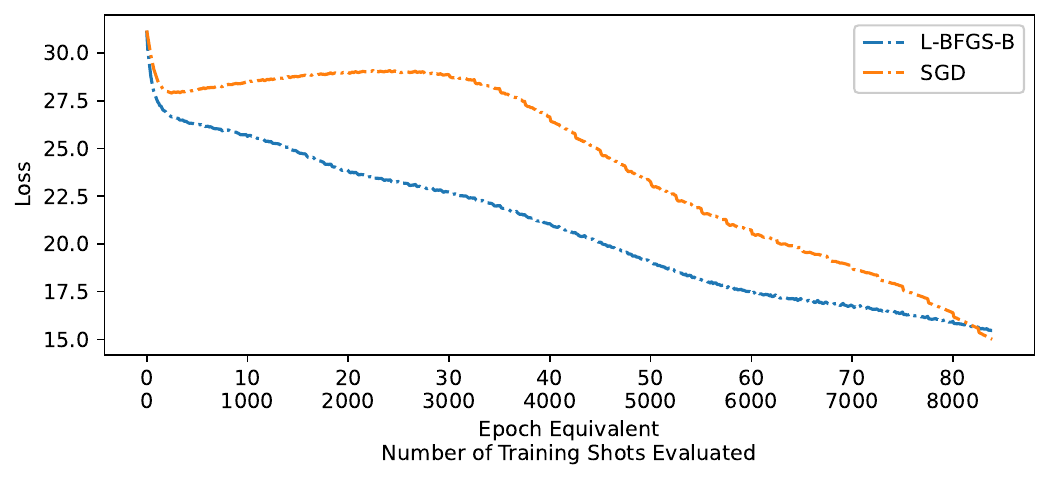}
        \caption[Classical FWI loss update.]{Classical FWI loss update for L-BFGS-B and Stochastic Gradient Descent. L-BFGS-B was a better loss optimizer than Stochastic Gradient Descent due to the monotonically decreasing loss. Stochastic Gradient Descent training should have been stopped at an earlier epoch due to the increase at 30 when compared to earlier epoches.}
        \label{fig:app_classical_fwi_loss}
\end{figure}

\begin{figure}[ht]
    \centering
    \includegraphics[width=0.44\textwidth]{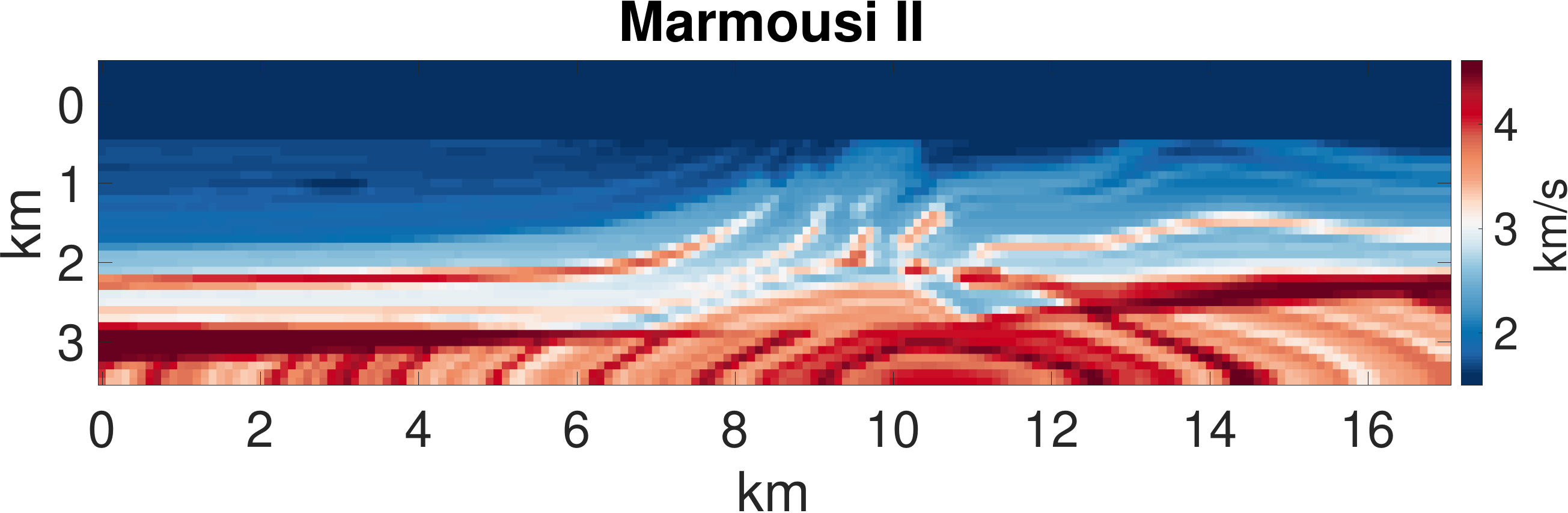}
    \includegraphics[width=0.44\textwidth]{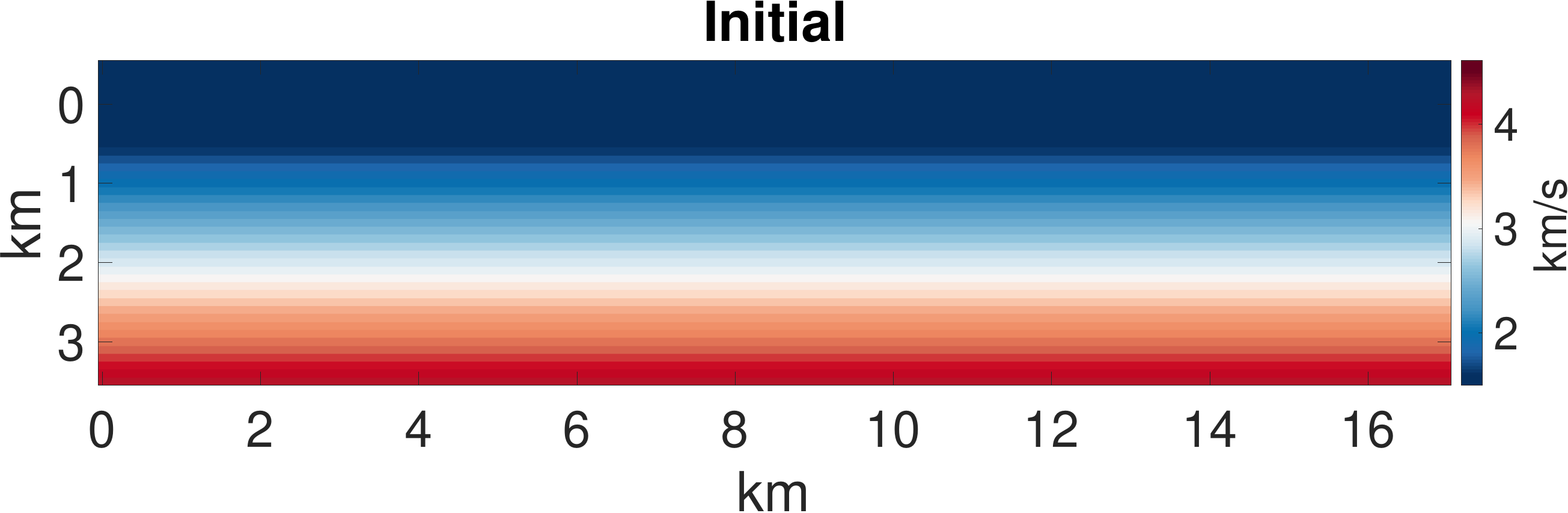}
    \includegraphics[width=0.44\textwidth]{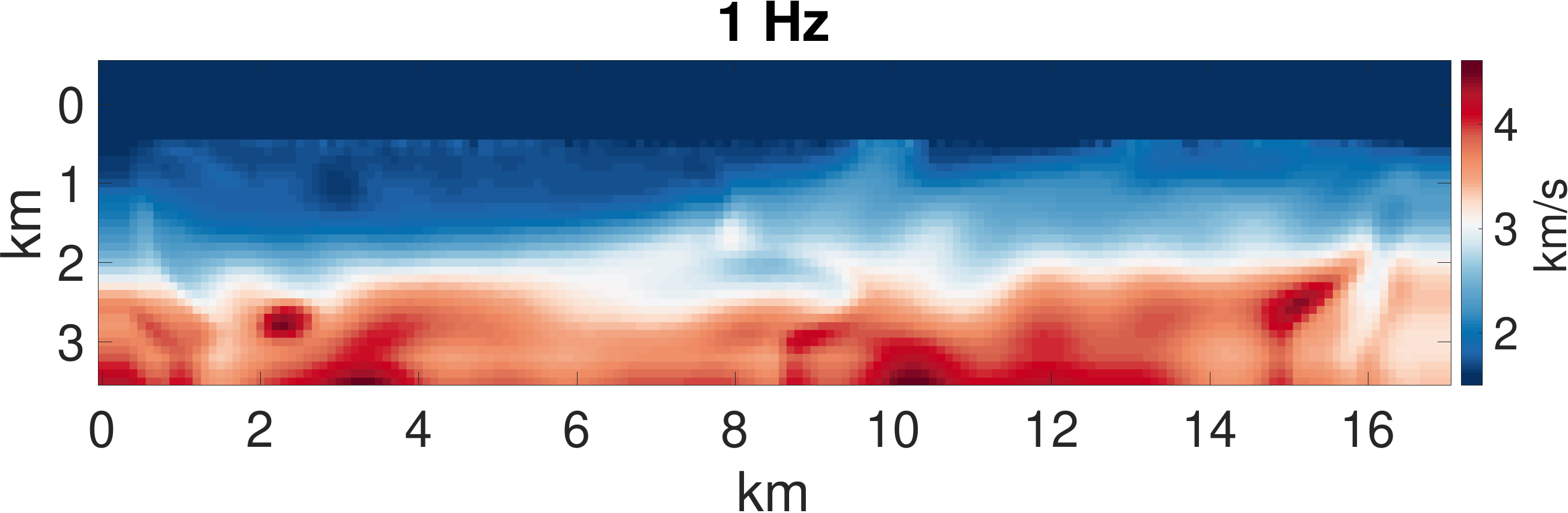}
    \includegraphics[width=0.44\textwidth]{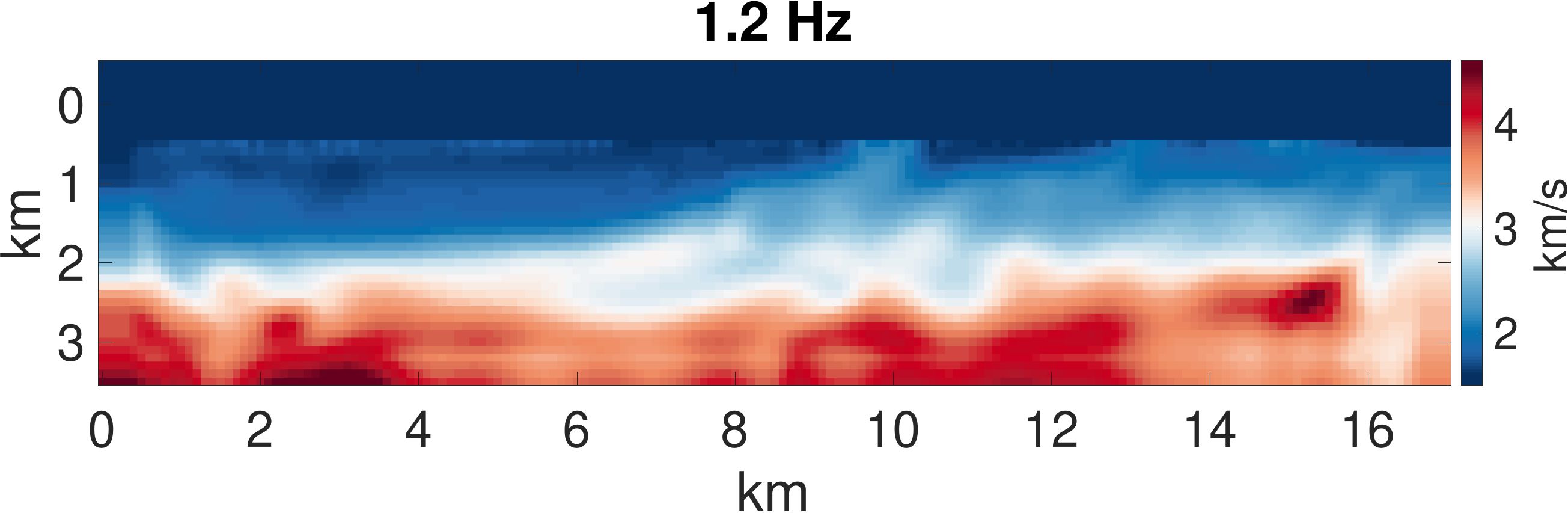}
    \includegraphics[width=0.44\textwidth]{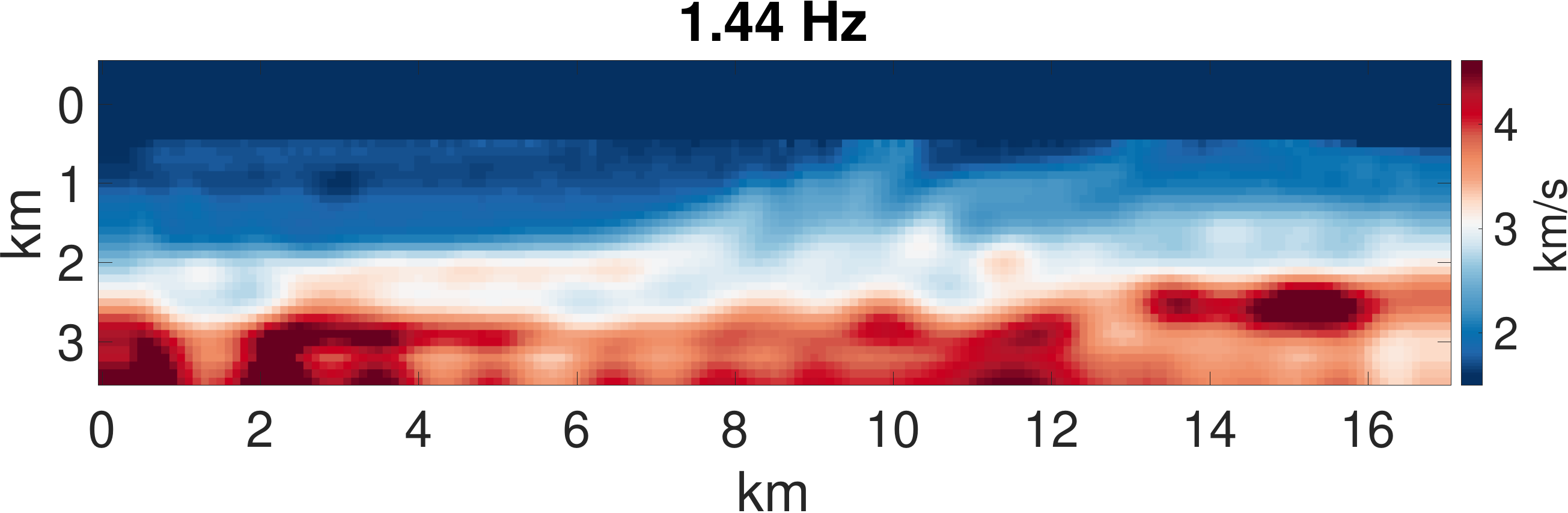}
    \includegraphics[width=0.44\textwidth]{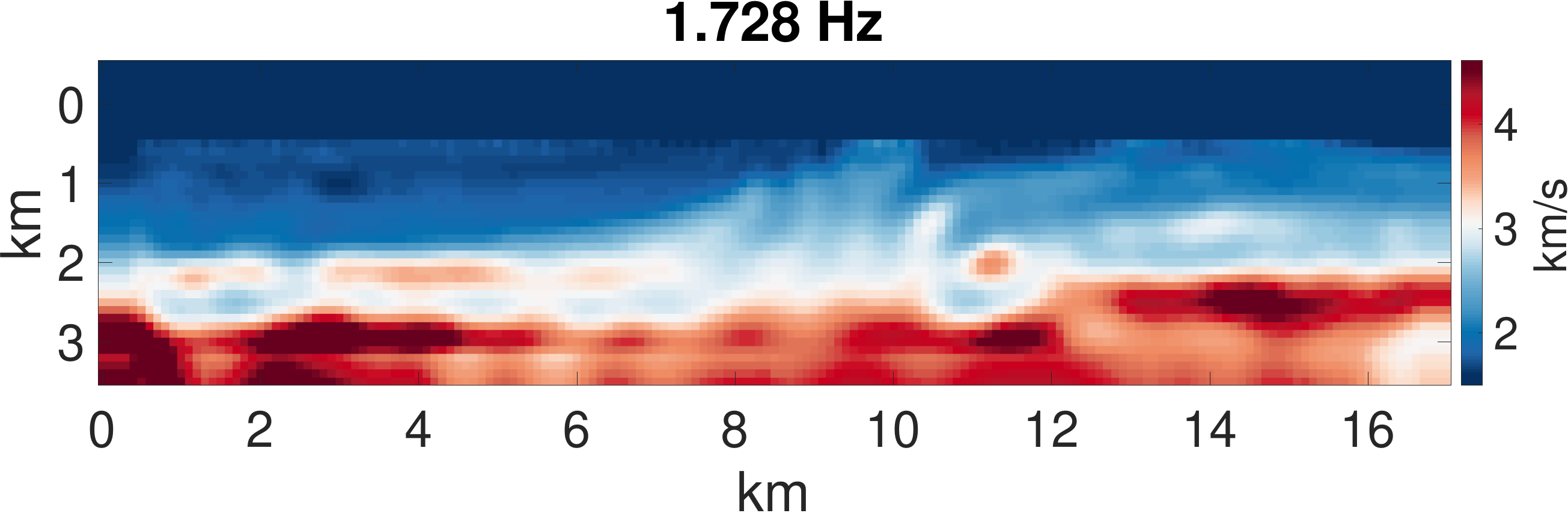}
    \includegraphics[width=0.44\textwidth]{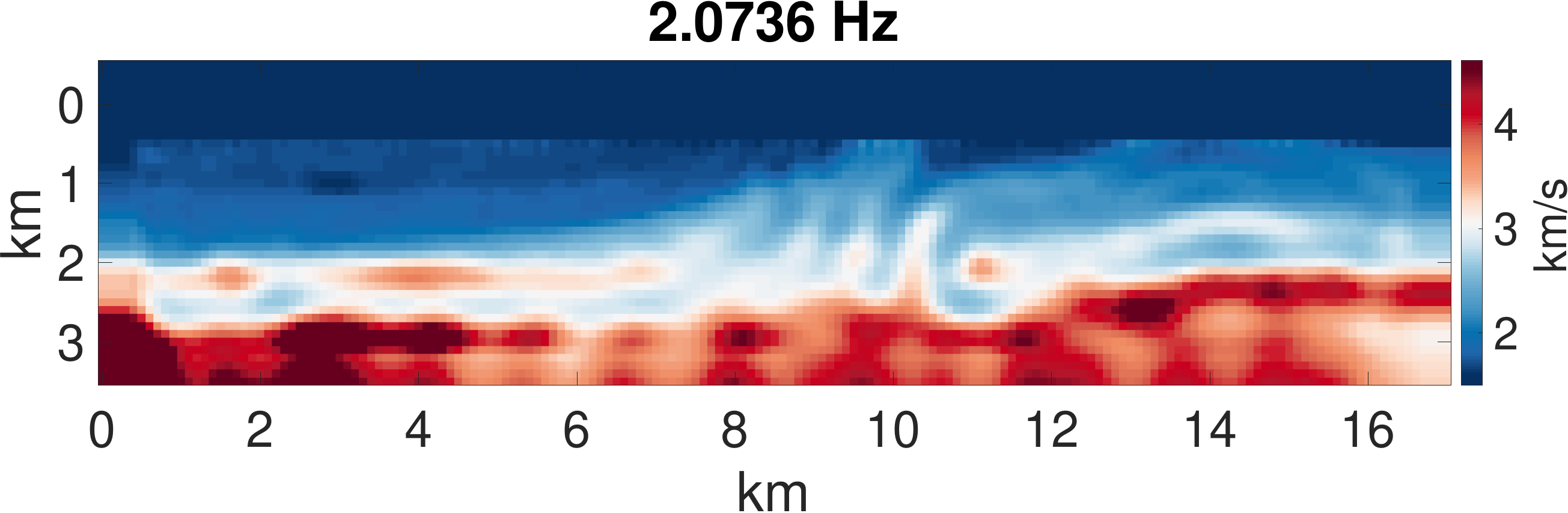}
    \includegraphics[width=0.44\textwidth]{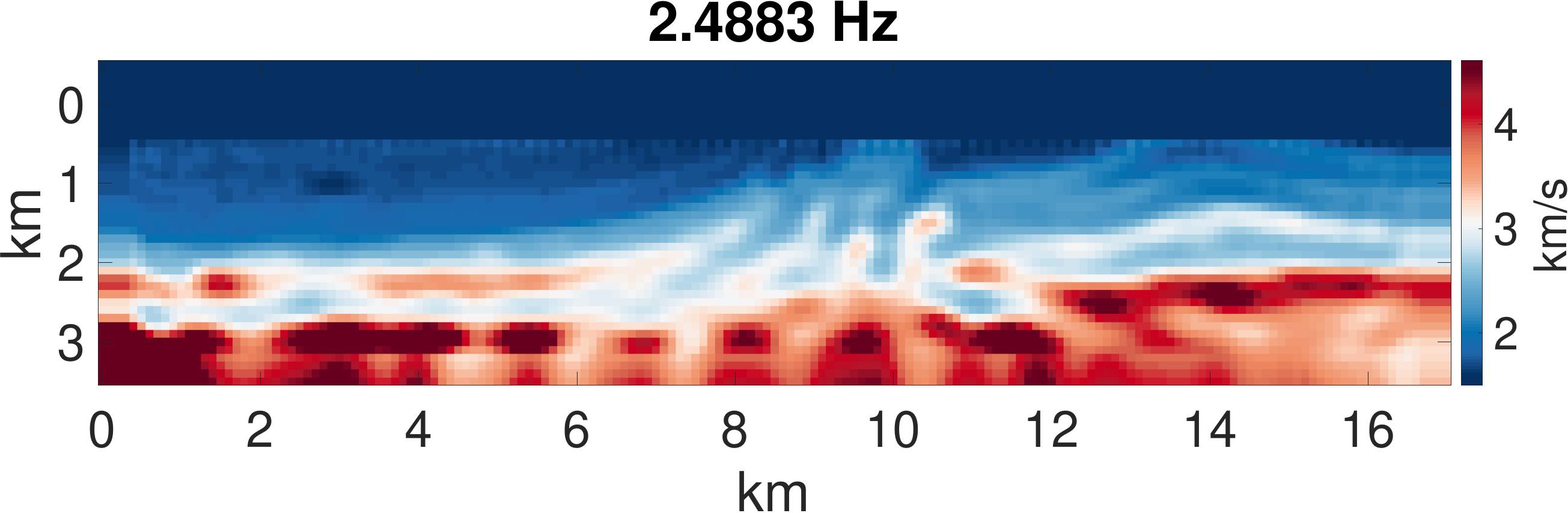}
    \includegraphics[width=0.44\textwidth]{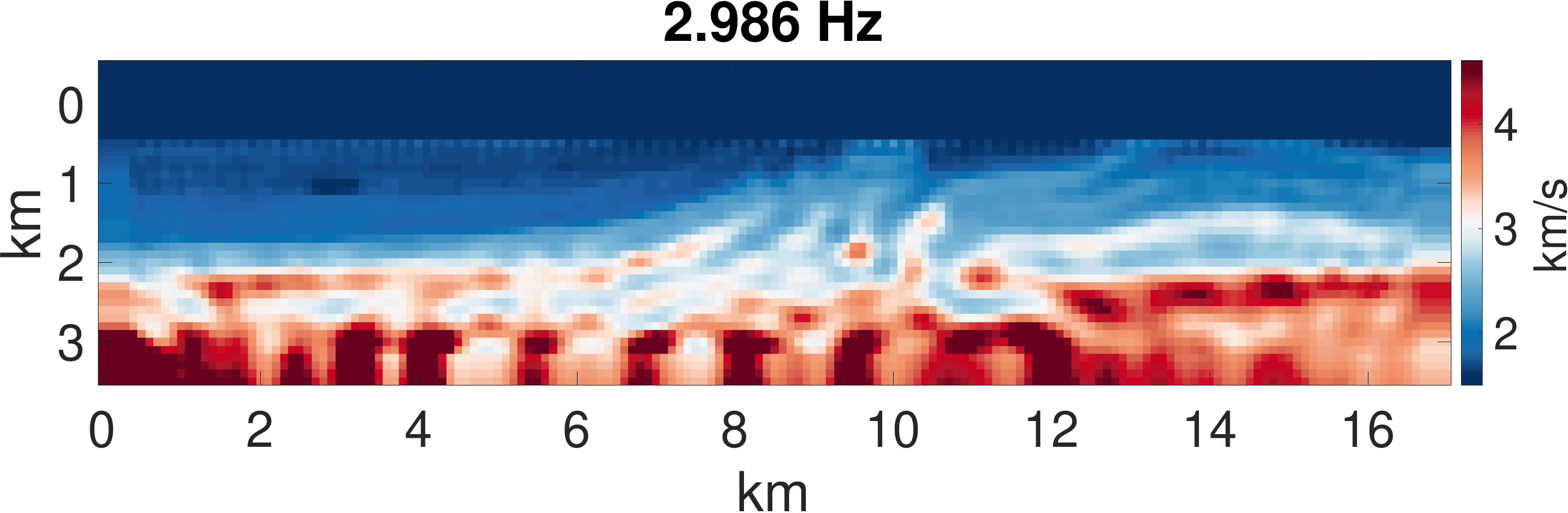}
    \includegraphics[width=0.44\textwidth]{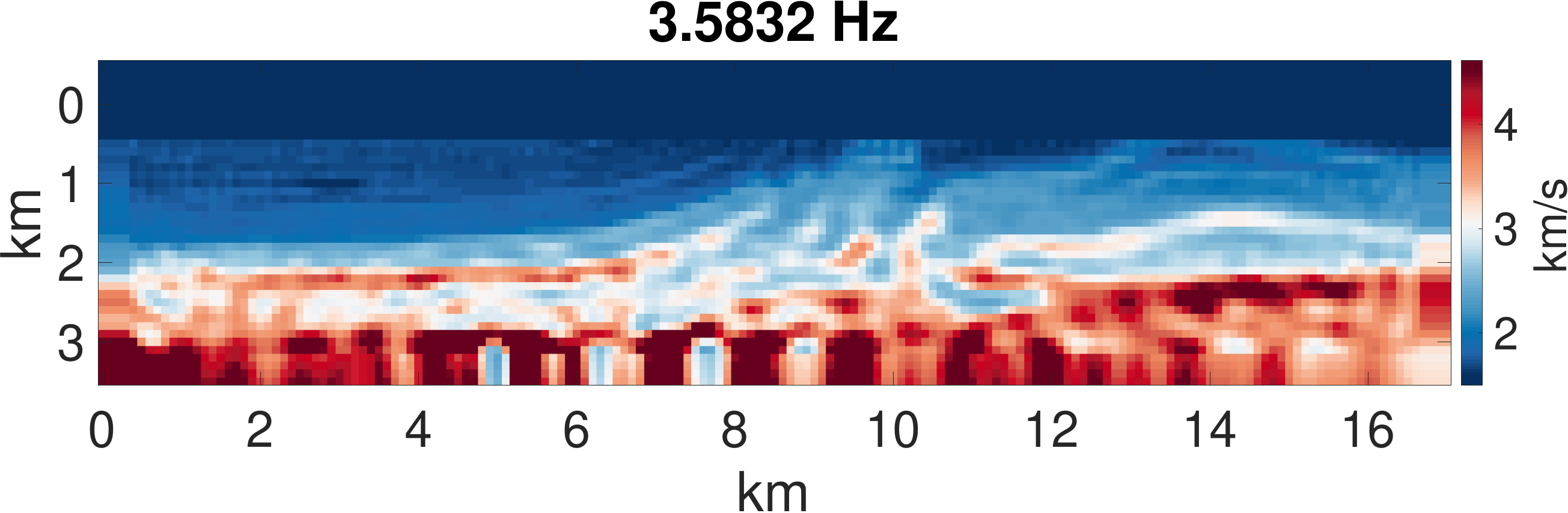}
    \includegraphics[width=0.44\textwidth]{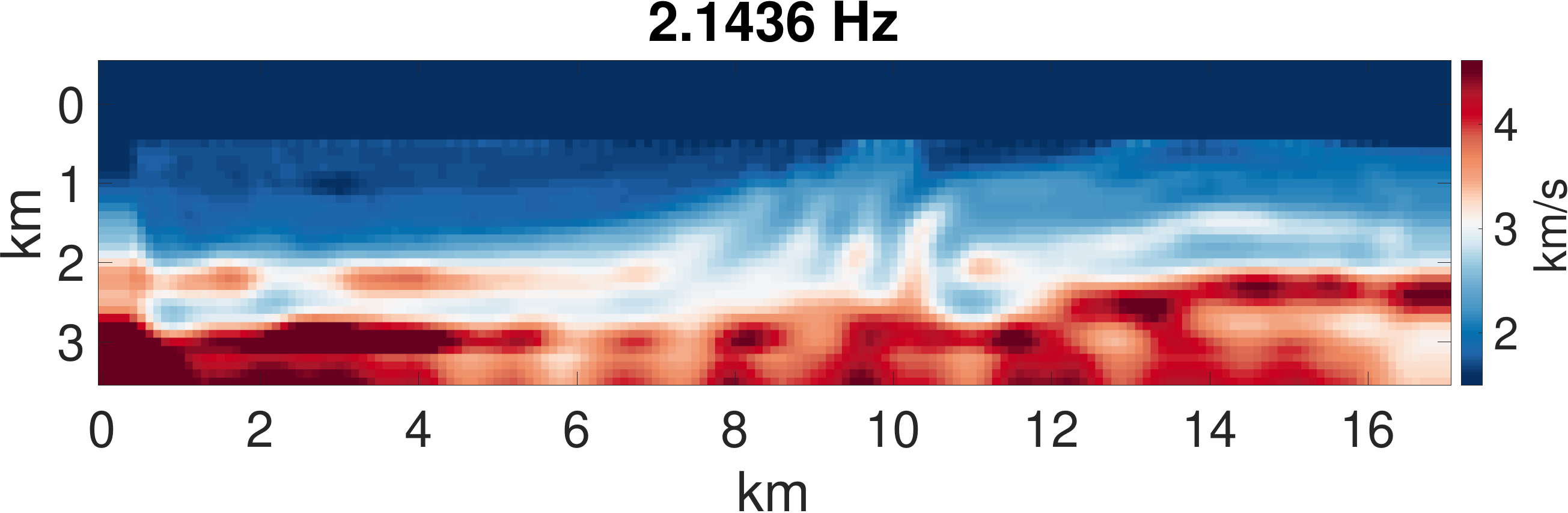}
    \includegraphics[width=0.44\textwidth]{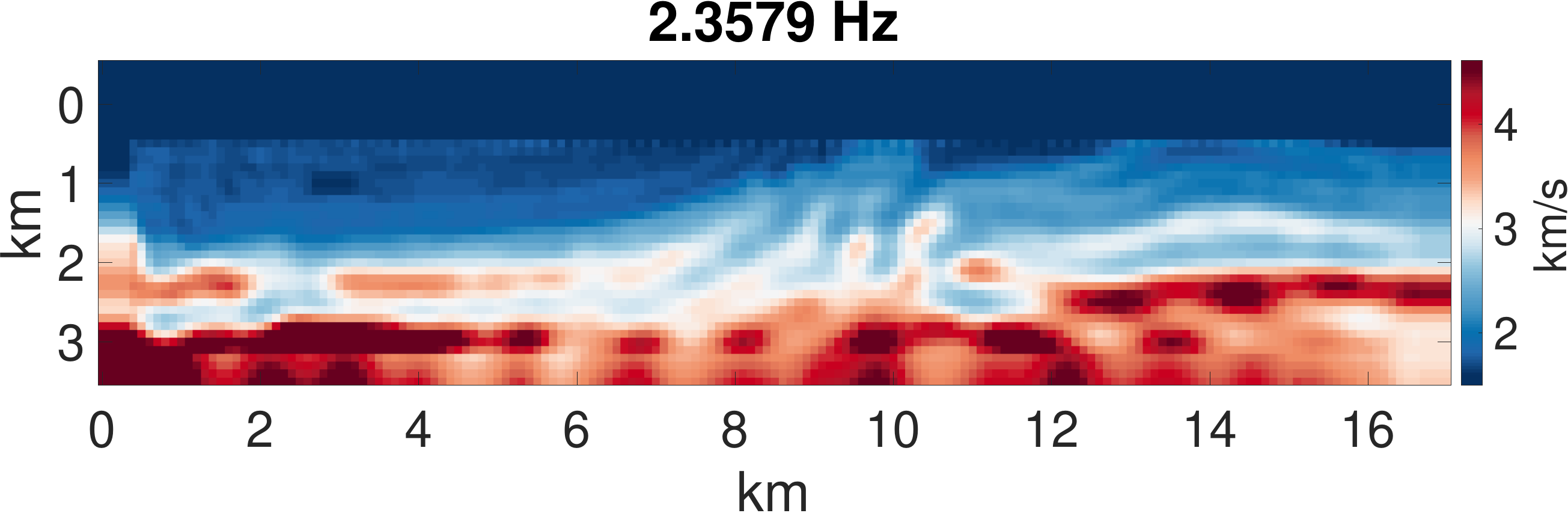}
    \includegraphics[width=0.44\textwidth]{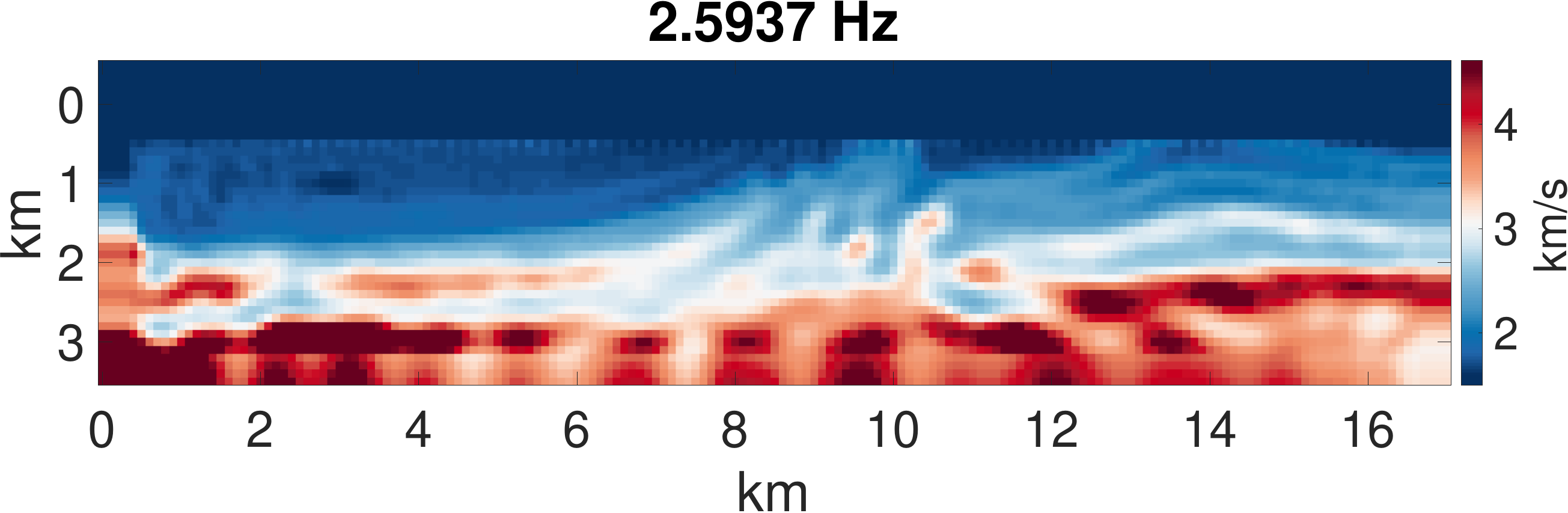}
    \includegraphics[width=0.44\textwidth]{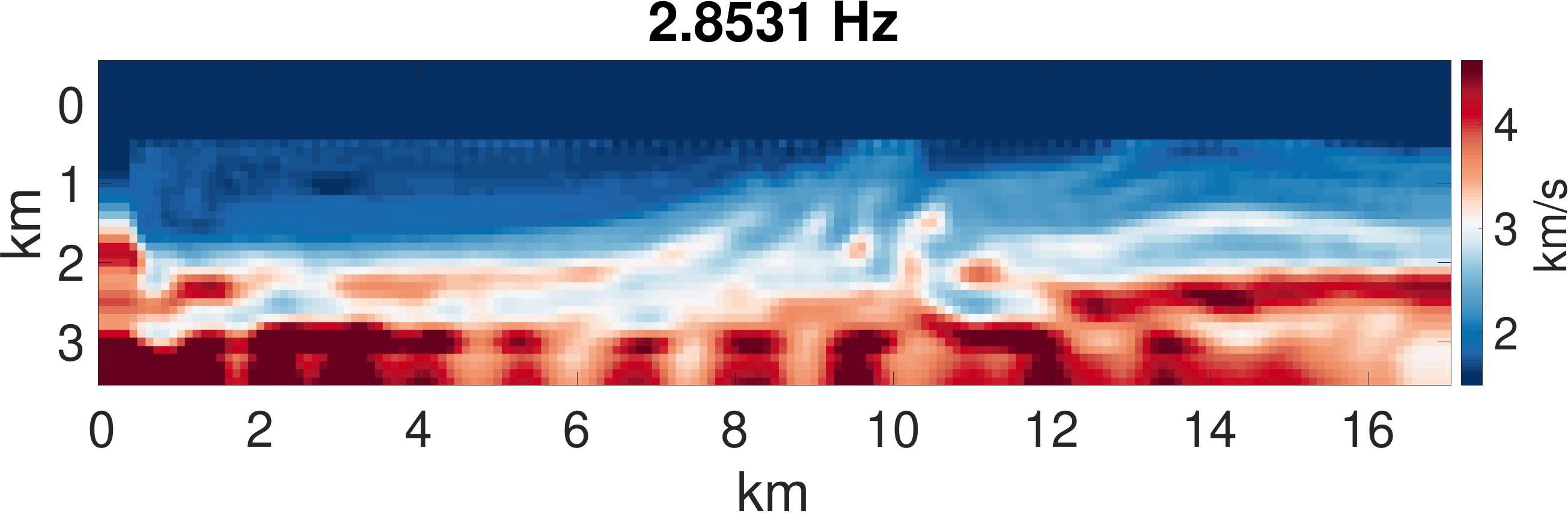}
    \includegraphics[width=0.44\textwidth]{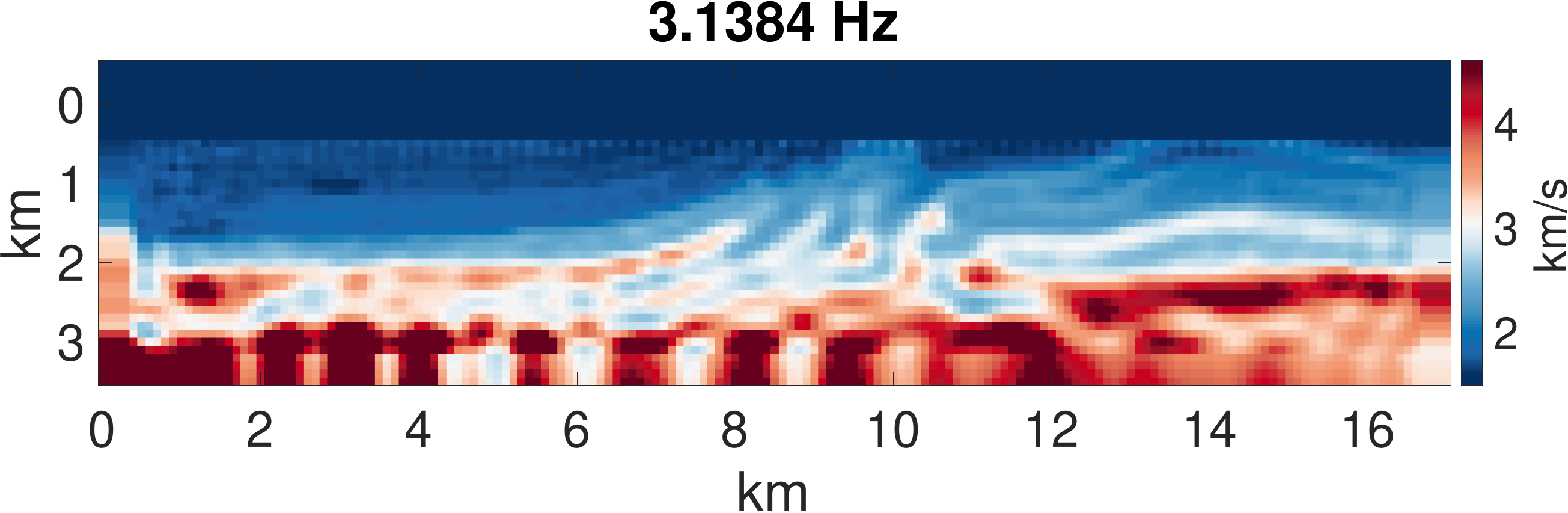}
    \includegraphics[width=0.44\textwidth]{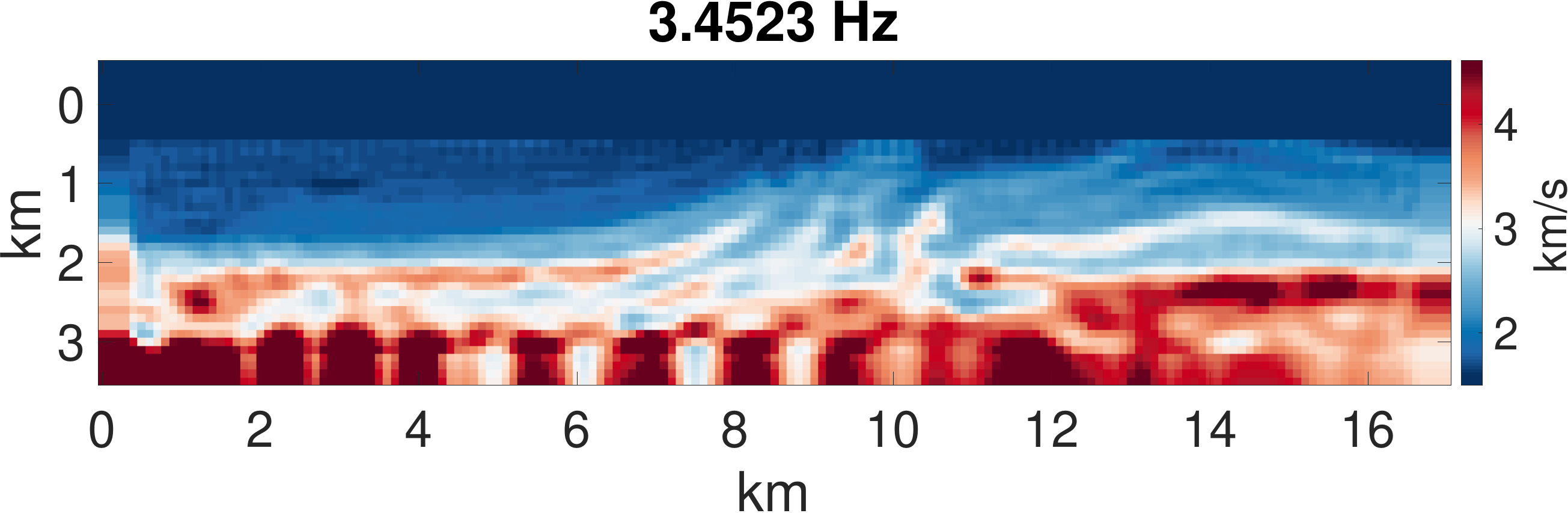}
    \includegraphics[width=0.44\textwidth]{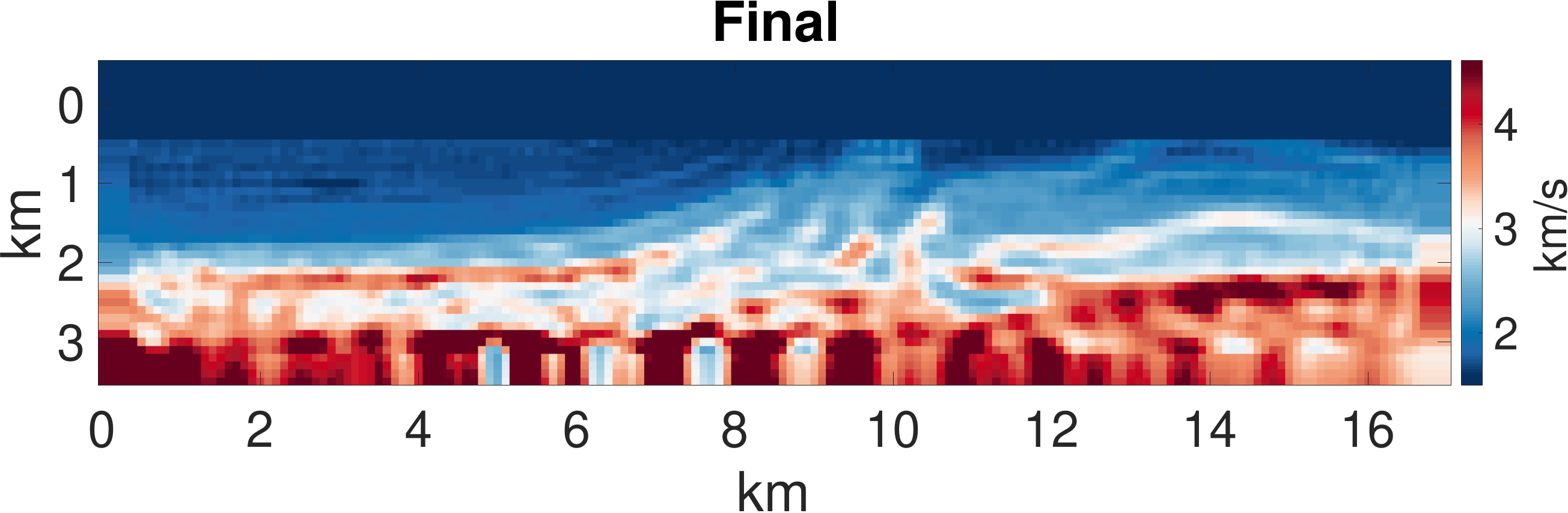}
    \caption[Classical FWI frequency updates.]{Classical FWI frequency updates. Starting from 1\si{Hz}, model update frequency was increased by a factor of 1.2 until a maximum frequency of 3.45\si{Hz}.The optimization algorithm was L-BFGS-B, with 50 iterations per step.}
    \label{fig:classical_fwi_progression}
\end{figure}

FWI with Sobolev space norm regularization was used as the deterministic version of FWI within this work. The maximum frequency of the inversion process was set to be 3.5\si{Hz}. The iterative update process started from frequency 1\si{Hz} and iteratively updated by a factor of 1.2 until reaching a maximum frequency of 3.45\si{Hz}. The optimization algorithm was L-BFGS-B, with 50 iterations per frequency. Figure~\ref{fig:app_classical_fwi_loss} is the loss update for L-BFGS-B and Stochastic Gradient Descent. Figure~\ref{fig:classical_fwi_progression} shows the progression of the frequency updates. Code for this implementation is available at \href{https://gitfront.io/r/zerafachris/fbeffbcbfa1a363bc271e3bcd3717a830d3bfe7f/academic/tree/PhD/code/classical_FWI/marmousi/*}{\url{./classical\_FWI/marmousi}}.

\subsection{Ray-Tracing}
\begin{figure}[ht]
        \centering
        \subbottom[Sample of ray-paths through Marmousi]{\includegraphics[width=0.65\textwidth]{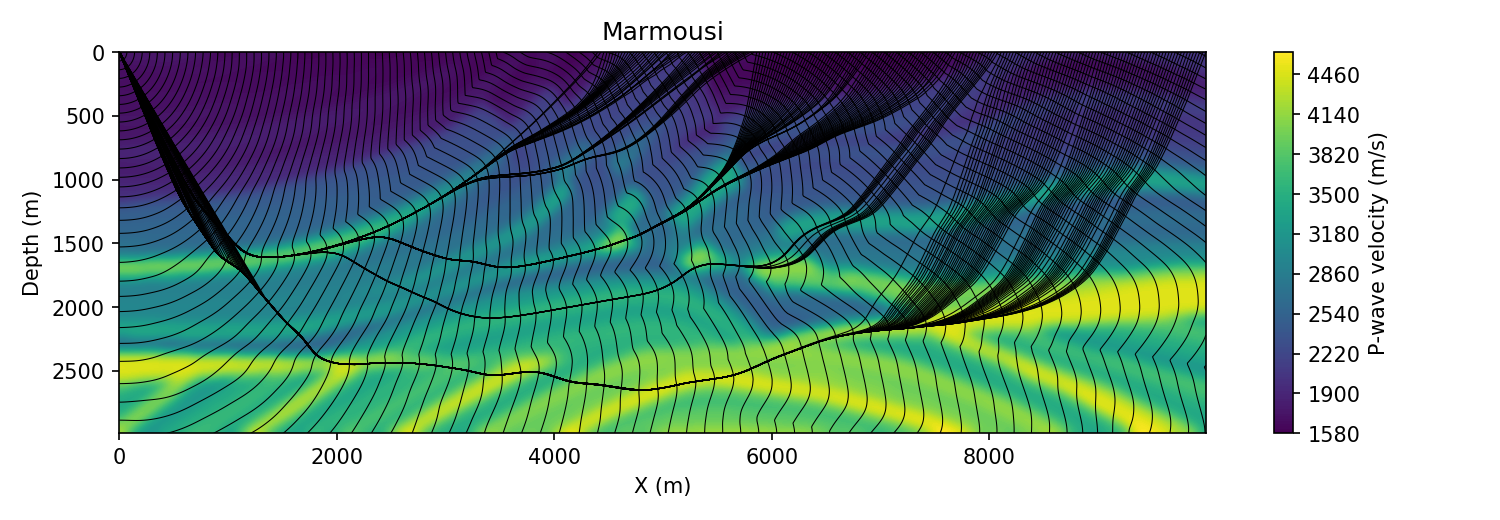}}

        \subbottom[Area of coverage intensity from ray-tracing.]{\includegraphics[width=0.65\textwidth]{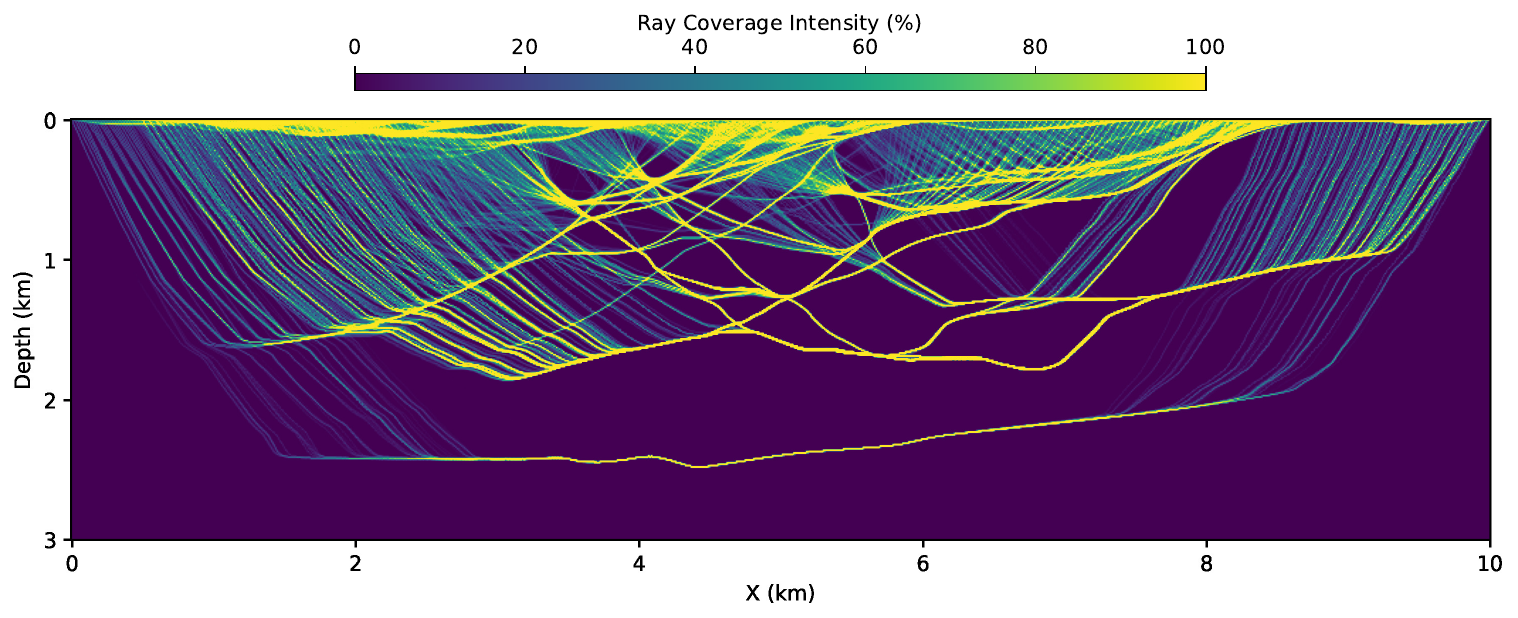}}
        \caption[Ray-tracing using \textbf{fteikpy}.]{Ray-tracing using \textbf{fteikpy}.}
        \label{fig:fteikpy_ray_tracing_marmousi}
\end{figure}

Pre-cursor to FWI is ray-tracing modelling to assess areas of update from standard FWI formulation. Open source version of \textbf{fteikpy} Python library provided by \citet{Noble2014} was adapted and utilized on the Marmousi-2 in §~\ref{sec:app_results_generation_marm}. This implementation computes accurate first arrival travel-times in 2D heterogeneous isotropic velocity models. The algorithm solves a hybrid Eikonal formulation with a spherical approximation near-source and a plane wave approximation in the far field. This reproduces properly the spherical behaviour of wave fronts in the vicinity of the source \citep{Noble2014}. Figure~\ref{fig:fteikpy_ray_tracing_marmousi} shows a sample of ray-paths for a source at 0km and depth 0km and ray coverage for the Marmousi model. The adapted code is available within \href{https://gitfront.io/r/zerafachris/52df30fb666ba880749c8e951a3d056ce628a6cd/PhD/blob/code/ray_tracing/marmousi/marmousi_ray_tracing.ipynb}{\url{./ray_tracing/marmousi/marmousi_ray_tracing.ipynb}}.
\clearpage

\section{Data-Driven FWI}\label{sec:app_results_summary_results_exp_1}
\subsection{DNN Architectures}\label{sec:app_results_architectural_summary_dnn_workflow}
Table~\ref{tab:app_results_dnn_architectures} lists DNN architectures used throughout Section~\ref{sec:results_FWI_as_a_Learned_Direct_Approximation}.
\begin{table*}[ht]
    \footnotesize
    \centering
    \begin{tabular}{@{}ll@{}}\toprule
    Architecture                           & Code Repository    \\ \hline
    Time to Pseudo-Spectral 1D             & \href{https://gitfront.io/r/zerafachris/52df30fb666ba880749c8e951a3d056ce628a6cd/PhD/blob/code/appendix/DNN_arch_time_pseudo.txt}{\url{./appendix/DNN\_arch\_time\_pseudo_1D.txt}}\\
    Time to Pseudo-Spectral 2D             & \href{https://gitfront.io/r/zerafachris/52df30fb666ba880749c8e951a3d056ce628a6cd/PhD/blob/code/appendix/DNN_arch_time_pseudo.txt}{\url{./appendix/DNN\_arch\_time\_pseudo_2D.txt}}\\
    Conv1D                                 & \href{https://gitfront.io/r/zerafachris/52df30fb666ba880749c8e951a3d056ce628a6cd/PhD/blob/code/appendix/DNN_arch_conv1d.txt}{\url{./appendix/DNN\_arch\_conv1d.txt}}\\
    Conv2D                                 & \href{https://gitfront.io/r/zerafachris/52df30fb666ba880749c8e951a3d056ce628a6cd/PhD/blob/code/appendix/DNN_arch_conv2d.txt}{\url{./appendix/DNN\_arch\_conv2d.txt}}\\
    VGG                                    & \href{https://gitfront.io/r/zerafachris/52df30fb666ba880749c8e951a3d056ce628a6cd/PhD/blob/code/appendix/DNN_arch_vgg.txt}{\url{./appendix/DNN\_arch\_vgg.txt}}\\
    ResNet                                 & \href{https://gitfront.io/r/zerafachris/52df30fb666ba880749c8e951a3d056ce628a6cd/PhD/blob/code/appendix/DNN_arch_resnet.txt}{\url{./appendix/DNN\_arch\_resnet.txt}}\\
    Marmousi - Time to Pseudo-Spectral     & \href{https://gitfront.io/r/zerafachris/52df30fb666ba880749c8e951a3d056ce628a6cd/PhD/blob/code/appendix/DNN_arch_marm_time_pseudo.txt}{\url{./appendix/DNN\_arch\_marm\_time\_pseudo.txt}}\\
    Marmousi - Pseudo-Spectral to Velocity & \href{https://gitfront.io/r/zerafachris/52df30fb666ba880749c8e951a3d056ce628a6cd/PhD/blob/code/appendix/DNN_arch_marm_pseudo_velocity.txt}{\url{./appendix/DNN\_arch\_marm\_pseudo\_velocity.txt}}\\
    \hline
    \end{tabular}
    \caption{Repositories defining different architectures used in Section~\ref{sec:results_FWI_as_a_Learned_Direct_Approximation}.}\label{tab:app_results_dnn_architectures}
\end{table*}







\subsection{Architecture and Loss Tuning}
Tables~\ref{tab:app_res1_duration}-\ref{tab:app_res1_inversion} show results for different architecture and loss optimizer combinations.
\begin{table*}[!ht]
    \footnotesize
    \centering
    \begin{tabular}{@{}llll@{}}\toprule
            Architecture & Loss Optimizer & Duration (Hours) & Rank         \\ \hline
            MLP          & Adagrad        & 45               & 18            \\
            MLP          & Adadelta       & 54               & 16            \\
            MLP          & RMSprop        & 31               & 20            \\
            MLP          & Adam           & 36               & 19            \\
            Conv1D       & Adagrad        & 49               & 17            \\
            Conv1D       & Adadelta       & 59               & 14            \\
            Conv1D       & RMSprop        & 59               & 14            \\
            Conv1D       & Adam           & 55               & 15            \\
            Conv2D       & Adagrad        & 66               & 12            \\
            Conv2D       & Adadelta       & 66               & 12            \\
            Conv2D       & RMSprop        & 67               & 10            \\
            Conv2D       & Adam           & 67               & 10            \\
            VGG          & Adagrad        & 75               & 8             \\
            VGG          & Adadelta       & 75               & 8             \\
            VGG          & RMSprop        & 177              & 1             \\
            VGG          & Adam           & 75               & 8             \\
            ResNet       & Adagrad        & 108              & 4             \\
            ResNet       & Adadelta       & 108              & 4             \\
            ResNet       & RMSprop        & 144              & 2             \\
            ResNet       & Adam           & 107              & 5             \\ \hline
    \end{tabular}
    \caption[Architecture and Loss comparison - Duration]{Architecture and Loss comparison - Duration. The shortest duration is better result.}
    \label{tab:app_res1_duration}
\end{table*}

\begin{table*}[!ht]
    \footnotesize
    \centering
    \begin{tabular}{@{}llll@{}}\toprule
            Architecture & Loss Optimizer & Train MSE & Rank         \\ \hline
            MLP          & Adagrad        & 6352.22549   & 4    \\
            MLP          & Adadelta       & 86460.7877   & 13   \\
            MLP          & RMSprop        & 8098.47514   & 7    \\
            MLP          & Adam           & 1369163.37   & 19   \\
            Conv1D       & Adagrad        & 5180.91202   & 2    \\
            Conv1D       & Adadelta       & 9913.78826   & 8    \\
            Conv1D       & RMSprop        & 14578.6939   & 9    \\
            Conv1D       & Adam           & 20305.0825   & 10   \\
            Conv2D       & Adagrad        & 6808.32294   & 5    \\
            Conv2D       & Adadelta       & 111152.159   & 15   \\
            Conv2D       & RMSprop        & 1618.25641   & 1    \\
            Conv2D       & Adam           & 5145821.2    & 20   \\
            VGG          & Adagrad        & 6139.20768   & 3    \\
            VGG          & Adadelta       & 93423.5512   & 14   \\
            VGG          & RMSprop        & 59200.3044   & 11   \\
            VGG          & Adam           & 78038.1544   & 12   \\
            ResNet       & Adagrad        & 7280.4593    & 6    \\
            ResNet       & Adadelta       & 131353.653   & 17   \\
            ResNet       & RMSprop        & 123707.716   & 16   \\
            ResNet       & Adam           & 232489.037   & 18   \\ \hline
    \end{tabular}
    \caption[Architecture and Loss comparison - Training MSE.]{Architecture and Loss comparison - Training MSE. The lowest MSE is the better result.}\label{tab:app_res1_train_MSE}
\end{table*}

\begin{table*}[!ht]
    \footnotesize
    \centering
    \begin{tabular}{@{}lll@{}}\toprule
            Architecture & Loss Optimizer & Rank         \\ \hline
            MLP          & Adagrad        & 20   \\
            MLP          & Adadelta       & 19   \\
            MLP          & RMSprop        & 9    \\
            MLP          & Adam           & 9    \\
            Conv1D       & Adagrad        & 15   \\
            Conv1D       & Adadelta       & 19   \\
            Conv1D       & RMSprop        & 9    \\
            Conv1D       & Adam           & 17   \\
            Conv2D       & Adagrad        & 15   \\
            Conv2D       & Adadelta       & 15   \\
            Conv2D       & RMSprop        & 9    \\
            Conv2D       & Adam           & 3    \\
            VGG          & Adagrad        & 15   \\
            VGG          & Adadelta       & 15   \\
            VGG          & RMSprop        & 9    \\
            VGG          & Adam           & 4    \\
            ResNet       & Adagrad        & 17   \\
            ResNet       & Adadelta       & 15   \\
            ResNet       & RMSprop        & 3    \\
            ResNet       & Adam           & 3   \\            \hline
    \end{tabular}
    \caption[Architecture and Loss comparison - Under-fitting/over-fitting]{Architecture and Loss comparison - Qualitative assessment of under-fitting/over-fitting and learning rate performance.}\label{tab:app_res1_validation_MSE}
\end{table*}

\begin{table*}[!ht]
    \footnotesize
    \centering
    \begin{tabular}{@{}llll@{}}\toprule
            Architecture & Loss Optimizer & Inversion RMSE (Hours) & Rank         \\ \hline
                MLP          & Adagrad        & 102.008808 & 11   \\
                MLP          & Adadelta       & 144.749442 & 8    \\
                MLP          & RMSprop        & 41.112304  & 15   \\
                MLP          & Adam           & 161.546795 & 7    \\
                Conv1D       & Adagrad        & 19.0347424 & 18   \\
                Conv1D       & Adadelta       & 15.4491243 & 19   \\
                Conv1D       & RMSprop        & 15.4070707 & 20   \\
                Conv1D       & Adam           & 20.1883355 & 17   \\
                Conv2D       & Adagrad        & 84.3773592 & 12   \\
                Conv2D       & Adadelta       & 252.899762 & 4    \\
                Conv2D       & RMSprop        & 45.4919989 & 14   \\
                Conv2D       & Adam           & 117.230985 & 10   \\
                VGG          & Adagrad        & 61.7665919 & 13   \\
                VGG          & Adadelta       & 135.254508 & 9    \\
                VGG          & RMSprop        & 23.2408697 & 16   \\
                VGG          & Adam           & 220.758795 & 5    \\
                ResNet       & Adagrad        & 358.686894 & 2    \\
                ResNet       & Adadelta       & 264.61373  & 3    \\
                ResNet       & RMSprop        & 162.649915 & 6    \\
                ResNet       & Adam           & 406.824565 & 1   \\            
             \hline
    \end{tabular}
    \caption[Architecture and Loss comparison - Inversion RMSE]{Architecture and Loss comparison - Inversion RMSE. RMSE for 100,000 validation velocity profile are compared to the true velocity.}\label{tab:app_res1_inversion}
\end{table*}

\subsection{Network Training Process}
For each of the Architecture-Loss combinations available, these networks were trained on an Intel i7-7800x X-series CPU workstation provided by the Department of Physics at the University of Malta. The script for this is available at \\
\href{https://gitfront.io/r/zerafachris/52df30fb666ba880749c8e951a3d056ce628a6cd/PhD/blob/code/appendix/ArchitectureComparison.py}{\url{./code/appendix/ArchitectureComparison.py}}. The initial maximum epoch was set to 20, with 5 epoch early stopping and 2 epoch learning rate reduction of 0.2 when reaching a plateau. The number of traces in the training and testing epoch generators were 1,000,000 and 100,000 respectively. The batch size was set to 100 for MLP and Conv1D, whilst 20 for the other networks due to ram size of the workstation.

\clearpage
\section[Theory-Guided FWI]{Theory-Guided FWI}
\subsection{1D Results}\label{sec:app_results_rnn_fwi_1d}
\begin{figure}[!ht]
        \centering
        \subbottom[Source for 1D experiments.\label{fig:app_rnn_1d_source}]{\includegraphics[width=0.48\textwidth, trim={0 0 19.5cm 0}, clip]{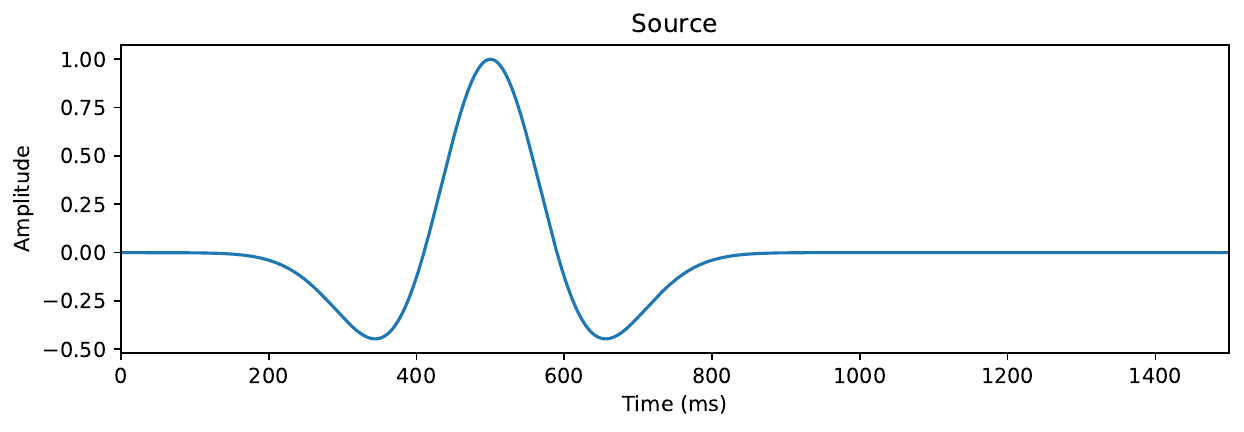}}
        \subbottom[1D constant 1500 m/s velocity model.\label{fig:app_rnn_1d_velocity}]{\includegraphics[width=0.45\textwidth, trim={21cm 0 0 0}, clip]{15_Dissertation_02_initial_experiment_1d_1d_direct_wave_velocity-eps-converted-to.pdf}}

        \subbottom[1D direct wave trace forward modelling comparison to analytical 1D Green’s function wavefield, RNN Time and RNN Freq showing no discrepancies.\label{fig:app_rnn_1d_direct_wave}]{\includegraphics[width=0.9\textwidth, trim={0 0 0 0.7cm}, clip]{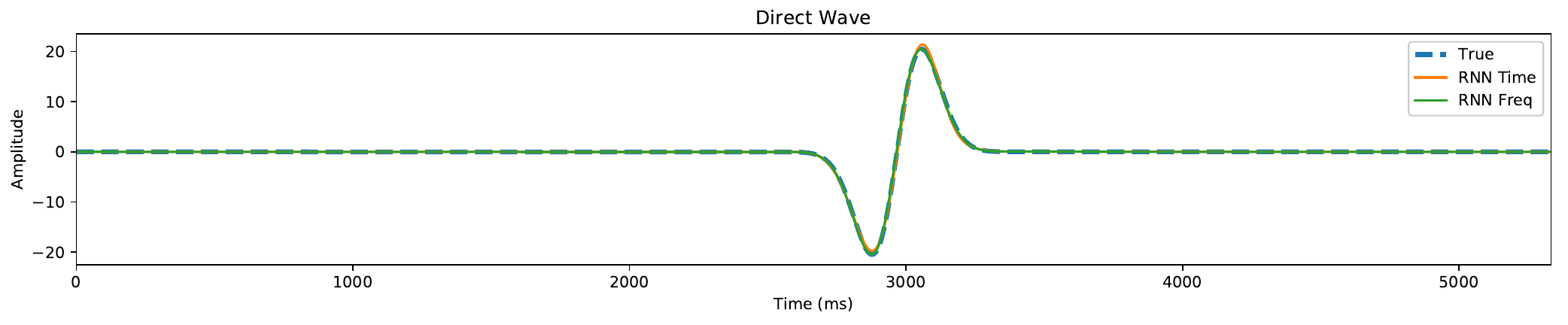}}
        \caption[1D Direct wave forward modelling comparison.]{1D Direct wave forward modelling comparison.}        
        \label{fig:app_rnn_1d_direct_results}
\end{figure}

Replacing the forward modelling component with RNN directly implies that the RNN should be able to retrieve all wavefield components. This was first tested by considering a 1D direct wave as shown in Figure~\ref{fig:app_rnn_1d_direct_results}. A 10Hz Ricker wavelet (Figure~\ref{fig:app_rnn_1d_source}) was injected into a 1D 1500 ms$^{-1}$ constant velocity model (Figure~\ref{fig:app_rnn_1d_velocity}) with a single source and single receiver. The wave was forward propagated for 5333 time-steps at 1ms, with a 10m grid spacing. The resulting direct wave is illustrated in Figure~\ref{fig:app_rnn_1d_direct_wave}, with True being the analytical solution calculated using a 1D Green’s function, RNN Time and RNN Freq are the RNN implementation for forward modelling using Time and Fourier spatial derivatives respectively. Qualitatively, there is no visible difference between either approach. Quantitative analysis shown in Table~\ref{tab:app_rnn_1d_direct} indicates that the pseudo-spectral approach is producing slightly better results with an improved error tolerance of 0.2 in amplitude, resulting in a 0.9\% improvement in the \ac{RPE}.

\begin{table*}[!ht]
        \footnotesize
        \centering
        \begin{tabular}{@{}lcc@{}}\toprule
Modelling  & Error Tolerance    & RPE (\%) \\ \hline
RNN Time   & 0.980 & 4.786                           \\
RNN Freq   & 0.780 & 3.809                           \\ \hline
        \end{tabular}
        \caption{Empirical comparison of 1D Direct wave modelling to 1D Green's function.}\label{tab:app_rnn_1d_direct}
\end{table*}

\begin{figure}[!ht]
	\centering
	\subbottom[1D step velocity model ranging from 1500 \si{ms^{-1}} to 2500 \si{ms^{-1}}.\label{fig:app_rnn_1d_direct_refl_transmitted_vel}]{\includegraphics[width=0.8\textwidth]{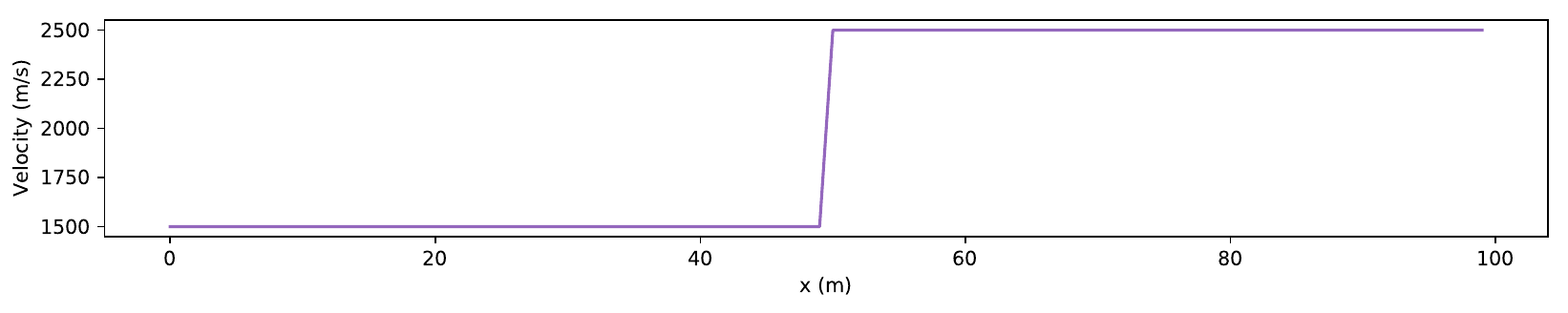}}
	\subbottom[Top: Full wavefield showing Direct and Reflected waves at about 20ms and Transmitted wave with peak at 96 ms. Middle and Bottom: Zoomed in sections from Full trace showing different components of the wavefield and near perfect reconstruction.\label{fig:app_rnn_1d_direct_refl_transmitted_wavefield}]{\includegraphics[width=0.9\textwidth]{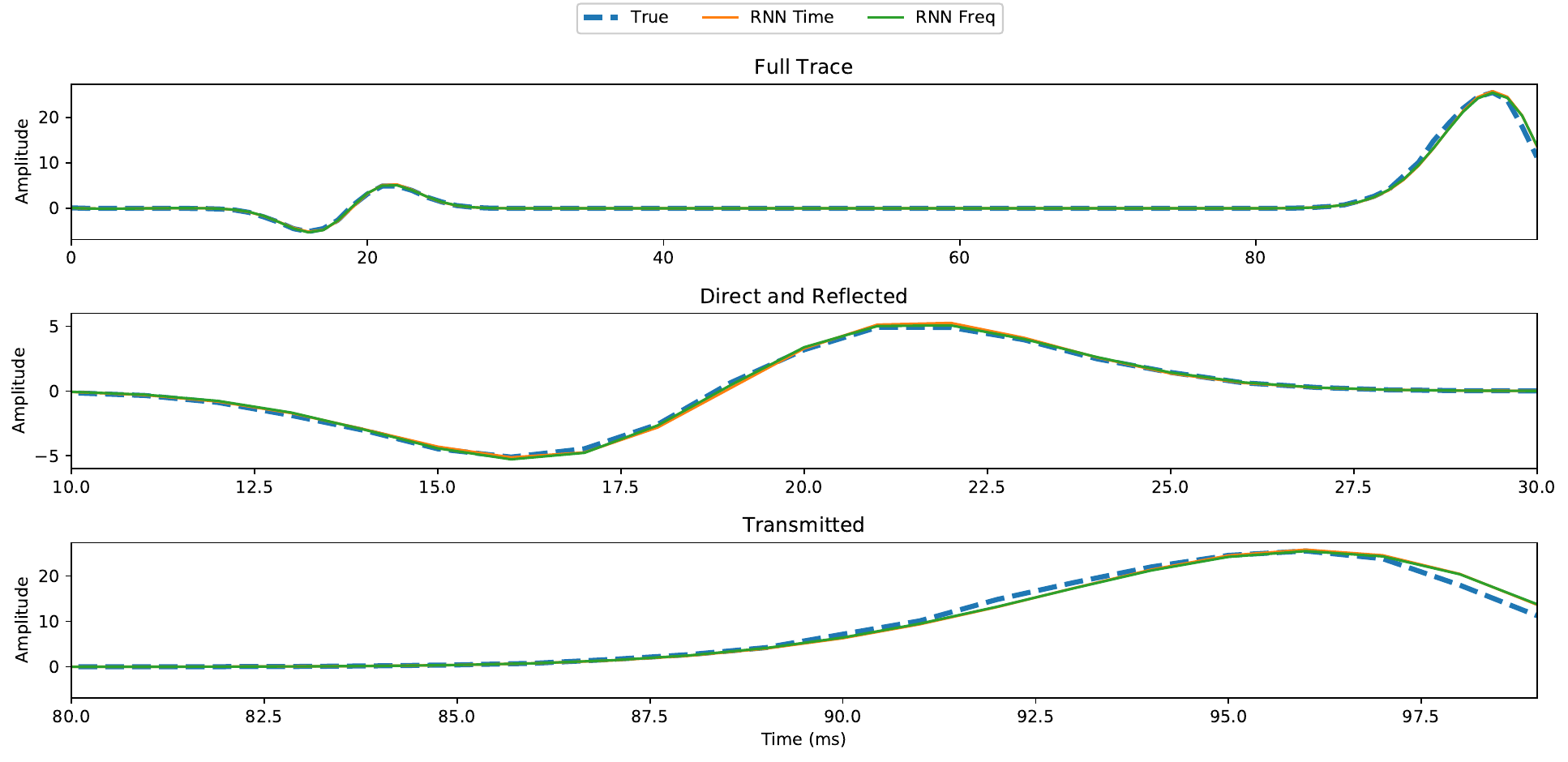}}
	\caption[1D direct, reflected and transmitted wave forward modelling comparison.]{1D direct, reflected and transmitted wave forward modelling comparison.}        
	\label{fig:app_rnn_1d_direct_refl_transmitted}
\end{figure}

An identical source was used to test for Reflected and Transmitted arrivals. Figure~\ref{fig:app_rnn_1d_direct_refl_transmitted_wavefield} is the 1D step velocity model ranging from 1500 \si{ms^{-1}} to 2500 \si{ms^{-1}} used to produce the forward propagated wavefields in Figure~\ref{fig:app_rnn_1d_direct_refl_transmitted_wavefield}. The top shows the full wavefield, with Direct and Reflected arrivals at about 20 ms and Transmitted wave with peak at 96 ms. Middle and bottom sections of Figure~\ref{fig:app_rnn_1d_direct_refl_transmitted_wavefield} show zoomed sections of the wavefield respectively. Either the Reflected and Transmitted component match near identically to the Green’s function formulation. Quantitatively, the RNN Time implementation was found to be slightly improved, with an improvement of 0.1 in error tolerance and 0.2\% RPE (Table~\ref{tab:app_rnn_1d_direct_refl_transmitted}).

\begin{table*}[!ht]
        \footnotesize
        \centering
        \begin{tabular}{@{}lcc@{}}\toprule
Modelling  & Error Tolerance    & RPE (\%) \\ \hline
RNN Time   & 2.460 & 9.637                           \\
RNN Freq   & 2.520 & 9.872                           \\ \hline
        \end{tabular}
        \caption{Empirical comparison of 1D direct, reflected and transmitted wave modelling.}\label{tab:app_rnn_1d_direct_refl_transmitted}
\end{table*}

\begin{figure}[!ht]
	\centering
	\subbottom[1D scattering velocity model ranging from 1500 \si{ms^{-1}} to 1550 \si{ms^{-1}}.\label{fig:app_rnn_1d_scattered_vel}]{\includegraphics[width=0.8\textwidth]{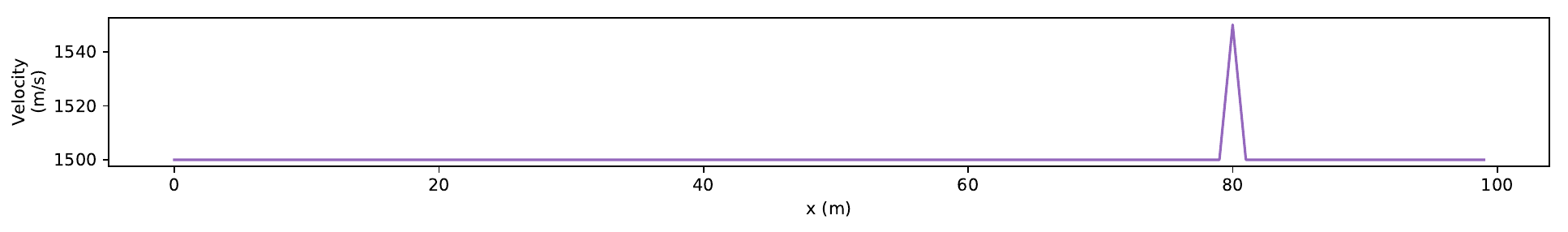}}
	\subbottom[Scattering wavefield modelling. Direct wavefield was excluded in the modelling.\label{fig:app_rnn_1d_scattered_wavefield}]{\includegraphics[width=0.9\textwidth]{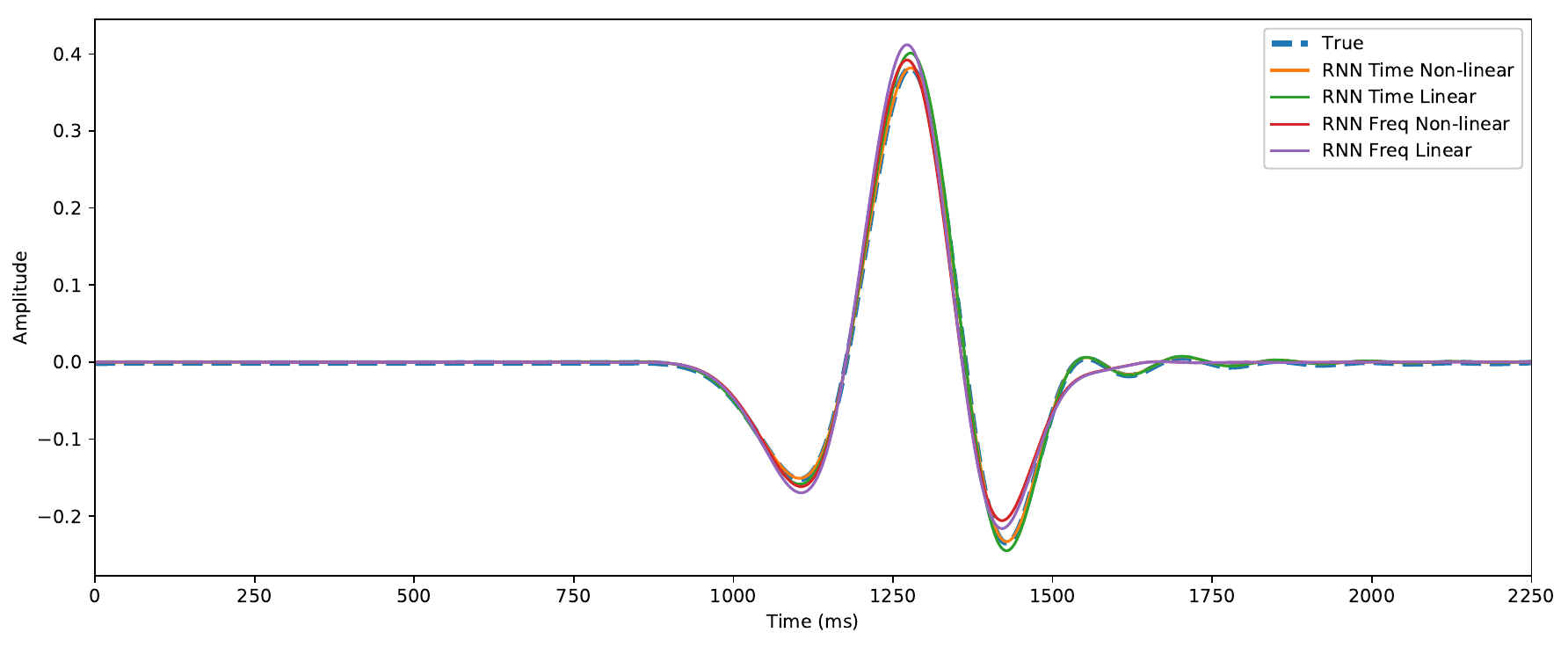}}
	\caption[1D scattering wave forward modelling comparison.]{1D scattering wave forward modelling comparison.}    
	\label{fig:app_rnn_1d_scattering}
\end{figure}

The remaining wavefield component to be modelled are Scattering waves. A constant velocity model with one-point scatterer was created with velocity ranging from 1500 \si{ms^{-1}} to 1550 \si{ms^{-1}} at the point scatterer (Figure ~\ref{fig:app_rnn_1d_scattered_vel}). The same source in the previous two wavefields was used. RNN implementations were modelled for the waveform at a point to be depended non-linearly on the scattering amplitude and then approximately linearised. The resulting wavefields are given in Figure~\ref{fig:app_rnn_1d_scattered_wavefield}. The Direct wave was not included in the Scattered wavefield reconstruction. From the error tolerance and RPE comparison (Table~\ref{tab:app_rnn_1d_scattering}), RNN Time produces marginally better results than RNN Freq.
This is due to the lack of ringing effect introduced due to discretisation beyond 1500ms which gets absorbed within the complex component of the pseudo-spectral approach. The RNN Frequency approach is thus a superior modelling approach in 1D. In the 2D case, RNN Time does not suffer from this effect.

\begin{table*}[!ht]
        \footnotesize
        \centering
        \begin{tabular}{@{}lcc@{}}\toprule
Modelling  & Error Tolerance    & RPE (\%) \\ \hline
RNN Time – Non-linear	& 0.009	& 0.031     \\
RNN Time – Linear 	& 0.010	& 2.493         \\
RNN Freq – Non-linear	& 0.030	& 7.649         \\
RNN Freq – Linear	& 0.040 & 9.711          \\ \hline
        \end{tabular}
        \caption{Empirical comparison of 1D scattering wave modelling.}\label{tab:app_rnn_1d_scattering}
\end{table*}

\subsection{RNN Hyper-Parameter Tuning}\label{sec:app_results_rnn_hp_tuning}
\begin{figure}[!ht]
	\centering
	\subbottom[RNN Time]{\includegraphics[width=0.95\textwidth]{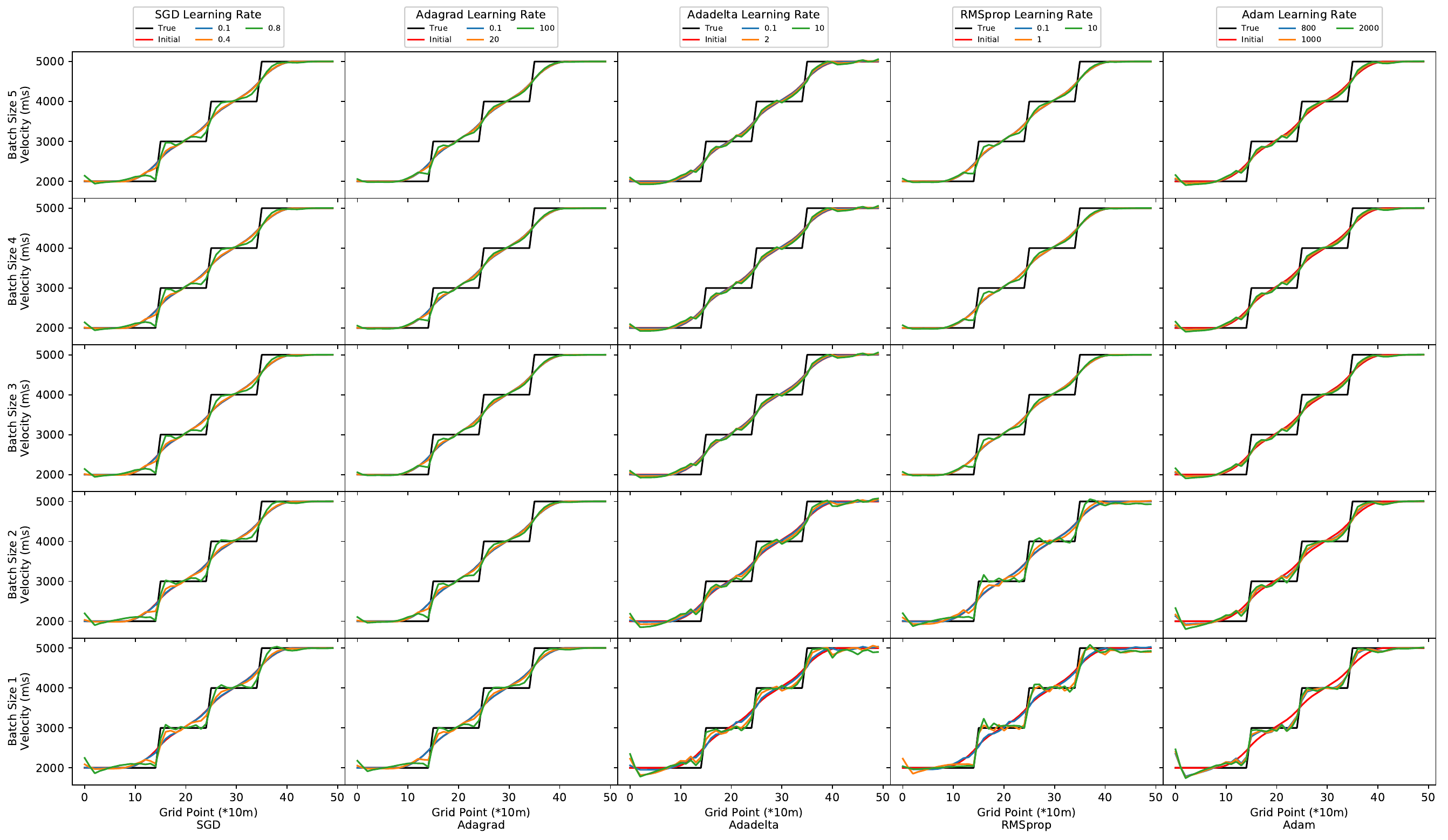}}
        \subbottom[RNN Freq]{\includegraphics[width=0.95\textwidth]{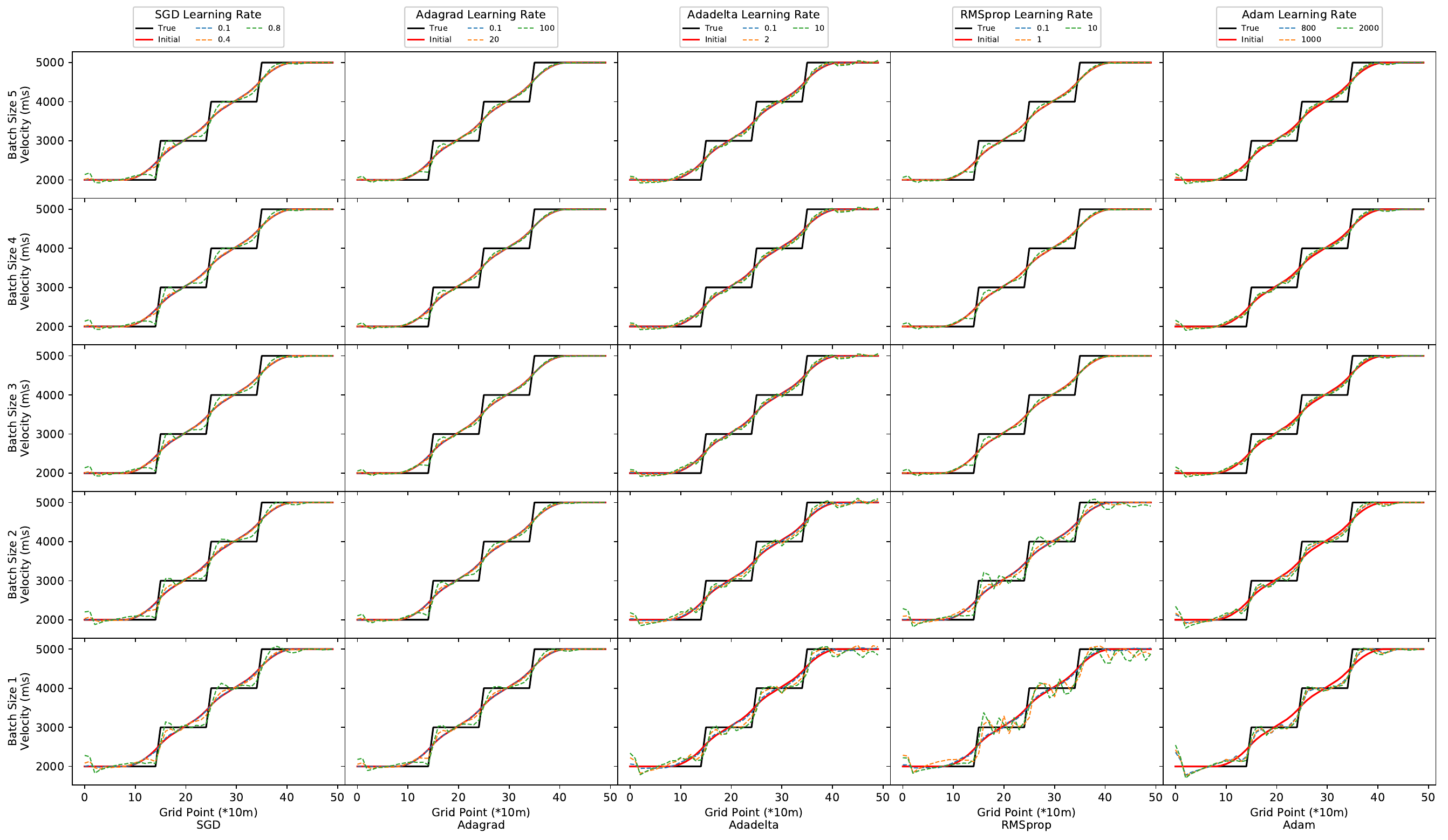}}
	\caption[Different Loss optimizer learning rate hyper-parameter tuning results.]{Loss optimizer learning rate hyper-parameter tuning results.}
	\label{fig:app_results_rnn_hp_tuning_inversion}
\end{figure}

Similarly to the approach shown in \cite{Sun2019}, a benchmark 1D 4-layer synthetic profile, with velocities [2, 3, 4, 5] \si{kms^{-1}}, was used to identify the ideal parameters for the RNN architecture. Classical 1D second-order \ac{FD} modelling was used to generate the required true receiver data. Batch size is used as a discriminator throughout Figure~\ref{fig:app_results_rnn_hp_tuning_inversion}. The results indicate that the larger the batch size used, the better the inversion as more data is being used. However, given fore-sight that this hyper-parameter tuning will be used a large dataset that might not fit in Graphical Processing Unit RAM, this was fixed at batch size one. 

\begin{figure}[!ht]
        \centering
        \includegraphics[width=0.99\textwidth]{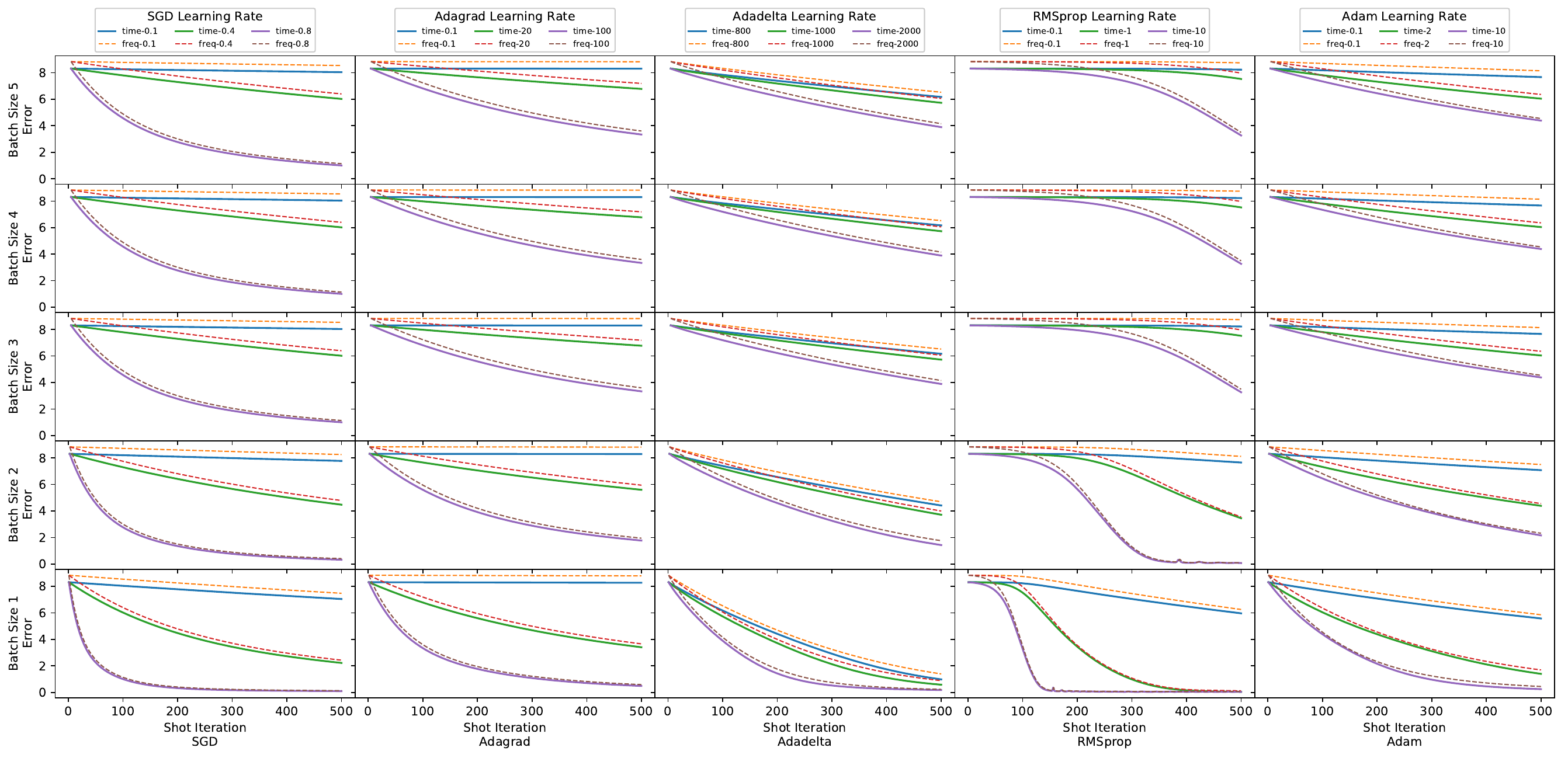}
        \caption[Losses for different loss optimizer learning rate hyper-parameter tuning.]{Losses for different loss optimizer learning rate hyper-parameter tuning.}
	\label{fig:app_results_rnn_hp_tuning_losses}
\end{figure}

\clearpage


\subsection{RNN Inversion Update Progress}\label{sec:app_results_rnn_update_progress}
\begin{figure}[!ht]
        \centering
        \includegraphics[width=0.9\textwidth]{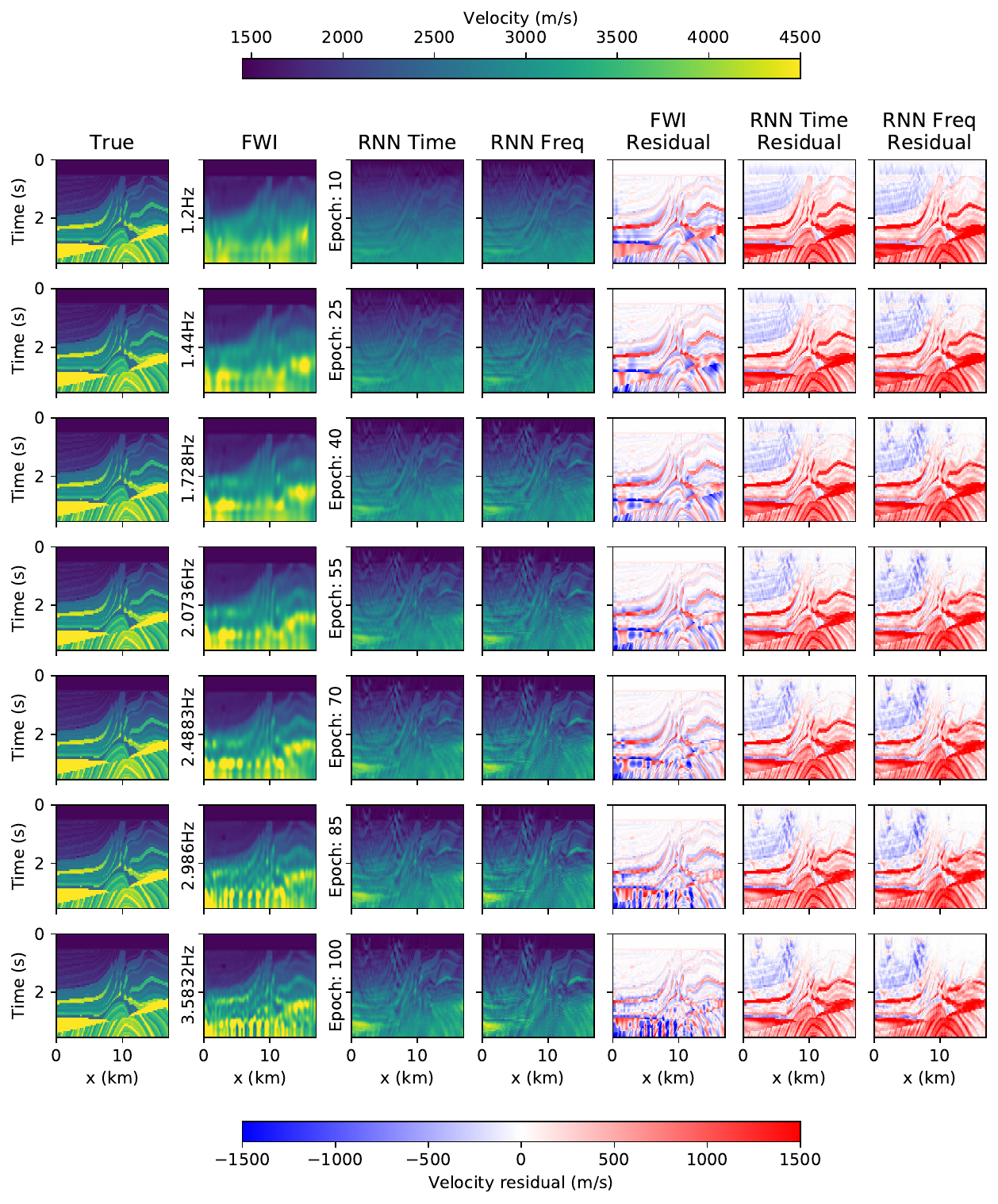}
        \caption[Velocity model inversion update progress.]{Velocity model inversion update progress for classical FWI, RNN Time and Freq, with residuals.}
	\label{fig:app_results_rnn_update_progress_model}
\end{figure}

\begin{figure}[!ht]
        \centering
        \includegraphics[width=0.95\textwidth]{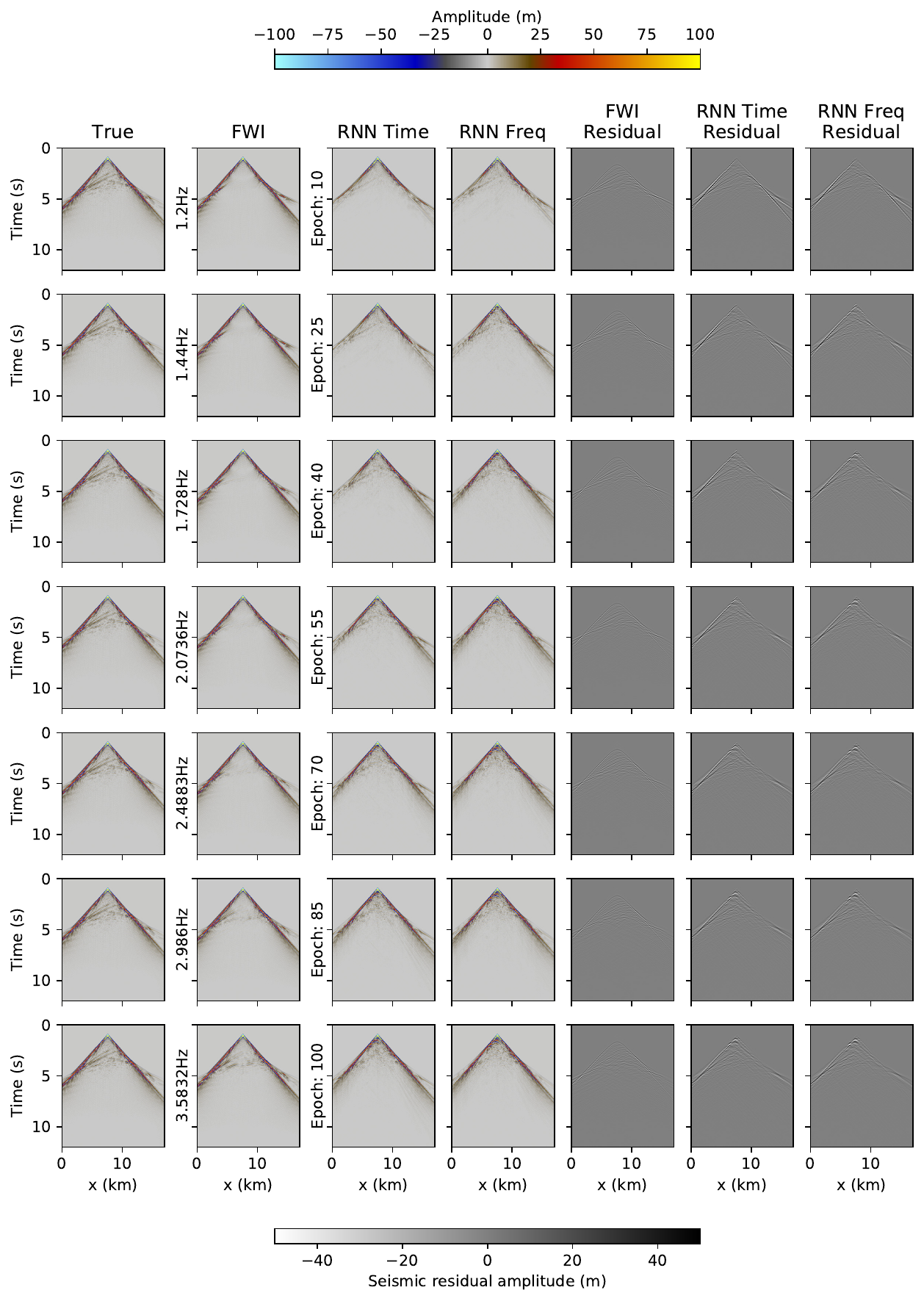}
        \caption[Receiver progress through model updates.]{Receiver progress through model updates for classical FWI, RNN Time and Freq, with residuals.}
	\label{fig:app_results_rnn_update_progress_rcv}
\end{figure}

Complementary to inverted Marmousi models in §~\ref{sec:results_rnn_comparison_to_FWI}, Figure~\ref{fig:app_results_rnn_update_progress_model} gives the update progress at epoch 10, 25, 40, 55, 70, 85 and 100 for RNN Time and RNN Freq, together with residual. Furthermore, classical FWI progress is included at different update frequency scales. In addition, receivers are provided in Figure~\ref{fig:app_results_rnn_update_progress_rcv}.

%% file: appC/appendix_c_ref_collab.tex
\chapter{Publications and Collaborations}

\section{Publications}
\begin{itemize}
    \item[] \href{https://www.xjenza.org/JOURNAL/OLD/7-1-2019/07.pdf}{Zerafa, C., Galea, P. \& Sebu, C. (2019). "Learning to Invert Pseudo-Spectral Data for Seismic Waveforms". \textit{Xjenza Online} \textbf{7}(1),3-17}.
    \item[] \href{https://www.earthdoc.org/content/papers/10.3997/2214-4609.201803015}{Zerafa, C. (2018) "DNN application of pseudo-spectral FWI". \textit{First EAGE/PESGB Workshop on Machine Learning: European Edition.}}
    \item[] \textbf{\textit{TBD}} \emph{Parts of this work are planned for publishing in \href{https://library.seg.org/journal/gpysa7}{SEG Geophysics} and \href{https://academic.oup.com/gji}{OUP Geophysical Journal International}}.
\end{itemize}

\section{Conferences \& Poster Presentations}
\begin{itemize}
    \item \textbf{Jul 2020} Zerafa, C. "Overview of Machine Learning Applications in Geophysics and Seismology". \textit{Department of Geosciences Summer Seminar, University of Malta}.
    \item \href{https://www.southampton.ac.uk/the-alan-turing-institute/news/events/2019/10/optimization-and-machine-learning.page}{\textbf{Nov 2019}. Zerafa, C., Galea, P. \& Sebu, C. "Learning to Invert Pseudo-Spectral Data for Seismic Waveforms". \textit{OptML: Optimization and Machine Learning, University of Southampton, UK}}.
    \item \textbf{Mar 2018}. Zerafa, C. "An Introduction to High Resolution Seismic Imaging". \textit{Scubed Annual Scientific Conference}.
\end{itemize}

\section{Appointments}
\begin{itemize}
    \item \textbf{Jan 2022} Post-Doc at Istituto Nazionale di Geofisica e Vulcanologia, Pisa.
    \newline Will be joining Istituto Nazionale di Geofisica e Vulcanologia for 2-year Post-Doc at Istituto Nazionale di Geofisica e Vulcanologia for project titled \emph{SOME Seismological Oriented Machine lEarning}. I will be involved in two main working groups:
    \begin{itemize}
        \item \emph{WP2}: Earthquake detection and characterization using large seismic datasets from tectonic/volcanic processes and hydrocarbon/geothermal exploitation. In this WP, we plan to extensively test published or newly developed supervised and unsupervised Machine Learning algorithms on the wealth of already available seismic datasets. We will use seismic datasets from regional (e.g., INSTANCE, AlpArray, Amatrice/Norcia) and local scale (Mugello, Irpinia fault system, Val d’Agri, Amiata, Larderello, Campi Flegrei, Etna).
        \item \emph{WP4}: Automatic extraction of phase and group velocity surface dispersion curves. We plan to build up on the recent work by \citet{Zhang2020}, applying the method to a various range of datasets already available at Istituto Nazionale di Geofisica e Vulcanologia \citep{Molinari2015,Molinari2020} or available in the near future as results of Department Projects at local and regional scales and in a wide frequency range.
    \end{itemize}
    \item \textbf{Apr 2020} - CA17137 WG2 - Co-Leader - \href{https://www.g2net.eu/wgs/wg2-machine-learning-for-low-frequency-seismic-measurement}{\url{www.g2net.eu/wgs/wg2}}. \newline I joined a leadership position within the working group together with Dr. Ilec. My involvement includes overseeing and organising the research initiatives within the group, introduced a common code repository to encourage collaboration and hold weekly meetings to steer the working group forward.
    \item \textbf{Sep 2018} - CA17137 University of Malta Representative - \url{https://www.g2net.eu/}. \newline Active member within CA17137 - g2net - A network for Gravitational Waves, Geophysics and Machine Learning, with particular focus to Working Group 2 - Machine Learning for low-frequency seismic measurement. Research deals with acquisition, processing and interpretation of seismic data, with the goal of combating the seismic influences at Gravitational Wave detector site, using multi-disciplinary research focussing on advanced techniques available from state of the art machine learning algorithms. 
\end{itemize}

\section{Collaborations}
\begin{itemize}
    \item \textbf{Apr 2020} Short Term Scientific Mission at Istituto Nazionale di Geofisica e Vulcanologia, Pisa. \newline In collaboration with Giunchi, C., De Matteo, G., Gaviano, S., developed two CNN approaches able to classify local earthquakes into 13 classes with different epicentre and magnitude characterizations.
\end{itemize}

\section{Organisation of Events and Guest Lecturing}
\begin{itemize}
    \item \textbf{Sep 2020} Kaggle Competition - \href{https://www.kaggle.com/c/g2net-gravitational-wave-detection}{Classification of Gravitational Wave Glitches}. Cuoco, E., Zerafa, C., Messenger, C., Williams, M. \newline Collaborating with Kaggle and other g2net members to host an international competition using gravitational wave data which could lead to novel was of automatic detection of gravitational wave detection. In turn, this could directly influence the next generation of gravitational wave research and better understanding of the known and unknown universe.
    
    \item \textbf{Mar 2020} CA17137 – g2net – 2nd Training School. \newline Hosted an international training school for CA17137 within the University of Malta. Was mainly responsible within the Scientific Organizing Committee, Local Organizing Committee and one of the lecturers. Classes though included a course in Machine Learning and hosted a hackathon for all participants. Testimonials, lecture notes, video recordings and code can  be found at \url{https://github.com/zerafachris/g2net_2nd_training_school_malta_mar_2020}.
    
    \item  \textbf{Sep 2019} PyMalta: First Steps towards Machine Learning \newline Hosted a two-part training introductory series on Machine Learning using Python. All code can be found in the repo \url{https://github.com/PyMalta/Introduction_to_ML_CZ}.
\end{itemize}

\section{Relevant Training \& Certification}
\begin{itemize}
    \item \textbf{Nov 2019} "Oberwolfach Graduate Seminar: Mathematics of Deep Learning", \textit{MFO}, \textit{Oberwalfach}, \textit{Germany}. Tutorial based course focussing on state-of-the-art mathematical analysis of deep learning algorithms. Training focussed on (i) approximation theory, (ii) expressivity, (iii) generalization, and (iv) interpretability. See~\href{https://www.mfo.de/occasion/1947a}{\url{MFO}}.
    
    \item \textbf{Mar 2019} "ML and Statistical Analysis", Distinction, \textit{The Data Incubator}. Tutorial based course on the use of ML on real world data set, with a heavy emphasis on creative use of different data science techniques to solve problems from multiple perspectives. Projects included NLP, clustering, Time series analysis and anomaly detection. See~\href{https://wqu.org/programs/data-science}{\url{DataIncubator}}.
    
    \item \textbf{Sep 2018} "Full Waveform Inversion: Maths and Geophysics", \textit{KIT, Karlsruhe, Germany}. Tutorials for high performance computing with specific focus on computational mathematics. See~\href{https://www.waves.kit.edu/summerschool2018.php}{\url{KIT2018}}.
    
    \item \textbf{Jul 2018} "International Summer School on Deep Learning", \textit{Gdansk University of Technology, Poland}. In-depth training and mini-projects on a range of learning methods, with particular focus on DNNs. See~\href{http://2018.dl-lab.eu/}{\url{ISSDL}}.
    
    \item \textbf{Jun 2018} "Scientific Computing and Python for Data Science", Distinction, \textit{The Data Incubator}. Tutorial based course covering linear regression and gradient descent techniques. Particular emphasis included cost function analysis, dimensionality reduction, regularization and feature engineering. See~\href{https://wqu.org/programs/data-science}{\url{DataIncubator}}.
    
    \item \textbf{Apr 2018} "Graphical Processing Unit-based analytics and data science, \textit{Vicomtech Research Center, Spain}. Intensive training on use of GPUs using PyCuda for big data analytics as applied to machine learning for image processing, segmentation, de-noising, filtering, interpolation and reconstruction. See~\href{https://bigskyearth.eu/apply-to-the-bigskyearth-training-school-2018-gpu-based-analytics-and-data-science/}{\url{BigSkyEarth}}.
    
\end{itemize}

%% file: main.bbl
\begin{thebibliography}{}

\bibitem[Abadi et~al., 2015]{Abadi2015}
Abadi, M., Agarwal, A., Barham, P., Brevdo, E., Chen, Z., Citro, C., Corrado, G., Davis, A., Dean, J., Devin, M., Ghemawat, S., Goodfellow, I., Harp, A., Irving, G., Isard, M., Jia, Y., Jozefowicz, R., Kaiser, L., Kudlur, M., Levenberg, J., Man{\'{e}}, D., Monga, R., Moore, S., Murray, D., Olah, C., Schuster, M., Shlens, J., Steiner, B., Sutskever, I., Talwar, K., Tucker, P., Vanhoucke, V., Vasudevan, V., Vi{\'{e}}gas, F., Vinyals, O., Warden, P., Wattenberg, M., Wicke, M., Yu, Y., and Zheng, X. (2015).
\newblock {TensorFlow: Large-Scale Machine Learning on Heterogeneous Systems}.
\newblock {\em Software available from tensorflow.org}.

\bibitem[Adler and {\"{O}}ktem, 2017]{Adler2017a}
Adler, J. and {\"{O}}ktem, O. (2017).
\newblock {Solving ill-posed inverse problems using iterative deep neural networks}.
\newblock {\em Inverse Problems}, 33(12):1--24.

\bibitem[Adler et~al., 2017]{Adler2017b}
Adler, J., Ringh, A., {\"{O}}ktem, O., and Karlsson, J. (2017).
\newblock {Learning to solve inverse problems using Wasserstein loss}.
\newblock {\em Iclr 2018}, pages 1--13.

\bibitem[Ali et~al., 2007]{Ali2007}
Ali, H. B.~H., Operto, S., Virieux, J., and Sourbier, F. (2007).
\newblock {3D acoustic frequency-domain full-waveform inversion}.
\newblock In {\em SEG Technical Program Expanded Abstracts 2007}, pages 1730--1734. Society of Exploration Geophysicists.

\bibitem[Anderson and Lucas, 2018]{Anderson2018}
Anderson, G.~J. and Lucas, D.~D. (2018).
\newblock {Machine Learning Predictions of a Multiresolution Climate Model Ensemble}.
\newblock {\em Geophysical Research Letters}, 45(9):4273--4280.

\bibitem[Andrychowicz et~al., 2016]{Andrychowicz2016}
Andrychowicz, M., Denil, M., Gomez, S., Hoffman, M.~W., Pfau, D., Schaul, T., Shillingford, B., and de~Freitas, N. (2016).
\newblock {Learning to learn by gradient descent by gradient descent}.
\newblock {\em Advances in Neural Information Processing Systems 29 (NIPS 2016)}.

\bibitem[Araya-Polo et~al., 2018]{Araya-Polo2018}
Araya-Polo, M., Jennings, J., Adler, A., and Dahlke, T. (2018).
\newblock {Deep-learning tomography}.
\newblock {\em The Leading Edge}, 37(1):58--66.

\bibitem[Arridge et~al., 2019]{Arridge2019}
Arridge, S., Maass, P., {\"{O}}ktem, O., and Sch{\"{o}}nlieb, C.-B. (2019).
\newblock {Solving inverse problems using data-driven models}.
\newblock {\em Acta Numerica}, 28:1--174.

\bibitem[Asnaashari et~al., 2013]{Asnaashari2013}
Asnaashari, A., Brossier, R., Garambois, S., Audebert, F.~F., Thore, P., and Virieux, J. (2013).
\newblock {Regularized seismic full waveform inversion with prior model information}.
\newblock {\em Geophysics}, 78(2):R25--R36.

\bibitem[Ben-Hadj-Ali et~al., 2008]{Ben-Hadj-Ali2008}
Ben-Hadj-Ali, H., Operto, S., and Virieux, J. (2008).
\newblock {Velocity model building by 3D frequency-domain, full-waveform inversion of wide-aperture seismic data}.
\newblock {\em Geophysics}, 73(5):VE101--VE117.

\bibitem[{Ben Hadj Ali} et~al., 2007]{BenHadjAli2007}
{Ben Hadj Ali}, H., Operto, S., Virieux, J., and Sourbier, F. (2007).
\newblock {3D acoustic frequency-domain full-waveform inversion}.
\newblock {\em Extended Abstracts}, pages 1730--1734.

\bibitem[Bengio et~al., 1994]{Bengio1994}
Bengio, Y., Simard, P., and Frasconi, P. (1994).
\newblock {Learning long-term dependencies with gradient descent is difficult}.
\newblock {\em IEEE transactions on neural networks}, 5(2):157--166.

\bibitem[Bergen et~al., 2019]{Bergen2019}
Bergen, K.~J., Johnson, P.~A., Maarten, V., and Beroza, G.~C. (2019).
\newblock {Machine learning for data-driven discovery in solid Earth geoscience}.
\newblock {\em Science}, 363(6433).

\bibitem[Berthelot et~al., 2018]{Berthelot2018}
Berthelot, D., Raffel, C., Roy, A., and Goodfellow, I. (2018).
\newblock {Understanding and improving interpolation in autoencoders via an adversarial regularizer}.
\newblock {\em arXiv preprint arXiv:1807.07543}.

\bibitem[Bishop, 2006]{Bishop2006}
Bishop, C.~M. (2006).
\newblock {\em {Pattern recognition and machine learning}}.
\newblock Springer.

\bibitem[Biswas and Sen, 2017]{Biswas2017}
Biswas, R. and Sen, M.~K. (2017).
\newblock {2D Full Waveform Inversion and Uncertainty Estimation using the Reversible Jump Hamiltonian Monte Carlo}.
\newblock {\em SEG Technical Program Expanded Abstracts 2017}, pages 1280--1285.

\bibitem[Biswas et~al., 2019]{Biswas2019}
Biswas, R., Sen, M.~K., Das, V., and Mukerji, T. (2019).
\newblock {Pre-stack inversion using a physics-guided convolutional neural network}.
\newblock In {\em SEG Technical Program Expanded Abstracts 2019}, pages 4967--4971. Society of Exploration Geophysicists.

\bibitem[Blanch et~al., 1995]{Blanch1995}
Blanch, J.~O., Robertsson, J. O.~A., and Symes, W.~W. (1995).
\newblock {Modeling of a constant Q: Methodology and algorithm for an efficient and optimally inexpensive viscoelastic technique}.
\newblock {\em Geophysics}, 60(1):176.

\bibitem[Bleibinhaus and Rondenay, 2009]{Bleibinhaus2009}
Bleibinhaus, F. and Rondenay, S. (2009).
\newblock {Effects of surface scattering in full-waveform inversion}.
\newblock {\em GEOPHYSICS}, 74(6):WCC69--WCC77.

\bibitem[Boehm et~al., 2015]{Boehm2015}
Boehm, C., Fichtner, A., de~la Puente, J., and Hanzich, M. (2015).
\newblock {Lossy Wavefield Compression for Full-Waveform Inversion}.
\newblock In {\em AGU Fall Meeting Abstracts}, volume 2015, pages S23C--2714.

\bibitem[Bogaerts et~al., 2020]{Bogaerts2020}
Bogaerts, T., Masegosa, A.~D., Angarita-Zapata, J.~S., Onieva, E., and Hellinckx, P. (2020).
\newblock {A graph CNN-LSTM neural network for short and long-term traffic forecasting based on trajectory data}.
\newblock {\em Transportation Research Part C: Emerging Technologies}, 112:62--77.

\bibitem[Born and Wolf, 1980]{Born1980}
Born, M. and Wolf, E. (1980).
\newblock {Principles of optics}.
\newblock {\em Pergamon Press}, 6:188--189.

\bibitem[Bottou et~al., 2018]{Bottou2018}
Bottou, L., Curtis, F.~E., and Nocedal, J. (2018).
\newblock {Optimization methods for large-scale machine learning}.
\newblock {\em Siam Review}, 60(2):223--311.

\bibitem[Boureau et~al., 2010]{Boureau2010}
Boureau, Y.-L., Ponce, J., and LeCun, Y. (2010).
\newblock {A theoretical analysis of feature pooling in visual recognition}.
\newblock In {\em Proceedings of the 27th international conference on machine learning (ICML-10)}, pages 111--118.

\bibitem[Boyd et~al., 2011]{Boyd2011}
Boyd, S., Parikh, N., and Chu, E. (2011).
\newblock {\em {Distributed optimization and statistical learning via the alternating direction method of multipliers}}.
\newblock Now Publishers Inc.

\bibitem[Bruna et~al., 2015]{Bruna2015}
Bruna, J., Sprechmann, P., and LeCun, Y. (2015).
\newblock {Super-Resolution with Deep Convolutional Sufficient Statistics}.

\bibitem[Bryson, 1961]{Bryson1961}
Bryson, A.~E. (1961).
\newblock {A gradient method for optimizing multi-stage allocation processes}.
\newblock In {\em Proc. Harvard Univ. Symposium on digital computers and their applications}, volume~72.

\bibitem[Bubba et~al., 2019]{Bubba2019}
Bubba, T.~A., Kutyniok, G., Lassas, M., Maerz, M., Samek, W., Siltanen, S., and Srinivasan, V. (2019).
\newblock {Learning the invisible: a hybrid deep learning-shearlet framework for limited angle computed tomography}.
\newblock {\em Inverse Problems}, 35(6):64002.

\bibitem[Bunks et~al., 1995]{Bunks1995}
Bunks, C., Saleck, F.~M., Zaleski, S., and Chavent, G. (1995).
\newblock {Multiscale seismic waveform inversion}.
\newblock {\em Geophysics}, 60(5):1457--1473.

\bibitem[Cai et~al., 2015]{Cai2015}
Cai, X.-H., Liu, Y., Ren, Z.-M., Wang, J.-M., Chen, Z.-D., Chen, K.-Y., and Wang, C. (2015).
\newblock {Three-dimensional acoustic wave equation modeling based on the optimal finite-difference scheme}.
\newblock {\em Applied Geophysics}, 12(3):409--420.

\bibitem[{Calde On-Mac{\'{i}}as} et~al., 1997]{CaldeOn-Macias1997}
{Calde On-Mac{\'{i}}as}, C., Sen, M.~K., and Stoffa, P.~L. (1997).
\newblock {Hopfield neural networks, and mean field annealing for seismic deconvolution and multiple attenuation}.
\newblock {\em GEOPHYSICS}, 62(3):992--1002.

\bibitem[Caruana et~al., 2000]{Caruana2000}
Caruana, R., Lawrence, S., and Giles, L. (2000).
\newblock {Overfitting in neural nets: Backpropagation, conjugate gradient, and early stopping}.
\newblock In {\em the 13th International Conference on Neural Information Processing Systems}, pages 402--408.

\bibitem[Chang et~al., 2017]{Chang2017}
Chang, J.~H., Li, C.-L., P{\'{o}}czos, B., Kumar, B. V.~K., and Sankaranarayanan, A.~C. (2017).
\newblock {One Network to Solve Them All---Solving Linear Inverse Problems using Deep Projection Models}.
\newblock {\em IEEE International Conference on Computer Vision (ICCV)}.

\bibitem[Chentouf, 1997]{Chentouf1997}
Chentouf, R. (1997).
\newblock {Construction de reseaux de neurones multicouches pour l'approximation}.
\newblock {\em Ph.D. thesis, Institut National Polytechnique, Grenoble}.

\bibitem[Chollet, 2015]{Chollet2015}
Chollet, F. (2015).
\newblock {Keras}.
\newblock {\em https://github.com/fchollet/keras}.

\bibitem[Ciresan et~al., 2011]{Ciresan2011}
Ciresan, D.~C., Meier, U., and Masci, J. (2011).
\newblock {A Committee of Neural Networks for Traffic Sign Classification}.
\newblock {\em International Joint Conference on Neural Networks (IJCNN-2011, San Francisco)}, 1(1).

\bibitem[Ciresan et~al., 2012]{Ciresan2012}
Ciresan, D.~C., Meier, U., Masci, J., and Schmidhuber, J. (2012).
\newblock {Multi-Column Deep Neural Network for Traffic Sign Classification.}
\newblock {\em Neural Networks}, pages 333--338.

\bibitem[Claerbout, 1971]{Claerbout1971}
Claerbout, J.~F. (1971).
\newblock {Toward a unified theory of reflector mapping}.
\newblock {\em Geophysics}, 36(3):467--481.

\bibitem[Claerbout, 1976]{Claerbout1976}
Claerbout, J.~F. (1976).
\newblock {\em {Fundamentals of geophysical data processing: McGraw-Hili Book Co}}.
\newblock Inc.

\bibitem[Cochrane and Cooper, 1991]{Cochrane1991}
Cochrane, G. and Cooper, A.~K. (1991).
\newblock {Sonobuoy seismic studies at ODP drill sites in Prydz Bay, Antarctica}.
\newblock pages 27--43.

\bibitem[Dadvand et~al., 2006]{Dadvand2006}
Dadvand, P., Lopez, R., and Onate, E. (2006).
\newblock {Artificial Neural Networks for the Solution of Inverse Problems A Variational Formulation for the Multilayer Perceptron}.
\newblock {\em Proceedings of the International Conference on Design Optimisation Methods and Applications ERCOFTAC}, 2006:1--10.

\bibitem[Dai and MacBeth, 1994]{Dai1994}
Dai, H. and MacBeth, C. (1994).
\newblock {Split shear-wave analysis using an artificial neural network}.
\newblock {\em First Break}, 12(12):605--613.

\bibitem[{De los Reyes} et~al., 2017]{DelosReyes2017}
{De los Reyes}, J.~C., Sch{\"{o}}nlieb, C.-B., and Valkonen, T. (2017).
\newblock {Bilevel parameter learning for higher-order total variation regularisation models}.
\newblock {\em Journal of Mathematical Imaging and Vision}, 57(1):1--25.

\bibitem[Deng and Yu, 2013]{Deng2013}
Deng, L. and Yu, D. (2013).
\newblock {Deep Learning: Methods and Applications}.
\newblock {\em Foundations and Trends{\textregistered} in Signal Processing}.

\bibitem[Dokmani{\'{c}} et~al., 2016]{Dokmanic2016}
Dokmani{\'{c}}, I., Bruna, J., Mallat, S., and de~Hoop, M. (2016).
\newblock {Inverse Problems with Invariant Multiscale Statistics}.
\newblock {\em CoRR}.

\bibitem[Dolenko et~al., 2015]{Dolenko2015}
Dolenko, S., Efitorov, A., Burikov, S., Dolenko, T., Laptinskiy, K., and Persiantsev, I. (2015).
\newblock {Neural Network Approaches to Solution of the Inverse Problem of Identification and Determination of the Ionic Composition of Multi-component Water Solutions}.
\newblock {\em CCIS}, 517:109--118.

\bibitem[Dowla and Rogers, 1996]{Dowla1996}
Dowla, F.~U. and Rogers, L.~L. (1996).
\newblock {\em {Solving Problems in Environmental Engineering and Geosciences with Artificial Neural Networks}}.
\newblock MIT Press.

\bibitem[Downton and Hampson, 2018]{Downton2018}
Downton, J.~E. and Hampson, D.~P. (2018).
\newblock {Deep neural networks to predict reservoir properties from seismic}.
\newblock {\em Presented at 6th GeoConvention 2018}.

\bibitem[Dreyfus, 1973]{Dreyfus1973}
Dreyfus, S. (1973).
\newblock {The computational solution of optimal control problems with time lag}.
\newblock {\em IEEE Transactions on Automatic Control}, 18(4):383--385.

\bibitem[Duchi et~al., 2011]{Duchi2011}
Duchi, J., Hazan, E., and Singer, Y. (2011).
\newblock {Adaptive subgradient methods for online learning and stochastic optimization.}
\newblock {\em Journal of machine learning research}, 12(7).

\bibitem[Elman, 1990]{Elman1990}
Elman, J.~L. (1990).
\newblock {Finding structure in time}.
\newblock {\em Cognitive science}, 14(2):179--211.

\bibitem[Elshafiey, 1991]{Elshafiey1991}
Elshafiey, I.~M. (1991).
\newblock {\em {Neural network approach for solving inverse problems}}.
\newblock PhD thesis.

\bibitem[Fahlman, 1991]{Fahlman1991}
Fahlman, S.~E. (1991).
\newblock {The recurrent cascade-correlation architecture}.
\newblock In {\em Advances in neural information processing systems}, pages 190--196.

\bibitem[Falsaperla et~al., 1996]{Falsaperla1996}
Falsaperla, S., Graziani, S., Nunnari, G., and Spampinato, S. (1996).
\newblock {Automatic classification of volcanic earthquakes by using multi-layered neural networks}.
\newblock {\em Natural Hazards}, 13(3):205--228.

\bibitem[Fletcher, 1987]{Fletcher1987}
Fletcher, R. (1987).
\newblock {Practical Methods of Optimization, Second Edition}.
\newblock pages 3--6.

\bibitem[Fukushima and Miyake, 1982]{Fukushima1982}
Fukushima, K. and Miyake, S. (1982).
\newblock {Neocognitron: A self-organizing neural network model for a mechanism of visual pattern recognition}.
\newblock In {\em Competition and cooperation in neural nets}, pages 267--285. Springer.

\bibitem[Galliani et~al., 2017]{Galliani2017}
Galliani, S., Lanaras, C., Marmanis, D., Baltsavias, E., and Schindler, K. (2017).
\newblock {Learned Spectral Super-Resolution}.

\bibitem[Gardner et~al., 1974]{Gardner1974}
Gardner, G. H.~F., Gardner, L.~W., and Gregory, A.~R. (1974).
\newblock {Formation Velocity and Density - The diagnostic basics for Stratigraphic Traps}.
\newblock {\em GEOPHYSICS}, 39(6):770--780.

\bibitem[Gauthier et~al., 1986]{Gauthier1986}
Gauthier, O., Virieux, J., and Tarantola, A. (1986).
\newblock {Two-dimensional nonlinear inversion of seismic waveforms: Numerical results}.
\newblock {\em Geophysics}, 51(7):1387--1403.

\bibitem[Gazdag, 1981]{Gazdag1981}
Gazdag, J. (1981).
\newblock {Modeling of the acoustic wave equation with transform methods}.
\newblock {\em Geophysics}, 46(6):854.

\bibitem[Gerdova et~al., 2002]{Gerdova2002}
Gerdova, I.~V., Churina, I.~V., Dolenko, S.~A., Dolenko, T.~A., Fadeev, V.~V., and Persiantsev, I.~G. (2002).
\newblock {New opportunities in solution of inverse problems in laser spectroscopy due to application of artificial neural networks}.
\newblock {\em SPIE Proceedings}, 4749(8):157--166.

\bibitem[Gers et~al., 1999]{Gers1999}
Gers, F.~A., Schmidhuber, J., and Cummins, F. (1999).
\newblock {Learning to forget: Continual prediction with LSTM}.

\bibitem[Gerstoft, 1994]{Gerstoft1994}
Gerstoft, P. (1994).
\newblock {Inversion of seismoacoustic data using genetic algorithms and a posteriori probability distributions}.
\newblock {\em The Journal of the Acoustical Society of America}, 95(2):770--782.

\bibitem[Goodfellow et~al., 2016]{Goodfellow2016}
Goodfellow, I., Bengio, Y., and Courville, A. (2016).
\newblock {\em {Deep Learning}}.
\newblock MIT Press, Cambridge MA.

\bibitem[Goodfellow et~al., 2015]{Goodfellow2015}
Goodfellow, I.~J., Shlens, J., and Szegedy, C. (2015).
\newblock {Explaining and harnessing adversarial examples}.
\newblock {\em CoRR}.

\bibitem[Graves, 2012]{Graves2012}
Graves, A. (2012).
\newblock {Supervised sequence labelling}.
\newblock In {\em Supervised sequence labelling with recurrent neural networks}, pages 5--13. Springer.

\bibitem[Grohs et~al., 2019]{Grohs2019}
Grohs, P., Perekrestenko, D., Elbr{\"{a}}chter, D., and B{\"{o}}lcskei, H. (2019).
\newblock {Deep Neural Network Approximation Theory}.

\bibitem[Hadamard, 1907]{Hadamard1907}
Hadamard, J. (1907).
\newblock {M{\'{e}}moire sur le probl{\`{e}}me d'analyse relatif {\`{a}} l'{\'{e}}quilibre des plaques {\'{e}}lastiques encastr{\'{e}}es}.
\newblock {\em Mem. Sav. Etrang.}, 33:515--641.

\bibitem[Hallstr{\"{o}}m, 2016]{Hallstrom2016}
Hallstr{\"{o}}m, E. (2016).
\newblock {Backpropagation from the beginning – Erik Hallstr{\"{o}}m}.
\newblock {\em Medium}.

\bibitem[Halpert, 2018]{Halpert2018}
Halpert, A.~D. (2018).
\newblock {Deep learning-enabled seismic image enhancement}.
\newblock In {\em SEG Technical Program Expanded Abstracts 2018}, pages 2081--2085. Society of Exploration Geophysicists.

\bibitem[Hammer, 2000]{Hammer2000}
Hammer, B. (2000).
\newblock {On the approximation capability of recurrent neural networks}.
\newblock {\em Neurocomputing}, 31(1-4):107--123.

\bibitem[Hamshaw et~al., 2018]{Hamshaw2018}
Hamshaw, S.~D., Dewoolkar, M.~M., Schroth, A.~W., Wemple, B.~C., and Rizzo, D.~M. (2018).
\newblock {A New Machine-Learning Approach for Classifying Hysteresis in Suspended-Sediment Discharge Relationships Using High-Frequency Monitoring Data}.
\newblock {\em Water Resources Research}, 54(6):4040--4058.

\bibitem[Haykin, 2009]{Haykin2009}
Haykin, S.~S. (2009).
\newblock {\em {Neural Networks and Learning Machines}}.
\newblock Number 3rd Edition. PEARSON, Prentice Hall.

\bibitem[He et~al., 2015]{He2015}
He, K., Zhang, X., Ren, S., and Sun, J. (2015).
\newblock {Spatial pyramid pooling in deep convolutional networks for visual recognition}.
\newblock {\em IEEE transactions on pattern analysis and machine intelligence}, 37(9):1904--1916.

\bibitem[He et~al., 2016a]{He2016IEEE}
He, K., Zhang, X., Ren, S., and Sun, J. (2016a).
\newblock {Deep residual learning for image recognition}.
\newblock {\em Proceedings of the IEEE}.

\bibitem[He et~al., 2016b]{He2016}
He, K., Zhang, X., Ren, S., and Sun, J. (2016b).
\newblock {Identity mappings in deep residual networks}.
\newblock In {\em European conference on computer vision}, pages 630--645. Springer.

\bibitem[Hebb, 1949]{Hebb1949}
Hebb, D.~O. (1949).
\newblock {The organization of behavior: A neuropsychological theory}.
\newblock {\em Brain Theory}.

\bibitem[Hecht-Nielsen, 1989]{Hecht-Nielsen1989}
Hecht-Nielsen, R. (1989).
\newblock {Theory of the Backpropagation Neural Network}.
\newblock {\em Proceedings Of The International Joint Conference On Neural Networks}, 1:593--605.

\bibitem[Hecht-Nielsen, 1990]{Hecht-Nielsen1990}
Hecht-Nielsen, R. (1990).
\newblock {\em {Neurocomputing}}.
\newblock Addison-Wesley Pub. Co.

\bibitem[Hicks and Pratt, 2001]{Hicks2001}
Hicks, G.~J. and Pratt, R.~G. (2001).
\newblock {Reflection waveform inversion using local descent methods: Estimating attenuation and velocity over a gas-sand deposit}.
\newblock {\em Geophysics}, 66(2):598--612.

\bibitem[Hinton et~al., 2012]{Hinton2012}
Hinton, G., Deng, L., Yu, D., Dahl, G.~E., Mohamed, A.-r., Jaitly, N., Senior, A., Vanhoucke, V., Nguyen, P., and Sainath, T.~N. (2012).
\newblock {Deep neural networks for acoustic modeling in speech recognition: The shared views of four research groups}.
\newblock {\em IEEE Signal processing magazine}, 29(6):82--97.

\bibitem[Hochreiter, 1991]{Hochreiter1991}
Hochreiter, J. (1991).
\newblock {Untersuchungen zu dynamischen neuronalen Netzen}.
\newblock {\em Master's thesis, Institut fur Informatik, Technische Universitat, Munchen}, pages 1--71.

\bibitem[Hochreiter and Schmidhuber, 1996]{Hochreiter1996}
Hochreiter, S. and Schmidhuber, J. (1996).
\newblock {\em {Bridging long time lags by weight guessing and “Long Short-Term Memory”}}.
\newblock IOP Press.

\bibitem[Hochreiter and Schmidhuber, 1997]{Hochreiter1997}
Hochreiter, S. and Schmidhuber, J. (1997).
\newblock {Long Short-term Memory}.
\newblock {\em Neural computation}, 9:1735--1780.

\bibitem[Hornik et~al., 1989]{Hornik1989}
Hornik, K., Stinchcombe, M., and White, H. (1989).
\newblock {Multilayer feedforward networks are universal approximator}.
\newblock {\em Neural Networks}, 2:359--366.

\bibitem[Hu et~al., 2017]{Hu2017}
Hu, J., Shen, L., and Sun, G. (2017).
\newblock {Squeeze-and-excitation networks}.
\newblock {\em arXiv preprint arXiv:1709.01507}, 7.

\bibitem[Hubel and Wiesel, 1959]{Hubel1959}
Hubel, D.~H. and Wiesel, T.~N. (1959).
\newblock {Receptive fields of single neurones in the cat's striate cortex}.
\newblock {\em The Journal of physiology}, 148(3):574.

\bibitem[Hubel and Wiesel, 1962]{Hubel1962}
Hubel, D.~H. and Wiesel, T.~N. (1962).
\newblock {Receptive fields, binocular interaction and functional architecture in the cat's visual cortex}.
\newblock {\em The Journal of physiology}, 160(1):106.

\bibitem[Hughes et~al., 2019]{Hughes2019}
Hughes, T.~W., Williamson, I. A.~D., Minkov, M., and Fan, S. (2019).
\newblock {Wave physics as an analog recurrent neural network}.
\newblock {\em Science advances}, 5(12):eaay6946.

\bibitem[Igel, 2017]{Igel2016}
Igel, H. (2017).
\newblock {\em {Computational seismology: a practical introduction}}.
\newblock Oxford University Press.

\bibitem[Innanen, 2014]{Innanen2014}
Innanen, K. (2014).
\newblock {Quantifying the incompleteness of the physics model in seismic inversion}.
\newblock {\em CREWES Research Report}, 26.

\bibitem[Ioffe and Szegedy, 2015]{Ioffe2015}
Ioffe, S. and Szegedy, C. (2015).
\newblock {Batch Normalization: Accelerating Deep Network Training by Reducing Internal Covariate Shift}.
\newblock 2017-Octob.

\bibitem[Jia et~al., 2014]{Jia2014}
Jia, Y., Shelhamer, E., Donahue, J., Karayev, S., Long, J., Girshick, R., Guadarrama, S., and Darrell, T. (2014).
\newblock {Caffe: Convolutional Architecture for Fast Feature Embedding}.
\newblock {\em arXiv preprint arXiv:1408.5093}, pages 1097--1105.

\bibitem[Jones and Davison, 2014]{Jones2014}
Jones, I.~F. and Davison, I. (2014).
\newblock {Seismic imaging in and around salt bodies}.
\newblock {\em Interpretation}, 2(4):SL1--SL20.

\bibitem[Jozinovi{\'{c}} et~al., 2020]{Jozinovic2020}
Jozinovi{\'{c}}, D., Lomax, A., {\v{S}}tajduhar, I., and Michelini, A. (2020).
\newblock {Rapid prediction of earthquake ground shaking intensity using raw waveform data and a convolutional neural network}.
\newblock {\em Geophysical Journal International}, 222(2):1379--1389.

\bibitem[Kalita and Alkhalifah, 2017]{Kalita2017}
Kalita, M. and Alkhalifah, T. (2017).
\newblock {Efficient full waveform inversion using the excitation representation of the source wavefield}.
\newblock {\em Geophysical Journal International}, 210(3):1581--1594.

\bibitem[Kamphorst and Ruelle, 1987]{Kamphorst1987}
Kamphorst, J.-P. E. S.~O. and Ruelle, D. (1987).
\newblock {Recurrence Plots of Dynamical Systems}.
\newblock {\em Europhysics Letters}, 4(9):17.

\bibitem[Kazei and Ovcharenko, 2019]{Kazei2019}
Kazei, V. and Ovcharenko, O. (2019).
\newblock {Simple frequency domain full-waveform inversion (FWI) regularized by Sobolev space norm}.

\bibitem[{Keivan Ekbatani} et~al., 2017]{Ekbatani2017}
{Keivan Ekbatani}, H., Pujol, O., and Segui, S. (2017).
\newblock {Synthetic Data Generation for Deep Learning in Counting Pedestrians}.
\newblock In {\em Proceedings of the 6th International Conference on Pattern Recognition Applications and Methods}, volume 2017-Janua, pages 318--323. SCITEPRESS - Science and Technology Publications.

\bibitem[Kelley, 1960]{Kelley1960}
Kelley, H.~J. (1960).
\newblock {Gradient Theory of Optimal Flight Paths}.
\newblock {\em ARS Journal}, 30(10):947--954.

\bibitem[Kelly et~al., 2017]{Kelly2017}
Kelly, B., Matthews, T.~P., and Anastasio, M.~A. (2017).
\newblock {Deep Learning-Guided Image Reconstruction from Incomplete Data}.
\newblock {\em arXiv preprint arXiv:1709.00584}.

\bibitem[Khan et~al., 2020]{Khan2020}
Khan, A., Sohail, A., Zahoora, U., and Qureshi, A.~S. (2020).
\newblock {A survey of the recent architectures of deep convolutional neural networks}.
\newblock {\em Artificial Intelligence Review}, pages 1--62.

\bibitem[Kingma and Ba, 2014]{Kingma2014}
Kingma, D.~P. and Ba, J. (2014).
\newblock {Adam: A method for stochastic optimization}.
\newblock {\em arXiv preprint arXiv:1412.6980}.

\bibitem[K{\"{o}}hler et~al., 2010]{Kohler2010}
K{\"{o}}hler, A., Ohrnberger, M., and Scherbaum, F. (2010).
\newblock {Unsupervised pattern recognition in continuous seismic wavefield records using Self-Organizing Maps}.
\newblock {\em Geophysical Journal International}, 182(3):1619--1630.

\bibitem[Kolmogoro and Tikhomirov, 1956]{Kolmogoro1956}
Kolmogoro, A.~N. and Tikhomirov, V.~M. (1956).
\newblock {On the representation of continuous functions of several variables as superpositions of functions of smaller number of variables}.
\newblock In {\em Soviet. Math. Dokl}, volume 108, pages 179--182.

\bibitem[Kolodzey, 1981]{Kolodzey1981}
Kolodzey, J. (1981).
\newblock {CRAY-1 Computer Technology}.
\newblock {\em IEEE Transactions on Components, Hybrids, and Manufacturing Technology}, 4(2):181--186.

\bibitem[Komatitsch and Tromp, 2002]{Komatitsch2002}
Komatitsch, D. and Tromp, J. (2002).
\newblock {Spectral-element simulations of global seismic wave propagation - I. Validation}.
\newblock {\em Geophysical Journal International}, 149(2):390--412.

\bibitem[Krizhevsky et~al., 2012]{Krizhevsky2012}
Krizhevsky, A., Sutskever, I., and Hinton, G.~E. (2012).
\newblock {ImageNet classification with deep convolutional neural networks}.
\newblock In Weinberger, F.~P., Burges, C. J.~C., Bottou, L., and Q., K., editors, {\em Advances in neural information processing systems}, volume~25, pages 1097--1105. Curran Associates, Inc.

\bibitem[Krizhevsky et~al., 2015]{Krizhevsky2015}
Krizhevsky, A., Sutskever, I., Hinton, G.~E., Tasci, T., and Kim, K. (2015).
\newblock {ImageNet Classification with Deep Convolutional Neural Networks}.
\newblock {\em Stanford cs231b}.

\bibitem[Kumar et~al., 2012a]{Kumar2012a}
Kumar, J., Ramrez, A., and Butt, S. (2012a).
\newblock {Preparing Data for Full Waveform Inversion: A Workflow for Free-surface Multiple Attenuation}.
\newblock {\em 74th EAGE Conference and Exhibition-Workshops}.

\bibitem[Kumar et~al., 2012b]{Kumar2012}
Kumar, M., Manral, D.~S., Banerjee, M.~K., Karmakar, K., Das, A., Reddy, B.~J., Dasgupta, R., and Singh, S.~N. (2012b).
\newblock {High Performance Computing in Geosciences: Promises Challenges}.
\newblock {\em 9th Biennial International Confrence Exposition on Petroleum Geophysics}.

\bibitem[Lailly and Bednar, 1983]{Lailly1983}
Lailly, P. and Bednar, J. (1983).
\newblock {The seismic inverse problem as a sequence of before stack migrations}.
\newblock {\em Conference on inverse scattering: theory and application}.

\bibitem[Lang and Hinton, 1988]{Lang1988}
Lang, K.~J. and Hinton, G.~E. (1988).
\newblock {The development of the time-delay neural network architecture for speech recognition}.
\newblock {\em Technical Report CMU-CS-88-152}.

\bibitem[Lang et~al., 1990]{Lang1990}
Lang, K.~J., Waibel, A.~H., and Hinton, G.~E. (1990).
\newblock {A time-delay neural network architecture for isolated word recognition}.
\newblock {\em Neural networks}, 3(1):23--43.

\bibitem[Langer et~al., 1996]{Langer1996}
Langer, H., Nunnari, G., and Occhipinti, L. (1996).
\newblock {Estimation of seismic waveform governing parameters with neural networks}.
\newblock {\em Journal Of Geophysical Research-Solid Earth}, 101(B9):20109.

\bibitem[Larsson et~al., 2016]{Larsson2016}
Larsson, G., Maire, M., and Shakhnarovich, G. (2016).
\newblock {Learning representations for automatic colorization}.
\newblock In {\em Lecture Notes in Computer Science (including subseries Lecture Notes in Artificial Intelligence and Lecture Notes in Bioinformatics)}, volume 9908 LNCS, pages 577--593.

\bibitem[Le et~al., 2017]{Le2017}
Le, T.~A., Baydin, A.~G., Zinkov, R., Wood, F., Atılım, G., Zinkov, R., and Wood, F. (2017).
\newblock {Using Synthetic Data to Train Neural Networks is Model-Based Reasoning}.
\newblock {\em ArXiv e-prints}, pages 3514--3521.

\bibitem[LeCun, 1989]{LeCun1989}
LeCun, Y. (1989).
\newblock {Generalization and network design strategies}.
\newblock {\em Connectionism in perspective}, 19:143--155.

\bibitem[LeCun et~al., 1990]{LeCun1990}
LeCun, Y., Boser, B.~E., Denker, J.~S., Henderson, D., Howard, R.~E., Hubbard, W.~E., and Jackel, L.~D. (1990).
\newblock {Handwritten digit recognition with a back-propagation network}.
\newblock In {\em Advances in neural information processing systems}, pages 396--404.

\bibitem[LeCun et~al., 2010]{LeCun2010}
LeCun, Y., Kavukcuoglu, K., and Farabet, C. (2010).
\newblock {Convolutional networks and applications in vision}.
\newblock In {\em Proceedings of 2010 IEEE international symposium on circuits and systems}, pages 253--256. IEEE.

\bibitem[Lee et~al., 2016]{Lee2016}
Lee, C.-Y., Gallagher, P.~W., and Tu, Z. (2016).
\newblock {Generalizing pooling functions in convolutional neural networks: Mixed, gated, and tree}.
\newblock In {\em Artificial intelligence and statistics}, pages 464--472.

\bibitem[Lee and Moloney, 2018]{Lee2018}
Lee, K. and Moloney, D. (2018).
\newblock {Evaluation of synthetic data for deep learning stereo depth algorithms on embedded platforms}.
\newblock In {\em 2017 4th International Conference on Systems and Informatics, ICSAI 2017}, volume 2018-Janua, pages 170--176. IEEE.

\bibitem[Lee et~al., 1996]{Lee1996}
Lee, M.~W., Hutchinson, D.~R., Collett, T.~S., and Dillon, W.~P. (1996).
\newblock {Seismic velocities for hydrate‐bearing sediments using weighted equation}.
\newblock {\em Journal of Geophysical Research: Solid Earth}, 101(B9):20347--20358.

\bibitem[Leshno et~al., 1993]{Leshno1993}
Leshno, M., Lin, V.~Y., Pinkus, A., and Schocken, S. (1993).
\newblock {Multilayer Feed Forward Networks With A Nonpolynomial Activation Function Can Approximate Any Function}.
\newblock {\em Neural Networks}, 6:861--867.

\bibitem[Levin, 1973]{Levin1973}
Levin, L. (1973).
\newblock {Universal sequential search problems}.
\newblock {\em Problemy Peredachi Informatsii}, 9(3):115--116.

\bibitem[Lewis and Vigh, 2017]{Lewis2017}
Lewis, W. and Vigh, D. (2017).
\newblock {Deep learning prior models from seismic images for full-waveform inversion}.
\newblock {\em SEG Technical Program Expanded Abstracts 2017}, pages 1512--1517.

\bibitem[Li et~al., 2018a]{Li2017}
Li, H., Xu, Z., Taylor, G., Studer, C., and Goldstein, T. (2018a).
\newblock {Visualizing the Loss Landscape of Neural Nets}.
\newblock In {\em Neural Information Processing Systems}.

\bibitem[Li et~al., 2018b]{Li2018a}
Li, H., Yang, W., and Yong, X. (2018b).
\newblock {Deep learning for ground-roll noise attenuation}.
\newblock In {\em SEG Technical Program Expanded Abstracts 2018}, pages 1981--1985. Society of Exploration Geophysicists.

\bibitem[Li et~al., 2019]{Li2019}
Li, S., Liu, B., Ren, Y., Chen, Y., Yang, S., Wang, Y., and Jiang, P. (2019).
\newblock {Deep-learning inversion of seismic data}.
\newblock {\em arXiv preprint arXiv:1901.07733}.

\bibitem[Lin et~al., 1996]{Lin1996}
Lin, T., Horne, B.~G., Tino, P., and Giles, C.~L. (1996).
\newblock {Learning long-term dependencies in NARX recurrent neural networks}.
\newblock {\em IEEE Transactions on Neural Networks}, 7(6):1329--1338.

\bibitem[Lines, 2014]{Lines2014}
Lines, L. (2014).
\newblock {FWI and the "Noise" Quandary}.
\newblock {\em CREWES Research Report}.

\bibitem[Linnainmaa, 1970]{Linnainmaa1970}
Linnainmaa, S. (1970).
\newblock {The representation of the cumulative rounding error of an algorithm as a Taylor expansion of the local rounding errors}.
\newblock {\em Master's Thesis (in Finnish), Univ. Helsinki}, pages 6--7.

\bibitem[Linnainmaa, 1976]{Linnainmaa1976}
Linnainmaa, S. (1976).
\newblock {Taylor expansion of the accumulated rounding error}.
\newblock {\em BIT}, 16(2):146--160.

\bibitem[Lippmann, 1987]{Lippmann1987}
Lippmann, R.~P. (1987).
\newblock {An Introduction to Computing with Neural Nets}.
\newblock {\em IEEE ASSP Magazine}, 4(April):4--22.

\bibitem[Liu et~al., 2020]{Liu2020}
Liu, Y., He, B., and Zheng, Y. (2020).
\newblock {Controlled-order multiple waveform inversion}.
\newblock {\em Geophysics}, 85(3):R243--R250.

\bibitem[Liu and Sen, 2011]{Liu2011}
Liu, Y. and Sen, M.~K. (2011).
\newblock {3D acoustic wave modelling with time-space domain dispersion-relation-based finite-difference schemes and hybrid absorbing boundary conditions}.
\newblock {\em Exploration Geophysics}, 42(3):176--189.

\bibitem[Lucas et~al., 2018]{Lucas2018}
Lucas, A., Iliadis, M., Molina, R., and Katsaggelos, A.~K. (2018).
\newblock {Using Deep Neural Networks for Inverse Problems in Imaging: Beyond Analytical Methods}.
\newblock {\em IEEE Signal Processing Magazine}, 35(1):20--36.

\bibitem[Malcolm and Willemsen, 2016]{Malcolm2016}
Malcolm, A. and Willemsen, B. (2016).
\newblock {Rapid 4D FWI using a local wave solver}.
\newblock {\em The Leading Edge}, 35(12):1053--1059.

\bibitem[Mangalathu et~al., 2020]{Mangalathu2020}
Mangalathu, S., Jang, H., Hwang, S.-H., and Jeon, J.-S. (2020).
\newblock {Data-driven machine-learning-based seismic failure mode identification of reinforced concrete shear walls}.
\newblock {\em Engineering Structures}, 208:110331.

\bibitem[Martens, 2010]{Martens2010}
Martens, J. (2010).
\newblock {Deep learning via Hessian-free optimization}.
\newblock {\em 27th International Conference on Machine Learning}, 951:735--742.

\bibitem[Martin et~al., 2002]{Martin2002}
Martin, G.~S., Marfurt, K.~J., and Larsen, S. (2002).
\newblock {Marmousi-2: An updated model for the investigation of AVO in structurally complex areas}.
\newblock In {\em SEG Technical Program Expanded Abstracts 2002}, pages 1979--1982. Society of Exploration Geophysicists.

\bibitem[Martin et~al., 2006]{Martin2006}
Martin, G.~S., Wiley, R., and Marfurt, K.~J. (2006).
\newblock {Marmousi2: An elastic upgrade for Marmousi}.
\newblock {\em The Leading Edge}, 25(2):156--166.

\bibitem[McCormack et~al., 1993]{McCormack1993a}
McCormack, M.~D., Zaucha, D.~E., and Dushek, D.~W. (1993).
\newblock {First-break refraction event picking and seismic data trace editing using neural networks}.
\newblock {\em Geophysics}, 58(1):67--78.

\bibitem[McCulloch and Pitts, 1943]{McCulloch1943}
McCulloch, W.~S. and Pitts, W. (1943).
\newblock {A logical calculus of the ideas immanent in nervous activity}.
\newblock {\em The Bulletin of Mathematical Biophysics}, 5(4):115--133.

\bibitem[Meinhardt et~al., 2017]{Meinhardt2017}
Meinhardt, T., Moeller, M., Hazirbas, C., and Cremers, D. (2017).
\newblock {Learning Proximal Operators: Using Denoising Networks for Regularizing Inverse Imaging Problems}.
\newblock In {\em Proceedings of the IEEE International Conference on Computer Vision}, volume 2017-Octob, pages 1799--1808.

\bibitem[Menke, 1989]{Menke1989}
Menke, W. (1989).
\newblock {Geophysical Data Analysis: Discrete Inverse Theory}.

\bibitem[Michaels and Smith, 1992]{Michaels1992}
Michaels, P. and Smith, R. B. R.~B. (1992).
\newblock {Recurrent Neural Network representation of the Inelastic Wave Equation and Full-Waveform inversion without local minima}.
\newblock {\em 62nd Ann. Internat. Mtg}, pages 22--25.

\bibitem[Minsky and Papert, 2017]{Minsky2017}
Minsky, M. and Papert, S.~A. (2017).
\newblock {\em {Perceptrons: An introduction to computational geometry}}.
\newblock MIT press.

\bibitem[Minsky and Papert, 1969]{Minsky1969}
Minsky, M.~L. and Papert, S.~A. (1969).
\newblock {Perceptrons: an introduction to computational geometry}.
\newblock {\em MA: MIT Press, Cambridge}.

\bibitem[Mispel et~al., 2019]{mispel2019high}
Mispel, J., Furre, A., Sollid, A., and Maa{\o}, F.~A. (2019).
\newblock {High Frequency 3D FWI at Sleipner: A Closer Look at the CO2 Plume}.
\newblock In {\em 81st EAGE Conference and Exhibition 2019}.

\bibitem[Molinari et~al., 2020]{Molinari2020}
Molinari, I., Obermann, A., Kissling, E., Het{\'{e}}nyi, G., and Boschi, L. (2020).
\newblock {3D crustal structure of the Eastern Alpine region from ambient noise tomography}.
\newblock {\em Results in Geophysical Sciences}, 1-4:100006.

\bibitem[Molinari et~al., 2015]{Molinari2015}
Molinari, I., Verbeke, J., Boschi, L., Kissling, E., and Morelli, A. (2015).
\newblock {Italian and A lpine three‐dimensional crustal structure imaged by ambient‐noise surface‐wave dispersion}.
\newblock {\em Geochemistry, Geophysics, Geosystems}, 16(12):4405--4421.

\bibitem[M{\o}ller, 1993]{Moller1993}
M{\o}ller, M. (1993).
\newblock {Exact Calculation of the Product of the Hessian Matrix of Feed-Forward Network Error Functions and a Vector in 0 (N) Time}.
\newblock {\em DAIMI Report Series}, page~14.

\bibitem[Mont{\'{u}}far et~al., 2014]{Montufar2014}
Mont{\'{u}}far, G., Pascanu, R., Cho, K., and Bengio, Y. (2014).
\newblock {On the Number of Linear Regions of Deep Neural Networks}.
\newblock {\em arXiv preprint arXiv:1402.1869}.

\bibitem[Morgan et~al., 2016]{Morgan2016}
Morgan, J., Warner, M., Arnoux, G., Hooft, E., Toomey, D., VanderBeek, B., and Wilcock, W. (2016).
\newblock {Next-generation seismic experiments - II: Wide-angle, multi-azimuth, 3-D, full-waveform inversion of sparse field data}.
\newblock {\em Geophysical Journal International}, 204(2):1342--1363.

\bibitem[Morgan et~al., 2013]{Morgan2013}
Morgan, J., Warner, M., Bell, R., Ashley, J., Barnes, D., Little, R., Roele, K., and Jones, C. (2013).
\newblock {Next-generation seismic experiments: wide-angle, multi-azimuth, three-dimensional, full-waveform inversion}.
\newblock {\em Geophysical Journal International}, 195(3):1657--1678.

\bibitem[Morgan et~al., 2009]{Morgan2009}
Morgan, J.~V., Christeson, G.~L., and Warner, M. (2009).
\newblock {Using swath bathymetry as an a priori constraint in a 3D full wavefield tomographic inversion of seismic data across oceanic crust}.
\newblock In {\em AGU Fall Meeting Abstracts}, volume~1, page~7.

\bibitem[Mothi et~al., 2013]{Mothi2013}
Mothi, S., Schwarz, K., and Zhu, H. (2013).
\newblock {Impact of full-azimuth and long-offset acquisition on Full Waveform Inversion in deep water Gulf of Mexico}.
\newblock {\em SEG Houston 2013 Annual Meeting}, (June 2013):924--928.

\bibitem[Mozer, 1992]{Mozer1992}
Mozer, M.~C. (1992).
\newblock {Induction of multiscale temporal structure}.
\newblock In {\em Advances in neural information processing systems}, pages 275--282.

\bibitem[Murat and Rudman, 1992]{Murat1992}
Murat, M.~E. and Rudman, A.~J. (1992).
\newblock {Automated First Arrival Picking: a Neural Network Approach}.
\newblock {\em Geophysical Prospecting}, 40(6):587--604.

\bibitem[Nangoo, 2013]{Nangoo2013}
Nangoo, T.~P. (2013).
\newblock {Seismic Full-Waveform Inversion of 3D Field Data – From the Near Surface to the Reservoir}.

\bibitem[Nath et~al., 1999]{Nath1999}
Nath, S.~K., Chakraborty, S., Singh, S.~K., and Ganguly, N. (1999).
\newblock {Velocity inversion in cross-hole seismic tomography by counter-propagation neural network, genetic algorithm and evolutionary programming techniques}.
\newblock {\em Geophysical Journal International}, 138(1):108--124.

\bibitem[{Nazari Siahsar} et~al., 2017]{NazariSiahsar2017}
{Nazari Siahsar}, M.~A., Gholtashi, S., Kahoo, A.~R., Chen, W., and Chen, Y. (2017).
\newblock {Data-driven multitask sparse dictionary learning for noise attenuation of 3D seismic data}.
\newblock {\em Geophysics}, 82(6):V385--V396.

\bibitem[Newton, 1687]{Newton1687}
Newton, I. (1687).
\newblock {Philosophiae Naturalis Principia Mathematica}.
\newblock {\em Pan}, page 510.

\bibitem[Nguyen and Mcmechan, 2015]{Nguyen2015}
Nguyen, B. and Mcmechan, G. (2015).
\newblock {Five ways to avoid storing source wavefield snapshots in 2D elastic prestack reverse time migration}.
\newblock {\em GEOPHYSICS}, 80:S1--S18.

\bibitem[Nicholson and Gibson, 2016]{Nicholson2016}
Nicholson, A. and Gibson, A. (2016).
\newblock {Deeplearning4j: Open-source distributed deep learning for the jvm}.
\newblock {\em Apache Software Foundation License}, 2.

\bibitem[Noble et~al., 2014]{Noble2014}
Noble, M., Gesret, A., and Belayouni, N. (2014).
\newblock {Accurate 3-D finite difference computation of traveltimes in strongly heterogeneous media}.
\newblock {\em Geophysical Journal International}, 199(3):1572--1585.

\bibitem[{\"{O}}ktem and Adler, 2018]{Oktem2018}
{\"{O}}ktem, O. and Adler, J. (2018).
\newblock {Mathematics of Deep Learning with an Emphasis on Inverse Problems}.

\bibitem[Olah, 2015]{Olah2015}
Olah, C. (2015).
\newblock {Understanding lstm networks}.

\bibitem[Operto et~al., 2015]{Operto2015}
Operto, S., Miniussi, A., Brossier, R., Combe, L., Haller, N., Kjos, E., and Metivier, L. (2015).
\newblock {Efficient 3D Frequency-domain Full-waveform Inversion of Ocean-bottom Cable Data - Application to Valhall in the Visco-ac}.
\newblock In {\em EAGE Extended Abstracts}.

\bibitem[Operto et~al., 2004]{Operto2004}
Operto, S., Ravaut, C., Improta, L., Virieux, J., Herrero, A., and Dell'Aversana, P. (2004).
\newblock {Quantitative imaging of complex structures from dense wide-aperture seismic data by multiscale traveltime and waveform inversions: a case study}.
\newblock {\em Geophysical Prospecting}, 52(6):625--651.

\bibitem[Operto et~al., 2007]{Operto2007}
Operto, S., Virieux, J., Amestoy, P., L'Excellent, J.-Y., Giraud, L., and Ali, H. B.~H. (2007).
\newblock {3D finite-difference frequency-domain modeling of visco-acoustic wave propagation using a massively parallel direct solver: A feasibility study}.
\newblock {\em Geophysics}, 72(5):SM195--SM211.

\bibitem[Operto et~al., 2006]{Operto2006}
Operto, S., Virieux, J., Dessa, J., and Pascal, G. (2006).
\newblock {Crustal seismic imaging from multifold ocean bottom seismometer data by frequency domain full waveform tomography: Application to the eastern Nankai trough}.
\newblock {\em Journal of Geophysical Research: Solid Earth}, 111(B9).

\bibitem[Parker, 1999]{Parker1999}
Parker, P.~B. (1999).
\newblock {\em {Genetic algorithms and their use in geophysical problems}}.
\newblock PhD thesis, University of California, Berkeley.

\bibitem[Paszke et~al., 2017]{Paszke2017}
Paszke, A., Chanan, G., Lin, Z., Gross, S., Yang, E., Antiga, L., and Devito, Z. (2017).
\newblock {Automatic differentiation in PyTorch}.
\newblock {\em Advances in Neural Information Processing Systems}, 30(Nips):1--4.

\bibitem[Patterson and Gibson, 2017]{Patterson2017}
Patterson, J. and Gibson, A. (2017).
\newblock {\em {Deep learning: A practitioner's approach}}.
\newblock " O'Reilly Media, Inc.".

\bibitem[Pearlmutter, 1989]{Pearlmutter1989}
Pearlmutter, B.~A. (1989).
\newblock {Learning state space trajectories in recurrent neural networks}.
\newblock {\em Neural Computation}, 1(2):263--269.

\bibitem[Peng et~al., 2018]{Peng2018}
Peng, C., Wang, M., Chazalnoel, N., and Gomes, A. (2018).
\newblock {Subsalt imaging improvement possibilities through a combination of FWI and reflection FWI}.
\newblock {\em The Leading Edge}, 37(1):52--57.

\bibitem[Petersen et~al., 2017]{Petersen2017}
Petersen, P.~C., B{\"{o}}lcskei, H., Grohs, P., and Kutyniok, G. (2017).
\newblock {Optimal approximation with sparse deep neural networks}.

\bibitem[Plate, 1993]{Plate1993}
Plate, T.~A. (1993).
\newblock {Holographic recurrent networks}.
\newblock In {\em Advances in neural information processing systems}, pages 34--41.

\bibitem[Plessix, 2009]{Plessix2009}
Plessix, R. (2009).
\newblock {Three-dimensional frequency-domain full-waveform inversion with an iterative solver}.
\newblock {\em Geophysics}, 74(6):WCC149--WCC157.

\bibitem[Plessix and Perkins, 2010]{Plessix2010}
Plessix, R.-E. and Perkins, C. (2010).
\newblock {Thematic Set: Full waveform inversion of a deep water ocean bottom seismometer dataset}.
\newblock {\em First Break}, 28(4):71--78.

\bibitem[Plessix et~al., 2014]{Plessix2014}
Plessix, R.-{\'{E}}., Stopin, A., Milcik, P., and Matson, K. (2014).
\newblock {Acoustic and anisotropic multi-parameter seismic full waveform inversion case studies}.
\newblock In {\em SEG Technical Program Expanded Abstracts 2014}, pages 1056--1060. Society of Exploration Geophysicists.

\bibitem[Poulton et~al., 1992]{Poulton1992}
Poulton, M.~M., Sternberg, B.~K., and Glass, C.~E. (1992).
\newblock {Location of subsurface targets in geophysical data using neural networks}.
\newblock {\em Geophysics}, 57(12):1534--1544.

\bibitem[Pratt, 1990]{Pratt1990b}
Pratt, R.~G. (1990).
\newblock {Inverse theory applied to multi-source cross-hole tomography - Part 2: Elastoc Wave-Equation Method}.
\newblock {\em Geophysical Prospecting}, 38(3):311--329.

\bibitem[Pratt, 1999]{Pratt1999}
Pratt, R.~G. (1999).
\newblock {Seismic waveform inversion in the frequency domain, Part 1: Theory and verification in a physical scale model}.
\newblock {\em GEOPHYSICS}, 64(3):888--901.

\bibitem[Pratt and Goulty, 1991]{Pratt1991}
Pratt, R.~G. and Goulty, N.~R. (1991).
\newblock {Combining wave-equation imaging with traveltime tomography to form high-resolution images from crosshole data}.
\newblock {\em Geophysics}, 56(2):208--224.

\bibitem[Pratt et~al., 1996]{Pratt1996}
Pratt, R.~G., Song, Z.-M., Williamson, P., and Warner, M. (1996).
\newblock {Two-dimensional velocity models from wide-angle seismic data by wavefield inversion}.
\newblock {\em Geophysical Journal International}, 124(2):323--340.

\bibitem[Pratt and Worthington, 1990]{Pratt1990a}
Pratt, R.~G. and Worthington, M.~H. (1990).
\newblock {Inverse theory applied to multi-source cross-hole tomography - Part 1: Acoustic Wave-Equation Method}.
\newblock {\em Geophysical prospecting}, 38(March 1989):287--310.

\bibitem[Press, 1968]{Press1968}
Press, F. (1968).
\newblock {Earth Models Obtained by Monte Carlo Inversion}.
\newblock {\em J. Geophys. Res.}, 73(16):5223--5234.

\bibitem[Pullammanappallil and Louie, 1994]{Pullammanappallil1994}
Pullammanappallil, S.~K. and Louie, J.~N. (1994).
\newblock {A generalized simulated-annealing optimization for inversion of first-arrival times}.
\newblock {\em Bulletin of the Seismological Society of America}, 84(5):1397--1409.

\bibitem[Puskorius and Feldkamp, 1994]{Puskorius1994}
Puskorius, G.~V. and Feldkamp, L.~A. (1994).
\newblock {Neurocontrol of nonlinear dynamical systems with Kalman filter trained recurrent networks}.
\newblock {\em IEEE Transactions on neural networks}, 5(2):279--297.

\bibitem[Raissi et~al., 2019]{Raissi2019}
Raissi, M., Perdikaris, P., and Karniadakis, G.~E. (2019).
\newblock {Physics-informed neural networks: A deep learning framework for solving forward and inverse problems involving nonlinear partial differential equations}.
\newblock {\em Journal of Computational Physics}, 378:686--707.

\bibitem[Randall, 2009]{Randall2009}
Randall, P. (2009).
\newblock {Pink Color of Seismic noise in the low frequency limit. [Accessed 10/8/2014].}

\bibitem[Ranzato et~al., 2007]{Ranzato2007}
Ranzato, M., Huang, F.~J., Boureau, Y.-L., and LeCun, Y. (2007).
\newblock {Unsupervised learning of invariant feature hierarchies with applications to object recognition}.
\newblock In {\em 2007 IEEE conference on computer vision and pattern recognition}, pages 1--8. IEEE.

\bibitem[Raschka and Mirjalili, 2017]{Raschka2017}
Raschka, S. and Mirjalili, V. (2017).
\newblock {\em {Python machine learning}}.
\newblock Packt Publishing Ltd.

\bibitem[Reading et~al., 2015]{Reading2015}
Reading, A.~M., Cracknell, M.~J., Bombardieri, D.~J., and Chalke, T. (2015).
\newblock {Combining Machine Learning and Geophysical Inversion for Applied Geophysics}.
\newblock In {\em ASEG-PESA 2015 24th International Geophysical Conference and Exhibition}, volume 2015, page~1.

\bibitem[Richardson, 2018]{Richardson2018}
Richardson, A. (2018).
\newblock {Seismic Full-Waveform Inversion Using Deep Learning Tools and Techniques}.
\newblock {\em arXiv preprint arXiv:1801.07232}.

\bibitem[Robinson and Fallside, 1987]{Robinson1987}
Robinson, A.~J. and Fallside, F. (1987).
\newblock {\em {The utility driven dynamic error propagation network}}.
\newblock University of Cambridge Department of Engineering Cambridge, MA.

\bibitem[Robinson, 1957]{Robinson1957}
Robinson, E.~A. (1957).
\newblock {PREDICTIVE DECOMPOSITION OF SEISMIC TRACES}.
\newblock {\em GEOPHYSICS}, 22(4):767--778.

\bibitem[Robinson, 1967]{Robinson1967}
Robinson, E.~A. (1967).
\newblock {Predictive decomposition of time series with application to seismic exploration}.
\newblock {\em Geophysics}, 32(3):418--484.

\bibitem[Romano et~al., 2016]{Romano2016}
Romano, Y., Elad, M., and Milanfar, P. (2016).
\newblock {The Little Engine that Could: Regularization by Denoising (RED)}.
\newblock {\em SIAM Journal on Imaging Sciences}, 10(4):1804--1844.

\bibitem[Romeo, 1994]{Romeo1994}
Romeo, G. (1994).
\newblock {Seismic signals detection and classification using artiricial neural networks}.
\newblock {\em Annals Of Geophysics}, 37(3).

\bibitem[Ronneberger et~al., 2015]{ronneberger2015u}
Ronneberger, O., Fischer, P., and Brox, T. (2015).
\newblock {U-net: Convolutional networks for biomedical image segmentation}.
\newblock In {\em International Conference on Medical image computing and computer-assisted intervention}, pages 234--241. Springer.

\bibitem[Rosenblatt, 1958]{Rosenblatt1958}
Rosenblatt, F. (1958).
\newblock {The perceptron: A probabilistic model for information storage and organization in the brain}.
\newblock {\em Psychological Review}, 65(6):386--408.

\bibitem[R{\"{o}}th and Tarantola, 1994]{Roth1994}
R{\"{o}}th, G. and Tarantola, A. (1994).
\newblock {Neural networks and inversion of seismic data}.
\newblock {\em Journal of Geophysical Research}, 99(B4):6753--6768.

\bibitem[Rothman, 1985]{Rothman1985}
Rothman, D.~H. (1985).
\newblock {\em {Large near-surface anomalies, seismic reflection data, and simulated annealing}}.
\newblock PhD thesis, Stanford University.

\bibitem[Ruder, 2016]{Ruder2016}
Ruder, S. (2016).
\newblock {An overview of gradient descent optimization algorithms}.
\newblock {\em arXiv preprint arXiv:1609.04747}.

\bibitem[Rumelhart et~al., 1986]{Rumelhart1986}
Rumelhart, D.~E., Hinton, G.~E., Williams, R.~J., Vinet, L., and Zhedanov, A. (1986).
\newblock {Learning representations by back-propagating errors}.
\newblock {\em Nature}, 323(6088):533--536.

\bibitem[Russakovsky et~al., 2015]{Russakovsky2015}
Russakovsky, O., Deng, J., Su, H., Krause, J., Satheesh, S., Ma, S., Huang, Z., Karpathy, A., Khosla, A., Bernstein, M., Berg, A.~C., and Fei-Fei, L. (2015).
\newblock {ImageNet Large Scale Visual Recognition Challenge}.
\newblock {\em International Journal of Computer Vision}, 115(3):211--252.

\bibitem[Russell and Norvig, 2008]{Russell2008}
Russell, S. and Norvig, P. (2008).
\newblock {\em {Artificial Intelligence: A Modern Approach}}.
\newblock Prentice Hall.

\bibitem[Rusu and Thompson, 2017]{Rusu2017}
Rusu, C. and Thompson, J. (2017).
\newblock {Learning fast sparsifying transforms}.
\newblock {\em IEEE Transactions on Signal Processing,}, 65(16):4376--4378.

\bibitem[Ryan, 1994]{Ryan1994}
Ryan, H. (1994).
\newblock {Ricker, Ormsby, Klauder, Butterworth - A choice of wavelets}.
\newblock {\em CSEG Recorder}, pages 8--9.

\bibitem[Santurkar et~al., 2018]{Santurkar2018}
Santurkar, S., Tsipras, D., Ilyas, A., and Madry, A. (2018).
\newblock {How does batch normalization help optimization?}
\newblock In {\em Advances in Neural Information Processing Systems}, pages 2483--2493.

\bibitem[Schakel and Mesdag, 2014]{Schakel2014}
Schakel, M.~D. and Mesdag, P.~R. (2014).
\newblock {Fully data-driven quantitative reservoir characterization by broadband seismic}.
\newblock In {\em 2014 SEG Annual Meeting}. OnePetro.

\bibitem[Schmidhuber, 1992a]{Schmidhuber1992}
Schmidhuber, J. (1992a).
\newblock {A fixed size storage O (n 3) time complexity learning algorithm for fully recurrent continually running networks}.
\newblock {\em Neural Computation}, 4(2):243--248.

\bibitem[Schmidhuber, 1992b]{Schmidhuber1992a}
Schmidhuber, J. (1992b).
\newblock {Learning complex, extended sequences using the principle of history compression}.
\newblock {\em Neural Computation}, 4(2):234--242.

\bibitem[Schmidhuber, 2015]{Schmidhuber2015}
Schmidhuber, J. (2015).
\newblock {Deep learing in neural networks: an overview}.
\newblock {\em Neural Networks}, 61.

\bibitem[Schraudolph, 2002]{Schraudolph2002}
Schraudolph, N. (2002).
\newblock {Fast curvature matrix-vector products for second-order gradient descent}.
\newblock {\em Neural Computation}, 14(7):1723--1738.

\bibitem[Schultz et~al., 1994]{Schultz1994}
Schultz, P.~S., Ronen, S., Hattori, M., and Corbett, C. (1994).
\newblock {Seismic-guided estimation of log properties (Part 1: A data-driven interpretation methodology)}.
\newblock {\em The Leading Edge}, 13(5):305--310.

\bibitem[Sen and Stoffa, 1995]{Sen1995}
Sen, M.~K. and Stoffa, P.~L. (1995).
\newblock {\em {Global optimization methods in geophysical inversion}}.
\newblock Elsevier.

\bibitem[Sever, 2015]{Sever2015}
Sever, A. (2015).
\newblock {An inverse problem approach to pattern recognition in industry}.
\newblock {\em Applied Computing and Informatics}, 11(1):1--12.

\bibitem[Shahnas et~al., 2018]{Shahnas2018}
Shahnas, M.~H., Yuen, D.~A., and Pysklywec, R.~N. (2018).
\newblock {Inverse Problems in Geodynamics Using Machine Learning Algorithms}.
\newblock {\em Journal of Geophysical Research: Solid Earth}, 123(1):296--310.

\bibitem[Shalova and Oseledets, 2020]{Shalova2020}
Shalova, A. and Oseledets, I. (2020).
\newblock {Tensorized Transformer for Dynamical Systems Modeling}.
\newblock {\em arXiv preprint arXiv:2006.03445}.

\bibitem[Shen and Clapp, 2015]{Shen2015}
Shen, X. and Clapp, R. (2015).
\newblock {Random boundary condition for memory-efficient waveform inversion gradient computation}.
\newblock {\em GEOPHYSICS}, 80:R351--R359.

\bibitem[Sheriff and Geldart, 1985]{Sheriff1985}
Sheriff, R.~E. and Geldart, L.~P. (1985).
\newblock {\em {Exploration seismology. Volume 2}}.
\newblock Cambridge University Press, New York, NY, United States.

\bibitem[Sherstinsky, 2020]{Sherstinsky2020}
Sherstinsky, A. (2020).
\newblock {Fundamentals of recurrent neural network (rnn) and long short-term memory (lstm) network}.
\newblock {\em Physica D: Nonlinear Phenomena}, 404:132306.

\bibitem[Shimshoni and Intrator, 1998]{Shimshoni1998}
Shimshoni, Y. and Intrator, N. (1998).
\newblock {Classification of seismic signals by integrating ensembles of neural networks}.
\newblock {\em IEEE Transactions on Signal Processing}, 46(5):1194--1201.

\bibitem[Shin et~al., 2010]{Shin2010}
Shin, C., Koo, N.-H.~H., Cha, Y.~H., and Park, K.-P.~P. (2010).
\newblock {Sequentially ordered single-frequency 2-D acoustic waveform inversion in the Laplace-Fourier domain}.
\newblock {\em Geophysical Journal International}, 181(2):935--950.

\bibitem[Siahkoohi et~al., 2019]{Siahkoohi2019}
Siahkoohi, A., Louboutin, M., and Herrmann, F.~J. (2019).
\newblock {The importance of transfer learning in seismic modeling and imaging}.
\newblock {\em GEOPHYSICS}, 84(6):A47--A52.

\bibitem[Simonyan et~al., 2014]{Simonyan2014}
Simonyan, K., Others, and Zisserman, A. (2014).
\newblock {Very Deep Convolutional Networks for Large-Scale Image Recognition}.
\newblock {\em arXiv preprint arXiv:1409.1556}, pages 1--14.

\bibitem[Singh, 2015]{Singh2015}
Singh, R.~V. (2015).
\newblock {ImageNet Winning CNN Architectures – A Review}.
\newblock pages 3--8.

\bibitem[Sirgue et~al., 2010]{Sirgue2010}
Sirgue, L., Barkved, O.~I., Dellinger, J., Etgen, J., Albertin, U., and Kommedal, J.~H. (2010).
\newblock {Thematic set: Full waveform inversion: The next leap forward in imaging at Valhall}.
\newblock {\em First Break}, 28(4):65--70.

\bibitem[Sirgue et~al., 2009]{Sirgue2009}
Sirgue, L., Barkved, O.~I., {Van Gestel}, J.~P., Askim, O.~J., and Kommedal, J.~H. (2009).
\newblock {3D waveform inversion on Valhall wide-azimuth OBC}.
\newblock In {\em 71st EAGE Conference and Exhibition incorporating SPE EUROPEC 2009}.

\bibitem[Sirgue et~al., 2007]{Sirgue2007}
Sirgue, L., Etgen, J., Albertin, U., and America, B.~P. (2007).
\newblock {3D full-waveform inversion: Wide-versus narrow-azimuth acquisitions}.
\newblock {\em SEG Technical Program Expanded Abstracts 2007}, pages 1760--1764.

\bibitem[Sirgue and Pratt, 2004]{Sirgue2004}
Sirgue, L. and Pratt, R.~G. (2004).
\newblock {Efficient waveform inversion and imaging: A strategy for selecting temporal frequencies}.
\newblock {\em Geophysics}, 69(1):231--248.

\bibitem[Srivastava et~al., 2014]{Srivastava2014}
Srivastava, N., Hinton, G., Krizhevsky, A., Sutskever, I., and Salakhutdinov, R. (2014).
\newblock {Dropout: A Simple Way to Prevent Neural Networks from Overfitting}.
\newblock {\em Journal of Machine Learning Research}, 15:1929--1958.

\bibitem[Stephens, 2006]{Stephens2006}
Stephens, A.~D. (2006).
\newblock {Measurable functions}.
\newblock {\em M3/M4S3 - Statistical Theory II Lecture Notes}.

\bibitem[Sun et~al., 2019a]{Sun2019}
Sun, J., Niu, Z., Innanen, K.~A., Li, J., and Trad, D.~O. (2019a).
\newblock {A theory-guided deep learning formulation of seismic waveform inversion}.
\newblock In {\em SEG Technical Program Expanded Abstracts 2019}, pages 2343--2347. Society of Exploration Geophysicists.

\bibitem[Sun et~al., 2019b]{Sun2019a}
Sun, J., Niu, Z., Innanen, K. A.~H., Li, J., and Trad, D. (2019b).
\newblock {A deep learning perspective of the forward and inverse problems in exploration geophysics}.
\newblock {\em CSEG Geoconvention, 2019}.

\bibitem[Sun et~al., 2018]{Sun2018}
Sun, Y., Xia, Z., and Kamilov, U.~S. (2018).
\newblock {Efficient and accurate inversion of multiple scattering with deep learning}.
\newblock {\em Optics Express}, 26(11):14678.

\bibitem[Sutton, 1988]{Sutton1988}
Sutton, R.~S. (1988).
\newblock {Learning to predict by the methods of temporal differences}.
\newblock {\em Machine learning}, 3(1):9--44.

\bibitem[Szegedy et~al., 2016]{Szegedy2016}
Szegedy, C., Ioffe, S., Vanhoucke, V., and Alemi, A. (2016).
\newblock {Inception-v4, Inception-ResNet and the Impact of Residual Connections on Learning}.

\bibitem[Szegedy et~al., 2014]{Szegedy2014}
Szegedy, C., Liu, W., Jia, Y., Sermanet, P., Reed, S., Anguelov, D., Erhan, D., Vanhoucke, V., Rabinovich, A., Hill, C., and Arbor, A. (2014).
\newblock {Going Deeper with Convolutions}.
\newblock {\em cv-foundation.org}, pages 1--9.

\bibitem[Tanaka, 2003]{Tanaka2003}
Tanaka, M.~M. (2003).
\newblock {\em {Inverse problems in engineering mechanics IV : International Symposium on Inverse Problems in Engneering Mechanics 2003 (ISIP 2003), Nagano, Japan}}.
\newblock Elsevier.

\bibitem[Tarantola, 1984a]{Tarantola1984a}
Tarantola, A. (1984a).
\newblock {Inversion of seismic reflection data in the acoustic approximation}.
\newblock {\em GEOPHYSICS}, 49(8):1259--1266.

\bibitem[Tarantola, 1984b]{Tarantola1984}
Tarantola, A. (1984b).
\newblock {Linearized inversion of seismic reflection data}.
\newblock {\em Geophysical prospecting}, 32(6):998--1015.

\bibitem[Tarantola, 1987]{Tarantola1987}
Tarantola, A. (1987).
\newblock {Inverse problem theory: Methods for data fitting and parameter estimation}.

\bibitem[Tarantola, 2005]{Tarantola2005}
Tarantola, A. (2005).
\newblock {\em {Inverse Problem Theory and Methods for Model Parameter Estimation}}.

\bibitem[Taylor et~al., 2016]{Taylor2016}
Taylor, G., Burmeister, R., Xu, Z., Singh, B., Patel, A., and Goldstein, T. (2016).
\newblock {Training neural networks without gradients: A scalable admm approach}.
\newblock In {\em International conference on machine learning}, pages 2722--2731. PMLR.

\bibitem[Thulasiraman and Swamy, 2011]{Thulasiraman2011}
Thulasiraman, K. and Swamy, M. N.~S. (2011).
\newblock {\em {Graphs: theory and algorithms}}.
\newblock John Wiley Sons.

\bibitem[Tikhonov, 1963]{Tikhonov1963}
Tikhonov, A.~N. (1963).
\newblock {On the Solution of Incorrectly Stated Problems and a Method of Regularization}.
\newblock {\em Dokl. Acad. Nauk SSSR}, 151:501--504.

\bibitem[Tikhonov and Arsenin, 1977]{Tikhonov1977}
Tikhonov, A.~N. and Arsenin, V.~Y. (1977).
\newblock {\em {Solutions of ill-posed problems}}.
\newblock Winston, New York.

\bibitem[T{\"{o}}rn and {\v{Z}}ilinskas, 1989]{Torn1989}
T{\"{o}}rn, A. and {\v{Z}}ilinskas, A. (1989).
\newblock {\em {Global optimization}}, volume 350.
\newblock Springer.

\bibitem[Tran and Hiltunen, 2011]{Tran2011}
Tran, K.~T. and Hiltunen, D.~R. (2011).
\newblock {Two-dimensional inversion of full waveforms using simulated annealing}.
\newblock {\em Journal of Geotechnical and Geoenvironmental Engineering}, 138(9):1075--1090.

\bibitem[Tran and Hiltunen, 2012]{Tran2012}
Tran, K.~T. and Hiltunen, D.~R. (2012).
\newblock {One-Dimensional Inversion of Full Waveforms using a Genetic Algorithm}.
\newblock {\em Journal of Environmental Engineering Geophysics}, 17(4):197--213.

\bibitem[Vaswani et~al., 2017]{Vaswani2017}
Vaswani, A., Shazeer, N., Parmar, N., Uszkoreit, J., Jones, L., Gomez, A.~N., Kaiser, L., and Polosukhin, I. (2017).
\newblock {Attention is all you need}.
\newblock {\em arXiv preprint arXiv:1706.03762}.

\bibitem[Vigh et~al., 2010]{Vigh2010}
Vigh, D., Starr, B., Kapoor, J., and Li, H. (2010).
\newblock {3D Full waveform inversion on a Gulf of Mexico WAZ data set}.
\newblock {\em SEG Technical Program Expanded Abstracts 2010}, pages 957--961.

\bibitem[Vigh and Starr, 2008]{Vigh2008}
Vigh, D. and Starr, E.~W. (2008).
\newblock {3D prestack plane-wave, full-waveform inversion}.
\newblock {\em GEOPHYSICS}, 73(5):VE135--VE144.

\bibitem[Virieux and Operto, 2009]{Virieux2009}
Virieux, J. and Operto, S. (2009).
\newblock {An overview of full-waveform inversion in exploration geophysics}.
\newblock {\em Geophysics}, 74(6):WCC1--WCC26.

\bibitem[Vito et~al., 2005]{Vito2005}
Vito, E.~D., Rosasco, L., Caponnetto, A., Giovannini, U.~D., and Odone, F. (2005).
\newblock {Learning from Examples as an Inverse Problem}.
\newblock {\em Journal of Machine Learning Research}, 6(May):883--904.

\bibitem[Waibel et~al., 1989]{Waibel1989}
Waibel, A., Hanazawa, T., Hinton, G., Shikano, K., and Lang, K.~J. (1989).
\newblock {Phoneme recognition using time-delay neural networks}.
\newblock {\em IEEE transactions on acoustics, speech, and signal processing}, 37(3):328--339.

\bibitem[Wang and Tsvankin, 2016]{Wang2016}
Wang, H. and Tsvankin, I. (2016).
\newblock {Feasibility of waveform inversion in acoustic orthorhombic media}.
\newblock {\em SEG Technical Program Expanded Abstracts 2016}, pages 311--316.

\bibitem[Wang and Mendel, 1992]{Wang1992}
Wang, L. and Mendel, J.~M. (1992).
\newblock {Adaptive minimum prediction‐error deconvolution and source wavelet estimation using Hopfield neural networks}.
\newblock {\em Geophysics}, 57(5):670--679.

\bibitem[Wang et~al., 2012]{Wang2012}
Wang, T., Wu, D.~J., Coates, A., and Ng, A.~Y. (2012).
\newblock {End-to-end text recognition with convolutional neural networks}.
\newblock In {\em Proceedings of the 21st international conference on pattern recognition (ICPR2012)}, pages 3304--3308. IEEE.

\bibitem[Wang, 2015]{wang2015frequencies}
Wang, Y. (2015).
\newblock {Frequencies of the Ricker wavelet}.
\newblock {\em Geophysics}, 80(2):A31----A37.

\bibitem[Wang and Oates, 2015]{Wang2015}
Wang, Z. and Oates, T. (2015).
\newblock {Encoding time series as images for visual inspection and classification using tiled convolutional neural networks}.
\newblock In {\em Workshops at the twenty-ninth AAAI conference on artificial intelligence}, volume~1.

\bibitem[Warner et~al., 2013]{Warner2013}
Warner, M., Ratcliffe, A., Nangoo, T., Morgan, J., Umpleby, A., Shah, N., Vinje, V., {\v{S}}tekl, I., Guasch, L., Win, C., Conroy, G., and Bertrand, A. (2013).
\newblock {Anisotropic 3D full-waveform inversion}.
\newblock {\em GEOPHYSICS}, 78(2):R59--R80.

\bibitem[Warner et~al., 2007]{Warner2007}
Warner, M., Stekl, I., and Umpleby, A. (2007).
\newblock {Full Wavefield Seismic Tomography–Iterative Forward Modelling in 3D}.
\newblock In {\em 69th EAGE Conference and Exhibition incorporating SPE EUROPEC 2007}.

\bibitem[Warner et~al., 2008]{Warner2008c}
Warner, M., Stekl, I., and Umpleby, A. (2008).
\newblock {3D wavefield tomography : synthetic and field data examples}.
\newblock {\em 2008 SEG Annual Meeting}, pages 3330--3334.

\bibitem[Webster, 1978]{Webster1978}
Webster, G.~M. (1978).
\newblock {\em {Deconvolution}}.
\newblock Society of Exploration Geophysicists, Tulsa, OK, geophysics edition.

\bibitem[Wei et~al., 2017]{Wei2017}
Wei, Q., Fai, K., and Carin, L. (2017).
\newblock {An Inner-loop Free Solution to Inverse Problems using Deep Neural Networks}.
\newblock {\em Advances in Neural Information Processing Systems}, pages 2370--2380.

\bibitem[Werbos, 1981]{Werbos81}
Werbos, P.~J. (1981).
\newblock {Applications of Advances in Nonlinear Sensitivity Analysis}.
\newblock In {\em Proceedings of the 10th IFIP Conference, 31.8 - 4.9, NYC}, pages 762--770.

\bibitem[Werbos, 1988]{Werbos1988}
Werbos, P.~J. (1988).
\newblock {Generalization of backpropagation with application to a recurrent gas market model}.
\newblock {\em Neural networks}, 1(4):339--356.

\bibitem[Williams and Zipser, 1989]{Williams1989}
Williams, R.~J. and Zipser, D. (1989).
\newblock {A learning algorithm for continually running fully recurrent neural networks}.
\newblock {\em Neural computation}, 1(2):270--280.

\bibitem[Wu et~al., 2018a]{Wu2018a}
Wu, G., Fomel, S., and Chen, Y. (2018a).
\newblock {Data‐driven time–frequency analysis of seismic data using non‐stationary Prony method}.
\newblock {\em Geophysical Prospecting}, 66(1):85--97.

\bibitem[Wu et~al., 2018b]{Wu2018}
Wu, Y., Lin, Y., and Zhou, Z. (2018b).
\newblock {InversionNet: Accurate and efficient seismic waveform inversion with convolutional neural networks}.
\newblock In {\em SEG Technical Program Expanded Abstracts 2018}, pages 2096--2100. Society of Exploration Geophysicists.

\bibitem[Xie et~al., 2017]{Xie2017}
Xie, Y., Zhou, B., Zhou, J., Hu, J., Xu, L., Wu, X., Lin, N., Loh, F.~C., Liu, L., and Wang, Z. (2017).
\newblock {Orthorhombic full-waveform inversion for imaging the Luda field using wide-azimuth ocean-bottom-cable data}.
\newblock {\em The Leading Edge}, 36(1):75--80.

\bibitem[Xu et~al., 2012]{Xu2012}
Xu, Q., Yu, H., Mou, X., Zhang, L., Hsieh, J., and Wang, G. (2012).
\newblock {Low-dose X-ray CT reconstruction via dictionary learning}.
\newblock {\em IEEE transactions on medical imaging}, 31(9):1682--1697.

\bibitem[Yadav et~al., 2015]{Yadav2015}
Yadav, N., Yadav, A., and Kumar, M. (2015).
\newblock {\em {An Introduction to Neural Network Methods for Differential Equations}}.

\bibitem[Yao et~al., 2020]{Yao2020}
Yao, G., Wu, D., and Wang, S.-X. (2020).
\newblock {A review on reflection-waveform inversion}.
\newblock {\em Petroleum Science}, 17(2):334--351.

\bibitem[Yao et~al., 2007]{Yao2007}
Yao, Y., Rosasco, L., and Caponnetto, A. (2007).
\newblock {On early stopping in gradient descent learning}.
\newblock {\em Constructive Approximation}, 26(2):289--315.

\bibitem[Yu and Principe, 2019]{Yu2019}
Yu, S. and Principe, J.~C. (2019).
\newblock {Understanding autoencoders with information theoretic concepts}.
\newblock {\em Neural Networks}, 117:104--123.

\bibitem[Zeiler, 2012]{Zeiler2012}
Zeiler, M.~D. (2012).
\newblock {Adadelta: an adaptive learning rate method}.
\newblock {\em arXiv preprint arXiv:1212.5701}.

\bibitem[Zeiler and Fergus, 2014]{Zeiler2014}
Zeiler, M.~D. and Fergus, R. (2014).
\newblock {Visualizing and understanding convolutional networks}.
\newblock In {\em European conference on computer vision}, pages 818--833. Springer.

\bibitem[Zhang et~al., 2013]{Zhang2013}
Zhang, D.~L., Dai, W., Ge, Z., and Schuster, G. (2013).
\newblock {Multiples waveform inversion}.
\newblock In {\em 75th EAGE Conference Exhibition incorporating SPE EUROPEC 2013}, pages cp--348. European Association of Geoscientists Engineers.

\bibitem[Zhang et~al., 2018]{Zhang2018}
Zhang, J., Lin, Y., Song, Z., and Dhillon, I.~S. (2018).
\newblock {Learning Long Term Dependencies via Fourier Recurrent Units}.

\bibitem[Zhang et~al., 2020]{Zhang2020}
Zhang, X., Jia, Z., Ross, Z.~E., and Clayton, R.~W. (2020).
\newblock {Extracting dispersion curves from ambient noise correlations using deep learning}.
\newblock {\em IEEE Transactions on Geoscience and Remote Sensing}, 58(12):8932--8939.

\bibitem[Zhang and Paulson, 1997]{Zhang1997}
Zhang, Y. and Paulson, K.~V. (1997).
\newblock {Magnetotelluric inversion using regularized Hopfield neural networks}.
\newblock {\em Geophys. Prosp.}, 45(05):725--743.

\bibitem[Zhou and Chellappa, 1988]{Zhou1988}
Zhou, Y.-T. and Chellappa, R. (1988).
\newblock {Computation of optical flow using a neural network.}
\newblock In {\em ICNN}, pages 71--78.

\bibitem[Zhu et~al., 1997]{Zhu1997}
Zhu, C.~Y., Byrd, R.~H., Lu, P.~H., and Nocedal, J. (1997).
\newblock {Algorithm 778: L-BFGS-B: Fortran subroutines for large-scale bound-constrained optimization}.
\newblock {\em Acm Transactions on Mathematical Software}, 23(4):550--560.

\bibitem[Zhu et~al., 2020]{Zhu2020}
Zhu, W., Xu, K., Darve, E., Biondi, B., and Beroza, G.~C. (2020).
\newblock {Integrating Deep Neural Networks with Full-waveform Inversion: Reparametrization, Regularization, and Uncertainty Quantification}.
\newblock {\em arXiv preprint arXiv:2012.11149}.

\bibitem[Zimmermann et~al., 2006]{Zimmermann2006}
Zimmermann, H.~G., Grothmann, R., Schaefer, A.~M., and Tietz, C. (2006).
\newblock {Identification and Forecasting of Large Dynamical Systems by Dynamical Consistent Neural Networks}.
\newblock {\em S. Haykin, J. Principe, T. Sejnowski, and J. McWhirter, editors, New Directions in Statistical Signal Processing: From Systems to Brain}, pages 203--242.

\end{thebibliography}
